\documentclass[a4paper,12pt]{book}
\usepackage[left=2.5cm,right=2.5cm,top=2.5cm,bottom=2.5cm]{geometry}

\usepackage[frenchb,english]{babel}
\usepackage[applemac]{inputenc}
\usepackage{enumerate}
\usepackage{amsmath}
\usepackage{amsthm}
\usepackage{amssymb}
\usepackage{graphicx}
\usepackage{floatflt}
\usepackage{fancybox}
\usepackage{stmaryrd}
\usepackage{appendix} 

\usepackage[Lenny]{fncychap}
\usepackage{fancyhdr}
\usepackage{minitoc}

\usepackage[backref=page]{hyperref} 
\renewcommand*{\backref}[1]{} 
\renewcommand*{\backrefalt}[4]{\ifcase #1  (Not cited.)
\else (Cited p.~#2.) 
\fi}

\usepackage{color}
\usepackage{sidecap}
\usepackage[width=.9\textwidth]{caption}
\usepackage{subcaption}

\usepackage{comment}
\includecomment{comment} 

\renewcommand{\Re}{\textrm{Re}}
\renewcommand{\Im}{\textrm{Im}}

\newcommand{\om}{\omega}
\newcommand{\Om}{\Omega}
\newcommand{\lam}{\lambda}
\newcommand{\Gam}{\Gamma}
\newcommand{\eps}{\epsilon}

\newcommand{\Obox}{\partial_t^2 - \partial_x^2}

\newcommand{\vac}[1]{|0_{#1} \rangle}
\newcommand{\vev}[2]{\langle 0_{#1} | #2 | 0_{#1} \rangle}
\newcommand{\M}{\mathcal M}
\newcommand{\vect}[1]{\overrightarrow{#1}}
\newcommand{\CM}{C^\infty(\mathcal M)}
\newcommand{\p}{\partial}
\newcommand{\g}{\mathbf{g}}
\newcommand{\grad}{\textrm{\bf grad}}
\newcommand{\scri}{\mathcal I}
\newcommand{\uf}{u}
\newcommand{\uc}{\underline{u}}
\newcommand{\vc}{\underline{v}}

\newcommand{\be} {\begin{equation}}
\newcommand{\ee} {\end{equation}}
\newcommand{\bsub}{\begin{subequations}}
\newcommand{\esub}{\end{subequations}}
\newcommand{\bea}{\begin{eqnarray}}
\newcommand{\eea}{\end{eqnarray}}
\newcommand{\bi} {\begin{itemize}}
\newcommand{\ei} {\end{itemize}}
\newcommand{\ben} {\begin{enumerate}}
\newcommand{\een} {\end{enumerate}}
\newcommand{\bmat} {\begin{pmatrix}}
\newcommand{\emat} {\end{pmatrix}} 
\newcommand{\bal} {\begin{aligned}}
\newcommand{\eal} {\end{aligned}}
\newcommand{\btab}{\begin{tabular}}
\newcommand{\etab}{\end{tabular}}
\newcommand{\Sec}[1]{Sec.\ref{#1}}

\newtheorem{theorem}{Theorem}[section]
\newtheorem{definition}{Definition}[section]
\newtheorem{lemma}{Lemma}[section]

\newtheorem{example}{Example}[section]
\newtheorem{convention}{Convention}[section]
\newtheorem{principle}{Principle}[section]

\newcommand{\eq}[1]{Eq.~\eqref{#1}}
\newcommand{\Sch}{Schwarzschild }
\newcommand{\tablespace}{\rule[-3mm]{0cm}{8mm}}

\begin{document}
\selectlanguage{english}
\frontmatter
\dominitoc

\thispagestyle{empty}

\noindent{\sc Université Paris-Sud 11
\hfill LPT Orsay}


\vspace{3cm}
\begin{center}
\textbf{\Huge On the phenomenology \\
of \\
black hole radiation: \\}
\vspace{13mm}
\textbf{\textit{\Large Stability properties of Hawking radiation in the presence of 
ultraviolet violation of local Lorentz invariance\\}}
\vspace{2.5cm}
{\large PhD thesis by \\}
\vspace{5mm}
\textbf{\Large Antonin Coutant}

\vfill
{\large Defended on October 1, 2012, in front of the jury\\
\vspace{5mm}
\btab{rcl}
\hline
Pr. Vitor Cardoso && Referee \\
Pr. Ted Jacobson && Referee \\
Pr. Vincent Rivasseau && Jury president \\
Pr. Roberto Balbinot && Jury member \\
Pr. Stefano Liberati && Jury member \\
Pr. Renaud Parentani && PhD advisor \\
\hline
\etab
}
\vspace{20pt}

\includegraphics[height=2.2cm,keepaspectratio]{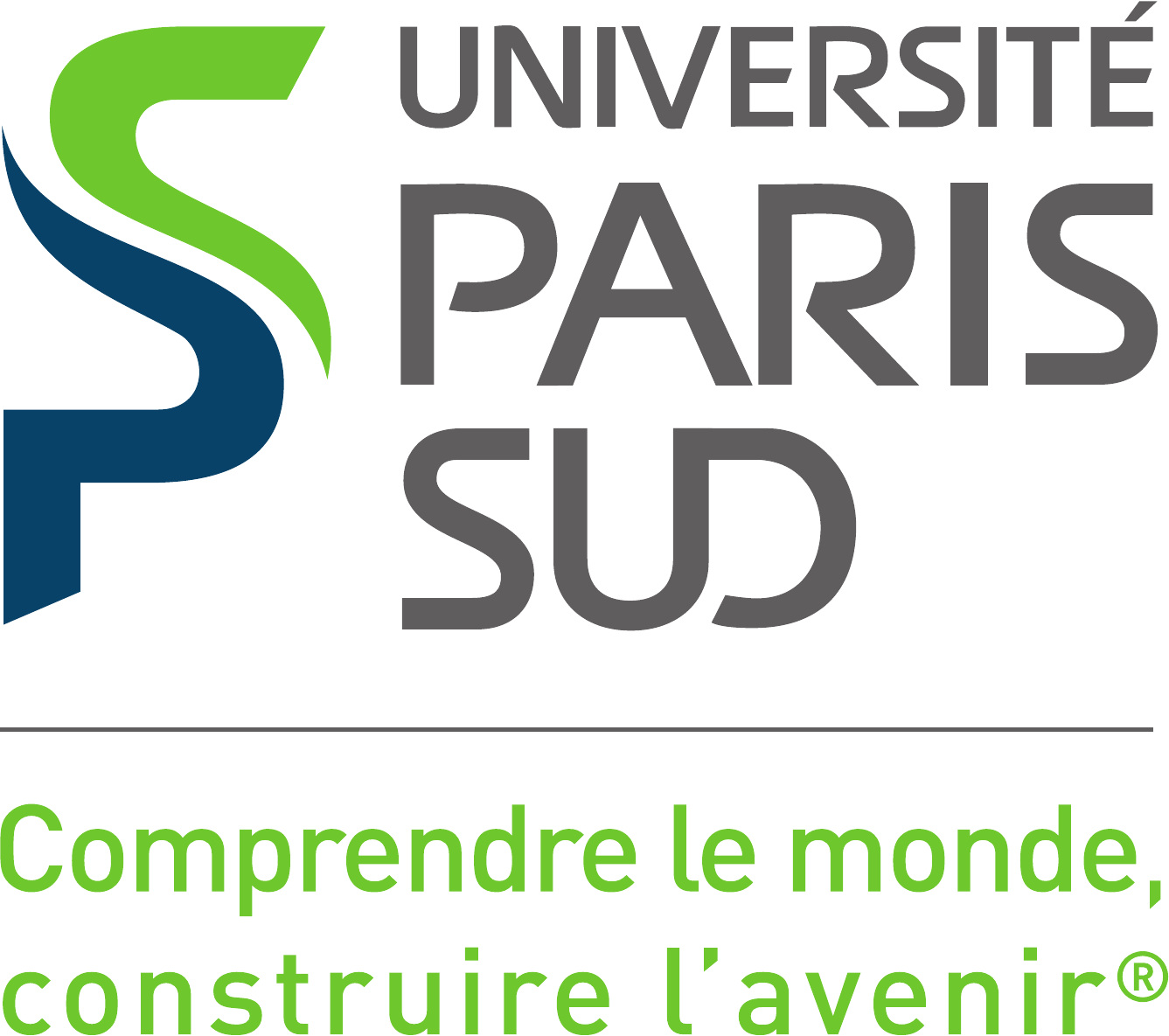}
\hspace{2cm}
\includegraphics[height=2.2cm,keepaspectratio]{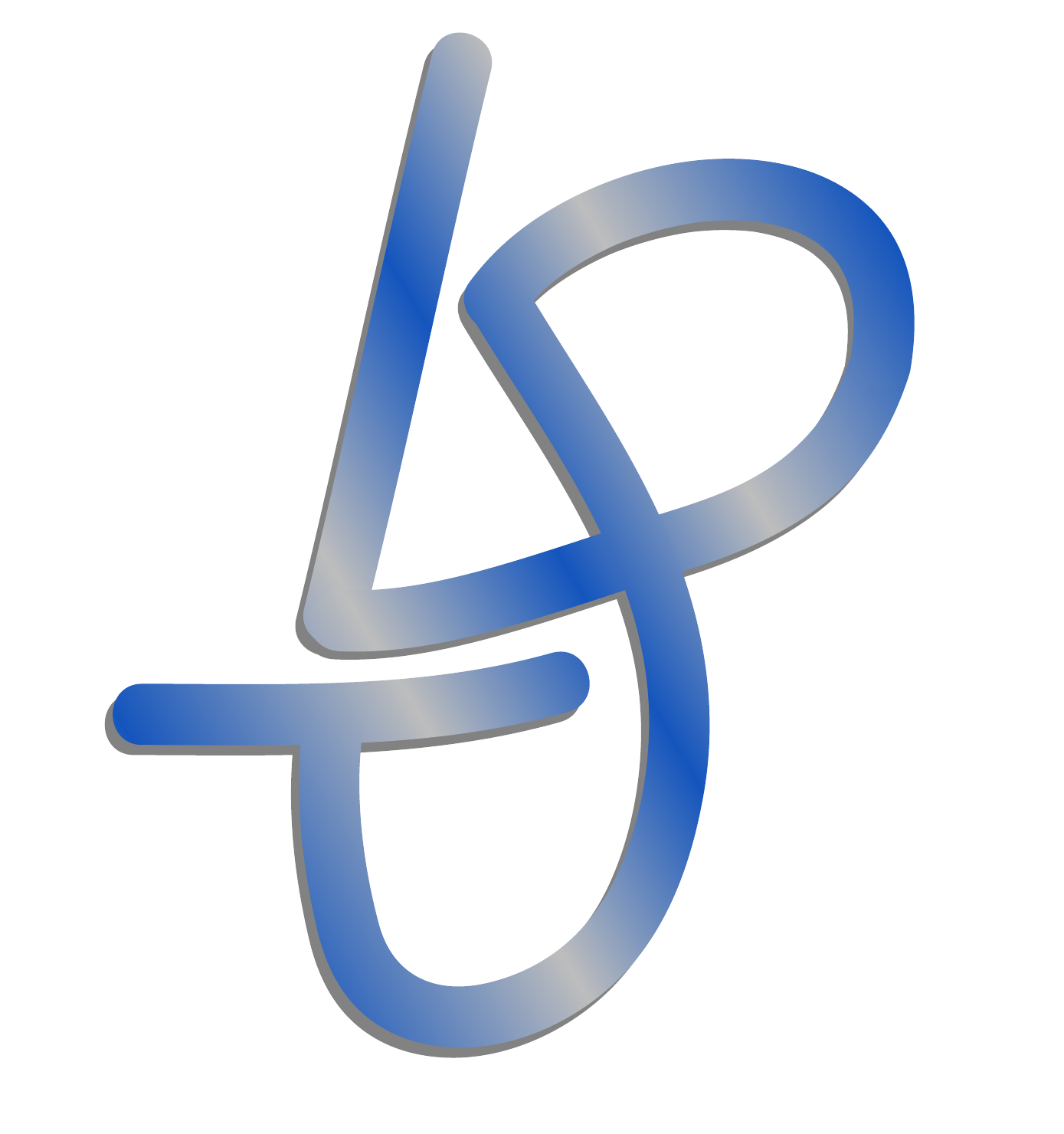}
\hspace{2cm}
\includegraphics[height=2.2cm,keepaspectratio]{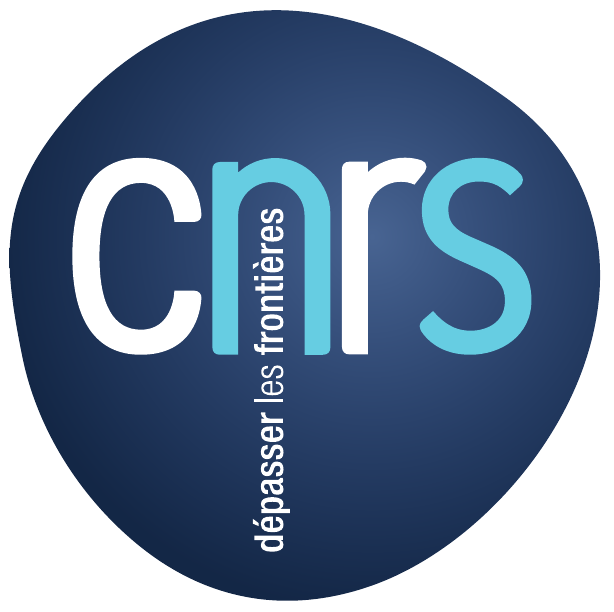}
\end{center}

\newpage
\thispagestyle{empty}
\cleardoublepage
\thispagestyle{plain}
\setcounter{page}{1}

\begin{center}
\textbf{Abstract (EN):}
\end{center}
In this thesis, we study several features of Hawking radiation in the presence of ultraviolet Lorentz violations. These violations are implemented by a modified dispersion relation that becomes nonlinear at short wavelengths. The motivations of this work arise on the one hand from the developing field of analog gravity, where we aim at measuring the Hawking effect in fluid flows that mimic black hole space-times, and on the other hand from the possibility that quantum gravity effects might be approximately modeled by a modified dispersion relation. We develop several studies on various aspects of the problem. First we obtain precise characterizations about the deviations from the Hawking result of black hole radiation, which are induced by dispersion. Second, we study the emergence, both in white hole flows or for massive fields, of a macroscopic standing wave, spontaneously produced from the Hawking effect, and known as `undulation'. Third, we describe in detail an instability named black hole laser, which arises in the presence of two horizons, where Hawking radiation is self-amplified and induces an exponentially growing in time emitted flux.\\

\vspace{0.3cm}
\noindent \textbf{Tags:} Hawking radiation, Analog gravity, Lorentz violation, Instabilities, Undulations

\vspace{1.2cm}

\begin{center}
\textbf{Résumé (FR):}
\end{center}
Dans cette thèse, nous étudions plusieurs aspects de la radiation de Hawking en présence de violations de l'invariance locale de Lorentz. Ces violations sont introduites par une modification de la relation de dispersion, devenant non-linéaire aux courtes longueurs d'onde. Les principales motivations de ces travaux ont une double origine. Il y a d'une part le développement en matière condensée de trous noirs analogues, ou l'écoulement d'un fluide est perçu comme une métrique d'espace-temps pour les ondes de perturbations et ou la radiation de Hawking pourrait être détectée expérimentalement. D'autre part, il se pourrait que des effets de gravité quantique puissent être modélisés par une modification de la relation de dispersion. En premier lieu, nous avons obtenu des caractérisations précises des conditions nécessaires au maintien de l'effet Hawking en présence de violation de l'invariance de Lorentz. De plus, nous avons étudié l'apparition d'une onde macroscopique de fréquence nulle, dans des écoulements de type trous blancs et également pour des champs massifs. Une autre partie de ce travail a consisté à analyser une instabilité engendrée par les effets dispersifs, ou la radiation de Hawking est auto-amplifiée, générant ainsi un flux sortant exponentiellement croissant dans le temps.\\

\vspace{0.3cm}
\noindent \textbf{Mots-clés:} Radiation de Hawking, Gravité analogue, Violation de Lorentz, Instabilités, Undulations\\

\vfill
\hfill
\begin{minipage}{10cm}
Thèse préparée dans le cadre de l'Ecole Doctorale 107 au 
Laboratoire de Physique Théorique d'Orsay (UMR 8627) 
Bât. 210, Université Paris-Sud 11, 91405 Orsay Cedex.
\end{minipage}


\newpage
\thispagestyle{plain}
\cleardoublepage
\thispagestyle{plain}

\noindent {\huge Acknowledgment}\\
 
First of all, I would like to thank my PhD advisor, Renaud Parentani. Working under his supervision was not only fruitful, it was also a great pleasure. I can't thank him enough for the countless conversations we had about various topics of physics, and for all the knowledge he shared with me. I particularly enjoyed our discussions on the interpretation of quantum mechanics or the origin of the second principle.\\

I am sincerely grateful to Vitor Cardoso and Ted Jacobson, who accepted to be my PhD referees. The same gratitude goes to the other members of my jury, Roberto Balbinot, Stefano Liberati and Vincent Rivasseau. They agreed to give some of their time to examine my work, and that was an honor for me. I also wish to thank my other collaborators  Stefano Finazzi, Alessandro Fabbri, and Paul R. Anderson, and I hope our work together was as pleasant for them as it was for me.\\

These three years in the LPT Orsay have been a real enjoyment. Working among the Cosmo group has been wonderful, and has only confirmed my willing to pursue my career as a physicist. I particularly thank my fellows from the SinJe, as much for the passionate scientific debates as for the human experience. A special thank goes to Xavier Bush for his careful reading of my manuscript. I am also very grateful to Patricia Dubois-Violette, Philippe Molle, and the rest of the administrative staff, whose efficiency and patience has been a precious help. \\

I am specially thankful to my friends and colleagues Baptiste Darbois-Texier, Marc Geiller, Sylvain Carrozza and Yannis Bardoux. The many conversations we had about physics has been, and continue to be, a great inspiration for me. I can only hope we will have many occasions to work together in the future.\\

I can't say how much I owe to my parents Isabelle and Bernard, my brother Balthazar and sister Bérénice, as well as my friends Grégoire and Ludovic, who has always been there for me. Their constant support has always been inflexible, even though they probably still wonder what my job really is.\\

Finally, my last words go to my dear Anne-Sophie. These last years has been constantly enlightened by her presence, support and love.

\newpage
\thispagestyle{plain}
\tableofcontents

\mainmatter

\chapter*{Introduction}
\markboth{Introduction}{Introduction}
\addstarredchapter{Introduction}
\phantomsection
\section*{The problem of quantum gravity}
\addcontentsline{toc}{section}{The problem of quantum gravity}

Modern physics relies in its fundamentals on two extremely well tested theories. The first one is Quantum Mechanics, which rules the microscopic world of atoms and elementary particles. At its antipodes, one finds General Relativity, the theory of gravitation describing the physics at large scales. Unfortunately, these two theories are intrinsically incompatible. \\

The logical inconsistency between both theories is in fact a consequence of a more general statement. For the axioms of quantum mechanics to be internally consistent, one needs to assume that \emph{all} physical degrees of freedom are quantum in nature. To understand this, we propose to go back to the beginning of the construction of the modern version of quantum mechanics. In 1926, Born published a remarkable paper, which was ultimately rewarded by a Nobel prize~\cite{Born26}. In his paper, Born showed that the wave equation proposed by Schrödinger was equally efficient to describe scattering processes as stationary states. More importantly, he underlined that the correct way to interpret the wave function was to see it as a probability density. This was the birth of the statistical interpretation of quantum mechanics. Shortly after, Heisenberg complemented this understanding by showing that there is an intrinsic uncertainty when trying to measure simultaneously the position and the momentum of a physical system~\cite{Heisenberg27}, {\it i.e.},
\be
\Delta x \Delta p \geqslant \frac\hbar2.
\ee
However, if one tries to couple a quantum system with classical degrees of freedom, the Heisenberg inequalities can be violated. This was pointed out in 1927 by Einstein. He proposed to Bohr a {\it gedanken experiment} to measure the position and the momentum of a system with an arbitrary precision. Bohr understood that the paradox could be resolved only if one assumed that the measuring device is also ruled by the quantum laws, and hence subjected to an uncertainty relation~\cite{Bohr49}\footnote{An english translation of the founding fathers papers have been published in a book by Wheeler and Zurek, which reviews the debates about measurements in quantum mechanics~\cite{Wheeler}.}. \\

\begin{figure}[!ht]
\begin{center} 
\includegraphics[scale=0.9]{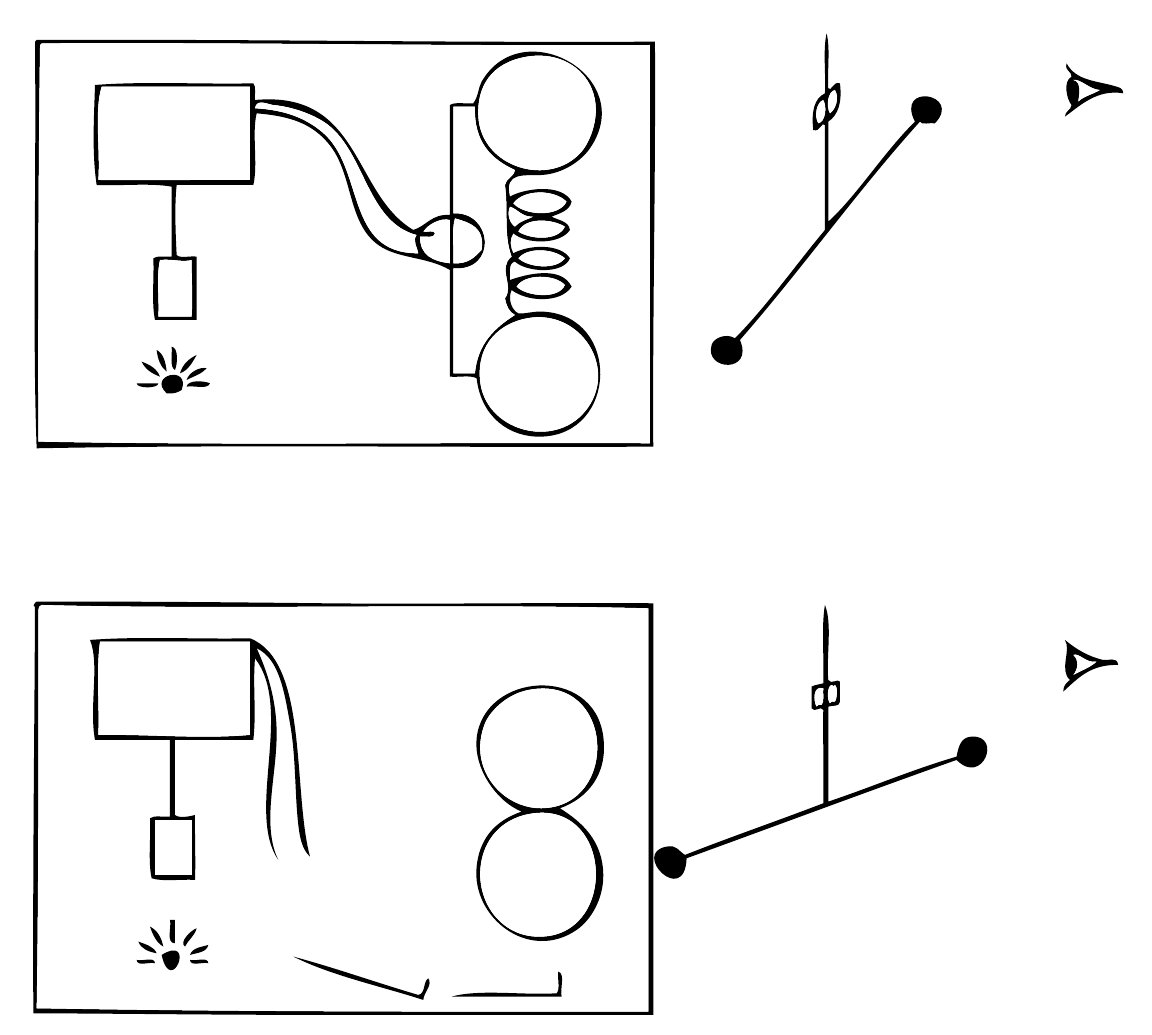}
\end{center}
\caption{In a box, we put two massive spheres, attached by a spring. A radioactive nucleus is connected to a mechanism that releases the spring. When the nucleus decays, the spheres are pulled together and the Cavendish balance switches position. If gravity is quantum, the balance will undergo a quantum jump when the nucleus decays. On the other hand, if the gravitational field is only sensitive to the `mean position' of the masses, the balance will move slowly following the law of radioactive decay. (Figure from~\cite{UnruhQG}.)}
\label{Cavendish_exp_fig} 
\end{figure}

In General Relativity, space-time itself is a dynamical system, giving rise to the gravitational field. Therefore, it does not escape the preceding argument. These degrees of freedom must be quantized as well. A modern version of the Einstein-Bohr debates were presented by Unruh~\cite{UnruhQG} in a book in honor of Bryce DeWitt~\cite{Christensen}. In particular, he considered a set up consisting of a Schrödinger's cat type of experiment, where the state of the cat is probed by a Cavendish balance, see Fig.\ref{Cavendish_exp_fig}. The quantum nature of the gravitational interaction would then be manifest at a macroscopic scale. 
\\

The necessity of building a quantum theory of gravity is thus established. The next question would be why did we succeed in quantizing all physical systems but gravity ? This is probably a much more delicate issue. We can at least compare it to the other interactions, namely the electroweak and strong forces. At the quantum level, these are described by bosonic particles. When considering particles in interactions, there is a huge number of virtual processes contributing to the transition amplitude. To compute physical observables, one must sum over these processes, but because arbitrary highly energetic ones are involved, we generically obtain diverging quantities. Properly removing these infinities leads to the theory of \emph{renormalization}, which is at the heart of modern high energy physics. Unfortunately, when one tries this to deal with the gravitational interaction, quantum general relativity is found to be \emph{non renormalizable}~\cite{tHooft74,Goroff85,Goroff85b}. This means that infinities can be removed, but they generate an infinite number of counter terms associated with an infinite number of new coupling constants. This does not mean that no predictions can be made~\cite{Donoghue97,Burgess}, but the theory can certainly not be trusted in the ultraviolet regime and therefore, is necessarily incomplete. Another proposition was made, to write down a wave equation for the gravitational field, that does not rely on a perturbative expansion. This is the well-known Wheeler-DeWitt equation~\cite{DeWitt67a} and its path integral approach~\cite{DeWitt67b,DeWitt67c}. Unfortunately, this equation is only formal and not mathematically well defined. \\

To tackle the puzzle of quantum gravity, several approaches can be followed. The first would be to address the problem frontally, that is to provide a new theory, reconciling gravity and quantum mechanics. Nowadays, there are two main candidates for such a theory. The first one is string theory~\cite{Zwiebach}, which relies on the assumption that the concept of point particle is only approximately valid, and fundamental objects are in fact extended. The second one is loop quantum gravity. This approach intends to quantize the gravitational field in a fully `background independent' manner~\cite{Thiemann,Perez12,Geiller11}. As a second approach, one can decide to stick to the theories we understand, and push them to their extreme limit, where quantum gravity presumably becomes a necessity. The main examples in that direction are most probably \emph{primordial cosmology} and \emph{black hole evaporation}. Their deeper understanding could provide us with crucial hints and guidelines about what the full quantum theory of gravity could or should be. The work presented in this thesis has been fundamentally motivated by this second line of thought. It has been devoted to the study of certain aspects of black hole radiation.

\section*{The background field approximation}
\addcontentsline{toc}{section}{The background field approximation}
In 1974 Hawking showed that black holes are not completely black but rather radiate as thermal objects~\cite{Hawking74}. To do so, he considered a relativistic quantum field, typically photons, in a black hole (classical) background. As we will discuss in Chapter \ref{HR_Ch}, this phenomenon has many crucial implications. However, it is legitimate to ask what is the validity of the approach. Can it be consistent to consider gravitation as a classical background, while matter fields are quantized ? As pointed out by Duff~\cite{Duff81}, when space-time geometry gives rise to quantum effects, such as particle creation, gravitons should be emitted as well, and hence gravity cannot be considered as classical. Following these lines, quantum field theory in curved background would be doomed to be either trivial or inconsistent. However, DeWitt answered to that worry by pointing out that gravitons can be included in the matter sector, and thus perfectly well described within the background field approximation~\cite{FullingQG}. It is only when including gravitational interactions that  standard quantum field theory methods fail. Therefore, quantum field theory in curved space-time seems to be a perfectly respectable physical theory, but whose range of validity is unclear.  

In quantum electrodynamics, the same question can be answered precisely. For example, to describe an electron orbiting around a nucleus, it is unnecessary to appeal for the full theory of quantum electrodynamics (QED), because the electron feels essentially a classical background electric field, given by the Coulomb potential of the nucleus. More generally, as explained in the thirteenth chapter of~\cite{Weinberg1}, a quantum particle will behave as if coupled to a classical external field if its \emph{sources are heavy enough} compared to the test particle. Unfortunately, for gravity, the answer most probably deeply entails the knowledge of the full quantum gravity theory. However, very interesting results have been obtained in a simplified context, known as `minisuperspace'. Indeed, in quantum cosmology, precise conditions concerning the validity of the background field approximation have been obtained~\cite{Parentani98,Massar98}, which essentially matches those derived in QED.

In the present work, we have reversed the philosophy. Namely, we assume that quantum field theory in curved space is a valid theory, and study the quantum effects in black hole physics. In chapter \ref{LIV_Ch}, we will \emph{model} the (potential) residual effects of quantum gravity at large scales by a modification of the dispersion relation. The spirit is really to make speculative assumptions about corrections to the semi-classical approach and to study their consequences.

\section*{What can we learn from black holes ?}
\addcontentsline{toc}{section}{What can we learn from black holes ?}
Why should one study Hawking radiation ? Has black hole physics something to tell about quantum gravity ? There are probably no definite answers to these questions. However, we see several points in black hole physics that might be related to or even lead to the full theory of quantum gravity. 

In parallel to the discovery of Hawking, and in fact shortly before, Bekenstein made the following proposition. According to what we know about the dynamics of black holes in General Relativity, combined with arguments from information theory, black holes must possess a proper entropy, proportional to their surface area~\cite{Bekenstein73}. The microscopic origin of this entropy is still an enigma. It is widely believed that its understanding will pass through a better knowledge of quantum gravity. The subject of black hole entropy, and more generally, black hole thermodynamics, is a rich domain of gravitational physics. In \Sec{BHthermo_Sec}, we shall say a few words about it. However, in our work, we have mainly focused on the process of Hawking radiation.

This radiation, leading to the evaporation of black holes, raises another deep question. What can happen at the end of the evaporation ? For large black holes, the background field approximation has chance to be valid, but when the black hole reaches a Planck size this is no longer true. In addition, this problem might be closely related to the well-known `information paradox'. Certainly this question appeals for quantum gravity to be answered. But even before reaching such a microscopic regime, the Hawking scenario of black hole radiation must be questioned. \\

Even for large black holes, the Hawking process seems to appeal to the full quantum gravity theory for a complete understanding. Indeed, in Chapter \ref{LIV_Ch}, we will see that the semi-classical derivation involves arbitrary high frequencies. In that sense, black holes act as `microscopes', since the ultraviolet features of the theory are naturally probed. In addition, these very high energetic fluctuations could lead to a complete invalidation of Hawking radiation. Following a proposition of Jacobson~\cite{Jacobson91}, we \emph{assume} that the resulting low energy effects are well modeled by a modification of the dispersion relation, {\it i.e.}, breaking local Lorentz invariance in the utlraviolet. The concern of Chapter \ref{LIV_Ch} will thus be the study of the modifications of Hawking radiation induced by dispersive effects. Moreover, the introduction of Lorentz violation is not free of consequences. In fact, the whole stability analysis must be reconsidered. Indeed, in Chapters \ref{mass_Ch} and \ref{laser_Ch}, we will see and analyze various unstable processes, induced by these violations. \\

In addition, our work is also motivated by an analogy discovered by Unruh~\cite{Unruh81}, between perturbations in a fluid flow and fields on a black hole geometry. This analogy will also be presented in Chapter \ref{LIV_Ch}. Because dispersive effects always exist in condensed matter, our results are of particular relevance in this analog context, where actual experiments can be performed, in contrast to the astrophysical case.

Before presenting the content of our work, we have devoted the first two chapters to a review of known material concerning black holes. In Chapter \ref{geometry_Ch}, we present classical features such as geometry, space-time in general relativity, and black hole properties. In Chapter \ref{HR_Ch}, we analyze quantum phenomena. After reviewing the canonical quantization of a field, we present the Unruh effect, followed by Hawking radiation.

\newpage
\section*{Preliminary remarks and conventions \markboth{Preliminaries}{Preliminaries}}
In this thesis, each chapter can be conceived as almost independent. In particular, a few notations may vary from one chapter to another, even though we tried to keep a general coherence. A consequent part of the manuscript consists in reviewing known materials relevant for our work. The new results are presented as an exposition of the studies realized in the following list of papers. \\

\bi
\item[$\star$] \cite{Coutant10} A.~Coutant and R.~Parentani, ``Black hole lasers, a mode analysis,'' Phys.\ Rev.\ D {\bf 81} (2010) 084042 \href{http://arxiv.org/abs/0912.2755}{[arXiv:0912.2755 [hep-th]]}. 

\item[$\star$] \cite{Coutant11} A.~Coutant, R.~Parentani and S.~Finazzi, ``Black hole radiation with short distance dispersion, an analytical S-matrix approach,'' Phys.\ Rev.\ D {\bf 85} (2012) 024021 \href{http://arxiv.org/abs/1108.1821}{[arXiv:1108.1821 [hep-th]]}. 

\item[$\star$] \cite{aQuattro} A.~Coutant, S.~Finazzi, S.~Liberati and R.~Parentani, ``Impossibility of superluminal travel in Lorentz violating theories,'' Phys.\ Rev.\ D {\bf 85} (2012) 064020 \href{http://arxiv.org/abs/1111.4356}{[arXiv:1111.4356 [gr-qc]]}. 

\item[$\star$] \cite{Coutant12} A.~Coutant, A.~Fabbri, R.~Parentani, R.~Balbinot and P.~Anderson, ``Hawking radiation of massive modes and undulations,'' Phys. Rev. D {\bf 86} (2012) 064022 \href{http://arxiv.org/abs/1206.2658}{arXiv:1206.2658 [gr-qc]}.
\ei

\vspace{1cm}
\noindent All along the thesis, we work with the following conventions:
\bi
\item We work in units where $c=1$, $\hbar = 1$ and $k_B = 1$. On the other hand, the Newton's constant $G$ will stay a dimensionfull quantity in order to keep track of the role of gravitation.
\item $M_{\odot} \simeq 2,0.10^{30}kg$ is the solar mass.
\item In $d+1$ dimensions, the signature of the metric is $+ \underbrace{-\ldots-}_{\times d}$.
\item We use $[ \mu \nu ]$ to symmetrize sums over indices
\be
T_{[\mu_1\ldots \mu_r]} = \frac1{r!} \sum_{\sigma \in S_r} T_{\sigma(\mu_1)\ldots \sigma(\mu_r)} \qquad {\rm but} \qquad T_{[\mu_1 | \mu_2 \ldots |\mu_r]} = \frac12 \left(T_{\mu_1 \mu_2 \ldots \mu_r} + T_{\mu_r \mu_2 \ldots \mu_1} \right) \nonumber.
\ee
\item We use the Einstein convention of repeated indices (exposed in \Sec{Cotangent_Sec}).
\item The identity operator is noted $\hat I$.
\ei

\chapter{Geometry of space-time and black holes}
\label{geometry_Ch}
\minitoc

\section{Fundamental principles of relativity}
\label{fundamentalGR_Sec}
In 1905, Albert Einstein published a revolutionary paper, which was the starting point of the theory of special relativity~\cite{Einstein05}. At that time, there was a well-known conflict between Maxwell theory of electrodynamics and the Newtonian laws of mechanics. This was due to the non invariance of Maxwell equation under Galilean transformations. The naive answer to that paradox was that Maxwell's equations are only valid in one, preferred frame, named the `luminiferous aether'. To resolve this conflict, Albert Einstein proposed a radically different solution. When passing from one Galilean frame to another, the speed of light is a universal quantity. Instead, time itself is relative. This implied modifying the old laws of Galilean transformations, and adopting new transformation laws, which would preserve the `interval' given by
\be
\Delta s^2 = c^2 \Delta t^2 - \Delta x^2 - \Delta y^2 - \Delta z^2. \label{SRds}
\ee
In fact, Einstein was already proven right by the experiments of Michelson and Morley, who measured the speed of light in different reference frames, and couldn't find any sensible difference. This led to the laws of Lorentz transformations, relating time and space coordinates between two Galilean frames. As Lorentz himself had already proven, this transformation group is a symmetry of Maxwell's equations. The success of Einstein's special theory of relativity ended the discussion about the existence of the aether. \\

Soon after 1905, Einstein realized that his new theory was incompatible with Newton universal theory of gravitation. It took him another 10 years to come up with a new theory, which was equally revolutionary in terms of new physical concepts. The starting point was the well known fact that the inertial mass $m_i$, appearing in Newton's law, is exactly equal to the gravitational mass $m_g$, relating the gravitational field to the gravitational force, {\it i.e.} 
\be
m_i = m_g .
\ee
Because of that, the gravitational field is indistinguishable from an acceleration. This was clearly illustrated by the famous {\it gedanken experiment} of Einstein where a man in an elevator cannot tell whether he feels a force due to a gravitational attraction, or because the elevator is accelerating. Einstein proposed to promote this simple observation to a fundamental postulate of gravitation theory. The second main point was the idea to extend what Einstein considered as the main lesson of special relativity. From special relativity, we know that not only no reference frame plays a privileged role in physics, but no clocks as well. Pushing that forward, Einstein postulated that no coordinate set, of any kind, should be preferred to another. These two observations led him to formulate the `equivalence principle', first physical postulate at the origin of the theory of general relativity.

\begin{principle}[Equivalence principle] We consider a point particle in an arbitrary gravitational field. At any space-time point $p$, there exist a set of coordinates $\xi^\mu$ such that in these coordinates, the laws of mechanics are Newton's inertial principle, {\it i.e.}
\be
\frac{d^2\xi^\mu}{d\tau^2}_{|p} = 0. \label{equiv_principle}
\ee
\end{principle}

The idea is then to build the theory of general relativity by formulating a coordinate independent theory, which looks like special relativity for infinitesimal space-time regions. The mathematical tool necessary to formulate such theory is differential geometry of Lorentzian manifolds. The next section is devoted to an introduction to these objects. Of course, this presentation has no ambition of being exhaustive. We shall instead focus on the notion of manifold and geometry. We provide definitions of tensors and some features of vector fields. This will turn out to be very useful to analyze the physical content of a specific geometry. However, we will be very brief, if not silent, concerning notions such as curvatures, tetrad or affine connection. The reason is that in our work, the geometry is mainly considered as a non dynamical background. Therefore it is crucial to properly interpret a geometry. Eluded notions are on the other hand more relevant to study the dynamics of space-time, something we shall barely speak about.

The following section is inspired mainly by references~\cite{Lafontaine,Straumann,Wald}, cited in order from the most mathematical to the most physical one. All along the presentation, we will try to present the various concepts using both abstract definitions and components in an arbitrary coordinate set. The first one presents the interest of being manifestly coordinate invariant, while the second one is often more handy to perform brut computations.

\section{Elements of geometry}
\label{math_Sec}
\subsection{Smooth manifolds}
\begin{figure}[!ht]
\begin{center} 
\includegraphics[scale=0.6]{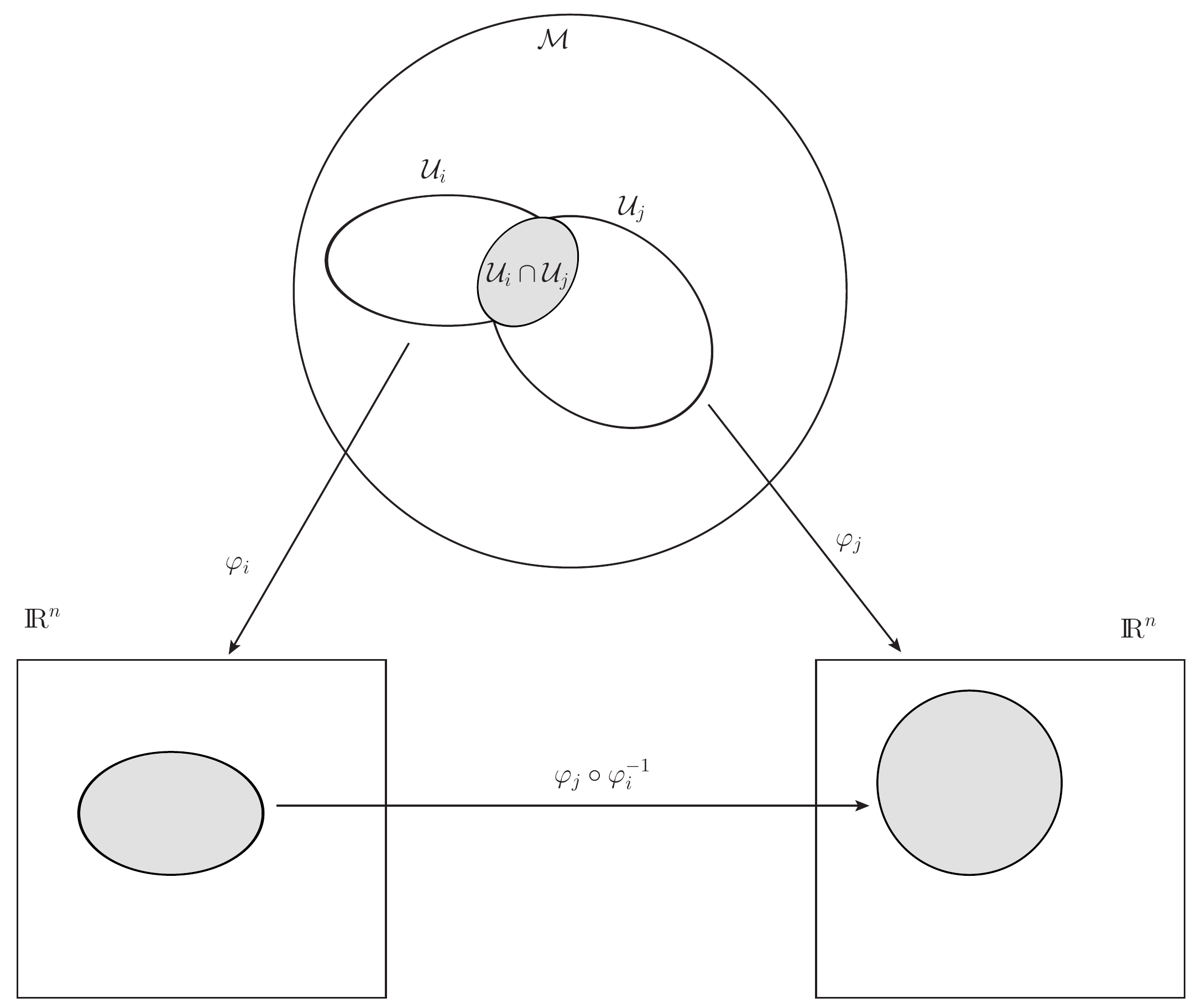}
\end{center}
\caption{Graphic representation of a smooth manifold.}
\label{manifold_fig} 
\end{figure}

\begin{definition}[differentiable manifold] A $C^\infty$-differentiable or smooth manifold $\M$ of dimension $n$ is a set of points, which is locally diffeomorphic to $\mathbb R^n$. More precisely, $\M$ is a separated topological set, such that there exist an open cover $(U_j)_{j \in I}$ and homeomorphisms $(\varphi_j)_{j \in I}$, which send the open sets $U_j$ to $\mathbb R^n$ (or an open subset of $\mathbb R^n$) and obey the compatibility condition
\bi
\item $\forall (i,j)$, the transition map $\varphi_i \circ \varphi_j^{-1}$ is a smooth function {\bf ($C^\infty$ compatibility)}.
\ei
The set of all couples $(U_j; \varphi_j)_{j \in I}$ is called an atlas of $\M$. 
\end{definition}
A couple $(U_j; \varphi_j)$ of the atlas is called a chart. In a physicist language, it is a coordinate patch. Indeed, one can parametrize the set $U_j$ using the mapping $\varphi_j$. For a point $m \in U_j$
\be
\varphi_j (m) = (x^1, x^2, \ldots x^n),
\ee
the $(x^1, x^2, \ldots x^n)$ are hence coordinates for the point $m$. On Fig.\ref{manifold_fig} we have drawn 2 charts and a transition map.

The differentiable structure of $\M$ is given by its atlas $(U_j; \varphi_j)_{j \in I}$. Roughly speaking, it allows to define the notion of derivatives on $\M$, by transporting the usual notion in $\mathbb R^n$ to $\M$, via $\varphi_j$, {\it i.e.}, deriving on $\M$ means deriving with respect to some coordinates. The $C^\infty$ compatibility is necessary to ensure that the notions of smoothness or differential are coordinate independent, {\it i.e.} independent of the map $\varphi_j$ we used. As a first example, we define the set of smooth and real or complex valued functions on $\M$.

\begin{definition} A continuous map $f : \M \to \mathbb R$ (or $\mathbb C$) is smooth if for any map $\varphi_j$ of the atlas, $f\circ \varphi_j^{-1} : \mathbb R^n \to \mathbb R$ (or $\mathbb C$) is smooth. The algebra of all smooth and real (complex) valued functions on $\M$ is noted $\CM$ (or $C_{\mathbb C}^\infty(\M)$). \label{smoothdef}
\end{definition}
One verifies that this definition is independent of the choice of $\varphi_j$, precisely because the transition maps are smooth.

A manifold possesses many global properties, like its topological structure. But locally, it looks like $\mathbb R^n$. For our purpose, we will be poorly interested into global properties. Therefore, we will often work in an open set $U_j$, with a set of coordinates $\varphi_j$. However, we will be very careful that the notions we define are independent of the coordinate choice.

To build differential calculus on a manifold, we need to define the notion of infinitesimal displacements. Figuratively, a point $p$ in an infinitesimal neighborhood of $m$ is 
\be
p \approx \underbrace{m}_{\in \M} + \underbrace{\epsilon}_{\ll 1} \underbrace{v}_{\in T_m\M}, \label{inftydisplace}
\ee
where $\epsilon$ is an infinitesimal quantity, and $v$ is a vector tangent to the manifold $\M$ at the point $m$. However, because a manifold is not naturally embedded into $\mathbb R^n$, the notion of `being tangent to' is ill defined. Therefore, we need to use an intrinsic definition for the tangent space. 

\begin{definition}[tangent space] Consider a point $m$ on a smooth manifold $\M$. A derivative at $m$ is a linear map $\delta: \CM \to \mathbb R$, which verifies the Leibniz rule
\be
\delta(fg) = f(m) \delta(g) + g(m) \delta(f),
\ee
and localized around $m$. This means that for any open set $U$ containing $m$, 
\be
\delta(f_{| U}) = \delta(f).
\ee
The set of all derivatives is called the tangent space and is noted $T_m\M$. \label{TmMdef}
\end{definition}

This definition can be shown to be equivalent to a more geometric one, {\it i.e.} to the set of derivatives of curves passing on $m$, where two curves are identified if they have the same derivative at $m$, when expressed in a certain set of coordinates. However, definition \ref{TmMdef} presents the interest of being manifestly coordinate invariant, since it does not refer to any $\varphi_j$ to be well defined. 
 
From the definition, it is immediate to see that $T_m\M$ is naturally a vector space. Moreover, if $m\in U_j$, using the coordinate patch $\varphi_j$, we can demonstrate that that $T_m\M$ is isomorphic to $\mathbb R^n$~\cite{Lafontaine}. More precisely 
\be
\delta(f_{|U_j}) = \sum_{\mu = 1}^n v^\mu \p_\mu(f\circ \varphi_j^{-1}), 
\ee
or simply   
\be
\delta = \sum_{\mu = 1}^n v^\mu \p_\mu ,\label{localvectdecomp}
\ee
where $(v^\mu)_{\mu=1..n} \in \mathbb R^n$ and $\p_\mu$ is the partial derivative with respect to the coordinate $x^\mu$. In particular, using a coordinate patch, we obtain an explicit basis of derivation, given by partial derivatives. In this construction, tangent vectors are identified with the directional derivatives of function. This definition of tangent vectors turns out to be more suitable for a generalization to vector fields. 

Having defined the tangent space, we can now build the differential of a function between two manifolds $f : \M \to \mathcal N$. With a definition similar to \ref{smoothdef}, we first define $f$ as being smooth if it is continuous and differentiable with respect to any coordinate system. 

\begin{definition} Let $f : \M \to \mathcal N$ be a smooth function. Its differential at a point $m$ is a linear map between tangent spaces defined as 
\be \bal
df_m : T_m\M &\to T_{f(m)}\mathcal N \\
v &\mapsto df_m \cdot v
\eal \ee
$df_m \cdot v$ is a derivation at $f(m)$ defined by 
\be
df_m \cdot v (g) = v(g\circ f).
\ee
\end{definition}

Figuratively, in the language of \eq{inftydisplace}, the differential of $f$ is the linear map such that  
\be
f(m + \epsilon v) \approx f(m) + \epsilon \ df_m \cdot v.
\ee
Using a coordinate patches around $m$ and $f(m)$, the function $f$ becomes
\be
f : x=(x^1,x^2,\ldots x^n) \mapsto (f^1(x),f^2(x),\ldots f^n(x)),
\ee
and we can express the latter definition in components
\be
[df_m \cdot v]^\nu = \sum_{\mu=1}^n \frac{\p f^\nu}{\p x^\mu} v^\mu.
\ee
A particular and fundamental class of smooth function is the set of diffeomorphisms.

\begin{definition}[diffeomorphism] Consider two manifolds $\M$ and $\mathcal N$, $f : \M \to \mathcal N$ is a diffeomorphism if it is a one-to-one mapping such that $f$ and $f^{-1}$ are smooth.

This implies that at each point $m\in \M$, $df_m$ is a linear isomorphism between the vector spaces $T_m\M$ and $T_{f(m)}\mathcal N$. 

The set of all diffeomorphisms of a manifold $\M$ into itself is noted $\mathrm{Diff}(\M)$.
\end{definition}

In particular, changes of coordinates, {\it i.e.} $\varphi_i \circ \varphi_j^{-1}$, are diffeomorphisms of $\mathbb R^n$. This notion is crucial since general relativity is a diffeomorphism invariant theory. In other words, observables never depend upon a choice of coordinate set. 

\subsection{Tangent bundle}
\subsubsection{Construction}
In this section, we would like to generalize the notion of tangent vector to that of vector field. This means a function $X$ which associates to each point $p$ of $\M$ a vector $X_p \in T_m\M$. The subtlety appear when we require it to be a smooth function, since every vector $X_p$ belongs to a different space. This is why we have to introduce the notion of tangent bundle.

\begin{definition}[tangent bundle] The tangent bundle of a smooth manifold $\M$ is the disjoint union of all the tangent spaces, {\it i.e.}
\be
T\M = \underset{m \in \M}{\amalg} T_m\M.
\ee
It possesses a canonical differential structure inherited from $\M$.  
\end{definition}

To build the differential structure of $T\M$, we use an atlas $(U_j; \varphi_j)_{j \in I}$ of $\M$. In the open set $U_j$, using the map $\varphi_j$, we show that
\be
T\M \simeq U_j \times \mathbb R^n.
\ee
The points of $T\M$ are the couples $(q,\dot q)$ where $q\in \M$ and $\dot q \in T_q\M$. In particular, if $\M$ is of dimension $n$, $T\M$ is of dimension $2n$. In classical mechanics, this is the configuration space over which the Lagrangian formalism is defined~\cite{Arnold}. 

\subsubsection{Vector fields}
It is now quite natural to define the notion of vector field. 

\begin{definition} A vector field $X$ of a manifold $\M$ is a smooth map $\M \to T\M$ such that at each point $m$, $X_m \in T_m\M$.
\end{definition}
The set of all vector field is noted $\Gamma(T\M)$. Of course, it is a vector space, since linear combinations of vector fields can naturally be defined pointwise.

A vector field also has a natural way to act on smooth functions $f\in C^{\infty}(\M)$. Indeed, to the function $f$, it associate the smooth function $L_X f$ defined as
\be
L_X f : m \mapsto df_m \cdot X_m.
\ee
The mapping $L_X : C^{\infty}(\M) \to C^{\infty}(\M)$ is the generalization of \ref{TmMdef} and is called a global derivation.

\begin{definition} A global derivation $\delta$ of a manifold $\M$ is a linear map of $\CM$ into itself that obeys the Leibniz rule 
\be
\delta(fg) = f \delta(g) + g \delta(f).
\ee
\end{definition}
What is remarkable, is that the correspondence between vector fields and global derivations, {\it i.e.} $X \leftrightarrow L_X$ is one-to-one. As we saw, to a vector field $X$ we associate the derivation $L_X$. Reciprocally, to a derivation $\delta$, we can always build a vector field $X$ such that $\delta = L_X$. To see this, we use a local coordinate patch, and show that 
\be
\delta(f)(m) = \sum_{\mu = 1}^n X^\mu(m) \p_\mu f,
\ee
where the $X^\mu$ are smooth functions. These are the components of $X$ in the local basis associated with the coordinates, as in \eq{localvectdecomp}. Therefore, using a coordinate patch, we often denote a vector field as  
\be 
X = \sum_{\mu = 1}^n X^\mu \p_\mu. \label{Xdecomp}
\ee
The representation of a vector field by a derivation turns out to be much more convenient to manipulate. As a first example, we define the transport of a vector field by a diffeomorphism.
\begin{definition} Let $X$ be a vector field on a manifold $\M$, and $\varphi$ a diffeomorphism from $\M$ to $\mathcal N$. We define a derivation on $\mathcal N$ as
\be
\delta : f \mapsto L_X(f\circ \varphi) \circ \varphi^{-1}.
\ee
The corresponding vector field $Y$ on $\mathcal N$ is the image of $X$ by $\varphi$ and is noted 
\be
Y = \varphi_* X.\label{phiXdef}
\ee
\end{definition}
Using the differential of the map $\varphi$, one can show
\be
(\varphi_*X)_y = d\varphi_{\varphi^{-1}(y)} \cdot X_{\varphi^{-1}(y)}.
\ee
A particular and crucial example is given when we use $(\mathcal U, \varphi)$ a chart of the manifold $\M$. Indeed, the decomposition of $X$ in local coordinates in \eq{Xdecomp} is an abuse of notation for $\varphi_*X$. Moreover, if one would like to make a change of coordinate one should use the preceding formula for the diffeomorphism
\be
\psi : (x^1,x^2,\ldots x^n) \mapsto (\tilde x^1,\tilde x^2,\ldots \tilde x^n).
\ee
Using the corresponding local basis, we obtain 
\be
(\psi_*X)^\nu = \sum_{\mu =1}^n \frac{\p \tilde x^\nu}{\p x^\mu} X^\mu.
\ee

\subsubsection{Commutator}
If $X$ and $Y$ are vector fields on $\M$, since $L_X$ and $L_Y$ are maps of $\CM$ in itself, it is natural to ask wether the composition is also a derivation. Since it is obviously linear, the only point to check is the Leibniz rule. Considering $f, g \in \CM$, a straightforward computation shows that
\be
L_X\circ L_Y (fg) = f L_X\circ L_Y(g) + L_X(f) L_Y(g) + g L_X\circ L_Y(f) + L_X(g) L_Y(f).
\ee
Therefore, the composition is not a derivation, however, the commutator
\be
[L_X,L_Y] = L_X\circ L_Y - L_Y\circ L_X
\ee
is. Hence, using the correspondence between derivations and vector fields, we define the commutator of two vector fields $[X,Y]$. This object shares the usual properties of commutators, since it is bilinear, antisymmetryc and satisfies the Jacobi identity
\be
[X,[Y,Z]] + [Y,[Z,X]] + [Z,[X,Y]] = 0.
\ee
Moreover, if $\psi$ is a diffeomorphism, using definition \ref{phiXdef}, we show 
\be
\varphi_*[X,Y] = [\varphi_*X, \varphi_*Y].
\ee
This means that the commutator can be computed in any coordinate system. Using such a chart, with 
\be
X = \sum_{\mu = 1}^n X^\mu \p_\mu,
\ee
and
\be
Y = \sum_{\mu = 1}^n Y^\mu \p_\mu,
\ee
we obtain 
\be
[X,Y] = \sum_{\mu = 1}^n \left( X^\nu \p_\nu Y^\mu - Y^\nu \p_\nu X^\mu \right) \p_\mu.
\ee
Physically, the commutator $[X,Y]$ is the directional derivative of $X$ in the direction of $Y$. This statement will be made clearer in a following section. For now, we can at least say that if $[X,Y]=0$, $X$ is independent of the direction pointed by $Y$. This is perfectly illustrated by the following theorem in the more general case of $p$ commuting vector fields.
\begin{theorem}
Let $(X_j)_{j=1..p}$ be $p \leqslant n$ vector fields on a $n$ dimensional manifold $\M$. We suppose that they commute among themselves, {\it i.e.}
\be
\forall i,j \ ,\ [X_i,X_j] = 0.
\ee
We suppose them to be non zero on a point $m\in \M$. Then, there exist a coordinate set $\psi$ in a neighborhood of $m$ such that 
\be
\psi_*X_j = \p_j.
\ee
\end{theorem}
This theorem shows the power of the concept of commutator. In fact, much more general results can be obtained with commuting conditions, which are not only useful in integrable system, but also in general relativity when one wants to implement symmetries on a space-time~\cite{Kundt66,Zegersthesis}.

\subsubsection{Flow of a vector field}
\label{XflowSec}
If $X$ is a vector field on $\M$, a way to represent it geometrically is to build  family of curves on $\M$ that are tangent to $X$ at each point. Those are the integral curves of $X$. As we shall see, this notion is crucial since it allows to build family of diffeomorphism on $\M$ induced by $X$. 

\begin{definition} Let's consider a vector field $X \in \Gamma(T\M)$ and a point $x \in \M$. 
Let $J$ be an open interval of $\mathbb R$ containing 0. A smooth curve $\gamma_x : J \to \M$ is called an integral curve of $X$ starting at $x$ if it satisfies the Cauchy problem
\be
\left\{ \bal
\dot \gamma_x(\lam) &= X_{\gamma(\lam)},\\
\gamma_x(0) &= x.
\eal \right.
\ee \label{intcurve}
\end{definition}
Using a local set of coordinates, the differential equation to solve reads 
\be
\frac{d x^\mu}{d\lam} = X^\mu(x^1(\lam), x^2(\lam), \ldots x^n(\lam)).
\ee
From the well-known result of nonlinear differential equations, and in particular the Cauchy-Lipschitz theorem, we know that $\gamma_x$ is uniquely defined on a maximal open interval $J_x \subset \mathbb R$ for every $x \in \M$. Moreover, the set $\mathcal D = \{(\lam,x) | x\in \M, \lam\in J_x\}$ is an open set of $\M$. This allows us to define the flow of $X$

\begin{definition}
The mapping 
\be \bal
\phi^X : \mathcal D &\to \M \\
(\lam,x) &\mapsto \gamma_x(\lam) 
\eal \ee
is smooth and is called the \emph{flow} of $X$.
\end{definition}
The term `flow' comes naturally from fluid mechanics, where $X$ is a velocity profile. In this context, the flow $\phi^X$ allows to go from the Eulerian description to the Lagrangian one. There, the parameter $\lam$ is denoted $t$ and represent the Newtonian time. Note that definition \ref{intcurve} is not the most general, since the velocity profile can be time dependent. This leads to the notion of time dependent vector field~\cite{Lafontaine}.

However, in General Relativity, the time is a coordinate, and the parameter $\lam$ has {\it a priori} no physical meaning. In particular, in this case $X$ never depends explicitly on $\lam$, and the differential equation defining the flow is autonomous. The most interesting way to manipulate the flow is to look at fixed $\lam$, and introduce a family of diffeomorphism. Indeed, one can show that $\mathcal D_\lam = \{x\in \M | \lam \in J_x \}$ is an open set of $\M$. Then, 
\be \bal
\phi^X_\lam : \mathcal D_\lam &\to \M \\
x &\mapsto \phi^X(\lam,x) 
\eal \ee
is a diffeomorphism from $\mathcal D_\lam$ to its image. Moreover, we have the fundamental identity
\be
\phi^X_\lam \circ \phi^X_\sigma = \phi^X_{\lam + \sigma}.
\ee
We point out that this is an abuse of notation, which means
\be
\phi^X_\lam(\phi^X_\sigma(x)) = \phi^X_{\lam+\sigma}(x),
\ee
which makes sense only if $\sigma \in J_x$ and $\lam \in J_{\phi^X_\sigma(x)}$. Because of this identity, we call $\lam \to \phi^X_\lam$ a \emph{local one parameter group}. The word `local' is here to recall that it is \emph{not} a group, because of the restriction concerning the domains of definition. Sometimes, there is no such restriction, {\it i.e.} $\mathcal D_\lam = \M$ for all $\lam$. In that case, $X$ is said to be complete.

\begin{lemma}
Let $X$ be a vector field on a manifold $\M$. If $\M$ is compact, or if $X$ vanishes outside a compact subset of $\M$, then $X$ is complete. 
\end{lemma} 
Unfortunately, in general, many vector fields do not have this property as one can see from the following example. 

\begin{figure}[!ht]
\begin{center} 
\includegraphics[scale=0.7]{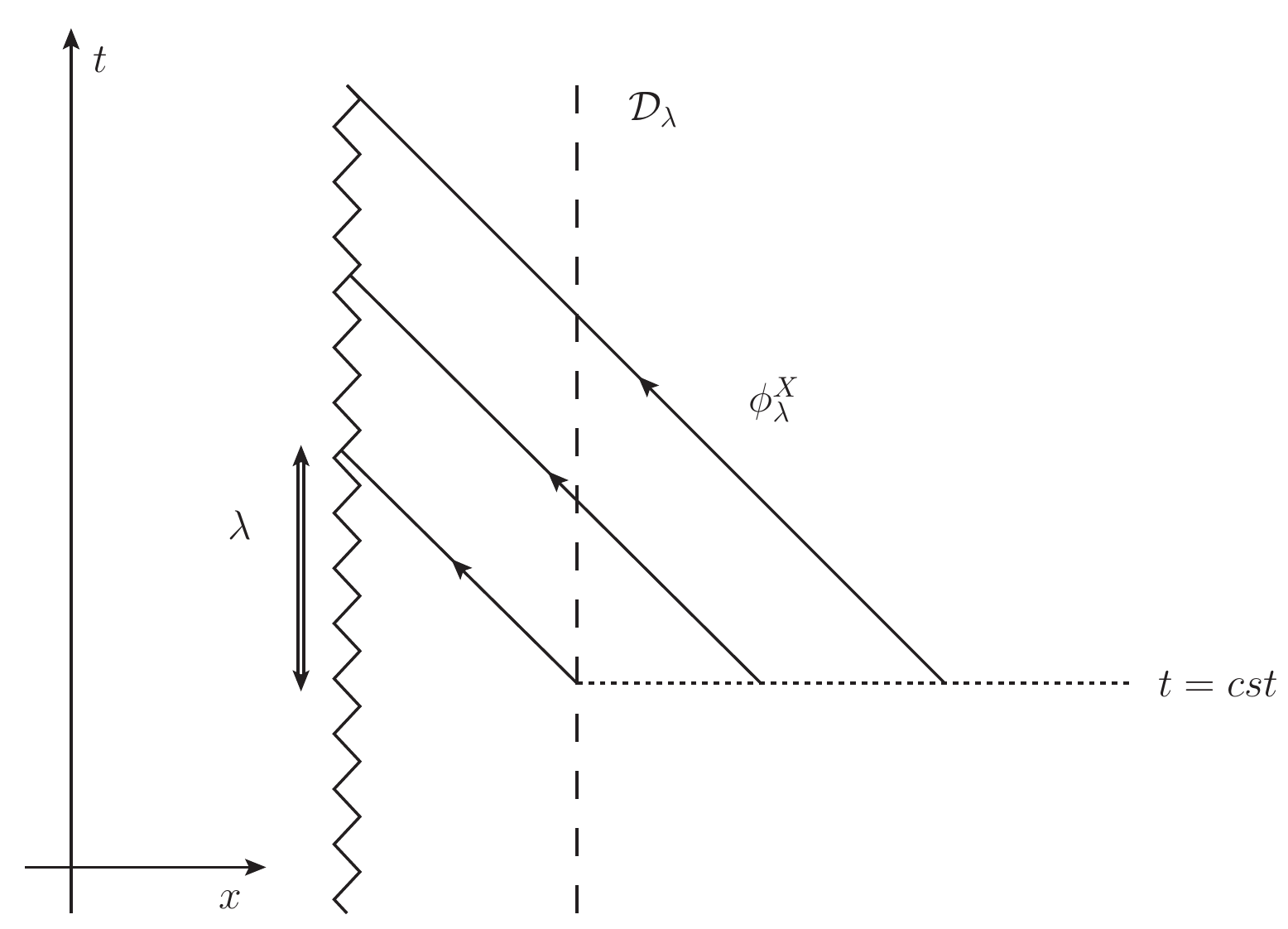}
\end{center}
\caption{Graphic representation of the domain $\mathcal D_\lam$, for a fixed value of $\lam$. After a time $\lam$, points that were at a distance $v\lam$ of $x=0$ (on the right of the dashed line) have reached the singularity.}
\label{singularityexample_fig} 
\end{figure}

\begin{example}
Let $\M = \{(t,x)\in \mathbb R^2 | x>0 \}$. We consider the vector field
\be
X = \p_t - v\p_x,
\ee
with $v>0$. A trajectory starting at $(t_0,x_0)$ will reach $x=0$ in a finite $\lam$ interval. The diffeomorphism $\phi^X_\lam$ is fairly simple to compute 
\be
\phi^X_\lam(t,x) = (t+\lam , x - v\lam).
\ee
Hence it is defined on 
\be
\mathcal D_\lam = \{(t,x) \in \M | x > v\lam \},
\ee
as one can see in Fig.\ref{singularityexample_fig}.
\end{example}

Of course, the latter example is quite artificial. However, it is representative of a generic feature that occurs in space-times containing a singularity. 

Going back to the local group $\phi^X_\lam$, we notice that by definition,
\be
\frac{d\phi^X_\lam}{d\lam} = X.
\ee
Hence, $X$ is often called the `infinitesimal generator' of the local group $\phi^X_\lam$. In particular, it allows us to define the `Lie derivative' with respect to $X$ of every object that can be transported by a diffeomorphism. As first examples, we have the results
\begin{lemma}
If $f \in \CM$, then
\be
\frac d{d\lam}( f\circ \phi^X_\lam) = L_X f.
\ee
\end{lemma}

\begin{lemma}
If $Y \in \Gamma(T\M)$, then
\be
\frac d{d\lam}( \phi^X_{\lam*}Y) = [X,Y].
\ee
\end{lemma}
In particular, we recover the interpretation of the commutator of the derivative of $Y$ in the direction $X$. 

In the following, this method will generalize the definition of the Lie derivative $L_X$ on any tensor field.

\subsection{Cotangent bundle and tensor fields}
\label{Cotangent_Sec}
\subsubsection{The cotangent bundle}
\label{1formSec}
When studying the vector space $T_m\M$, it is interesting not to look solely at the vectors, but also the linear form on this space, {\it i.e.} the dual $T^*_m\M$. Furthermore, to also consider the dependence on $m$, we follow the construction of the tangent bundle to build the cotangent bundle.

\begin{definition}[cotangent bundle] The cotangent bundle of a smooth manifold $\M$ is the disjoint union of all the dual tangent spaces, {\it i.e.}
\be
T^*\M = \underset{m \in \M}{\amalg} T^*_m\M.
\ee
\end{definition}
Just like $TM$, it possesses a canonical differential structure inherited from $\M$. The notion dual to vector fields is the 1-form field, or simply 1-form.

\begin{definition} A 1-form $\om$ of a manifold $\M$ is a smooth map $\M \to T^*\M$ such that at each point $m$, $\om_m \in T_m^*\M$. 
\end{definition}

When using a chart $(U_j,\varphi_j)$, a 1-form decomposes
\be
\om = \sum_{\mu =1}^n \om_\mu dx^\mu,
\ee
where $dx^\mu$ is the basis dual to partial derivatives
\be
dx^\mu(\p_\nu) = \delta^\mu_{\ \nu}.
\ee
One can also notice that $dx^\mu$ is the differential of the coordinate function $(x^1,x^2,\ldots x^n) \mapsto x^\mu$. When one disposes of a 1-form $\om$ and a vector field $X$, one can cannocially build the smooth function 
\be
\om(X) : m \mapsto \om_m(X_m).
\ee
In components, this reads
\be
\om(X) = \sum_{\mu = 1}^n \om_\mu X^\mu.
\ee
$\om(X)$ is called the \emph{contraction} of $\om$ and $X$. 

There is a fundamental class of 1-forms, which are given by the differential of functions. Indeed, if $f\in \CM$, 
\be
\om : m \mapsto df_m
\ee
is a 1-form. However, one can show that conversely, not all 1-form is the differential of a function. In fact, $\om$ can be \emph{locally} written as a differential if and only if
\be
\p_\mu \om_\nu - \p_\nu \om_\mu = 0.
\ee
The globalization of this property, and its generalization to $p$-forms lead to the theory of de Rham Cohomology~\cite{Lafontaine}. 

We want now to transport $\om$ using a diffeomorphism, which in particular will give us coordinate change formulae. The definition is quite intuitive to build, indeed, if $f\in \CM$ and $\psi$ is a diffeomorphism, $f$ is changed into $f\circ \psi$. Hence, it is natural to define the image of its differential by $\psi$ through 
\be
\psi^*df = d(f\circ \psi).
\ee
This leads to the following definition of the image of $\om$ by $\psi$, or the \emph{pull back} of $\om$ by $\psi$.

\begin{definition} Let $\om$ be a 1-form on $\M$, and $\psi : \mathcal N \to \M$ a diffeomorphism. We define the pull back $\psi^*\om$ as a 1-form on $\mathcal N$ by 
\be
(\psi^*\om)_m(v) = \om_{\psi(m)}(d\psi_m \cdot v),
\ee
for all $v\in T_m\mathcal N$.
\end{definition}

When one wants to make a coordinate change 
\be
\om = \sum_{\mu =1}^n \om_\mu dx^\mu \quad \to \quad \psi^*\om = \sum_{\mu =1}^n \tilde \om_\mu d\tilde x^\mu,
\ee
one should use the map $\psi : (\tilde x^1,\tilde x^2,\ldots \tilde x^n) \mapsto (x^1, x^2,\ldots x^n)$. The components of $\om$ in the local basis become 
\be
(\psi^*\om)_\nu = \sum_{\mu =1}^n \frac{\p x^\mu}{\p \tilde x^\nu} \om_\mu.
\ee
We notice that in this formula, the role of $x^\mu$ and $\tilde x^\mu$ are \emph{exchanged} with respect to \eq{Xdecomp}. This is why $X$ is said to be \emph{contravariant}, while $\om$ is \emph{covariant}. We keep trace of this property by noting the components of $X$ with upper indices and those of $\om$ with lower indices. 

\subsubsection{Tensor fields}
After having defined vector fields and 1-forms, the natural generalization is to build smooth fields of tensors of rank $(r,s)$ on $T_m\M$. 

\begin{definition} A tensor field $T$ of rank $(r,s)$ is a smooth map from $\M$ to the set 
\be
T(\M)^r_s = \underbrace{T\M \otimes \ldots \otimes T\M}_{r \text{ times}} \otimes \underbrace{T^*\M \otimes \ldots \otimes T^*\M}_{s \text{ times}},
\ee
such that for each point $m\in \M$, $T_m$ is a tensor of rank $(r,s)$ on the vector space $T_m\M$. More explicitly, $T_m$ is a $r+s$ linear map that takes as arguments $r$ elements of $T^*_m\M$ and $s$ elements of $T_m\M$
\be
T_m : \underbrace{T_m^*\M \otimes \ldots \otimes T_m^*\M}_{r \text{ times}} \otimes \underbrace{T_m\M \otimes \ldots \otimes T_m\M}_{s \text{ times}} \to \mathbb R.
\ee
The set of all tensor field of rank $(r,s)$ is noted $\mathcal T^r_s(\M)$. To consider arbitrary ranks, we define
\be
\mathcal T(\M) = \bigoplus_{r,s} \mathcal T^r_s(\M).
\ee
\end{definition}
When using a coordinate patch, we already know a basis for $T\M$ and $T^*\M$. Hence, using basic properties of tensor product, we decompose a tensor field $T$ into a local basis 
\be
T = \sum_{\mu_1,\ldots \mu_r = 1}^n \sum_{\nu_1,\ldots \nu_s = 1}^n T^{\mu_1\ldots \mu_r}_{\ \ \ \ \ \ \ \nu_1 \ldots \nu_s} \p_{\mu_1} \otimes \ldots \otimes \p_{\mu_r} \otimes dx^{\nu_1} \otimes \ldots \otimes dx^{\nu_s}.
\ee
We see here that when decomposing a tensor on a local basis, there is many indices to sum over. Hence, to enlight the writing, we shall adopt the convention of repeated indices.

\begin{convention}[Einstein's convention] When writing a tensor in components, it is assumed that any indices appearing twice is summed over.
\end{convention}

The next step is to extend the definition of a pull back by a diffeomorphism $\psi$ for any tensor field $T$. When the tensor field is completely covariant, it is straightforward, by imposing that for any tensor fields $T$ and $S$, we have 
\be
\psi^*(S\otimes T) = \psi^*S \otimes \psi^*T \label{vectprodPB}.
\ee
However, as we saw in \Sec{1formSec}, the transformation of a contravariant tensor is the inverse of a covariant one. Hence, the pull back is extended to all tensor field, by making the convention that a vector field $X$, {\it i.e.} a $(1,0)$ tensor field, is pulled back through 
\be
\psi^*X = (\psi^{-1})_*X. \label{XPB}
\ee

\begin{definition} Let $T$ be a tensor field of rank $(r,s)$ on $\M$, and $\psi : \mathcal N \to \M$ a diffeomorphism. We define the pull back of $T$ by $\psi$ by 
\be
(\psi^*T)_m = T_{\psi(m)}(\om^1(d\psi_m^{-1} \cdot), \ldots \om^r(d\psi_m^{-1} \cdot), d\psi_m \cdot v_1, \ldots d\psi_m \cdot v_s).
\ee
\end{definition}

The latter definition is quite abrupt, however, it presents the interests of being manifestly coordinate independent. In practice, the main properties given by \eq{vectprodPB} and \eqref{XPB} are much more useful. 

When $\psi$ is the coordinate change $(\tilde x^1,\tilde x^2,\ldots \tilde x^n) \mapsto (x^1, x^2,\ldots x^n)$, using a local basis we derive the transformation law 
\be
T^{\mu'_1 \ldots \mu'_r}_{\ \ \ \ \ \ \ \nu'_1 \ldots \nu'_s} =  T^{\mu_1 \ldots \mu_r}_{\ \ \ \ \ \ \ \nu_1 \ldots \nu_s}  \frac{\p \tilde x^{\mu'_1}}{\p x^{\mu_1}} \cdots \frac{\p \tilde x^{\mu'_r}}{\p x^{\mu_r}} \frac{\p x^{\nu_1}}{\p \tilde x^{\nu'_1}} \cdots \frac{\p x^{\nu_s}}{\p \tilde x^{\nu'_s}} .
\ee
When we know how to transport tensor fields with diffeomorphisms, we can apply it to a family of diffeomorphism, and in particular when it is the flow of some vector field $X$. The infinitesimal version of the pull back of a tensor field $T$ by the flow $\phi^X_\lam$ gives the Lie derivative of $T$ with respect to $X$.

\begin{definition} Let's consider a tensor field $T$, a vector field $X$ and its flow $\phi^X_\lam$. The Lie derivative of $T$ with respect to $X$ is given by 
\be
L_XT = \frac{d}{d\lam}(\phi^{X*}_\lam T)_{|\lam = 0}.
\ee
\end{definition}

Note that this definition is slightly abusive, since $\phi_\lam$ is not a diffeomorphism on the full manifold, but only on the open set $\mathcal D_\lam$. However, for any point $x\in \M$, one can consider $\lam$'s small enough so that $x \in \mathcal D_\lam$ and the latter definition is fine at fixed $x$.

Probably the most useful property of the Lie derivative $L_X$ is that it defines a derivation on $\mathcal T(\M)$, {\it i.e.}, for $T$ and $S$ tensor fields
\be
L_X(T\otimes S) = (L_XT) \otimes S + T \otimes (L_XS).
\ee

\subsection{Space-time as a Lorentzian manifold}
\subsubsection{Metric structure}
\begin{definition}
A pseudo-Riemannian manifold $(\M, \g)$ is a $n$ dimensional smooth manifold endowed with a tensor $\g \in \mathcal T^0_2(\M)$, such that at each point $m$, $\g_m$ is a symmetric and non-degenerated bilinear form. 

\bi
\item When its signature is $(+,-,\ldots -)$, it is a Lorentzian manifold.
\item When its signature is $(+,\ldots +)$, it is a Riemannian manifold.
\ei
\end{definition}

As usual, one can use a coordinate set to write the metric tensor in a basis 
\be
\g = g_{\mu \nu} dx^\mu dx^\nu,
\ee
where $dx^\mu dx^\nu$ is the \emph{symmetric product} of the 1-forms $dx^\mu$ and $dx^\nu$. If $\theta_1$ and $\theta_2$ are 1-forms on $\M$, we define their symmetric product by 
\be
\theta_1 \theta_2 = \theta_1 \cdot \theta_2 = \frac12 (\theta_1 \otimes \theta_2 + \theta_2 \otimes \theta_1).
\ee
This tool turns out to be quite convenient to decompose the metric $\g$ in a simple way. Note that a $(0,2)$ tensor can be equivalently seen as a (non degenerated) bilinear form or a quadratic form . In order to make the distinction, we shall call $\g$ the bilinear form and $ds^2$ the associated quadratic form.

\begin{principle} In general relativity, space-time is a $3+1$ Lorentzian manifold $(\M,\g)$. At each point $m$, $T_m\M$ is the Minkowski space-time of special relativity.
\end{principle}

This definition is the local version of \eq{SRds}, it translates the equivalence principle \eqref{equiv_principle} in geometrical terms. A point in the manifold is an \emph{event}, with definite time and position. A smooth curve on $\M$ represents a \emph{world line}, that is a trajectory in space-time.

Because the metric can take any sign, we divide the tangent space $T_m\M$ into 3. Let $v\in T_m\M$
\bi
\item if $\g(v,v) > 0$, $v$ is {\bf time-like},
\item if $\g(v,v) = 0$, $v$ is {\bf null} or {\bf light-like},
\item if $\g(v,v) < 0$, $v$ is {\bf space-like}.
\ei
When $v$ is either time-like or null, it is said {\bf causal}. We extend this definition to any curve $\gamma : \lam \to \gamma(\lam)$ if all its tangent vectors $\dot \gamma(\lam)$ are of the same type.

\begin{principle} A massive observer in a space-time $(\M, \g)$ has to follow a {\bf time-like} trajectory. If the observer goes from an event $E_1$ to an event $E_2$ through the world line $\gamma$, then 
\be
\Delta \tau = \int_{E_1}^{E_2} \sqrt{ds^2} = \int_{\lam_1}^{\lam_2} \sqrt{\g_{\gamma(\lam)}(\dot \gamma(\lam), \dot \gamma(\lam))} d\lam, \label{taudef}
\ee
is the time elapsed as measured by the observer. A massless object will instead follow a \emph{null} world line.
\end{principle}

This principle is simply the translation into curved space that a physical object cannot travel faster than light.

\subsubsection{Raising and lowering indices}
If $(\M,\g)$ is a pseudo-Riemannian manifold, the metric tensor induces an isomorphism between the tangent bundle and the cotangent bundle. Indeed, the morphism 
\be \bal
\Theta : T\M &\to T^*\M \\
(x,v) &\mapsto \g_x(v, \cdot )
\eal \ee
is smooth and induces for each $x$ an isomorphism between the vector spaces $T_x\M$ and $T_x^*\M$. To enlighten the notations, we introduce 
\bsub \bea
\Theta_x : &v \mapsto v^\flat, \\
\Theta_x^{-1} : &\alpha \mapsto \alpha^\sharp .
\eea \esub
We now use a local basis and work with components
\be
v = v^\mu \p_\mu.
\ee
By definition, 
\be
(v^\flat)_\mu = g_{\mu \nu} v^\mu.
\ee
Conversely, if $\alpha = \alpha_\mu dx^\mu$,
\be
(\alpha^\sharp)^\mu = g^{\mu \nu} \alpha_\nu,
\ee
where $g^{\mu \nu}$ is the inverse matrix of $g_{\mu\nu}$. In the following, when working in components, we omit the $\flat$ and $\sharp$, since the presence of the index up or down indicates whether we consider an element of $T\M$ or $T^*\M$. This mechanism can be extended to tensor fields without pain. Indeed, to $T \in \mathcal T^r_s(\M)$, we associate $\tilde T \in \mathcal T^{r-1}_{s+1}$ by 
\be
\tilde T(\om^1, \ldots \om^{r-1}, v_1, \ldots v^{s+1}) = T(\om^1, \ldots \om^{r-1}, (v_1)^\flat, \ldots v^{s+1}).
\ee
In components, this reads
\be
T^{\mu_1 \ldots \mu_{r-1}}_{\ \ \ \ \ \ \ \nu_1 \ldots \nu_{s+1}} = g_{\nu_1 \mu_r} T^{\mu_1 \ldots \mu_r}_{\ \ \ \ \ \ \ \nu_2 \ldots \nu_{s+1}}.
\ee
Therefore, in tensor calculus, $g_{\mu \nu}$ is used to lower indices, and conversely, $g^{\mu \nu}$ raises indices. Note also that $g^{\mu \nu}$ defined as the inverse of $g_{\mu \nu}$ is the same as defined by raising the two indices of $g_{\mu \nu}$, making our notations consistent. 

This isomorphism has many other consequences. For example, it allows us to define the \emph{gradient} of a function.

\begin{definition}
Consider a smooth function $f\in \CM$, we define its gradient as the vector field
\be
\grad(f) = (df)^\sharp.
\ee
\end{definition}

\subsubsection{family of trajectories}
Let's consider a vector field $\uf$ on a space-time. As we saw in \Sec{XflowSec}, it generates a family of curve through its flow $\phi$. To an initial event $x$, we associate a world line $\lam \to \phi_\lam(x)$ which starts at $x$ and follows $\uf$. If the vector field is time-like, that is $\g(\uf,\uf) > 0$ everywhere, it represent a field of 4-velocities. This is what is needed to describe a fluid in space-time for example. Moreover, the norm of the field relates the parameter $\lam$ to the proper time of an observer following the flow $\uf$, as we see from \eq{taudef}. In particular, if the vector field is unitary, $\tau = \lam$.

One can make the last discussion using components of $u$
\be
\uf = u^\mu \p_\mu.
\ee
Then the flow $x^\mu(\lam)$ is such that 
\be
d\tau^2 = ds^2(\dot x^\mu) = g_{\mu \nu}\frac{dx^\mu}{d\lam}\frac{dx^\nu}{d\lam} d\lam^2 = (u^\mu u_\mu) d\lam^2.
\ee
If $u^\mu u_\mu = 1$, then $d\tau = d\lam$. Moreover, using Leibniz rule, one derive the identity 
\be
u = u^\mu \p_\mu = \frac{dx^\mu}{d\lam} \p_\mu = \p_\lam.
\ee
This last equality is in fact abusive since $\lam$ is not a coordinate (so far). However, this identity can be made rigorous when one looks at the action of $u$ on a smooth function $f$
\be
L_u f = \p_\lam f = \frac{d}{d\lam}(f\circ \phi_\lam).
\ee
In particular, when $u$ is unitary, its Lie derivative represent the time derivative as measured by observers following the flow of $u$.

\subsubsection{Geodesics}
Among all the allowed trajectories, there is a special class, the one which minimizes the proper time needed to go from an event $E_1$ to an event $E_2$. Such trajectory is called a geodesic. 

\begin{principle} A point particle, undergoing no external force follows a geodesic curve.
\end{principle}

By the definition we gave, a geodesic trajectory can be obtained from the Euler-Lagrange equations, starting with the action proportional to the proper time. Hence, 
\be
\mathcal S[q(\lam)] = \int \sqrt{ds^2(\dot q(\lam))} d\lam.
\ee
This action is manifestly reparametrization invariant. However, if in addition we impose the parameter to be the proper time, the equation of motion are equally derived from 
\be
\mathcal S[q(\tau)] = -\frac12 \int ds^2(\dot q(\tau)) d\tau, \label{geodesicaction}
\ee
where the prefactor is chosen in order to find back the Newtonian action for low velocities. For practical purposes, the expression \eqref{geodesicaction} for the action is often more convenient.

\subsubsection{Killing field}
\begin{definition} Let $(\M,\g)$ be a pseudo-Riemannian manifold, and $\varphi$ a diffeomorphism of $\M$. 
\bi
\item If $\varphi^*\g = \g$, then $\varphi$ is an isometry of the metric $\g$.
\item If $\varphi^*\g = \Om^2 \g$, with $\Om$ a non vanishing smooth function on $\M$, then $\varphi$ is a conformal transformation.
\ei
\end{definition}
In that definition, only the first line corresponds to a real symmetry of the space-time. However, the second one is very useful because it preserves the \emph{causal structure} of space-time. Indeed, a null curve or geodesic stays a null curve or geodesic after a conformal transformation. Using this, we can embed many space-times into compact ones, without altering the causal structure and bringing infinity at a finite distance. This is at the heart of the Penrose-Carter diagrams~\cite{Wald}.

\begin{definition} Consider a manifold $\M$ endowed with a metric $\g$. The vector field $K$ is called a Killing field if and only if
\be
L_K \g = 0.
\ee
\end{definition}

If $\phi$ is the flow of $K$, it is equivalent to say that for all $\lam$ small enough, 
\be
\phi_\lam^*\g = \g.
\ee
This means that the local one parameter group is an isometry of $\g$. In other words, when one follows the integral curves of $K$, one sees always the same metric. Killing fields are thus the infinitesimal expression of a \emph{symmetry} of the metric.

\section{General relativity}
\label{RG_Sec}
\subsection{Einstein's equations}
\label{Eequ_Sec}
In general relativity, the geometry is not a static background. It is dynamical. Matter moves along curves in space-time, and space-time geometry is modified by the presence of matter. In fact, the absence of background structure is one of the main features of general relativity. To describe the dynamics of gravity, Einstein built a set of equation under a tensorial form. This ensure the coordinate independence of the dynamics. The idea is to couple the geometry to the energy of matter, given by the stress energy tensor $T_{\mu \nu}$. 

As we said in the introduction, we shall not provide a precise definition of curvature tensors. Instead, we will rapidly present the main ideas that lead to Einstein's equation. Developing further the mathematics of Lorentzian geometry, one can show that there exists a tensor of rank $(1,3)$ that characterizes the geometry, in the sense that it vanishes if and only if the geometry is flat (at least locally, see~\cite{Straumann} for a precise proof). This tensor is called the Riemann tensor $R^\mu_{\ \nu \rho \sigma}$, and is obtained as a nonlinear combination of $g_{\mu \nu}$ and its first two derivatives. With its help, one builds the Einstein tensor as 
\be
G_{\mu \nu} = R_{\mu \nu} - \frac12 R g_{\mu \nu}, \label{Gmunu_def}
\ee
where $R_{\mu \nu} = R^\rho_{\ \mu \rho \nu}$ is the Ricci tensor and $R = R^\mu_{\ \mu}$ the scalar curvature. The $G_{\mu \nu}$ tensor is of rank $(0,2)$. It allows us to define the Einstein equation as 
\be
G_{\mu \nu} = 8\pi G T_{\mu \nu}. \label{Einsteinequ}
\ee
This contains the fundamental idea of general relativity, that matter is coupled to geometry via this equation. It is characterized by several key properties. Firstly it is tensorial and is a second order differential equation for the metric $g_{\mu \nu}$. Secondly, because the specific combination \eqref{Gmunu_def} satisfies the identity $\nabla^\mu G_{\mu \nu} = 0$ (contracted Bianchi identity), Einstein's equation implies the energy conservation 
\be
\nabla^\mu T_{\mu \nu} = 0.
\ee
Finally, the proportionality factor between the geometric tensor and the stress-energy one, $8\pi G$, is chosen in order to recover Newton's theory of gravitation in the limit of static and weak gravitational fields.

\subsection{Black hole space-times}
\label{BHST_Sec}
\subsubsection{Event horizon}
A black hole is a region of space-time where the gravitational field is so strong, that no physical object can escape from it. In order to formulate this mathematically, one needs to define an `outside region', where observers can probe whether or not some region of space-time can emit a signal. Asymptotic infinity will play such a role. Roughly speaking, we define asymptotic infinity by looking at the locus of all causal curves when time goes to infinity. However, this notion is meaningful if space-time is asymptotically flat, that is, it looks like Minkowski at infinity, {\it i.e.}, 
\be
\g_x \underset{x\to \infty}{\longrightarrow} \eta.
\ee
More precisely, infinity is decomposed into 
\bi
\item Time-like future (resp. past) infinity, noted $\imath^+$ (resp. $\imath^-$), is the \emph{future} (resp. \emph{past}) infinity of all time-like curves,
\item Null future (resp. past) infinity, noted $\scri^+$ (resp. $\scri^-$), is the \emph{future} (resp. \emph{past}) infinity of all null curves,
\item Space infinity, noted $\imath^0$, is the infinity of all space-like curves.
\ei
To obtain a detailed construction of these notions, we refer the reader to~\cite{Frauendiener00,Wald}. For the present purpose, the above intuitive definition shall be enough. 

\begin{definition} Let $(\M,\g)$ be a Lorentzian manifold. An event $p_1$ is in the past (resp. future) of an event $p_2$ is there exist a causal curve, future (resp. past) oriented that goes from $p_1$ to $p_2$. 

We call chronological past of a region $S \subset \M$ the set of all points that are in the past of a point in $S$, we note it $I^-(S)$. Similarly, we define the chronological future $I^+(S)$.
\end{definition}

The black hole region of a space-time $\M$ is then defined as 
\be
\mathcal B = \M / I^-(\scri^+).
\ee
By definition, the region inside a black hole is not in causal contact with infinity, this means that asymptotically, one receives no signal from the black hole region. The \emph{event horizon} is then the boundary of the black hole
\be
\mathcal H = \p \mathcal B.
\ee
Beyond the event horizon, no one can ever escape or send a signal to the outside.

It is important to notice that the entire future history of the space-time must be known to define the event horizon. This means that the event horizon is not defined as a \emph{local} concept~\cite{Wald99}. 

\subsubsection{Killing horizon}
When considering a black hole space-time, we say that it is stationary, or `at equilibrium' if the metric is invariant under `time translation'. More precisely, if it possesses a Killing field $K$ which is time-like asymptotically. The Killing field can be used to define another notion of horizon. Indeed, if it is not time-like on the whole space-time, then its norm $K^2 \dot = ds^2(K)$ must vanish somewhere. We define a hypersurface $\Sigma$, such that 
\be
K^2_{| \Sigma} = K_\mu K^\mu_{| \Sigma} = 0.
\ee
Beyond this surface, $K$ becomes space-like. This means that a physical object cannot stay `at rest', because the orbits of $K$ are no longer causal. However, it does not mean that it cannot escape from this region. Around a rotating black hole, there is a region where all observers are dragged and must corotate with the hole. This region is called an \emph{ergoregion}. Such phenomenon occurs for example around a Kerr black hole~\cite{Visser07}. 

For the sake of simplicity, we shall consider only non rotating black holes. Equivalently, we require that the space-time is not only stationary, but also static. For this, the Killing field must also satisfy (everywhere) the so called `Frobenius condition'
\be
K^\flat \wedge dK^\flat = K_{[ \mu} \p_\nu K_{\rho]} = 0,
\ee
where $d$ is the exterior derivative and $\wedge$ is the wedge product of differential forms~\cite{Lafontaine,Straumann,Wald}. In case of spherical symmetry, this condition is automatically fulfilled. 

We now assume that this surface $\Sigma$ coincide with the event horizon $\mathcal H$. By construction, $\mathcal H$ is a null surface. This means that the induced metric $\g_{| \mathcal H}$ is degenerated. Since $K$ is orthogonal to this surface, and $K^\mu K_\mu$ is constant on it, its gradient must be proportional to $K$. The coefficient of proportionality defines what we call $\kappa$, the `surface gravity' of the horizon
\be
\grad(K^2) = -2\kappa K, \label{kappa_def}
\ee
or in local coordinates
\be
\p^\nu(K^\mu K_\mu) = -2 \kappa K^\nu.
\ee
A priori, $\kappa$ depends on the point on $\mathcal H$. One can show, under general asumptions~\cite{Wald}, that it is in fact constant. This constitutes the `zeroth law of black hole thermodynamics'. The surface gravity is a geometric invariant that characterizes a horizon. 

Up to this point, the notion of Killing horizon and surface gravity seems fully local. However, if $K$ is a time-like Killing field, so is $\lam K$ for any real $\lam$. This introduces an ambiguity to the definition that cannot be solved locally. If the space-time is asymptotically flat, then the normalization is fixed by requiring that 
\be
K \to \p_t,
\ee
where $\p_t$ is the asymptotic (Killing) Minkowski time derivative. When this identification is not possible, the Killing field is ambiguously defined. An instructive counter example is found in Minkowski space. One can consider the boost Killing field, which defines a horizon. But its surface gravity cannot be defined universally (with the dimension of a frequency). However, if one considers an accelerated trajectory, and normalizes the boost Killing on it, then the surface gravity will coincide with Unruh temperature, see \Sec{unruheffect_Sec}.

We mention that there exists another definition of horizon as `apparent horizon'. This last definition is local, and does not need stationarity of space-time. However, it depends on which `time slicing' is used. 

Unfortunately, in general, these three notions of horizon do \emph{not} coincide. In~\cite{Wald99}, it is discussed the precise relation between a Killing horizon and an event horizon. Let's also mention the result of Hawking that establishes the equivalence between event and Killing horizons for all \emph{stationary} black holes in the vacuum or electrovacuum in general relativity~\cite{HawkingEllis}.  Moreover, as discussed in~\cite{Poisson}, the apparent horizon and event horizon coincide also when space-time is \emph{stationary}. When non stationary effects are included, these definitions differ. This is the case in particular when a black hole evaporates (see Chapter \ref{HR_Ch}, in \Sec{Observables_Sec}), which makes the geometry slightly non stationary. We refer to~\cite{YorkQG} for a very interesting discussion about this effect during the evaporation process.

\section{Spherically symmetric Black holes}
\label{BHinPG_Sec}
\subsection{Geodesic flow}
\label{PGgeod_Sec}
\subsubsection{Geodesic equation}
In general relativity, there are a few black hole solutions. In fact, when looking for spherically symmetric black holes, the unique solution is the well-known Schwarzschild metric. This result is the Birkhoff theorem. More generally, in $3+1$ dimensions, when solving Einstein's equation coupled to Maxwell, the general stationary black hole solution is fully characterized by its mass $M$, its charge $Q$ and its angular momentum $J$. It is given by the Kerr-Newman metric. Of course, one can obtain much more complicated black hole solutions when considering modified theories of gravity~\cite{Emparan02}, or exotic matter~\cite{Bardoux12}. In this work, for the sake of simplicity, we shall mainly work with non rotating black holes. However, the physics of Hawking radiation (chapter \ref{HR_Ch}) easily generalizes to the general case. We also notice that rotating bodies display interesting physics, and in particular concerning instabilities, as briefly discussed in chapter \ref{laser_Ch}.

In the following, we consider space-times described by the metric
\be
ds^2 = f(r) dt_s^2 - \frac{dr^2}{f(r)} - r^2 d\Om_{S^2}^2. \label{Schw}
\ee
Even though it is not the general spherically symmetric case, it covers a large enough class of metric for our purpose. Among others, it includes  the Schwarzschild, Reissner-Nordström metrics and their de Sitter or Anti de Sitter extensions. A full description of that class of metrics can be found {\it e.g.} in~\cite{Jacobson07}. In the case of a Schwarzschild black hole, the function $f$ is given by
\be
f(r) = 1 - \frac{2GM}r. \label{Schwf}
\ee
However, to keep the discussion general, we let $f$ arbitrary. We only assume that the geometry is asymptotically flat, {\it i.e.}, $f(r) \underset{\infty}{\to} 1$, and contains a horizon at some location $r=r_{\mathcal H}$, where $f(r_{\mathcal H})=0$.

We now want to determine the geodesics of this geometry. From spherical symmetry, we deduce that the motion is planar, in the sense that one can parametrize the sphere $S^2$ by angles $(\theta, \varphi)$ such that the trajectory stays at colatitude $\theta = \frac\pi2$. Moreover, we have a constant of motion $\ell$
\be
\ell = r^2 \dot \varphi.
\ee
This is nothing but Kepler's second law. In addition to the spherical symmetry, the geometry \eqref{Schw} is stationary. Indeed, there is a time-like Killing field given by
\be
K = \partial_{t_s}.
\ee
This generates another conserved quantity 
\be
E = \g(K,\dot x^\mu) = g_{\mu\nu} K^\mu \dot x^\mu.
\ee
We are now left with a $1+1$ dimensional problem with the equation of motion 
\be
\left\{ \bal
&f(r) \dot t_s = E,\\
&\dot{r}^2 + f(r) \left( 1 + \frac{\ell^2}{r^2} \right) = E^2. \label{Schgeod}
\eal \right.
\ee
This is the geodesic equation of a massive particle in the geometry \eqref{Schw}, with $E$ the energy per unit of mass and $\ell$ the angular momentum per unit of mass. When used in the Schwarzschild metric, these equations are essential for the solar system tests of general relativity. 

\subsubsection{Painlevé-Gullstrand coordinate set}

We are interested here in the study of a black hole horizon, that is across $f(r) = 0$, where the coordinates are singular. To build a new set of coordinates, we follow radially in-falling geodesics to use their clocks as a new time variable. Therefore, we consider $\ell = 0$ and $\dot r < 0$. Moreover, we replace the parameter $E$ using the asymptotic value of the velocity $v_\infty = \underset{r \to \infty}{\lim}\frac{dr}{dt_s}$, that is
\be
E^2 = \frac1{1-v_\infty^2}.
\ee
In particular, we assume $E>1$, which means that we consider only trajectories that were at spatial infinity at $t\to -\infty$. 
Following~\cite{Poisson00}, we build the corresponding family of geodesic, solutions of \eq{Schgeod}, as the flow of the vector field
\be
\uf_{v_\infty} = \frac1{f} \p_{t_s} - \sqrt{1 -(1- v_\infty^2)f} \p_r.
\ee
Because the norm of $\uf_{v_\infty}$ is constant, the relation between the parameter and the proper time is simple
\be
d\lam = \sqrt{1-v_\infty^2} d\tau.
\ee
The idea of the Painlevé-Gullstrand coordinate set is to use the proper time of one of this families of geodesic as a new time coordinate. To obtain the simplest expression, we choose a vanishing asymptotic velocity, $v_\infty=0$, and define $\uf = \uf_0$. Its proper time is easily obtained by 
\be
dt_{\rm PG} = \uf^\flat = dt_s + \frac{\sqrt{1 -f}}{f} dr.
\ee
Using this new time coordinate, the metric reads
\be
ds^2 = dt_{\rm PG}^2 - \left( dr + \sqrt{1- f(r)} dt_{\rm PG}\right)^2 - r^2 d\Om_{S^2}^2.
\ee
This metric is cast under the canonical Painlevé-Gullstrand form when we introduce 
\be
v(r) = -\sqrt{1-f(r)}, \label{PGvdef}
\ee
leaving the metric
\be
ds^2 = dt_{\rm PG}^2 - \left( dr - v(r) dt_{\rm PG}\right)^2 - r^2 d\Om_{S^2}^2. \label{PGds3}
\ee
In the sequel, we shall often forget the subscript PG, and work with $t$ being the Painlevé-Gullstrand time. We point out that this metric is still stationary, as $K = \p_{t_s} = \p_t$ is still a Killing field, but it is not reversible. Indeed, it is not invariant under $t \to -t$. This comes from the fact that \eq{PGds3} describes a black hole, which dynamics is irreversible since objects can fall in but not come out. In the language of the full analytical extension of Schwarzschild~\cite{Wald,Straumann}, it only describes the two quadrant of the black hole, {\it i.e.}, only the future horizon, not the past one.

We underline that even though this construction gives the most compact expressions, there is nothing special about the choice of $\uf$. As explained in~\cite{Poisson00}, one could have chosen any of the $\uf_{v_\infty}$ to build a new time coordinate. In Painlevé-Gullstrand coordinates, this family reads
\be
\uf_{v_\infty} = \frac{1+ v \sqrt{1 - (1-v_\infty^2)(1-v^2)}}{1-v^2} \p_t - \sqrt{1 - (1-v_\infty^2)(1-v^2)} \p_r. \label{framefamily}
\ee
The fact that none of the $\uf_{v_\infty}$ plays a privileged role is the translation of the local Lorentz invariance of general relativity, and especially the boost symmetry. In particular, at spatial infinity, we pass from $\uf$ to $\uf_{v_\infty}$ by applying the usual Minkowski boost of parameter $v_\infty$.

Moreover, when taking the limit $v_\infty \to 1$, one obtains a null family of falling in geodesics. The corresponding parameter is no longer a proper time, but is still affine. Using it as a new coordinate, one obtains the well-known Eddington-Finkelstein coordinate set~\cite{Poisson00}.

\subsubsection{Null geodesics}

To understand the causal structure of the geometry, we focus now on null geodesics. Forgetting about the non radial part of \eq{PGds3}, the metric can be written
\be
ds^2 = (1-v^2) \left(dt - \frac{dr}{1+v}\right)\left(dt + \frac{dr}{1-v}\right).
\ee
We now define a pair of null coordinates
\be
\left\{ \bal
d\uc &= dt - \frac{dr}{1+v}, \\
d\vc &= dt + \frac{dr}{1-v}. 
\eal \right. \label{PGlightconecoord}
\ee
In order not to confuse these coordinates with the profile $v$ and the freely falling frame $u$, we note the advanced (resp. retarded) $\vc$ (resp. $\uc$) with an underline. Using them, the metric becomes
\be
ds^2 = (1-v^2) d\uc d\vc.
\ee
Under this form, it is straightforward to see that the null geodesics are simply $\uc = $ cste and $\vc = $ cste. Moreover, because of the form of the metric above, the function $1-v^2$ will often be referred as the conformal factor.

\subsection{Near horizon region}
\label{HRNHR_Sec}
\subsubsection{Surface gravity}

The stationary Killing field $K$ has the same expression in Painlevé-Gullstrand coordinate, {\it i.e.}, 
\be
K = \p_t.
\ee
Its norm is easily obtained
\be
K^2 = 1-v^2(r).
\ee
Therefore, when $v=\pm 1$, there is a Killing horizon. In the case of Schwarzschild (see \eq{Schwf}), we see that there is only one horizon located at $r_{\mathcal H} = 2GM$. Moreover, we can compute its surface gravity in full generality. We note $v' = \p_x v$, and obtain 
\be
\grad(1-v^2) = -2 v^2 v' \p_t + 2vv' (1-v^2) \p_x.
\ee
This is equal to $-2\kappa K$ on the horizon ($1-v^2 = 0$) and gives for the surface gravity 
\be
\kappa = v'_{|\mathcal H}. \label{PGsurfgrav}
\ee
For Schwarzschild, we recover the well known result
\be
\kappa = \frac1{4GM}.
\ee
In particular, $\kappa > 0$. More generally, unless specified otherwise, we assume $\kappa >0$. As we will see in \Sec{WHundul_Sec}, the case $\kappa<0$ corresponds to a white hole horizon, the time reverse of a black hole.

\subsubsection{Light cones}
\begin{figure}[!ht]
\begin{center} 
\includegraphics[scale=0.7]{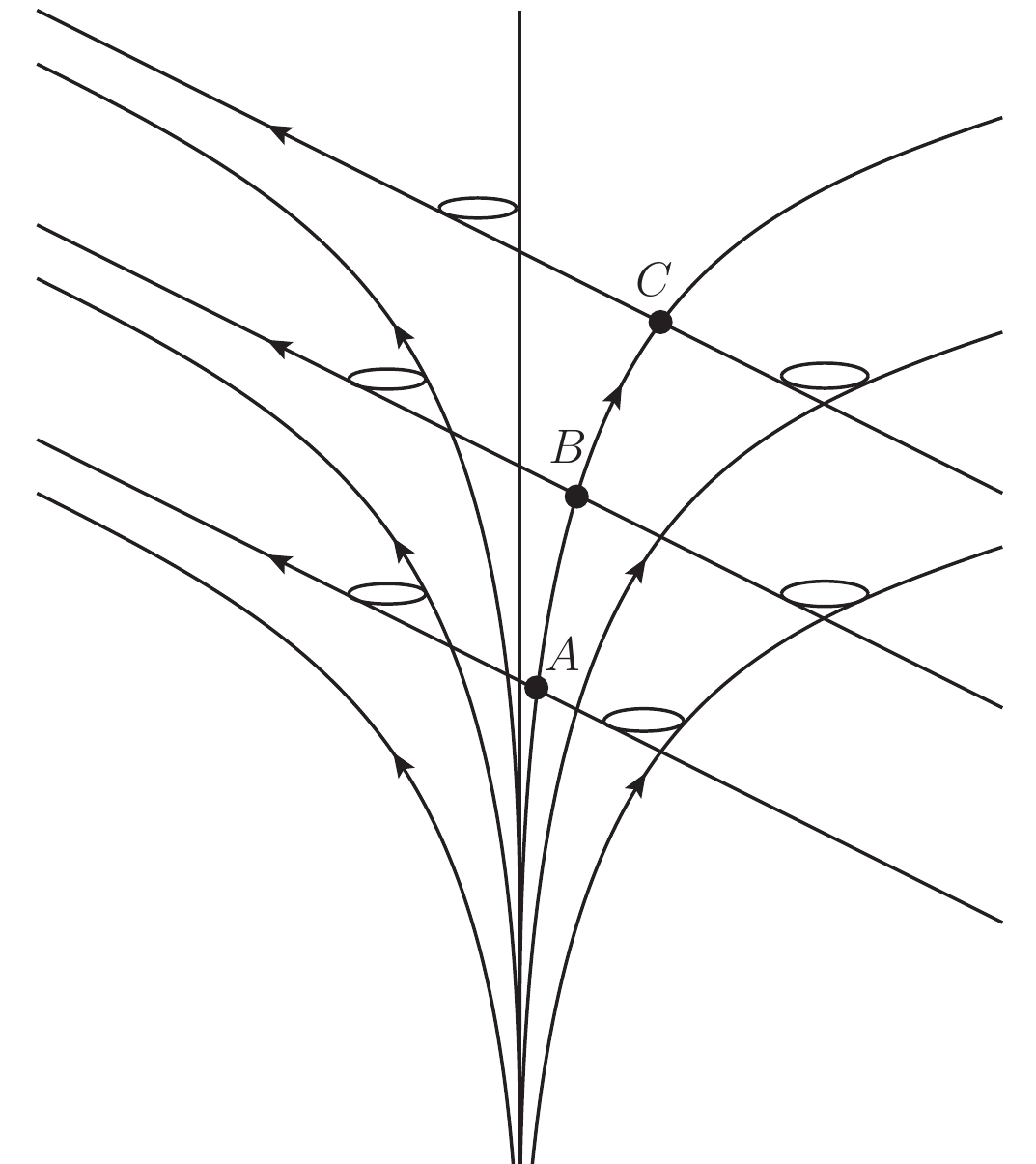}
\end{center}
\caption{Space-time diagram of null geodesics around a horizon. If a family of observers fall in with almost the speed of light, they essentially follow $v = $ cste trajectories. When they cross an outgoing geodesic, in $A$, $B$ and $C$, they can measure the freely falling frequency. Comparing their results at regular time interval will give the redshift law of \eq{Carterkappa}.}
\label{lightconeSTdiag_fig} 
\end{figure}

To focus on the near horizon region, we introduce the coordinate $x = r-r_{\mathcal H}$, so that the horizon is located at $x=0$, where $v=-1$. In its close vicinity, the profile $v$ is approximately linear and reads
\be
v = -1 + \kappa x + O(x^2). \label{vNHRapprox}
\ee
Integrating \eq{PGlightconecoord}, we obtain the trajectories of $u$ and $v$ null geodesics
\be
\left\{ \bal
t^{\rm u} &= \frac{1}{\kappa} \ln|x| + {\rm cste}, \\
t^{\rm v} &= - \frac{x}{2} + {\rm cste}' .
\eal \right. \label{NHRtraj}
\ee
These trajectories are represented in a space-time diagram in Fig.\ref{lightconeSTdiag_fig}. We see explicitly that at $x=0$ there is a change of regime. Indeed, on the right, the $u$ is right moving and the $v$ left moving. But in the left region, both trajectories are left moving. Any physical, {\it i.e.} causal, trajectory must lie inside the light cone, and therefore, on the left region, all trajectories are dragged toward the left. This means in particular that once $x=0$ is crossed, one cannot go back to the region of positive $x$. This is the characteristic behavior near a horizon. 

If one imagines a photon, following a null trajectory, one would like to characterize it not only by its motion, but also by its frequency. In relativity, the frequency is an observer dependent quantity. It is obtained as the time component of the 4-momentum vector, with respect to some observer. We first recall how to build the 4-momentum vector. We start with the Lagrangian of a single particle in an arbitrary geometry obtained from \eq{geodesicaction}, 
\be
\mathcal L = - \frac12 g_{\mu \nu} \dot x^\mu \dot x^\nu.
\ee
The conjugate momentum is then obtained by Legendre transform\footnote{Note that the sign conventions are such that spatial momentum shares the same direction as the spatial velocity.}.
\be
p_\mu = - g_{\mu \nu} \dot x^\nu.
\ee
Using $p_\mu$, we define several notions of frequency, depending on which time is used. The first that could come to mind is by using the Killing field $K$. This gives the Killing frequency $\om = - K^\mu p_\mu$, which is the one measured by static observers, {\it i.e.}, such that $r = $ cste. Because $K$ is a Killing, this frequency is conserved along geodesics. Moreover, asymptotically, it coincides with the usual Minkowski notion of frequency. The main drawback of this frequency, is that it is not a frequency everywhere. Indeed, inside the black hole, $K$ is space-like, and thus $\om$ is a momentum. 

On the other hand, we can also consider frequencies as measured by freely falling observers. For example, following the integral curves of $\uf$, we define $\Om = -\uf^\mu p_\mu$. Unlike $\om$, $\Om$ is not a conserved quantity. When one follows an outgoing geodesic, the freely falling frequency is redshifted. In the near horizon region, it follows the law 
\be
\Om(t) = \Om_0 e^{-\kappa t}. \label{Carterkappa}
\ee
This exponential redshift, and the fact that it is governed by the surface gravity $\kappa$, is a characteristic of Killing horizons. The slightly tricky point is that a single freely falling observer cannot test this law by itself. To obtain it, one should compare the frequencies measured by several freely falling observers, when crossing the same light-like geodesic. On Fig.\ref{lightconeSTdiag_fig}, we represented three crossing points $A$, $B$ and $C$. The $\Om$ measured on each of these by freely falling observers must follow the law \eqref{Carterkappa}.

We conclude this section by noting that the discrepancy between the two inequivalent notions of frequency we looked at, $\om$ and $\Om$, is a key point to understand Hawking radiation. This last statement will be made clearer in \Sec{eternalHR_Sec}.

\subsection{Field propagation around a black hole}
\label{fieldprop_Sec}
\subsubsection{General metric}
We consider here a scalar field $\phi$ propagating freely in a curved space-time $(\M,\g)$. 
\be
S[\phi] = \frac12 \int \left[g^{\mu\nu}\p_\mu\phi \p_\nu\phi - m^2 \phi^2 - \xi R \phi^2\right] \sqrt{-g} d^4x^\mu, 
\ee
where $g = \det{\g}$. By comparing this action to the one in Minkowski space \eqref{MinkAction}, we see that the first term corresponds to the kinetic term, and the second one to the mass term. However, the last one has no equivalent in flat space. It is the only extra term that is diffeomorphism invariant, and of the same dimension as the kinetic term, {\it i.e.} such that $\xi$ is dimensionless. In $3+1$ dimensions, one distinguishes 2 peculiar values of $\xi$
\bi
\item For $\xi = \frac16$, the coupling of the scalar field is \emph{conformal}. This means that any conformal transformation is a symmetry of the equation of motion.
\item For $\xi = 0$, the coupling is called \emph{minimal}.
\ei
In the sequel, we shall focus on the minimal coupling case. Note however that in 1+1 dimensions, then $\xi = 0$ both correspond to minimal and conformal coupling. This property will turn out to be very useful, since we shall often consider 1+1 problems, either to start with, or by symmetry reduction of a 3+1 problem. 

Sticking to the 3+1 dimensional and minimally coupled problem, we derive the equation of motion
\be
\frac1{\sqrt{-g}} \p_\mu \left( g^{\mu \nu} \sqrt{-g} \p_\nu \phi\right) + m^2 \phi = 0. \label{gmunuwavequ}
\ee

\subsubsection{Spherically symmetric black hole}
We apply this to the Painlevé-Gullstrand metric of \eq{PGds3}. The action reads 
\be
S[\phi] = \frac12 \int \left[[(\p_t + v\p_r)\phi]^2 - (\p_r\phi)^2 - \frac{(\p_\theta \phi)^2}{r^2} - \frac{(\p_\varphi \phi)^2}{r^2 \sin^2\theta} - m^2 \phi^2 \right] r^2\sin \theta dt drd\theta d\varphi . 
\ee
This gives the equation of motion
\be
(\p_t + \p_r v)r^2(\p_t + v\p_r)\phi - \p_r(r^2 \p_r \phi) - \hat L^2 \phi + m^2 \phi = 0, \label{EOM31}
\ee
where $\hat L^2$ is the Laplacian of the sphere $S^2$
\be
\hat L^2 = \frac1{\sin \theta} \p_\theta \sin \theta \p_\theta + \frac1{\sin^2 \theta} \p^2_\varphi.
\ee
This operator not only appears naturally in \eq{EOM31}, it also commutes with the differential operator defining \eq{EOM31}. This is the manifestation of the spherical symmetry of the problem. Hence, one can decompose the field into a sum of spherical harmonics as 
\be
\phi = \frac1{\sqrt{4\pi}r} \sum_{\ell, n} \phi_{\ell, n}(r,t) Y_{\ell, n}(\theta, \varphi). \label{Yharmonics}
\ee
Exceptionally, we call $n$ the vertical angular momentum instead of the standard $m$~\cite{Gottfried}, so as not to confuse it with the mass of the field $m$. The reduced equation of motion then reads
\be
(\p_t + \p_r v)(\p_t + v\p_r)\phi_{\ell, n} - \p_r^2\phi_{\ell, n} + \left(\frac{\ell (\ell+1)}{r^2} - \frac{2vv'}{r} + m^2 \right) \phi_{\ell, n} = 0. \label{fullPGwavequ}
\ee
The full equation of motion is quite hard to solve. However, one needs not to solve the general case in order to understand the physics of Hawking radiation. We can neglect the last term, called the `gravitational potential'. To do so, we should first consider modes for which it is minimal. For this reason, we now consider massless $s$-waves, {\it i.e.} $\ell = 0$ and $m=0$. As explained in~\cite{Page76,Page76b}, $s$-waves contribution is about $90\%$ of the Hawking radiation flux. Moreover, for solar mass black holes and above, massive fields barely radiate~\cite{Page77}. For such fields, neglecting the residual potential $- \frac{2vv'}{r}$, the equation of motion reduces to 
\be
(\p_t + \p_r v)(\p_t + v\p_r)\phi_{0, 0} - \p_r^2\phi_{0, 0} = 0. \label{PGwavequ}
\ee
In that case, we drop the indices $(\ell, n)$ and simply name the mode $\phi$. We see that this corresponds to the propagation of a field in the $1+1$ reduction of the Painlevé-Gullstrand metric. In \Sec{Observables_Sec} we shall further discuss the effect of the gravitational potential, and in \Sec{massfields_Sec} the effect of a mass is studied with care. 

\section{Hints of black hole thermodynamics}
\label{BHthermo_Sec}
In general relativity, black holes are perfect absorbers, since nothing can come out from them. However, it is still possible to extract energy from a black hole. A classical example is given by the well-known Penrose process~\cite{Straumann,Wald,Poisson}. Also, when two or more black holes collide and coalesce to form a bigger black hole, a lot of energy is radiated away through gravitational waves. The dynamics of black holes, dictated by Einstein's equation is rather complicated. However, in 1973, Bardeen, Carter and Hawking developed an astonishing result~\cite{Bardeen73}. The mechanics of black holes are governed by four elementary laws, closely analogous to the laws of thermodynamics. Among them, the first two are of primary importance

\bi
\item {\it First law} : for any physical process involving a change on a black hole state, its mass $M$, angular momentum $J$ and area $\mathcal A$ must follow the law 
\be
\underbrace{\delta M - \Om \delta J}_{\textrm{Internal energy}} = \underbrace{\frac{\kappa}{8\pi} \delta \mathcal A}_{\text{heat flow}}. \label{firstlaw}
\ee 
In that formula, $\Om$ is the angular velocity of the horizon, and $\kappa$ is the surface gravity, as defined by \eq{kappa_def}.

\begin{figure}[!h]
\begin{center} 
\includegraphics[scale=0.7]{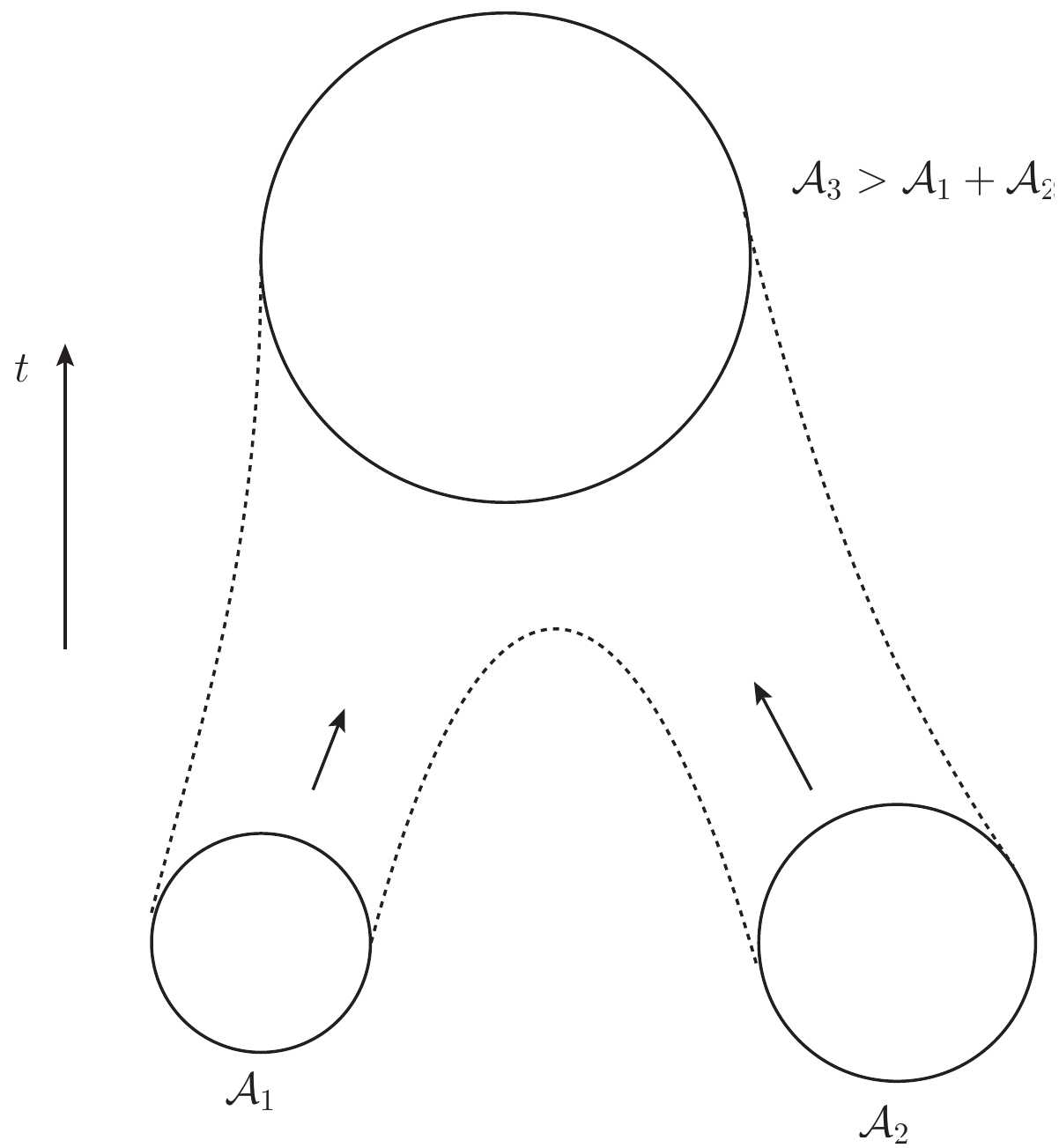}
\end{center}
\caption{Schematic representation of the coalescence of two black holes into a bigger one. A lot of energy is lost by gravitational radiation. But the losses are limited by the second law, which states that the final area $\mathcal A_3$ must be bigger than the initial area $\mathcal A_1 + \mathcal A_2$.}
\label{BHsecondlaw_fig} 
\end{figure}

\item {\it Second law} : during any black hole transformation, we have 
\be
\delta \mathcal A \geqslant 0. \label{secondlaw}
\ee
Historically, motivated by a study of black hole collisions, Hawking showed this inequality (see Fig.\ref{BHsecondlaw_fig}), but without interpreting it as a second principle~\cite{Hawking71}.
\ei

If the mass $M$ clearly constitutes the rest energy of the black hole, the identification of its area with an entropy is much less trivial. In fact, in~\cite{Bardeen73}, the authors refused to see in these laws more than a formal analogy. On the contrary, Bekenstein, who was working in parallel on that matter, claimed that this is more than a naive analogy~\cite{Bekenstein73,Bekenstein74}. He proposed to attribute to all black holes, an entropy $S_{\rm BH}$ proportional to the area, motivated both by Hawking area law and arguments from information theory. Pushing this idea forward, he conjectured a `generalized second law', unifying the thermodynamic and black hole one. When coupling an ordinary physical system to a black hole, the total entropy of the system can only grow, {\it i.e.},
\be
\Delta(S_{\rm BH} + S_{\rm ordinary}) \geqslant 0.
\ee
In this context, the phenomenon of Hawking radiation is a great leap toward the Bekenstein interpretation of black hole entropy. Indeed, if the entropy is proportional to the area, then from \eq{secondlaw}, the temperature of the black hole must be proportional to its surface gravity. In 74, Hawking~\cite{Hawking74} showed, by including quantum effects, that a black hole spontaneously emits a flux of particles, exactly as a black body does at temperature 
\be
T_H = \frac{\kappa}{2\pi}. \label{T_H}
\ee
This computation supports the idea that a black hole is indeed a thermodynamical object. In fact, not only does it support it, it is necessary that a black hole radiates in order not to violate the generalized second law~\cite{Unruh82,Unruh83}. This also fixes the proportionality factor between the entropy and the area of the black hole
\be
S_{\rm BH} = \frac{\mathcal A}{4G}. \label{S_BH}
\ee
Chapter \ref{HR_Ch} is devoted to a detailed description of this phenomenon. \\

Black hole entropy is also one of the main achievements to expect from a quantum theory of gravity. Indeed, one would hope to obtain the entropy expression \eqref{S_BH} from a counting of \emph{microscopic} states. Up to now, there has been interesting results in string theory~\cite{Callan96} or in loop quantum gravity~\cite{Engle09}. We also point out interesting attempts using general relativity in 2+1 dimensions~\cite{Carlip96,Carlip93}, or in the 't Hooft $S$-matrix approach~\cite{tHooft96}. More recently, there has been an even more ambitious proposal, which generalizes the notion of black hole entropy to entropy of any local causal horizons. Gravitation dynamics would then be the thermodynamics equilibrium condition of underlying degrees of freedom~\cite{Jacobson95}. Despite all these approaches, it is fair to say that a complete microscopic understanding of black hole entropy has not been obtained so far.

\chapter{Quantum field theory in curved space-time}
\label{HR_Ch}
\minitoc

\section{Quantum field in flat space}
Quantum field theory was the main achievement of high energy physics during the second half of the twentieth century. It arises from the reconciliation between quantum mechanics and special relativity. One can basically divide this field in two main parts. The first one is the theory of fundamental interactions, where particles collide in an empty Minkowski background. The second one would be the effects of quantum fields coupled with a classical background field. What interests us in the following is the theory of quantum fields in curved space-time background, which obviously fits in the second approach. However, we shall first present the main features of quantum field theory in Minkowski space, which is essential to understand the general case. As we shall see with the Unruh effect, the physics of non inertial systems in Minkowski is already non trivial and highly valuable for the black hole problem.

\label{QFTfst_Sec}
\subsection{Field quantization}
\label{canonicalquantize_Sec}
\subsubsection{Classical phase space}
In this section, we review the canonical quantization of a field in Minkowski space. For the sake of simplicity, we only consider a real scalar field in 1+1 dimensions. The generalization of this to higher dimensions is straightforward. But the 2 dimensional case is not only the simplest, it is also an approximation that arises naturally when considering a highly symmetric problem, as the spherically symmetric black hole of \Sec{BHinPG_Sec}. For higher spins the discussion is similar, but it displays several technicalities that are irrelevant for our purpose. Hence, the dynamics of the field we consider is obtained by the action
\be
\mathcal S[\phi] = \int \mathcal L dxdt = \frac12 \int [(\p_t \phi)^2 - (\p_x \phi)^2 - m^2 \phi^2] dxdt. \label{MinkAction}
\ee
Minimizing $\mathcal S[\phi]$, we derive the equation of motion
\be
(\Obox + m^2 )\phi = 0. \label{Minkwavequ}
\ee
The aim is now to describe this equation in the Hamiltonian formalism. The advantages are multiple. 
Firstly, it reveals the canonical structure, which dictates the quantization procedure. 
Secondly, it recasts the wave equation as a first order in time equation, adapted to the study of the Cauchy problem. Moreover, since we shall mainly consider linear wave equations, we will be able to solve it by exploiting results from spectral theory.

We build the conjugate momentum
\be
\pi = \frac{\delta \mathcal L}{\delta \p_t \phi} = \p_t \phi.
\ee
The phase space consists in the set of pair of fields $(\phi_t(x), \pi_t(x))$. In this section, following the notations of~\cite{Delduc}, we write $\phi_t(x)$ instead of $\phi(t,x)$ to underline the fact that we are propagating a field in $x$ space through time. The phase space is naturally structured with a symplectic 2-form. However, in the context of field equations, we prefer to view it as a pseudo scalar product
\be
(\phi_1, \phi_2)_{\rm KG} = i \int_{-\infty}^{+\infty} (\phi_1^* \pi_2 - \pi_1^* \phi_2) dx \label{MinkPS}.
\ee
This scalar product is non positive definite, and hence phase space possesses only the structure of a Krein space. This structure is equivalently characterized by the equal-time Poisson brackets
\bsub \bea
\{ \phi_t(x), \phi_t(x') \} &=& \{ \pi_t(x), \pi_t(x') \} = 0,\\
\{ \phi_t(x), \pi_t(x') \} &=& \delta (x - x').
\eea \esub
Moreover, the dynamics is encoded into the Hamiltonian. Its density is obtained by a Legendre transform of the Lagrangian density
\be
\mathcal H = \pi \p_t \phi - \mathcal L,
\ee
giving the Hamiltonian functional
\be
H[\phi, \pi] = \frac12 \int_{-\infty}^{+\infty} (\pi^2 + (\p_x \phi)^2 + m^2 \phi^2) dx. \label{MinkHam}
\ee
The evolution equation on phase space now reads
\be
\p_t \bmat \phi \\ \pi \emat = B \cdot \bmat \phi \\ \pi \emat = \bmat 0 & 1 \\ \p_x^2 - m^2 & 0 \emat \cdot \bmat \phi \\ \pi \emat .
\ee
Because the Klein-Gordon scalar product of \eq{MinkPS} is conserved, the operator $iB$ is self-adjoint in phase space. Moreover, the Hamiltonian of \eq{MinkHam} is also conserved. Usually, like in Minkowski space, this Hamiltonian is positive definite. This allows to endow the phase space with a Hilbert structure, and leads to the usual spectral theorems that guarantee the existence of an eigenbasis of normal modes. In the case of \eq{Minkwavequ}, such a decomposition is quite easy to obtain, using Fourier transform. Indeed, solutions of \eq{Minkwavequ} are superpositions of plane waves
\be
\varphi_\om(x) = e^{-i (\om t \pm k_\om x)},
\ee
with
\be
k_\om = \sqrt{\om^2 - m^2}.
\ee
Computing the group velocity
\be
v_g  =(\p_\om k)^{-1},
\ee
we distinguish left moving modes, denoted with a $v$ and right moving modes, denoted with a $u$. Imposing that the field is real, we obtain its decomposition
\be
\phi(t,x) = \int_m^{+\infty} \left[ a_\om^u e^{-i (\om t - k_\om x)} + a_\om^v e^{-i (\om t + k_\om x)} + (a_\om^u)^* e^{i (\om t - k_\om x)} + (a_\om^v)^* e^{i (\om t + k_\om x)} \right] d\om \label{Minkbasis},
\ee
Using this decomposition together with \eq{MinkHam}, we express the Hamiltonian as
\be
H = \int_m^{\infty} \frac{4\pi \om^2}{|\p_\om k|} \left( |a_\om^u|^2 + |a_\om^v|^2 \right) d\om.
\ee
It is also interesting to notice that the Hamiltonian is simply expressed in term of the scalar product \eqref{MinkPS} 
\be
H = \frac12 (\phi | i\p_t \phi)_{\rm KG}.
\ee
This relation rely on the fact that the Hamiltonian is quadratic in the field $\phi$ and its conjugate momentum $\pi$.

\subsubsection{Canonical quantization}
The quantization of a field is obtained by promoting $\phi$ and $\pi$ as operators acting on some Hilbert space $\mathcal F$, which is the set of quantum states of the theory. Their action is specified by the equal-time commutation relations, obtained from the Poisson brackets
\be
i \hbar \{\cdot , \cdot \} \rightarrow [\cdot, \cdot].
\ee
In the following, we adopt the convention that $\hbar = 1$. For the real scalar field, this reads
\be \bal
\left[ \hat \phi(t,x), \hat \phi(t,x') \right] &= \left[ \hat \pi(t,x), \hat \pi(t,x') \right] = 0,\\
\left[ \phi(t,x), \pi(t,x') \right] &= \delta (x - x') \hat I. \label{ETC}
\eal \ee
In the Heisenberg representation, states in $\mathcal F$ do not evolve, but the field operator does, following \eq{Minkwavequ}. Hence, it admits the same decomposition as the classical field, since it obeys the same equation \eqref{Minkwavequ}. Moreover, the real character of the field imposes that $\hat \phi$ be self-adjoint
\be
\hat \phi(t,x) = \int_m^{+\infty} \left[ \tilde a_\om^u e^{-i (\om t - k_\om x)} + \tilde a_\om^v e^{-i (\om t + k_\om x)} + (\tilde a_\om^u)^\dagger e^{i (\om t - k_\om x)} + (\tilde a_\om^v)^\dagger e^{i (\om t + k_\om x)} \right] d\om .
\ee
The only difference with respect to \eq{Minkbasis} is that the coefficients are now operators. What remains to be determined is their action on the space $\mathcal F$. For this, we express them in terms of the field operator
\be
\tilde a_\om^j = \frac{|\p_\om k|}{4\pi \om} (e^{-i(\om t \pm k_\om x)}, \hat \phi)_{\rm KG},
\ee
where we choose the + if $j = v$ and - for $j=u$. This equation gives an explicit expression of $\tilde a$ in terms of $\hat \phi$ and $\hat \pi$
\be
\tilde a_\om^j = \frac{|\p_\om k|}{4\pi \om} i \int e^{i(\om t \pm k_\om x)} \left( \hat \pi - i\om \hat \phi \right) dx.
\ee
We deduce the commutation relation between the $\tilde a$ operators
\be
\left[ \tilde a_\om^j , \tilde a_{\om'}^{j'\dagger} \right] = \frac{4\pi \om}{|\p_\om k|} \delta(\om - \om') \delta_{jj'},
\ee
while the other commutators vanish. Because $\om>0$, $\tilde a$ plays the role of a creation operator, while $\tilde a^\dagger$ the role of a destruction operator. To properly interpret them, it is convenient to normalize their commutation relation. For this, we define the normalized operators $\hat a_\om^j$, by decomposing the field operator as
\be
\hat \phi(t,x) = \int_m^{+\infty} \sqrt{\frac{|\p_\om k|}{4\pi \om}} \left[ \hat a_\om^u e^{-i (\om t - k_\om x)} + \hat a_\om^v e^{-i (\om t + k_\om x)} + (\hat a_\om^u)^\dagger e^{i (\om t - k_\om x)} + (\hat a_\om^v)^\dagger e^{i (\om t + k_\om x)} \right] d\om . \label{phidecomp}
\ee
The new commutation relations now read
\be
\left[ \hat a_\om^j , \hat a_{\om'}^{j'\dagger} \right] = \delta(\om - \om') \delta_{jj'}. \label{ETCa}
\ee
To construct the full space $\mathcal F$, we first define the vacuum state, as the normalized state annihilated by all $\hat a_\om^j$, {\it i.e.}, 
\be
\hat a_\om^j \cdot \vac{} = 0,
\ee
and
\be
\langle 0 \vac{} = 1.
\ee
States containing particles are now built by acting with creation operators on the vacuum. The one particle vector space is thus defined as
\be
\mathcal F_1 = \left\{ \int f(\om) \hat a_\om^{u\dagger} \cdot \vac{} +  \int g(\om) \hat a_\om^{v\dagger} \cdot \vac{} \ / \ f, g \in L^2 \right\}.
\ee
Because we consider bosonic fields, the many particle vector spaces are obtained by symmetric products\footnote{For fermionic fields, it would be an antisymmetric product~\cite{Weinberg1}.} of the one particle space~\cite{Weinberg1}. Therefore, the full Hilbert space\footnote{The bar stands for Cauchy completion~\cite{Delduc}.} is 
\be
\mathcal F = \overline{\text{vect}\{\vac{}\} \oplus \mathcal F_1 \oplus (\mathcal F_1)^{\odot 2} \oplus (\mathcal F_1)^{\odot 3}...} \label{Fockdecomp}
\ee
This decomposition is the Fock representation of the Hilbert space. In terms of $\hat a$ and $\hat a^\dagger$ operators, the Hamiltonian has a quite simple expression  
\be
\hat H = \int_m^{\infty} \om \left( \hat a_\om^{u\dagger} \hat a_\om^u + \hat a_\om^{v\dagger} a_\om^v \right) d\om.
\label{MinkQHam}
\ee
This shows that $\hat a_\om^\dagger$ creates a quanta of energy $\om$. There are several interesting points to underline concerning the Hamiltonian. The first one is that there is an ambiguity when one defines it as an operator, starting from its classical expression \eqref{MinkHam}. Indeed, since $\hat a$ and $\hat a^\dagger$ do not commute, there is a choice of ordering to make. Here, we used the convention of normal ordering. This is the unique choice that ensure a vanishing expectation value in the vacuum $\vev{}{\hat H} = 0$. Normal ordering is specified by putting the expression between double dots, {\it e.g.}, 
\be
\hat H = \frac12 \int_{-\infty}^{+\infty} : \hat \pi^2 + (\p_x \hat \phi)^2 + m^2 \hat \phi^2 : dx . 
\ee
Moreover, using the commutation relation of \eq{ETCa}, we see that any other choice of ordering shifts the Hamiltonian of a multiple of the identity. The normal order procedure equally applies to all observables that are non linear in the field operators, but whose expectation value is supposed to vanish in the vacuum, such as the momentum current.

The second fundamental point about the Hamiltonian is that it is bounded from below, and thus has a ground state. Moreover, this ground state is unique and hence unambiguously defines the vacuum state of the theory.

\subsection{Green functions}
\label{Greenfunctions_Sec}
In quantum field theory, it is possible to describe the field and the quantum state using only complex valued functions. These functions are called Green functions. There exist many different Green functions, and the study of their discrepancies and relations is an interesting but long topic. However, this discussion is crucial for quantum field in curved spaces. Indeed, in arbitrary space-times, we no longer have the Fock decomposition of \eq{Fockdecomp}. Therefore, we will appeal much more to the Green functions in order to analyze the physics taking place. In particular, in \Sec{UnruhDetector_Sec} we show that the two point function encodes the response of a particle detector to a quantum field. This will allow us to analyze the Unruh effect (\Sec{unruheffect_Sec}) and Hawking radiation (\Sec{collapseHR_Sec}) with the help of Green functions. In this section, we focus on the main ingredients characterizing the various Green functions. For a detailed discussion, we send the reader to the literature~\cite{Delduc,BirrellDavies}, and especially~\cite{Fulling}, where this points are treated with care. 

\subsubsection{Classical Green functions}
We start by computing the commutator of fields at arbitrary times. Using the equation of motion for the field operator, we show
\be
(\square +m^2) \left[ \hat \phi(t,x) , \hat \phi(t',x') \right] = \left[ (\square +m^2)\hat \phi(t,x) , \hat \phi(t',x') \right] = 0.
\ee
Moreover, at equal time $t=t'$, this operator is proportional to the identity. Hence, we can propagate it from $t=t'$ to $t\neq t'$ using the equation of motion, and show that it stays proportional to the identity at all time. Another way to see this is to compute it explicitly using the decomposition \eqref{phidecomp} and the commutation relation \eqref{ETCa}. This defines the Green function $G_c$ as
\be
\left[\hat \phi(t,x) , \hat \phi(t',x') \right] = i G_c(t,x;t',x') \hat I.
\ee
Note that calling $G_c$ a Green function is slightly abusive since it satisfies the homogeneous equation of motion, rather than sourced by a Dirac distribution. Using the equal time commutation relations \eqref{ETC} we characterized $G_c$ as the solution of a Cauchy problem 
\be
\left\{ \bal
&(\Obox +m^2) G_c(t,x;t',x') = 0,\\
&G_c(t=t';x,x') = 0,\\
&\p_t G_c(t=t';x,x') = \delta(x-x').
\eal \right.
\ee

\subsubsection{Quantum Green functions}
The commutator Green function, defined in the preceding paragraph, is also obtained as a vacuum expectation value
\be
i G_c(t,x;t',x') = \vev{}{\left[\hat \phi(t,x) , \hat \phi(t',x') \right]}.
\ee
We now follow the opposite path, and define a function as a vacuum expectation value of product of fields. We construct this way the Wightman function 
\be
G_+(t,x;t',x') = \vev{}{\hat \phi(t,x) \hat \phi(t',x')}.
\ee
It is straightforward to see that this Green function is also a solution of the equation of motion \eqref{Minkwavequ}. However, unlike $G_c$, the full Cauchy problem is not easy to formulate. On the other hand, by explicitly computing the expectation value, using the field decomposition \eqref{phidecomp}, we show that
\be
G_+(t,x;t',x') = \int_m^{\infty} \sqrt{\frac{|\p_\om k|}{4\pi \om}} \left[ e^{i k_\om (x-x'))} + e^{-i k_\om (x-x')} \right] e^{-i \om (t-t')}  d\om.
\ee
$G_+$ is not characterized by initial conditions, but by its Fourier content. In particular, it contains only positive frequencies. More precisely, it is the same function as $G_c$, but where we kept only the positive frequencies, {\it i.e.},
\be
\text{if } G_c = \int G_\om e^{-i \om (t-t')}  d\om \quad \text{then} \quad G_+ = \int \Theta(\om) G_\om e^{-i \om (t-t')}  d\om.
\ee
$G_c$ and $G_+$ are members of two fundamentally different classes of Green functions. Indeed, $G_c$ is defined as the vacuum expectation value of an operator that is proportional to the identity. Therefore, it does not depend on the choice of the state in which the expectation value is computed. Moreover, it is a real function, as one can see from the self-adjointness of the field operator. On the other hand, $G_+$ is a complex-valued function, and its definition does depend on the choice of a state. Hence, $G_c$ encodes only the {\bf dynamics}, while $G_+$ also specifies the nature of the {\bf quantum state}. For this reason, the first is referred as a classical Green function, while the second as a quantum Green function.

A characteristic feature of linear field theory, is that any state initially gaussian stays gaussian at all times. This means that the knowledge of $G_+$ is enough to compute all possible observables. For example, all the $n$ point functions are obtained as products of the two-point function $G_+$, through the identity 
\be
\vev{}{\hat \phi(x^\mu_1) \hat \phi(x^\mu_2) ... \hat \phi(x^\mu_{2n})} = \sum_{(p_j,q_j)} \prod_{j = 1}^n G_+(x^\mu_{p_j}; x^\mu_{q_j}),
\ee
where the sum is over all partitions of $\llbracket 1, 2n \rrbracket$ into an unordered set of ordered pairs $(p_j,q_j)$, $p_j > q_j$~\cite{Fulling}. In particular, the ground state of the Hamiltonian \eqref{MinkHam}, the vacuum state, is a gaussian state. But it is not the only one, {\it e.g.}, any squeezed vacuum will also be gaussian~\cite{Leonhardt}. This property is specially relevant for our derivation of the Hawking effect, because we start in a vacuum state and propagate it on a collapsing geometry, which forms a black hole. Therefore, the late time state, even though it is no longer the vacuum, is still gaussian. In table \ref{Greenfunction_tab}, we present the usual zoology of classical and quantum Green functions, and their relation to $G_+$.

\begin{table}[!ht]
\btab{|c|c|c|c|}
\hline
Green function & Vacuum expectation value & Equation of motion & Relation to $G_+$ \\
\hline
Classical &&& \\
\hline
$G_c(t,x;t',x')$\tablespace & $ -i \vev{}{\left[\hat \phi(t,x) , \hat \phi(t',x') \right]}$ & $(\square + m^2)G_{\rm adv} = 0 $ & $2\Im(G_+)$ \\
\hline
$G_{\rm ret}(t,x;t',x')$\tablespace & - & $(\square + m^2)G_{\rm ret} = \delta^{(4)}$ & $2 \Theta(t-t') \Im(G_+)$ \\
\hline
$G_{\rm adv}(t,x;t',x')$\tablespace & - & $(\square + m^2)G_{\rm adv} = \delta^{(4)}$ & $-2\Theta(t'-t) \Im(G_+)$ \\
\hline
$\overline G(t,x;t',x')$\tablespace & - & $(\square + m^2)G_{\rm adv} = \delta^{(4)}$ & $\frac12 (G_{\rm adv} + G_{\rm ret})$ \\
\hline
Quantum &&& \\
\hline
$G_-(t,x;t',x')$\tablespace & $\vev{}{\hat \phi(t',x') \hat \phi(t,x)}$ & $(\square + m^2)G_- = 0 $ & $G_+^*$ \\
\hline
$G^{(1)}(t,x;t',x')$\tablespace & $\vev{}{ \{ \hat \phi(t,x) , \hat \phi(t',x') \}} $ & $(\square + m^2)G_{\rm F} = 0 $ & $ 2\Re(G_+) $ \\
\hline
$G_{\rm F}(t,x;t',x')$\tablespace & $\vev{}{T \hat \phi(t,x) \hat \phi(t',x')}$ & $(\square + m^2)G_{\rm F} = i \delta^{(4)}$ & $\overline G + \frac i2 G^{(1)}$ \\
\hline
\etab
\caption{Two point functions of a free field in flat space.}
\label{Greenfunction_tab}
\end{table}

\section{The Unruh-DeWitt particle detector}
\label{UnruhDetector_Sec}
\subsection{Particle detector model}
\label{detectmodel_Sec}
To describe the physical content of a quantum field in more physical terms, we propose here to study a model of particle detector. This one consists in a system with several internal energy levels, coupled to the field $\hat \phi$. The idea is that whenever it encounters an excitation of the field, a particle, it will click, {\it i.e.}, pass from one energy level to another, exactly like a Geiger counter does when it detects an $\alpha$-particle. The particle detector model not only deepens our understanding of a field in Minkowski, but it is also a convenient tool in curved space, where the Fock decomposition of the Hilbert space and the particle interpretation become fuzzy. This kind of model was first discussed in the context of quantum field theory in curved space-time by Unruh and DeWitt.

The proposed model is inspired from many references~\cite{Unruh76,BirrellDavies,Primer} and has been extensively studied for many purposes, including Unruh effect and Hawking radiation~\cite{Unruh76}, but also decoherence and thermalization dynamics~\cite{Unruh89} (see~\cite{Massar06} and references therein). It consists in a point-like detector, probing the field only along its trajectory $x^\mu(\tau)$. This idealization is convenient but by no means necessary~\cite{Unruh76,Parentani95}. The detector interacts with the field through the interaction Hamiltonian
\be
H_I(\tau) = g\ \hat m(\tau) \hat \phi(x^\mu(\tau)),
\ee
where $\tau$ is the proper time of the detector. We work here in the interaction picture, that is, operators evolve as if the coupling were not there ($g=0$), but states do according to the interacting Hamiltonian~\cite{Gottfried}. The Hilbert space of this bipartite system is simply the tensor product
\be
\mathcal H = \mathcal H_{\rm detect} \otimes \mathcal H_{\phi}.
\ee
When the interaction is absent, the eigenstates of energy $E$ of the detector are noted $|E\rangle$. The aim is now to compute the transition probabilities for the detector to go from an initial eigenstate $|E_i\rangle$ to a final one $|E_f \rangle$, {\it i.e.}, the transition
\be
|E_i\rangle \otimes \vac{} \longrightarrow |E_f\rangle \otimes |\psi_n \rangle. \label{transit}
\ee
The field is assumed to be initially in the vacuum, and $(\psi_n)_{n \in \mathbb N}$ denotes any basis of out coming states for the field. The probability of this transition is given by the well-known Fermi golden rule~\cite{Gottfried}. We propose here to review its derivation, and in particular to make an explicit link with the preceding discussion about Green functions. In the interaction picture, we define $U_I(\tau)$, the evolution operator which brings states from $\tau=0$ to $\tau$. This one is related to the interaction Hamiltonian through 
\be
U_I(\tau) = T\exp \left(i \int_0^\tau H_I(\tau') d\tau' \right).
\ee
At first order in perturbation theory, {\it i.e.} for $g\to 0$, this becomes
\be
U_I(\tau) \simeq \hat I + i \int_0^\tau H_I(\tau') d\tau'.
\ee
Therefore, the probability amplitude of the transition \eqref{transit} reads
\be \bal
A_n(\tau) &= \langle E_f | \otimes \langle \psi_n | U_I(\tau) |E_i\rangle \otimes \vac{} ,\\
&= i g \int_0^\tau \langle E_f | \hat m(\tau) | E_i \rangle \langle \psi_n | \hat \phi(x^\mu(\tau)) \vac{} d\tau.
\eal
\ee
By definition, $\hat m(\tau)$ is the free evolution of $\hat m$, and $|E\rangle$ are eigenstates of the free Hamiltonian, hence
\be
\langle E_f | \hat m(\tau) | E_i \rangle = e^{-i (E_i - E_f) \tau} \langle E_f | \hat m(0) | E_i \rangle.
\ee
The probability to make a transition is the square modulus of $A_n$ 
\be
|A_n(\tau)|^2 = g^2 \iint_0^\tau e^{-i (E_i - E_f) (\tau_2 - \tau_1)} |\langle E_f | \hat m(0) | E_i \rangle|^2 \langle 0 | \hat \phi(x^\mu(\tau_1)) |\psi_n \rangle \langle \psi_n | \hat \phi(x^\mu(\tau_2)) \vac{} d\tau_1 d\tau_2.
\ee
Because we only care about the detector transitions, we trace over all the possible out coming states of the field to obtain the transition probabilities of the detector only 
\be
\mathcal P_{i\rightarrow f}(\tau) = \sum_n |A_n(\tau)|^2.
\ee
Using the identity $\sum_n |\psi_n \rangle \langle \psi_n | = \hat I$, we get
\be
\mathcal P_{i\rightarrow f}(\tau) = g^2 \iint_0^\tau e^{-i (E_i - E_f) (\tau_2 - \tau_1)} |\langle E_f | \hat m(0) | E_i \rangle|^2 \vev{}{\hat \phi(x^\mu(\tau_1)) \hat \phi(x^\mu(\tau_2))} d\tau_1 d\tau_2.
\ee
We see that transition probabilities are governed by the two-point function $G_+$ of the field. The matrix elements of $\hat m(0)$ simply tell how strongly the two levels are coupled to the field. To cast this last formula into a more elegant form, we introduce the interval of time $\Delta \tau = \tau_1 - \tau_2$ and the mean time $T = (\tau_1 + \tau_2)/2$. We estimate the transition probability at late time, $\tau \gg |E_i - E_f|^{-1}$
\be
\mathcal P_{i\rightarrow f}(\tau) = g^2|\langle E_f | \hat m(0) | E_i \rangle|^2  \int_0^\tau \int_{-\infty}^{+\infty} e^{ i (E_i - E_f) \Delta \tau } G_+(x^\mu(\tau_1) ; x^\mu(\tau_2)) d\Delta \tau dT. \label{GoldenRule}
\ee
In several configurations, the trajectory and the state of the field are stationary, {\it i.e.} $G_+$ only depends on the interval of time $\Delta \tau$. In that case, it is more relevant to consider the transition rate, which corresponds to the  probability per unit of proper time, {\it i.e.} 
\be
\dot{\mathcal P}_{i\rightarrow f} = \frac{d \mathcal P_{i\rightarrow f}}{d\tau}. \label{rate}
\ee
This suppresses the integral $\int_0^\tau$ in \eq{GoldenRule} and gives a constant transition rate.

\subsection{Inertial detector}
\label{inertialdetect_Sec}
We suppose here that the detector is inertial. Without loss of generality, we assume it at rest with respect to the $(t,x)$ frame. The trajectory is therefore simply
\be
\left\{ \bal
t(\tau) &= \tau, \\
x(\tau) &= x_0 .
\eal \right.
\ee
Therefore, the transition probability is the inverse Fourier transform of $G_+$ evaluated at $E_i - E_f$. But, by definition of the vacuum state, $G_+$ contains only positive frequencies, therefore 
\be
\int_{-\infty}^{+\infty} e^{ i (E_i - E_f) \Delta \tau } G_+(x^\mu(\tau_1) ; x^\mu(\tau_2)) d\Delta \tau \propto \Theta(E_i - E_f),
\ee
and thus
\be
\dot{\mathcal P}_{i\rightarrow f} \propto \Theta(E_i - E_f).
\ee
This means that if the detector is initially in its ground state, it undergoes no transition, since by definition $E_0 - E_f < 0$. On the other hand, if it is initially excited, it will end up transiting to a lower energy state, by emitting a $\phi$-particle. The transition rate hence gives the life-time of the excited state. This is exactly what one would expect in the vacuum state of the field, and is what occurs for any atomic system coupled with the electromagnetic field. In 1+1 dimension, the Wightman function of a massless scalar field admits a rather simple expression. Hence, we perform the discussed computation explicitly. We use 
\be
G_+(t,x;t',x') = -\frac1{4\pi} \ln \left( (\Delta t - i \epsilon)^2 - \Delta x^2 \right), \label{G1+1}
\ee
with $\Delta t = t-t'$ and $\Delta x = x-x'$. Along a trajectory at rest, this gives 
\be
G_+(x^\mu(\tau_1) ; x^\mu(\tau_2)) = -\frac1{2\pi} \ln( \Delta \tau - i \epsilon),
\ee
giving a transition rate
\be
\dot{\mathcal P}_{i\rightarrow f} = -\frac{g^2}{2\pi} |\langle E_f | \hat m(0) | E_i \rangle|^2 \int_{-\infty}^{+\infty} e^{ i (E_i - E_f) \Delta \tau } \ln( \Delta \tau - i \epsilon) d\Delta \tau.
\ee
We integrate this expression by part, with the usual prescription to set to zero infinitely oscillating functions\footnote{In fact, their contributions give Dirac distributions and derivatives, which have a support on $E_i - E_f = 0$. Since we consider transitions for $E_i \neq E_f$, these contributions are irrelevant.}.
\be
\dot{\mathcal P}_{i\rightarrow f} = -i \frac{g^2}{2\pi (E_i - E_f)} |\langle E_f | \hat m(0) | E_i \rangle|^2 \int_{-\infty}^{+\infty} e^{ i (E_i - E_f) \Delta \tau }\frac1{\Delta \tau - i \epsilon} d\Delta \tau. \label{rateIPP}
\ee
And by residue theorem, we conclude
\be
\dot{\mathcal P}_{i\rightarrow f} = \frac{g^2}{(E_i - E_f)} |\langle E_f | \hat m(0) | E_i \rangle|^2 \Theta(E_i - E_f).
\ee
The essential ingredient to retain from this computation is the consequence of the analytic properties of $G_+$. Indeed, this Green function is a distribution defined as the boundary value of an analytic function, {\it i.e.} this is why the $i\epsilon$ prescription is essential. We see that the stability of the detector's ground state, and hence the vacuum character of the field state, is one to one related to the analytic behavior of $G_+$ in the half plane $\Im(\Delta t) < 0$. This characterization of the vacuum state through analytic properties of the Wightman function is a very general feature in quantum field theory. This is at the origin of the Damour-Ruffini derivation of the Hawking effect in black holes~\cite{Damour76}, a point that will be detailed in \Sec{eternalHR_Sec}. In the next section, we will see that for the Unruh effect, the thermal behavior is encoded by the KMS condition, which is a periodicity in imaginary time. 

\subsection{Unruh effect}
\label{unruheffect_Sec}
\subsubsection{Thermalization of the detector}

\begin{figure}[!ht]
\begin{center} 
\includegraphics[scale=0.7]{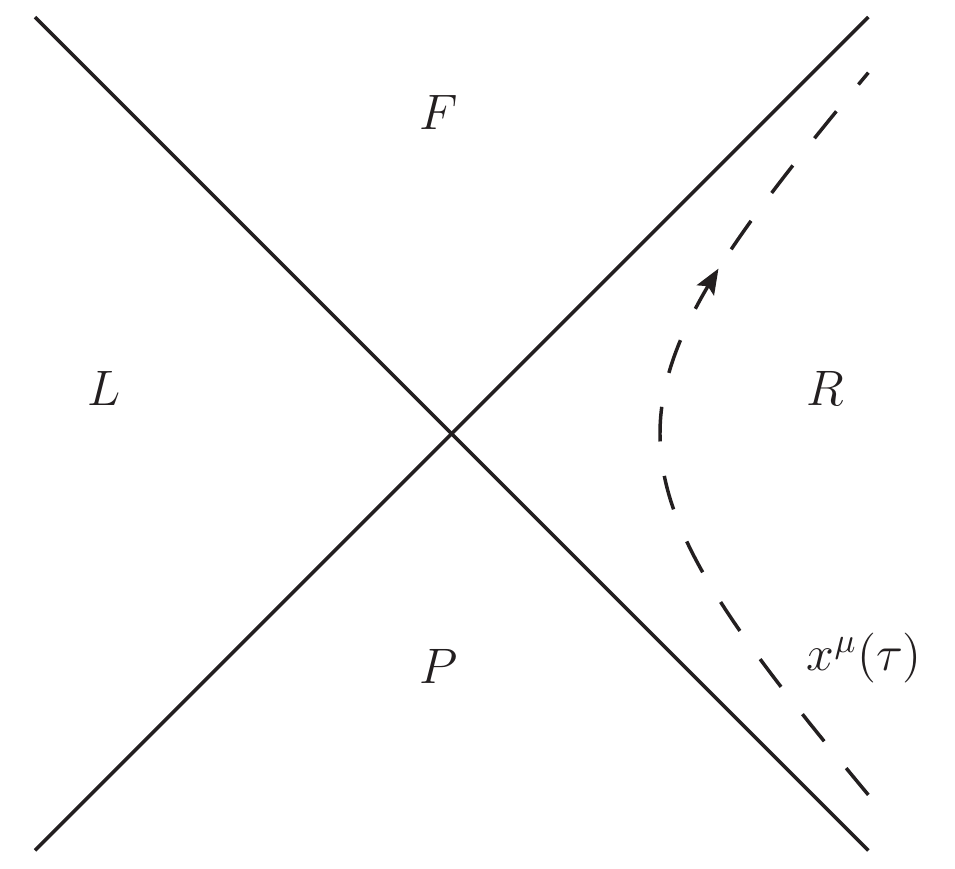}
\end{center}
\caption{Space-time diagram of a uniformly accelerated body. On this trajectory, it is causally disconnected from the region $L$, and cannot send signals to region $P$ or receive from region $F$.}
\label{Rindlerdiag_fig} 
\end{figure}

We now consider a detector moving along a uniformly accelerated trajectory. By a proper choice of time and space origin, this motion reads~\cite{dInverno}
\be
\left\{ \bal
t(\tau) &= \frac1a \sinh(a \tau), \\
x(\tau) &= \frac1a \cosh(a \tau) .
\eal \right. \label{acc_traj}
\ee
In Fig.\ref{Rindlerdiag_fig}, we have represented this trajectory.

From what precedes, and in particular from \eqref{G1+1} the expression of $G_+$, we see that all we need is the proper interval of time along the trajectory.
\be
(\Delta t - i\epsilon)^2 - \Delta x^2 = \frac4{a^2} \sinh^2\left(a (\Delta \tau - i\epsilon)/2\right). \label{trajTau}
\ee
We observe that this interval does not depend on $\tau_1$ and $\tau_2$ separately, but only through the difference $\Delta \tau$. This means that a uniformly accelerated detector perceives the vacuum as a stationary state, as we discussed before \eq{rate}. As pointed out in~\cite{Primer}, the origin of this miracle comes from the fact that this trajectory is the orbit of an element of the Lorentz group, in this case a boost. Inserting this result into the formula for the transition rate, and by performing an integration by part as for \eq{rateIPP}, we show 
\be \bal
\dot{\mathcal P}_{i\rightarrow f} &= -\frac{g^2}{4\pi} |\langle E_f | \hat m(0) | E_i \rangle|^2 \int_{-\infty}^{+\infty} e^{ i (E_i - E_f) \Delta \tau } \ln\left(\sinh^2\left(a (\Delta \tau - i\epsilon)/2\right) \right) d\Delta \tau,\\
&= - i \frac{g^2 a}{4\pi} |\langle E_f | \hat m(0) | E_i \rangle|^2 \int_{-\infty}^{+\infty} \frac{e^{ i (E_i - E_f) \Delta \tau}}{E_i - E_f} \frac{\cosh\left(a (\Delta \tau)/2\right)}{\sinh\left(a (\Delta \tau - i\epsilon)/2\right)} d\Delta \tau . 
\eal \label{rateUnruh} \ee
This integral can be evaluated using a residue theorem. The integrand possesses an infinite sequence of poles for $\Delta \tau = 2in \pi/a$ $(n\in \mathbb Z)$. Since they lie both on the upper and lower half plane, the result will be non zero for both signs of $E_i - E_f$. In particular, the ground state will have a probability to be spontaneously excited. To understand the physics, the full evaluation of \eqref{rateUnruh} is not necessary. What matters is the ratio between excitation and desexcitation transition rates, since they will determine the equilibrium state of the detector. Let's consider the ground state of energy $E_0$, and an excited state of energy $E_e > E_0$. The rate of excitation is given by 
\be
\dot{\mathcal P}_{g \rightarrow e} = \frac{g^2}{E_e - E_0} |\langle E_f | \hat m(0) | E_i \rangle|^2 \sum_{n = 1}^{\infty} e^{ - \frac{2\pi (E_e - E_0) n }a} ,
\ee
while the desexcitation rate reads
\be
\dot{\mathcal P}_{e \rightarrow g} = \frac{g^2}{E_e - E_0} |\langle E_f | \hat m(0) | E_i \rangle|^2 \sum_{n = 0}^{\infty} e^{ - \frac{2\pi (E_e - E_0) n }a} .
\ee
The only difference being that only the second one picks up the pole at $\Delta \tau = i \epsilon$. Hence, the ratio is exactly a Boltzmann factor
\be
\frac{\dot{\mathcal P}_{g \rightarrow e}}{\dot{\mathcal P}_{e \rightarrow g}} = e^{- \frac{2\pi (E_e - E_0)}a}.
\ee
As understood by Einstein in the context of interaction between light and matter~\cite{Einstein17}\footnote{On the subject, we also point out reference~\cite{TerHaar}, where an english translation of Einstein's papers about quantum theory can be found.}, this means that after a short time, the detector will be in a thermal state at the temperature
\be
k_B T = \frac{a}{2\pi}.
\ee
This is what is called the Unruh effect: a uniformly accelerating detector perceives the Minkowski vacuum as a thermal bath. By reading carefully the proof we just gave, we see that the main ingredient is the fact that $G_+(x^\mu(\tau_1) ; x^\mu(\tau_2))$ is periodic in the imaginary time $\Delta \tau$. This is the essence of the KMS condition. Moreover, this periodicity is already visible at the level of the trajectory, {\it i.e.}, in \eq{trajTau}. Therefore, sufficient ingredients are the Lorentz invariance of the field, which implies that $G_+$ will be a function of $(\Delta t - i\epsilon)^2 - \Delta x^2$ and the structure of the trajectory of the detector, {\it i.e.}, the orbit of a boost. Hence, Poincaré invariance is at the heart of this effect.

\subsubsection{The Rindler horizon}
As we saw on Fig.\ref{Rindlerdiag_fig}, a uniformly accelerated detector sees both a past and a future horizon. We focus here on the future horizon, but the discussion is the same for the past one. The future horizon shares many features with that of a black hole, as described in \Sec{BHST_Sec}. This horizon can be define almost as an event horizon, as the boundary of the causal past of the trajectory of the detector. Let's call $\mathcal T$ the trajectory given by \eq{acc_traj}, the future horizon is then 
\be
\mathcal H_F = \p I^-(\mathcal T),
\ee
where the index $F$ stands for `future'. More interestingly, this horizon is also a Killing horizon. For this, we consider the Killing field associated with a boost symmetry of Minkowski space, namely
\be
\tilde B = x \p_t + t \p_x.
\ee
The trajectory of \eq{acc_traj} is an orbit of that vector field. Moreover, we compute its norm
\be
\tilde B^2 = x^2 - t^2,
\ee
which vanishes exactly on the horizon (both past and future in fact). As discussed in \Sec{BHST_Sec}, in order to compute the surface gravity, we must first normalize the Killing field $\tilde B$. However, unlike in a black hole geometry, there is \emph{no} universal choice for its normalization, since no equivalent of $\scri^+$ is available. What one should do instead, is normalize $\tilde B$ on the trajectory of the detector. This gives
\be
B = a (x \p_t + t\p_x),
\ee
meaning that $B = \p_\tau$, the derivative with respect to the proper time of the detector. In that case, the surface gravity of the Rindler horizon is $\kappa = a$. Therefore, we see that the Unruh temperature follows the Hawking temperature announced in \eq{T_H},
\be
T_U = \frac{\kappa}{2\pi} = \frac{a}{2\pi}.
\ee
This shows that the phenomenon of Unruh effect can be attributed to the Rindler horizon. In fact, this statement can be made more explicit. Indeed, the field operator admits a decomposition $\hat \phi = \hat \phi_L + \hat \phi_R$, where $\hat \phi_L$ contains the degrees of freedom living on the Left quadrant, and $\hat \phi_L$ those living on the Right quadrant~\cite{Unruh76} (see Fig.\ref{Rindlerdiag_fig}). Thus, we can express the Minkowski vacuum as an entangled state between two Rindler wedges. Because an accelerating observer only has access to one, the perceived state is the trace over the second wedge, which turns out to be exactly a thermal state~\cite{WaldQ}. However, we insist on the fact that the Rindler horizon, and its surface gravity, are \emph{observer dependent}. An inertial detector in Minkowski space sees no horizon, as it is well-known. This is consistent with the fact that Unruh effect is \emph{locally} identical to the Hawking effect, as we shall explain in \Sec{collapseHR_Sec}. Only when considering the \emph{global} geometry, and in particular the asymptotic regions as $\scri^+$, one can tell whether the horizon is a Rindler one or a black hole one.

\section{Hawking radiation}
\label{HRrelat_Sec}

Classically, as we explained in \Sec{BHST_Sec}, no signal can come out from a black hole. Therefore, it was a huge surprise when Hawking, including quantum effects, predicted that black holes should emit a thermal flux of particles, just like a black body~\cite{Hawking74,Hawking75}. In this section, we shall derive the Hawking result in details. Understanding the features and implications of Hawking radiation is essential to motivate our work presented in the next chapters. Following the historical discovery, we shall first present the computation from the formation of the black hole in \Sec{collapseHR_Sec}. In \Sec{eternalHR_Sec}, we will show that the dynamics of the formation is in fact irrelevant, and the Hawking effect can be obtained by considering only the stationary geometry of a black hole. In \Sec{Observables_Sec}, we discuss a few important features of the manifestation of Hawking radiation. The discussions along this chapter are mainly based on references~\cite{Unruh76,Primer,Massar96,Macherthesis}.

\subsection{A collapsing model}
\label{collapseHR_Sec} 

In the universe, a black hole arises as the final state of a heavy star, which collapses on itself due to strong gravitational interactions. When nuclear reactions terminate inside a star, the gravitational force makes the star contract. If the residual pressure is high enough, the star reaches an equilibrium as a white dwarf or a neutron star. On the other hand, if the pressure is lower that the gravitational forces, as it is for heavy stars, the collapse continues until the star fully disappears inside a black hole~\cite{Straumann}. The dynamic of a collapsing star is a very complicated phenomenon, but as far as Hawking radiation is concerned, we don't need to consider sophisticated models. Hence, we shall mainly focus on a very simple toy model, where the computations can be carried out without much difficulties. In the end of this section, we consider a more general model of collapsing star, and establish that the conclusions drawn in the toy model stay valid. This toy model was first considered by Unruh~\cite{Unruh76}, in a long paper were it was presented together with the Unruh effect and the understanding of Hawking radiation for eternal black holes (see \Sec{eternalHR_Sec}). \\

\begin{figure}[!ht]
\begin{center} 
\includegraphics[scale=0.7]{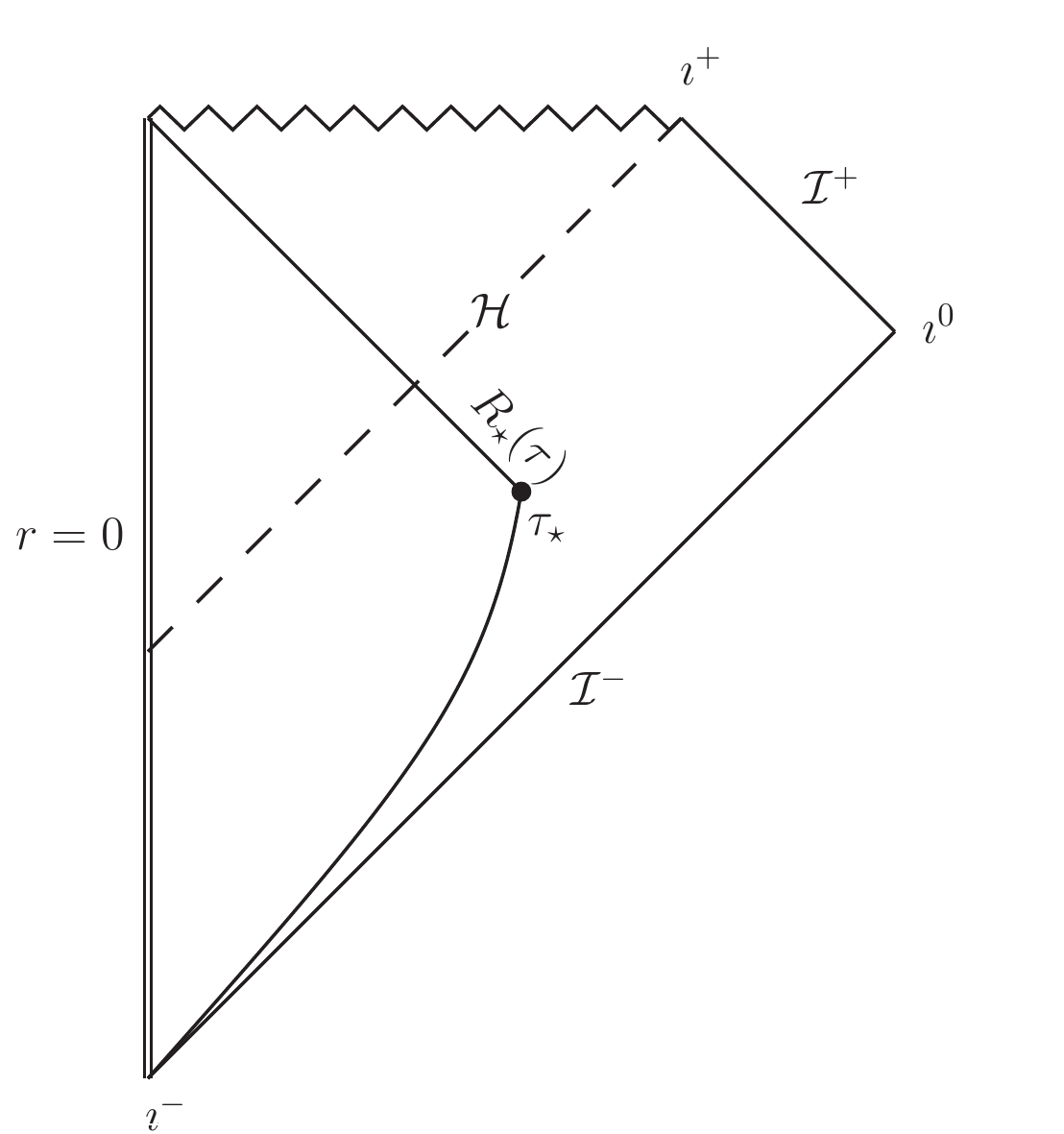}
\end{center}
\caption{Penrose diagram of a collapsing star.}
\label{collapse_fig} 
\end{figure}

The collapsing star is modeled by a spherical shell of matter, falling in very quickly, at almost the speed of light. Inside the star, space-time is empty and flat, outside, its metric is the \Sch one. In addition, we consider the geometry to be reduced to a 1+1 dimensional one. As explained in \Sec{fieldprop_Sec}, even though it is not the most realistic case, it is enough to capture the essential physics for Hawking radiation. The only difference with a pure 1+1 dimensional case is that, in order to mimic the potential barrier of a 3+1 model, we shall impose a boundary condition for the radiating field, $\phi(r=0) = 0$. This ensures that the 3+1 field of \eq{Yharmonics} is regular at $r=0$. In this model, the collapsing metric reads
\be
ds^2 = \left\{ 
\bal &d\tau^2 - dr^2, \quad r < R_\star(\tau),\\
&\left(1- \frac{2GM}{r} \right)dt^2 - \frac{dr^2}{1- \frac{2GM}r}, \quad r>R_\star(\tau).
\eal \right. \label{Unruhcollapse}
\ee
$R_\star$ stands for the radius of the shell. Its equation of motion is assumed to be
\be
R_\star(\tau) = \left\{ 
\bal &R_0, \quad \tau < \tau_\star,\\
&R_0 - v_\star \tau \quad \tau > \tau_\star. 
\eal \right. \label{shelltraj}
\ee
We assume that $v_\star \simeq 1$, so that after $\tau_\star$, the surface of the star follows a time-like trajectory, which is almost light-like. Moreover, we assume that the star is not too compact, {\it i.e.}, $R_0\gg 2GM$, thus before $\tau_\star$, space-time is essentially flat. In Fig.\ref{collapse_fig}, we show the Penrose diagram of our model.  
The coordinate $r$ is the same both inside and outside. This can be made unambiguously because our model represent the reduction of a 3+1 spherically symmetric. $r$ is thus defined so that $4\pi r^2$ is the area of $SO(3)$ orbits~\cite{Straumann,Primer}. This intrinsic and covariant definition for $r$ ensures that we can use the same inside and outside the star. On the other hand, the time coordinate is a priori different in both regions. To relate them, we assume that the induced metric on the surface of the star is the same on both sides. In other words, we assume that the proper time measured on the surface of the star is the same when computed inside or outside. In the language of \Sec{Cotangent_Sec}, this means  
\be
R_\star^*ds^2_- = R_\star^*ds^2_+. \label{firstJC}
\ee
In our case, it gives
\be
\frac{dt}{d\tau} = \frac1{1 - \frac{2GM}{R_\star(\tau)}} + O(1-v_\star). \label{junction1}
\ee
\eq{firstJC} is called the first junction condition. It ensures that the geometry is regular across the star's surface~\cite{Poisson}. The second junction condition would give the stress tensor of the shell as the discontinuity of the derivatives of the metric. However, in the present case, we are not interested in the properties of the matter constituting the collapsing shell, we only need to assume that it follows a quasi light-like trajectory. 
On top of this collapsing geometry, we consider a scalar field initially in the vacuum state. The aim is to compute the late time state of this field, after the star has collapsed into a black hole. Because we work in 1+1 dimension, the scalar field equation of motion is straightforward to solve. Indeed, since it is conformally invariant, and all geometries are conformally flat, the wave equation always reduces to 
\be
\p_{\uc} \p_{\vc} \phi = 0,
\ee
where $\uc$, $\vc$ are light-cone coordinates, {\it i.e.}, such that $ds^2 = \Om^2 d\uc d\vc$. In the present case, we write the geometry under the following form 
\be
ds^2 = \left\{ 
\bal &dUd\vc, \qquad (\text{inside}),\\
&\left(1- \frac{2GM}{r} \right)d\uc d\vc, \quad (\text{outside}).
\eal \right. \label{UnruhcollapseUV}
\ee
These new coordinates are related to the old ones by 
\be \left\{ \bal
U &= \tau - r +4GM - 2R_0, \\
\uc &= t - r - 2GM \ln(r-2GM), \\
\vc &= \tau + r - R_0, \qquad (\text{inside}), \\
\vc &= t + r + 2GM \ln(r-2GM), \qquad (\text{outside}). 
\eal \right. \label{coordUVSch} \ee
The various origins are chosen such that the star surface is at $\vc=0$, and the horizon at $U=0$. The continuity equation across the star's surface, or \eq{junction1}, gives the link between $U$ and $\uc$. Before the collapse, for $\tau < \tau_\star$, we have
\be
U = \left(1- \frac{2GM}{R_0}\right) \uc \simeq \uc, \label{Uu-}
\ee
while for $\tau >\tau_\star$, we obtain 
\be
\uc = f(U) = U - 4GM \ln\left(- \frac U2\right) \label{Uu+}
\ee
The coefficient in front of the logarithm is nothing else than the surface gravity $\kappa = 1/4GM$. This is not a coincidence. It comes from the redshift factor near the horizon of a black hole, which is universally given by $e^{-\kappa t}$, as explained in \Sec{HRNHR_Sec}. The appearance of $\kappa$ will be made even more explicit in the end of the section, when studying general black holes. Using the functional inverse $f^{-1}$ of $f$, we obtain the general solution for $\phi$
\be
\phi(\uc,\vc) = \left\{ 
\bal &F(U) + G(\vc), \qquad (\text{inside}),\\
& F( f^{-1}(\uc)) + G(\vc), \quad (\text{outside}).
\eal \right. 
\ee
Imposing the boundary condition $\phi(r=0) =0$, it reduces to 
\be
\phi(u,v) = \left\{ 
\bal &F(U) - F(\vc), \qquad (\text{inside}),\\
& F( f^{-1}(\uc)) - F(\vc), \quad (\text{outside}).
\eal \right.
\ee
In second quantization, the field operator obeys the same equation. Thus the general solution is equally valid, except that now $F$ is an operatorial function $\hat F$. In order to compare it to the well understood Minkowski case, we develop $\hat F$ in Fourier modes, and use the real character of $\hat \phi$ to relate positive and negative frequencies. This gives
\be
\hat \phi(\uc,\vc) = \int_0^{+\infty} \left[ \hat a_\om \left( e^{-i\om \vc} - e^{-i\om f^{-1}(\uc)}\right) + \hat a_\om^\dagger \left( e^{i\om \vc} - e^{i\om f^{-1}(\uc)}\right)\right] \frac{d\om}{\sqrt{4\pi \om}}. \label{phidecompHRin}
\ee
At early times, that is for $t\to -\infty$ , space-time looks very much like Minkowski space. From Eqs. \eqref{coordUVSch} and \eqref{Uu-}, we see that in this limit, $\uc \to -\infty$, and $U \sim \uc$. Hence, the preceding equation reduces to its flat space expression of \eq{phidecomp} and $\hat a_\om$ and $\hat a_\om^\dagger$ have the usual particle interpretation. Because they describe the {\it in}-going state, we refer to them as $\hat a_\om^{\rm in}$ and $\hat a_\om^{{\rm in} \dagger}$. They are the creation and annihilation operators in the Fock state of quanta defined on $\scri^-$, {\it i.e.}, for $t\to -\infty$ (see Fig.\ref{collapse_fig}). To impose that the field state is initially the vacuum, we define 
\be
\hat a_\om^{\rm in} \vac{\rm in} = 0, \label{invacdef}
\ee
meaning that $\vac{\rm in}$, the {\it in}-vacuum, contains no particle at early time. In particular, before the collapse and the formation of the black hole the field is in its ground state. We now look at the late time state of the field. When $t\to +\infty$, we have
\be
f^{-1}(\uc) \sim -2 e^{- \kappa \uc}. \label{latetimef}
\ee
Because this relation is non linear, the decomposition of $\hat \phi$ is not the canonical one for the asymptotic region $t\to +\infty$, in other words $\hat a_\om^{\rm in} \neq \hat a_\om^{\rm out}$. To understand the nature of this late time state, we shall use different methods. In the next paragraph, we compute the two-point function, and see the response of an Unruh detector to it. This presents the interest of introducing no extra formalism, but also to analyze what an actual, local observer would see. In a second time, we will decompose the field in terms of late time, asymptotic Minkowski particles. 

Using Eqs. \eqref{phidecompHRin} and \eqref{invacdef}, the two-point function is fairly simple to calculate. For all $\uc, \vc, \uc', \vc'$, one finds 
\be
G_+(\uc,\vc;\uc',\vc') = -\frac1{4\pi} \ln(\vc -\vc' -i\epsilon) + \frac1{4\pi} \ln\left( f^{-1}(\uc) - f^{-1}(\uc') - i\epsilon \right). \label{G+Unruh}
\ee
At late times, this becomes
\be
G_+(\uc,\vc;\uc',\vc') = -\frac1{4\pi} \ln |\vc -\vc' -i\epsilon| + \frac1{4\pi} \ln \left|\sinh\left( \frac{\kappa}2(\uc-\uc') - i\epsilon \right)\right| + \frac1{4\pi} \ln \left( e^{-\frac{\kappa}2(\uc+\uc')} \right) .
\ee
Using the results of \Sec{UnruhDetector_Sec}, this function is simple to interpret. We first notice that the last term will play no role. Indeed, as we discussed before \eq{rateIPP}, it contributes to the two point function as a polynomial, which has no impact on the transition amplitudes of the detector. The first term indicates that $v$-modes, {\it i.e.}, those falling toward the black hole, are in their vacuum state. On the other hand, the second term is periodic in $(\uc-\uc')_{\mathbb C}$ of imaginary period $2i \pi /\kappa$. This means that an (intertial) particle detector will perceive the state of $u$-modes as a thermal state at temperature 
\be
T_H = \frac{\kappa}{2\pi}. \label{TH}
\ee
This is the well-known result established by Hawking in 74~\cite{Hawking74}: a black hole spontaneously emits a thermal flux of outgoing particles at the Hawking temperature $T_H$. 
More precisely, one could imagine coupling the detector only to $u$-modes, {\it e.g.}, by using the Hamiltonian
\be
H_I(\tau) = g\ \hat m(\tau) \p_{\uc} \hat \phi(x^\mu(\tau)).
\ee
In that case, the detector will thermalize with $u$-modes and end up at equilibrium at temperature $T_H$. Of course, such a coupling is somehow artificial, since it is not Lorentz invariant, but it sheds some light on the discrepency between $u$ and $v$-modes. On the other hand, if it couples to all modes, it will not reach the temperature, since the probability of desexcitation is increased (with respect to the thermal case) by the chance of emitting a $v$-mode. 

One can also imagine sending a detector into the black hole. What happens if it falls freely across the horizon ? To see this, we deduce from \eq{coordUVSch} that the detector crosses the horizon for $\uc \to +\infty$, and $\vc>0$. In a small neighborhood of the horizon, $\vc \sim \vc_0$ barely varies, and the metric felt by the detector is approximately 
\be \bal
ds^2 &\simeq 2\kappa e^{\kappa \vc_0} e^{-\kappa \uc}d\uc d\vc, \\
&\simeq e^{\kappa \vc_0} dUd\vc. \label{ds2NHR_detect}
\eal \ee
Therefore, $\vc$ and $U = f^{-1}(\uc)$ are the cartesian light-cone coordinates of the local Minkowski patch, which means that the detector sees a two point function $G_+$ looking like 
\be
G_+(\uc,\vc;\uc',\vc') = -\frac1{4\pi} \ln(\vc -\vc' -i\epsilon) + \frac1{4\pi} \ln\left(U-U' - i\epsilon \right). \label{G+NHRFF}
\ee
In particular, the state of the field, as seen by the detector is the vacuum. Indeed, by applying the computation of \Sec{UnruhDetector_Sec} we show that it will not click while crossing the horizon. To be exact, this is only true for high enough energies, {\it i.e.}, in the ultraviolet sector. Indeed, the internal energy $\Delta E = E_e - E_g$ of the detector must be large enough so that during the time necessary to click $\Delta \tau \sim \Delta E^{-1}$, the detector stays in a close vicinity of the horizon, {\it i.e.}, the metric is well approximated by \eq{ds2NHR_detect}. Explicitly, this requires $\Delta E \gg \kappa$. 

As a last {\it gedanken experiment}, we consider the detector in the near horizon region, but accelerated such that it doesn't fall into the black hole, but stays at constant $r$. Such a detector will perceive the same metric as \eq{ds2NHR_detect}, but its proper time will be proportional to $\uc$ rather than $U$. More precisely, the interval of proper time reads
\be
\Delta \tau = \sqrt{2\kappa (r - 2GM)} \Delta \uc.
\ee
Therefore, for the same reason as the asymptotic one, the detector will see a thermal bath, but at a different temperature 
\be
T(r) = \frac{T_H}{\sqrt{2\kappa (r - 2GM)}}.
\ee
This is not surprising, when noticing that the ratio $T/T_H$ is nothing else than the gravitational redshift $(\sqrt{g_{00}})^{-1}$. However, it becomes more interesting when we compute the proper acceleration $a$ needed for the detector to stay on a trajectory $r = $ cste. Indeed, in the near horizon region, one obtains 
\be
a = \sqrt{-a^\mu a_\mu} \sim \frac{\kappa}{\sqrt{2\kappa (r - 2GM)}} = 2\pi T(r).
\ee
This means that a detector accelerating close to the horizon sees a thermal state at exactly its Unruh temperature. 

With the preceding analysis, we have shown that the late time state of the field is the local vacuum on the (future) horizon. It is the state that resembles the most of Minkowski vacuum in the near horizon region. In particular, locally, the Hawking effect is indistinguishable from the Unruh effect, as mentioned in Sec.\ref{unruheffect_Sec}. This is a manifestation of the equivalence principle. Near the horizon, one cannot say if one sees a Rindler horizon, or that of a black hole. It is only asymptotically, that the emitted particles become on shell, and are detected as in a thermal bath by all inertial observers. This was first understood by Unruh~\cite{Unruh76}, and hence, this particular state of the field, described by \eq{G+Unruh}, is called the \emph{Unruh vacuum}. It is the final state reached by the field after a black hole has formed. In Table \ref{HRdetector_tab}, we summarize how various detectors perceive the Unruh vacuum, depending on their trajectories. 

\begin{table}[!ht]
\centering
\btab{|c|c|c|} 
\hline
location & acceleration & Temperature \\
\hline
$\imath^+$ & 0 & $T_H$ \\
\hline
$\mathcal H$ & 0 & 0 \\
\hline
$\mathcal H$ & $\kappa/\sqrt{g_{00}}$ & $T_H/\sqrt{g_{00}}$ \\
\hline
\etab
\caption{Detector's response when coupled to a field in the Unruh vacuum.}
\label{HRdetector_tab}
\end{table}

There exists another stationary state, where both $u$-modes and $v$-modes are seen as being in a thermal bath by an asymptotic detector. This state is the \emph{Hartle-Hawking vacuum}~\cite{Hartle76}. It is easily obtained by coupling $u$ and $v$ modes, initially in the Unruh vacuum, so that the $v$-modes thermalize and end up at the same temperature. The easiest way to do that would be to put the black hole in a box surrounded by perfect mirrors~\cite{BirrellDavies}. In such a state, a detector will this time exactly thermalize at the Hawking temperature $T_H$. This state is particularly relevant when considering the analytic extension of the \Sch black hole. It is the only state which is regular everywhere, in particular on both the past and future horizons. On the contrary, the Unruh vacuum is regular on the future horizon, but singular on the past one. However, in astrophysics, the Unruh vacuum is the most natural, because a past horizon is never formed after a collapse, and as we saw, the field's state relaxes toward the Unruh vacuum. 

\subsubsection{{\it in} and {\it out} basis}
\label{inout_Sec}

There is an alternative way to analyze the final state of the field. One can try to decompose the field operator in order to recognize {\it out}-modes, that is a decomposition analogous to \eq{phidecompHRin} but where modes are plane waves with respect to asymptotic late time observers. Asymptotically, $\uc$ and $\vc$ are cartesian Minkowski coordinates, therefore such a decomposition would read
\be
\hat \phi(\uc,\vc) = \int_0^{+\infty} \left[ \hat a_\lam^{\rm out} \left( e^{-i\lam \vc} - e^{-i\lam \uc}\right) + \hat a_\lam^{{\rm out} \dagger} \left( e^{i\lam \vc} - e^{i\lam \uc}\right)\right] \frac{d\lam}{\sqrt{4\pi \lam}}. \label{phidecompHRout}
\ee
All we need to know is thus the relation between {\it in} and {\it out} modes. For $\om, \lam >0$, this relation is written 
\be
\frac{e^{-i \om f^{-1}(\uc)}}{\sqrt{4\pi \om}} = \int_0^{+\infty} \left[ \alpha_{\lam \om} \frac{e^{-i \lam \uc}}{\sqrt{4\pi \lam}} + \beta_{\lam \om} \frac{e^{i \lam \uc}}{\sqrt{4\pi \lam}}  \right] d\lam. \label{modeBogo}
\ee
This together with the complex conjugated relation gives us the link between {\it in} and {\it out} creation and annihilation operators. This is obtained by uniqueness of the field operator, {\it i.e.}, by identifying decompositions \eqref{phidecompHRin} and \eqref{phidecompHRout}. 
\be \left\{ \bal
\hat a_\om^{\rm out} &= \int_0^{+\infty} \left[ \alpha_{\om \lam} \hat a_\lam^{\rm in} + \beta_{\om \lam}^* \hat a_\lam^{{\rm in} \dagger} \right] d\lam,\\
\hat a_\om^{{\rm out} \dagger} &= \int_0^{+\infty} \left[ \beta_{\om \lam} \hat a_\lam^{\rm in} + \alpha_{\om \lam}^* \hat a_\lam^{{\rm in} \dagger} \right] d\lam.
\eal \right. \label{FullcollapseBogo} 
\ee
Such a transformation is called a Bogoliubov transformation. Note that the matrix relation between modes is the transposed of the one between operators. Only the second one coincide with the $S$-matrix of scattering theory~\cite{Gottfried,Weinberg1}. The Bogoliubov transformation is a crucial tool for quantum field theory in non trivial backgrounds. Indeed, it allows us to analyze the content of the {\it in} vacuum (or any {\it in} Fock state) in terms of {\it out}-going quanta. In particular, the number of quanta present in the {\it in} vacuum is 
\be
n_\om^{\rm out} = \vev{\rm in}{\hat a_\om^{{\rm out} \dagger} \hat a_\om^{\rm out}},
\ee
which gives 
\be
n_\om^{\rm out} = \int_0^{+\infty} |\beta_{\om \lam}|^2 d\lam. \label{HRmeannumber}
\ee
To obtain an explicit expression, one can write $\alpha$ and $\beta$ as overlaps of {\it in} and {\it out} modes~\cite{Primer}. In the present case, a simple Fourier transform is enough to obtain \eq{modeBogo} explicitly. Moreover, we shall write the Bogoliubov only at late time, which is the regime we are interested in, hence the function $f^{-1}$ is given by \eq{latetimef}. Therefore, for $\lam, \om >0$,  
\be \bal
\alpha_{\om \lam} &= \sqrt{\frac\om\lam} \int_{-\infty}^{+\infty} e^{2i\lam e^{-\kappa \uc}} e^{i\om \uc} \frac{d\uc}{2\pi},\\
&= \sqrt{\frac\om\lam} \int_0^{+\infty} e^{2i\lam X} X^{-i\frac{\om}\kappa } \frac{dX}{2\pi \kappa X}, \\
&= \frac{e^{\frac{\pi \om}{2\kappa}}}{2\pi \kappa} (2\lam)^{i \frac\om\kappa} \sqrt{\frac\om\lam} \Gamma \left(-i \frac\om\kappa \right).
\eal \ee
$\Gamma$ is the Euler function, whose properties are recalled in App.\ref{Specialfunction_App}. Similarly
\be \bal
\beta_{\om \lam} &= \sqrt{\frac\om\lam} \int_{-\infty}^{+\infty} e^{2i\lam e^{-\kappa \uc}} e^{-i\om \uc} \frac{d\uc}{2\pi},\\
&= \frac{e^{-\frac{\pi \om}{2\kappa}}}{2\pi \kappa} (2\lam)^{-i \frac\om\kappa} \sqrt{\frac\om\lam} \Gamma \left(i \frac\om\kappa \right),\\
&= e^{-\frac{\pi \om}{\kappa}} \alpha_{\om \lam}^*.
\eal \ee
When computing the mean number of quanta at fixed frequency $\om$, using \eq{HRmeannumber}, one finds an infinite result. This comes from the stationary character of Hawking radiation. The total number of emitted particle is infinite, but the rate, {\it i.e.}, the number of emitted quanta per unit of time, remains finite and constant through time. From now on, we shall refer to the flux as $n_\om$ but keeping in mind that it does not corresponds to a mean particle number. To evaluate it properly, we compute the number of quanta emitted during the interval $u_1 \leqslant \uc \leqslant u_1 + \Delta u$. The range of (ingoing) frequencies $\lam$ that contributes to the (outgoing) flux satisfies 
\be
\om e^{\kappa u_1} \leqslant \lam \leqslant \om e^{\kappa (u_1+\Delta u)}. \label{inoutredshift}
\ee
This statement follows from the redshift factor that {\it in} modes undergo when leaving the horizon, see \Sec{HRNHR_Sec}, \eq{Carterkappa}. From this, the flux emitted during the above interval reads 
\be \bal
n_\om^{\rm out} &= \frac1{\Delta u} \langle \hat a_\om^{{\rm out} \dagger} \hat a_\om^{\rm out} \rangle ,\\
&= \frac1{\Delta u} \int_{\om e^{\kappa u_1}}^{\om e^{\kappa (u_1+\Delta u)}} |\beta_{\om \lam}|^2 d\lam ,\\
&= \frac1{\Delta u} \int_{\om e^{\kappa u_1}}^{\om e^{\kappa (u_1+\Delta u)}} \frac1{e^{\frac{2\pi \om}{\kappa}}-1} \frac{d\lam}{2\pi \kappa \lam},
\eal \label{fluxderivation_HR}
\ee
which finally gives
\be
n_\om^{\rm out} = \frac1{2 \pi} \frac1{e^{\frac{2\pi \om}{\kappa}}-1}.
\ee
this result is exactly the thermal flux emitted by a black body at the Hawking temperature $T_H = \kappa/2\pi$.

\subsubsection{Collapse of arbitrary black hole}
\begin{figure}[!ht]
\begin{center} 
\includegraphics[scale=0.7]{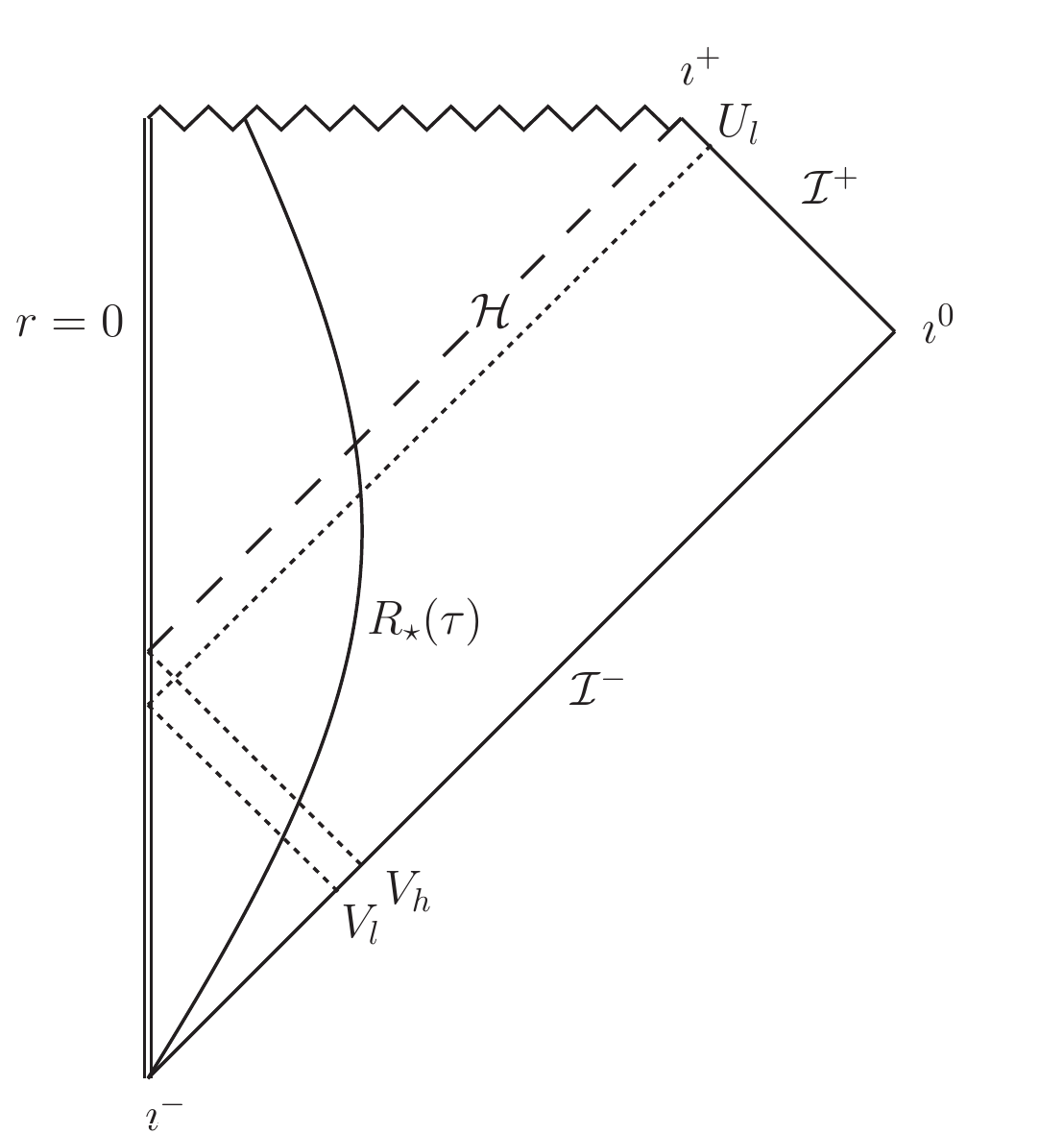}
\end{center}
\caption{Penrose diagram of a general collapse.}
\label{collapse_G_fig} 
\end{figure}

In what follows, we study a more general model of collapsing star forming a black hole. The aim is to understand that the details of the collapse are irrelevant for what concerns the late time state of the quantum field. For the sake of simplicity, we shall keep the assumption of spherical symmetry, and thus work within the 1+1 reduced geometry. However, we will make no assumptions concerning the matter content of the star, or the exact motion of its surface during the collapse. 
\be
ds^2 = \left\{ 
\bal & A(\tau,r) (d\tau^2 - dr^2), \quad r < R_\star(\tau),\\
& dt^2 - (dr - v(r) dt)^2, \quad r>R_\star(\tau).
\eal \right. \label{Generalcollapse}
\ee
Just like for the preceding case, we go straight to light cone coordinates. 
\be
ds^2 = \left\{ 
\bal & A\, dUdV, \quad r < R_\star(\tau),\\
&(1-v^2)d\uc d\vc, \quad r>R_\star(\tau).
\eal \right. \label{GeneralcollapseUV}
\ee
Since the star's surface does not {\it a priori} follow a light-like trajectory, we can hardly impose the advanced coordinate $\vc$ to be the same on both sides. However, we can assume without loss of generality that $U$ depends only on $\uc$ and $V$ on $\vc$. Doing so, what we need to do is to impose the continuity of the metric across the star's surface, which gives 
\be
(1-v^2)\frac{d\uc}{dU}\frac{d\vc}{dV} = A.
\ee
Moreover, since the star motion is expressed through the coordinate $r$, we shall also impose its continuity. The relation between $r$, $\uc, \vc$ and $U,V$ are still given by Eqs.~\eqref{coordUVSch}. In addition, we have at the star's surface \be
\frac{dV}{d\tau} = 1+ \dot R_\star \quad {\rm and } \quad \frac{dU}{d\tau} = 1- \dot R_\star.
\ee
Therefore, the continuity condition on $r$ reads
\be
2\dot R_\star = (1-v^2)(1+\dot R_\star) \frac{d\vc}{dV} - (1-v^2)(1-\dot R_\star)\frac{dU}{d\uc}.
\ee
Solving these equations, we extract the derivatives of $u$ and $v$
\bsub \bea
\frac{d\uc}{dU} &=& \frac{\left(\dot R_\star^2 + A(1-v^2)(1-\dot R_\star^2) \right)^{\frac12} - \dot R_\star}{(1-v^2)(1-\dot R_\star)},\\
\frac{d\vc}{dV} &=& \frac{\left(\dot R_\star^2 + A(1-v^2)(1-\dot R_\star^2) \right)^{\frac12} + \dot R_\star}{(1-v^2)(1+\dot R_\star)} .
\eea \esub
As in the preceding case, to obtain the late time state of the field, all one needs to do is evaluate $u$ and $v$ as functions of $U$ and $V$ when $t\to +\infty$. As we saw for \eq{latetimef} and is reviewed in Fig.\ref{collapse_G_fig}, this is done when the function $R_\star$ approaches $2GM$, {\it i.e.}, when the horizon is formed. Let's say this happens at a time $\tau_h$. When $\tau \to \tau_h$, the surface of the star falls in, following a time-like trajectory, thus 
\be
R_\star \sim 2GM - v_\star (\tau - \tau_h),
\ee
with $0< v_\star < 1$. In this regime, the derivative of the $\uc$ coordinate is given by
\be
\frac{d\uc}{dU} = - \frac1{\kappa(U-U_0)},
\ee
where $U_0$ is a constant that can be removed by a proper choice of origin for $U$. We deduce the relation $U(\uc)$
\be
U = U_0 \underbrace{- ae^{-\kappa \uc}}_{<0}. \label{Uu_G}
\ee
To obtain the above equation, we exploited the fact that $v'_{\mathcal H} = \kappa$, as we derived in \eq{PGsurfgrav}. This relation \eqref{Uu_G} is crucial, and contains the main physical information. First, the relation is exponential. Exactly as \eq{latetimef}, this is the key point which leads to a thermal state for $u$-modes at late times. Moreover, we see here that the coefficient in the exponential is exactly the surface gravity $\kappa$, which therefore governs the temperature through \eq{TH}. This is a major point of black hole radiation. The only information that the field retains at late time, is the value of the surface gravity, and no other detail of the geometry, since the function $v(r)$ can be arbitrary without altering our conclusion. Furthermore, following the path of \Sec{inout_Sec}, we compute the Bogoliubov coefficient, and obtain
\be
\beta_{\om \lam} = \frac{e^{-\frac{\pi \om}{2\kappa}}}{2\pi \kappa} (a\lam)^{-i \frac\om\kappa} \sqrt{\frac\om\lam} \Gamma \left(i \frac\om\kappa \right).
\ee
We first notice that the coefficient $a>0$ that appears in \eq{Uu_G} only contributes to $\beta$ as a phase. This phase is free of physical consequences, since it can always be gauged away by a proper change of origin for $\uc$. Moreover, the condition 
\be
\left| \frac{\beta_{\om \lam}}{\alpha_{\om \lam}} \right| = e^{-\frac{\pi \om}{\kappa}},
\ee
holds, which is the characteristic of a thermal state at late times. Concerning $v$-modes, the computation is equally simple. Near the horizon, one gets
\be
\vc = \frac{A_0(1+v_\star)}{2v_\star} V + {\rm cste},
\ee
where $A_0 = A(v=v_h) >0$ is approximately constant. Indeed, as we see on Fig.\ref{collapse_G_fig}, only a finite and small interval of $\vc$ describes rays reaching asymptotic infinity. The above equation means that a positive frequency $v$-mode stays a positive frequency $v$-mode, and hence the late time state of $v$-modes is the vacuum. This confirms that the field $\hat \phi$ ends up in the Unruh vacuum, long enough after the black hole has formed.

\subsection{Radiation of eternal black holes}
\label{eternalHR_Sec}
Shortly after the discovery of black hole radiation, an alternative derivation was proposed by Unruh~\cite{Unruh76}. As we noticed in the preceding section, the late time state of the field is independent of the detail of the collapse dynamics. Therefore, it should be possible to characterize this state by considering only the stationary black hole geometry obtained after the black hole has formed. Unruh showed that this is possible if one looks for a vacuum state which is regular across the (future) horizon. More precisely, the criterion is that any observer freely falling into the hole would not detect any quanta while crossing the horizon, at least at high energies, see the discussion after \eq{G+NHRFF}. To understand this, we study again the dynamics of the field on the reduced 1+1 geometry obtained from \eq{PGwavequ}. Before going to the details of the computation, we focus first on the main points of the canonical analysis, as performed in flat space in \Sec{canonicalquantize_Sec}. The dynamics of a massless scalar field in this geometry is derived from the action
\be
\mathcal S[\phi] = \frac12 \int [(\p_t \phi + v \p_x \phi)^2 - (\p_x \phi)^2 ] dxdt. 
\ee 
Hence, the canonical momentum reads $\pi = \p_t \phi + v \p_x \phi$, which allows us to build the conserved scalar product
\be
(\phi_1|\phi_2) = i \int \left[ \phi_1^*(\p_t \phi_2 + v \p_x \phi_2) - \phi_2 (\p_t \phi_1 + v \p_x \phi_1)^* \right] dx. \label{HRscalprod}
\ee
This scalar product is crucial, because it determines whether a mode is associated with an annihilation or creation operator, as in \Sec{canonicalquantize_Sec}. 
Because of conformal invariance, the $u$ and $v$ sectors exactly decouple, and the mode equation is easy to solve. However, it is of pedagogical value here not to solve it completely, but to present the separate dynamics of both sectors. Indeed, any solution of \eqref{PGwavequ} can be written as a sum $\phi = \phi^{u} + \phi^{v}$. Each term obeys an independent equation of motion 
\be
\left\{ \bal
(\p_t + v(x) \p_x) \phi^v &= \p_x \phi^v,\\
(\p_t + v(x) \p_x) \phi^u &= -\p_x \phi^u. 
\eal \right. \label{PGdecoupledwavequ}
\ee
Because $v(x) = -1$ on the horizon, the first equation is regular, while the second is singular. This is the origin of the dissymmetrical character of the Unruh vacuum regarding $u$ and $v$ sectors, as described in the preceding section. For the sake of simplicity of the discussion, we focus on the non trivial sector, {\it i.e.}, we consider only $u$-modes. We first decompose them in a superposition of single Killing frequency modes. Hence 
\be
\phi^u(t,x) = \int_{\mathbb R} \phi_\om^u(x) e^{-i\om t} d\om.
\ee
Because $K = \p_t$ is a Killing field, it commutes with the wave operator, and thus the dynamics can be studied at a fixed frequency $\om$. Moreover, asymptotically, $K$ is the Minkowskian time derivatives, and thus the corresponding eigen-modes are associated with the usual particle interpretation developed in \Sec{canonicalquantize_Sec}. The mode equation obeyed by $\phi_\om^u$ reads
\be
(\om + i (1+v) \p_x)\phi_\om^u = 0. \label{uPGwavequ}
\ee
To characterize the frequency as seen by a freely falling observer, we use the vector field $\uf$ of \Sec{BHinPG_Sec}. The freely falling frequency $\Om$ is thus defined as the eigen-value of the operator $i u^\mu \p_\mu$. Moreover, on the $u$-sector, we have the convenient property that
\be
u^\mu \p_\mu = \p_t + v \p_x = - \p_x,
\ee
as we see from \eq{PGdecoupledwavequ}. This means that for these modes, the freely falling frequency coincides with the spatial momentum. Not only this correspondence is convenient, but as noticed in~\cite{Brout95} it will turn out to be equally valid when introducing dispersion, in \Sec{SecpWKB}. Moreover, diagonalizing the operator $-i\p_x$ is straightforward using Fourier transform. The key point of the present derivation of Hawking radiation, is to characterize the Unruh vacuum using modes containing only \emph{positive} freely falling frequencies in the near horizon region. Hence, by construction, a detector crossing the horizon geodesically would detect no quanta. Such modes have the form of 
\be
\phi_\om^{u, {\rm in}}(x) = \int_0^{+\infty} f(\Om) e^{i \Om x} \frac{d\Om}{\sqrt{2\pi}}. \label{FFOmdecomp}
\ee
As we see, they are characterized by being \emph{analytic} on the upper complex $x$-plane. In addition, we build the corresponding, {\it i.e.}, with the same $\om$, negative norm modes, which contain only negative freely falling frequency $\Om$. These are analytic on the lower half plane. Combining Eqs. \eqref{uPGwavequ} and \eqref{FFOmdecomp}, we see that they are easily expressed as $(\phi_{-\om}^{u, {\rm in}})^*$. We underline that this construction puts no restrictions on the Killing frequency $\om$, therefore, the {\it in}-modes exist for both signs of frequency $\om$, and the field operator, restricted to $u$-sector, reads
\be \bal
\hat \phi^u(t,x) = \int_0^{+\infty} \Big( &\left[\hat a_\om^{\rm in} \phi_\om^{u, {\rm in}}(x) + (\hat a_{-\om}^{\rm in})^{\dagger} (\phi_{-\om}^{u, {\rm in}}(x))^*\right] e^{-i\om t} \\
&+ \left[ \hat a_{-\om}^{\rm in} \phi_{-\om}^{u, {\rm in}}(x) + (\hat a_{\om}^{\rm in})^{\dagger} (\phi_\om^{u, {\rm in}}(x))^*\right] e^{i\om t} \Big) d\om 
\eal \label{uBHdecomp}
\ee
We emphasize that the {\it in} modes we have constructed here do not correspond with single particle modes in the local Minkowski patch around the horizon. However, because they contain only positive frequencies, they characterize univocally the local vacuum. In other words, the prescription $a_\om^{\rm in} \vac{} = 0$ picks up the state that contains no quanta in the sense of local Minkowski, or freely falling, observers. This state is, as we already understood, the Unruh vacuum. Hence, because of that characterization, the modes $\phi_\om^{u, {\rm in}}$ are often referred as `Unruh modes'.

At this level, it is already possible to obtain the thermal character of black hole radiation. Indeed, the general $u$-solution of the mode equation \eqref{uPGwavequ} reads 
\be
\phi_\om^u(x) = (A \Theta(x) + B \Theta(-x)) e^{i \int^x \frac{\om}{1+v(x')} dx'}. \label{PGoutmodes}
\ee
The appearance of the Heaviside $\Theta$ function comes from the singular character of the mode equation \eqref{uPGwavequ} at $x=0$. In the near horizon region, one has $v(x) = -1 +\kappa x$, with $\kappa$ the surface gravity. Thus the modes look like 
\be
\phi_\om^u(x) = (A \Theta(x) + B \Theta(-x)) |x|^{i \frac{\om}{\kappa}}.
\ee
To be analytic for $\Im(x) >0$, this mode must be proportional to $(x+i\epsilon)^{i\frac{\om}\kappa}$. Therefore
\be
\phi_\om^{u, {\rm in}}(x) = \underbrace{A \Theta(x) |x|^{i \frac{\om}{\kappa}}}_{\rm Hawking\; quantum} + \underbrace{A e^{- \frac{\om\pi}{\kappa}} \Theta(-x) |x|^{i \frac{\om}{\kappa}}}_{\rm partner}.
\ee
By extending this for all values of $x$ using \eq{PGoutmodes}, we see that the first terms represent an outgoing particle escaping toward asymptotic infinity, while the second, of opposite norm, accounts for its partner, falling in the black hole toward the central singularity. Because {\it in} modes contain both values of the frequency $\om$, there is a mixing between creation and annihilation operators, and hence, the {\it in} vacuum contains outgoing particles. Moreover, we see that the outgoing modes are always accompanied by a partner, living inside the horizon. This is in fact necessary by conservation of the Killing energy. It shows that the process of Hawking radiation really consists in pair creation~\cite{Primer}. In addition, the relative weight of outgoing modes gives us 
\be
\left|\beta_\om / \alpha_\om \right| = e^{- \frac{\om\pi}{\kappa}},
\ee
which is characteristic of a thermal spectrum. To confirm the above analysis, and obtain the full Bogoliubov transformation, we solve \eq{uPGwavequ} in Fourier space, so that $-i \p_x \to p$. \eq{uPGwavequ} becomes 
\be
(\om - i\kappa -i\kappa p\p_p )\tilde \phi_\om(p) = 0,
\ee
whose general solution is
\be
\tilde \phi_\om(p) = \tilde A \Theta(p) |p|^{-i\frac\om\kappa - 1} + \tilde B \Theta(-p) |p|^{-i\frac\om\kappa - 1}.
\ee
To normalize the {\it in} and {\it out} modes, we use the canonical scalar product \eqref{HRscalprod}. Exploiting again the wave equation restricted to the $u$-sector \eqref{PGdecoupledwavequ}, a straightforward computation leads to a much simpler expression for the scalar product
\be
(\phi_\om^u |\phi_{\om'}^u) = -i \int \left[ \phi_\om^{u*}\p_x \phi_{\om'}^u - \phi_{\om'}^u \p_x \phi_\om^{u*} \right] dx. \label{NHR_PS}
\ee
In $p$-space, this simply reads
\be
(\tilde \phi_\om^u |\tilde \phi_{\om'}^u) = 2 \int p \phi_\om^{u*} \phi_{\om'}^u dp. 
\ee
Therefore, the normalized {\it in} mode of positive norm reads 
\be
\tilde \phi_\om^{u, {\rm in}}(p) = \Theta(p) \frac{p^{-i\frac\om\kappa - 1}}{\sqrt{4\pi \kappa}}.
\ee
Going back to $x$-space, we obtain 
\be
\phi_\om^{u, {\rm in}}(x) = \frac1{\sqrt{4\pi \kappa}} \int_0^{+\infty} p^{-i\frac\om\kappa -1} e^{ipx} \frac{dp}{\sqrt{2\pi}}.
\ee
This integral is easily computed using the Euler $\Gam$ function, and we derive 
\be
\phi_\om^{u, {\rm in}}(x) = \sqrt{\frac{\om}{2\pi \kappa}} e^{\frac{\om \pi}{2\kappa}} \Gamma\left(-i\frac\om\kappa\right) \times \Bigg[\underbrace{\Theta(x) \frac{|x|^{i\frac\om\kappa}}{\sqrt{4\pi \om}}}_{\phi_\om^{\rm out}} +e^{- \frac{\om \pi}\kappa} \underbrace{\Theta(-x) \frac{|x|^{i\frac\om\kappa}}{\sqrt{4\pi \om}}}_{(\phi_{-\om}^{\rm out})^*} \Bigg].
\ee
The {\it out} modes appearing in this formula were normalized using \eq{NHR_PS}. A similar computation leads to the decomposition of $(\phi_\om^{u, {\rm in}})^*$ into {\it out} modes. Transposing these relations, we obtain the Bogoliubov transform relating {\it in} and {\it out} creation and annihilation operators 
\be
\bmat \hat a_\om^{\rm out} \\ (\hat a_{-\om}^{\rm out})^\dagger \emat = \bmat \alpha_\om & \beta_\om \\ \tilde \beta_\om & \tilde \alpha_\om \emat \cdot \bmat \hat a_\om^{\rm in} \\ (\hat a_{-\om}^{\rm in})^\dagger \emat,
\ee
with
\be
\bmat \alpha_\om & \beta_\om \\ \tilde \beta_\om & \tilde \alpha_\om \emat = \sqrt{\frac{\om}{2\pi \kappa}} e^{\frac{\om \pi}{2\kappa}} \bmat \Gamma\left(-i\frac\om\kappa\right) & \Gamma\left(i\frac\om\kappa\right) e^{- \frac{\om \pi}\kappa}  \\ \Gamma\left(-i\frac\om\kappa\right) e^{- \frac{\om \pi}\kappa}  & \Gamma\left(i\frac\om\kappa\right) \emat. \label{relat_HR_Bogo}
\ee
This equation shows one of the main advantages of working with Unruh modes. Indeed, in that case, the Bogoliubov transformation is diagonal in $\om$, compare \eq{FullcollapseBogo} and \eqref{relat_HR_Bogo}. We conclude this derivation by showing that we obtain once more a Planckian spectrum for the flux of emitted quanta in the Unruh vacuum, since 
\be
\vev{\rm in}{\hat a_{\om'}^{{\rm out}\dagger} \hat a_\om^{\rm out}} = |\beta_\om|^2 \delta(\om - \om').
\ee
As in Eqs. \eqref{inoutredshift} and \eqref{fluxderivation_HR}, the $\delta$ function is here to encode a stationary flux of particles, which reads 
\be
n_\om^{\rm u} = \frac1{2\pi} \frac1{e^{\frac{2\pi \om}{\kappa}}-1}. \label{HRrelat_flux}
\ee
For convenience, we shall often forget the $2\pi$, and work with the shortcut definition $n_\om^{\rm u} = |\beta_\om|^2$.

\subsection{Observables}
\label{Observables_Sec}
\subsubsection{Asymptotic energy flux}
To compute the mean energy flux emitted by the hole, we simply need to compute $\vev{\rm in}{\hat T_{\mu \nu}}$. Unfortunately, computing naively this expression leads to an infinite result. This is because $\hat T_{\mu \nu}$ is a quadratic expression of the field operators, as discussed after \eq{MinkHam}. The renormalization of the stress-energy tensor in quantum field theory in curved space-time is a central but technical topic. The procedure to follow is to subtract the (infinite) contribution of the local Minkowski divergence. More precisely, in a small vicinity around each $x\in \M$, we define a local vacuum $|I(x)\rangle$ with the help of modes that resemble the most Minkowski plane waves. This can be done because the divergence is in the ultraviolet regime, which appeal for very short distances around $x$, in a neighborhood that is essentially Minkowski. The renormalized stress tensor is obtained by subtracting this (infinite) mean value
\be
\hat T^{\rm ren}_{\mu \nu} = \hat T_{\mu \nu} - \langle I(x) | \hat T_{\mu \nu} |I(x) \rangle \hat I.
\ee
Of course, the latter equation is only formal since both terms on the right side are infinite. One needs first to regularize it. the most powerful method is the point splitting approach, which is essentially base on the structure of Green function, where we subtract the Minkowski divergence before taking the coincident point limit. All this procedure can be shown to be perfectly consistent and leads to a finite result for a wide class of space-times. For a more detailed discussion we refer to~\cite{Fulling,BirrellDavies}. Here, it is unnecessary to develop the full technology. Instead, we shall exploit the fact that the geometry is asymptotically flat. Let's consider as an example the momentum flux. Asymptotically we are in flat space, hence~\cite{Delduc},
\be
\hat J = \hat T_{tr} =  \p_t \hat \phi \, \p_r \hat \phi.
\ee
Therefore, it is tempting to prescribe a normal order procedure. However, this is not as easy as in flat space, since we dispose of several inequivalent notions of creation and annihilation operators. To overcome this ambiguity, we prescribe the {\it out}-normal order, which means that we put all annihilation operators on the right when using a decomposition with {\it out}-operators. This is physically reasonable, because if the state of the field was the {\it out} vacuum, we would expect, by definition, to detect no energy. Moreover, in asymptotically flat spaces, it can be shown that it is equivalent to the point splitting procedure. 
\be
\hat T^{\rm ren}_{\mu \nu} =\ : \hat T_{\mu \nu} :_{\rm out}.
\ee
Because the state of the field is the {\it in} rather than the {\it out} vacuum, the mean energy flux will be non trivial. Using Eqs. \eqref{uBHdecomp} and \eqref{invacdef}, 
\be
\vev{\rm in}{\hat J} = \int_0^{+\infty} \om |\beta_\om|^2 \frac{d\om}{2\pi},
\ee
which gives exactly a Planckian thermal flux in 1+1 dimensions
\be
\langle \hat J \rangle = \frac{\pi}{12} T_H^2. \label{idealHRflux}
\ee
To obtain the full 3+1 result, we just recall that the field $\phi$ used below is related to the 3+1 field by \eq{Yharmonics}. Transposed to the stress-energy tensor, it gives
\be
\langle \hat T_{\mu \nu}^{4D} \rangle = \frac1{4\pi r^2} \langle \hat T_{\mu \nu}^{2D} \rangle,
\ee
where the equality holds for $\mu, \nu = t,r$ and the other $4D$ components vanish. This integrated on a sphere gives again the result \eqref{idealHRflux}. Note that this equation is exact only because we neglected the gravitational potential in \eq{fullPGwavequ}. We now discuss its effects.

\subsubsection{Greybody factors}
In fact, the expression \eqref{idealHRflux} overestimates the flux. Indeed, for a real $3+1$ black hole, not all the radiation emitted by the horizon reaches the asymptotic region. There is a loss due to the potential barrier appearing in \eq{fullPGwavequ}. However, this effect has no conceptual consequences on the Hawking radiation, since it affects only the infrared. Indeed, the scattering matrix relating {\it in} and {\it out} modes admits the exact factorization 
\be
S = S_{\rm NHR} \cdot S_{\rm gf}.
\ee
This factorization of the $S$-matrix will be explained in more details in \Sec{StrucSec}. For now, we just expose its signification. $S_{\rm NHR}$ is the Hawking $S$-matrix found in \eq{relat_HR_Bogo}. It relates {\it in} and {\it out} creation and annihilation operators. The {\it in} vacuum, or Unruh vacuum is defined by a prescription which concerns the ultraviolet regime. Indeed, from equation \eqref{FFOmdecomp}, we see that we can add many low negative frequency modes without altering the analytic characterization of Unruh vacuum. On the other hand, $S_{\rm gf}$ concerns only the infrared, and becomes trivial when $\om \to \infty$. It is obtained by solving an usual scattering problem, which is \eq{fullPGwavequ} written in tortoise coordinates
\be
\p_t^2 \phi - \p_{r^*}^2 \phi + V_{\rm Grav}(r) \phi = 0. 
\ee
We emphasize that this extra scattering is purely \emph{elastic}, {\it i.e.}, it does not mix annihilation and creation operators. Hence, $S_{\rm gf} \in U(2)$ while $S_{\rm NHR} \in U(1,1)$\footnote{In fact, it is slightly more complicated, since they must be $3\times 3$, see \Sec{StrucSec}.}. In particular, the temperature is unaffected by the greybody factors, and is still defined by $S_{\rm NHR}$. This can be seen when the field is in the Hartle-Hawking vacuum. In that case, the black hole is in equilibrium with the walls, and the only parameter describing the state is the temperature, {\it i.e.}, greybody factors disappear. On the other hand, the outgoing flux in the Unruh vacuum is reduced. In \eq{idealHRflux}, the $\pi/12$ will be replaced by a numerical factor $\xi$ depending on the nature of the field and the full propagation from the horizon to infinity. As far as the flux is concerned, this numerical factor can be pretty large, reducing the flux up to $\sim 99\%$~\cite{Page76,Page76b,Page77}.
\begin{figure}[!ht]
\begin{center} 
\includegraphics[scale=0.7]{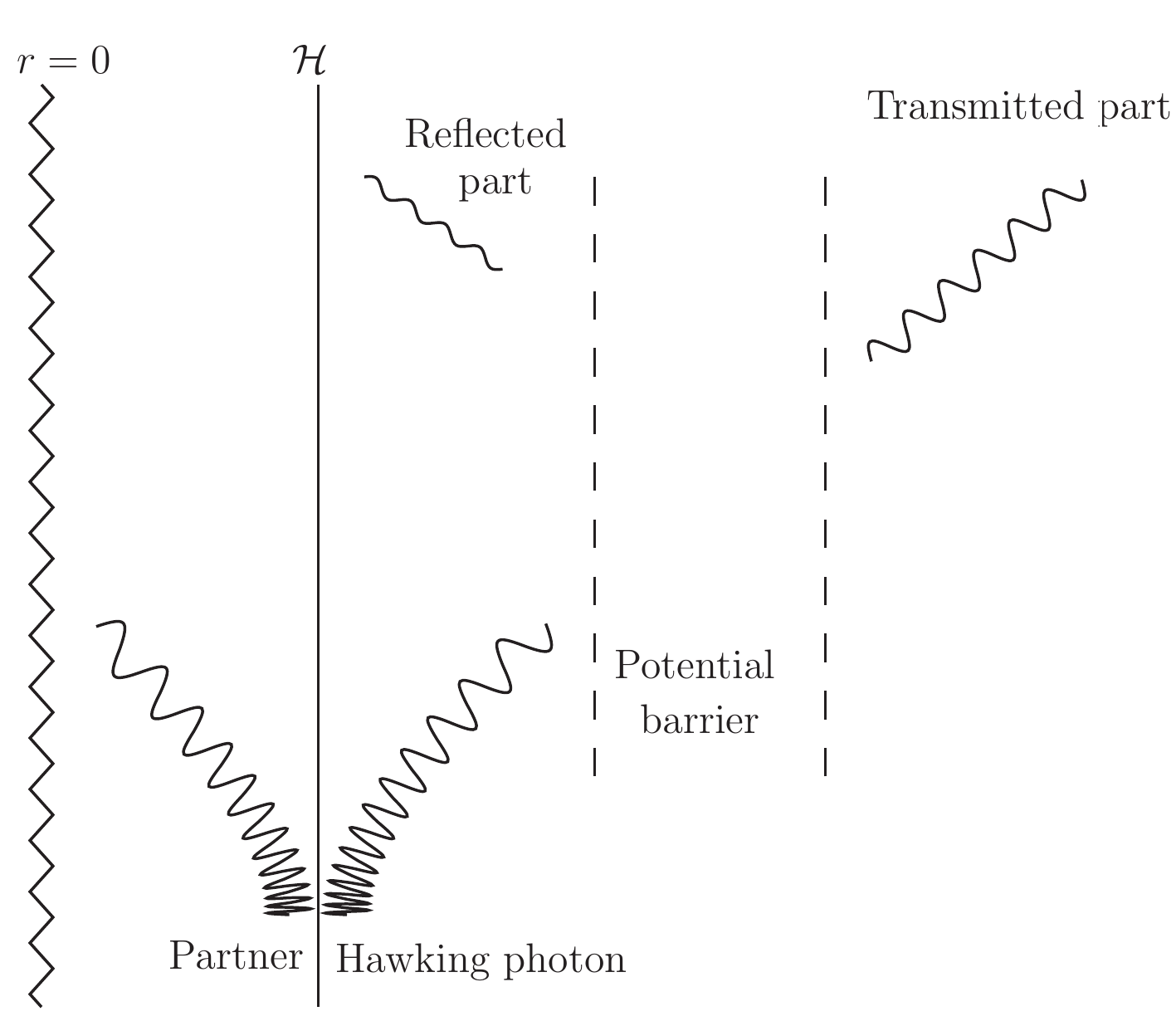}
\end{center}
\vspace{-0.5cm}
\caption{Space-time diagram of the scattering of Hawking quanta near a black hole. The near horizon behavior, where the redshift is depicted, is independent of the elastic exterior scattering, which is responsible for the grey body factors corrections.}
\label{GFscattering_fig} 
\end{figure}

Unfortunately, even in the optimistic estimate, without greybody factors, the flux emitted by a black hole is very small. Indeed, the temperature takes the value 
\be
T_H \approx \left( \frac{M_{\odot}}{M}\right) 6.10^{-8}\ K,
\ee
In particular, for a solar mass black hole, the temperature is 7 order of magnitude smaller than the cosmic microwave background $\sim 2,7 K$, erasing all hopes of observing Hawking radiation in any near future. This is why this phenomenon is mainly a playground for theoretical work, and speculations about the quantum theory of gravity.

\subsubsection{The back reaction}
To conclude the chapter, we shall say of few words concerning the back reaction of Hawking radiation on the black hole. When quantum fluctuations of gravity can be neglected, the black hole is perturbed by the $T_{\mu \nu}$ of its own radiation. Hence, its dynamics is given by the Einstein equation sourced by the (renormalized) stress energy tensor of the radiation 
\be
G_{\mu \nu} = 8\pi G \langle \hat T_{\mu \nu}^{\rm ren} \rangle. \label{backreacequ}
\ee
This equation can also be obtained from energy conservation. The energy flux emitted to infinity reduces the mass of the black hole so that the total energy black hole + radiation is conserved. Such equation follows from \eqref{backreacequ} evaluated asymptotically and integrated on a sphere. It gives
\be
\frac{dM}{dt} = - \xi \, T_H^2(M). \label{evapoequ}
\ee
Therefore, the black hole slowly evaporates as $M - M_0 \propto - t^{1/3}$. Of course, to obtain the life-time one should precisely incorporate the effects of greybody factors, that is compute $\xi$, as discussed in the preceding paragraph and in~\cite{Page76}. However, this does not affect the conclusion that a black hole evaporates, by losing its mass through Hawking radiation. \\

\begin{figure}[!ht]
\begin{center} 
\includegraphics[scale=0.7]{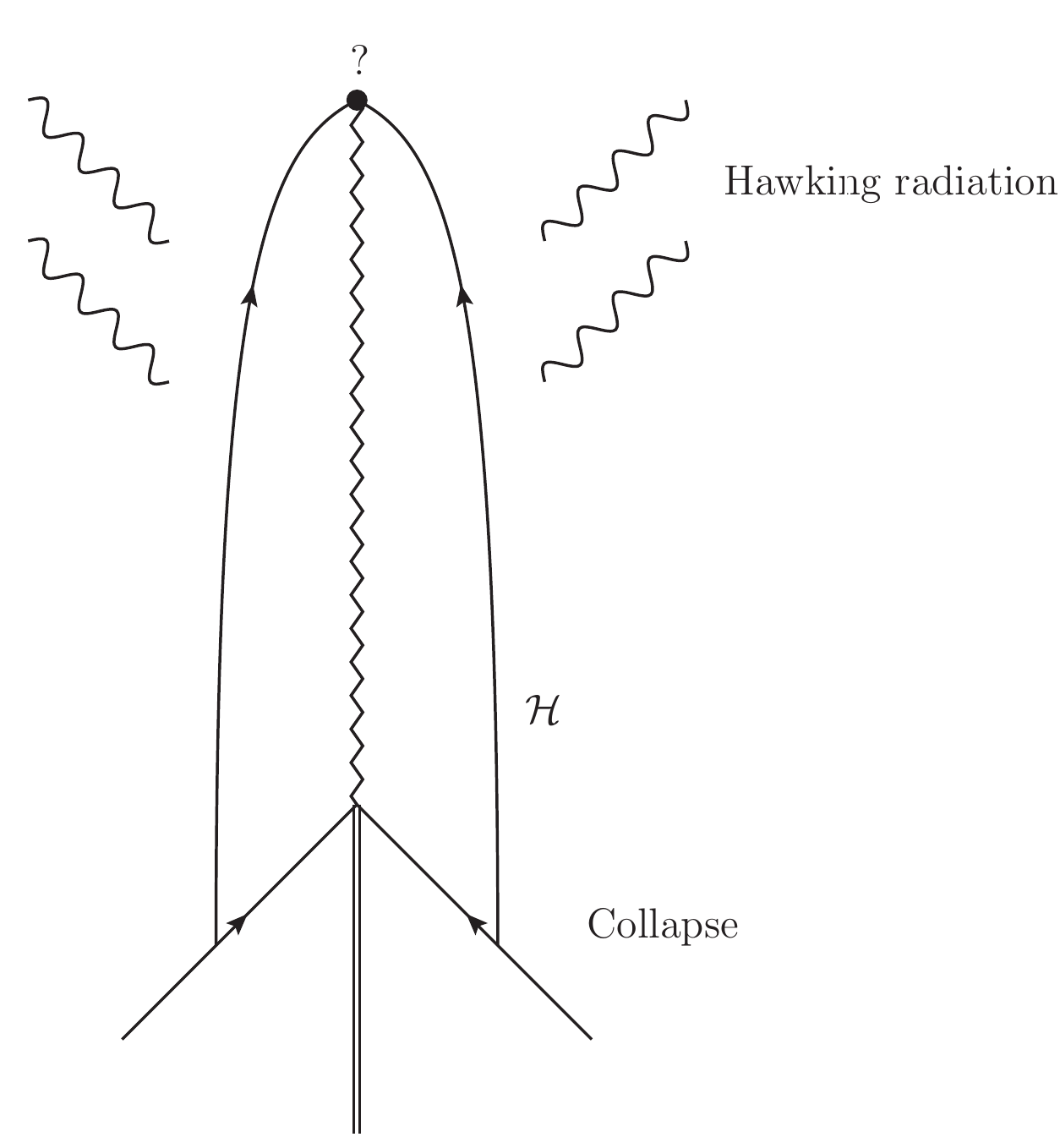}
\end{center}
\caption[width=5cm]{Space-time diagram of an evaporating black hole. The black hole slowly loses its mass through Hawking radiation, until it reaches a final state, unknown from the semi-classical theory.}
\label{evaporation_fig} 
\end{figure}

Despite the intensive work on the matter during the last 30 years, the scenario of black hole evaporation still raises many open questions. One of these concerns the final stage of the evaporation of a black hole, where the semi-classical equation can certainly not be trusted (see Fig.\ref{evaporation_fig}). Another one is the validity of the semi-classical approximation, a point that we shall discuss in the beginning of chapter \ref{LIV_Ch}. In both cases, our lack of answers is deeply entailed to the full problem of quantum gravity. 

\chapter{Breaking Lorentz invariance}
\label{LIV_Ch}
\minitoc

\section{Dispersive field theories}
\subsection{The transplanckian question}
\label{transpl_Sec}
Hawking radiation, as described in Chapter \ref{HR_Ch}, is one of the main predictions of quantum field theory in curved backgrounds. 
As we discussed in Sec.~\ref{BHthermo_Sec}, it plays a crucial role in black hole thermodynamics and in the attempts to 
construct theories of quantum gravity. Unfortunately, such theories are up to now still incomplete, and no complete microscopic or quantum description of Hawking radiation has been provided. 
The semi-classical treatment stays the most relevant model. Therefore, it is crucial to determine under which conditions this treatment is valid, 
and how its properties depend or not on hypotheses concerning the ultra-violet behavior of the theory.\\

One of the most puzzling features of Hawking radiation concerns the role of ultra high frequencies in its derivation. This can be seen as follows. Imagine that a Hawking quantum is detected at infinity, with some frequency $\om_0$. What happens if one traces back in time this particle ? It goes back closer and closer to the horizon. Its Killing frequency is unchanged, but its freely falling frequency $\Om$, which is the one as defined in some small Minkowski patch, gets blueshifted. Since null geodesics undergo an infinite focusing (see \Sec{HRNHR_Sec}), this blueshift never stops. Therefore, we can question the validity of the semi-classical approach in Hawking's derivation~\cite{Jacobson91}. Can we trust a theory where gravity is treated classically ? Because of this infinite blueshift, outgoing quanta could interact gravitationally very strongly, and hence invalidate the approximation~\cite{tHooft85,tHooft96}. Of course this blueshift is in fact never infinite. One can only trace back to the time the horizon was formed, from the collapse of some star. To put some numbers in the discussion, we consider a solar mass black hole, $M = M_{\odot}$. $1$s after the black hole has formed, the blueshift factor is of the order of $e^{10^5}$. As Unruh often says, `{\it this number is so huge, that you can put any units [for the frequency], it will only make a trivial change in the exponent}'. In particular, when one detects a particle at a typical frequency of the order of the Hawking temperature $T_H$, it emerges from quanta with freely falling frequency way above the Planck scale, where the semi-classical treatment can certainly not be trusted. This puzzle is often referred as the `transplanckian problem'. However, following Jacobson, we shall rather refer to it as a `transplanckian question'. Indeed, at late time, no observable explicitly shows these transplanckian frequencies, neither the fluxes, nor the mean value of the (renormalized) stress energy tensor $\langle \hat T_{\mu \nu}^{\rm ren} \rangle$. Hence, it is only by knowing the full quantum dynamics that we would be able to see whether this redshift invalidates or not the Hawking scenario.\\

As mentioned in~\cite{Primer}, the infinite redshift on the horizon is probably not relevant as far as standard model particle interactions are concerned, since these are presumably asymptotically free. However, this transplanckian question can certainly not be washed out when it comes to gravitational interactions, because they tend to increase for higher energies. There have been several attempts to take into account gravitational interactions in the process of Hawking radiation, mainly by 't Hooft~\cite{tHooft85,tHooft96}, but we also mention~\cite{Parentani02}, where a potential spontaneous Lorentz symmetry breaking is considered. The main conclusion is probably that transplanckian frequencies always induce strong interactions near the horizon. For example, when considering an ingoing spherical shell of matter interacting with an outgoing one, of respective frequency $\om_{\rm in}$ and $\lam_{\rm out}$, the interaction matrix elements essentially go like 
\be
\langle \om_{\rm in} |\hat S_{\rm int} | \lam_{\rm out} \rangle \approx 4G \frac{\om \lam r_{\mathcal H}}{r - r_{\mathcal H}}.
\ee
Since this blows up on the horizon, it could drastically alter the semi-classical picture. However, up to now, no fully satisfactory computations have been obtained. This is mainly due to the fact that every perturbative treatment seems to break down due to the presence of this transplanckian frequencies. In the work presented in this chapter, the philosophy is thus to determine under which conditions Hawking radiation is a robust scenario, and how deeply it relies on the ultraviolet behavior of the full theory. To address this question, Jacobson first proposed to introduce a modification of the dispersion relation~\cite{Jacobson91} that would emerge from the microscopic structure of the theory. This idea was inspired by a very nice analogy found by Unruh~\cite{Unruh81}, between sound propagation on moving flows, and light on a black hole geometry. We shall first present this analogy and then introduce the Jacobson-Unruh dispersive model.

\subsection{Analog models}
\label{analogmodel_Sec}
\subsubsection{The Unruh analogy}
\begin{figure}[!ht]
\begin{center} 
\includegraphics[trim= 0 3cm 13mm 0,clip]{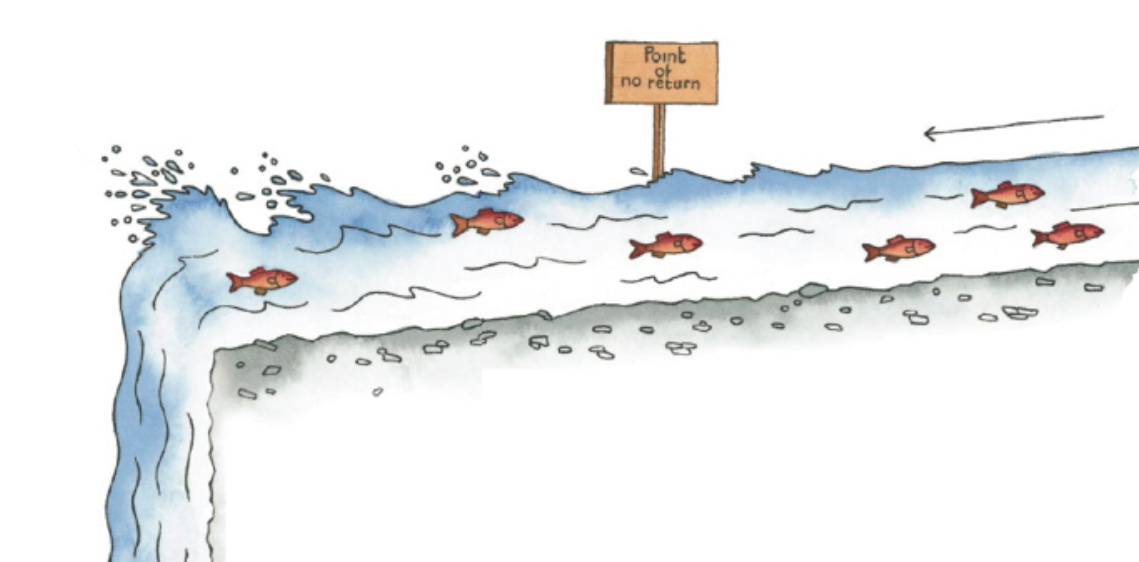}
\end{center}
\caption{Artist view of an analogue black hole in a flow. Fishes that have passed the point of no return can no longer communicate with those upstream. [Credit: Yan Nascimbene]}
\label{sonicBHpic_fig}
\end{figure}

In 1981, while preparing a lecture on fluid mechanics, Unruh noticed that sound waves in a moving fluid obey a wave equation which is exactly the one on a space-time geometry~\cite{Unruh81}. Indeed, in the hydrodynamic regime, {\it i.e.} for long wavelength, sound waves see an effective metric characterized by the background flow. More precisely, the velocity profile is decomposed into a background flow and fluctuations around it 
\be
\vect v = \vect{v_0} + \delta \vect{v}.
\ee
Whenever the background flow becomes faster that the speed of sound $c$, {\it i.e.}, propagation speed of the fluctuations, there is a horizon. Indeed, let's assume for simplicity that $\vect v$ is unidimensional in the $x$ direction and that $v<0$, so that the current is flowing to the left. In addition, suppose that $v=-c$ at $x=x_h$. On the right of this point, sound waves can be emitted toward both sides. However, as soon as we get on the left side of $x_h$, all sound waves are dragged by the current, and propagate to the left (in the lab frame). Anyone located in this region ($x<x_h$) wouldn't be able to communicate to the right region. The interior region effectively behaves as a `dumb hole'\footnote{Note that in this context, `dumb' stands for `mute', not `stupid'. Because of this amusing {\it quiproquo}, Unruh was forced to change the name of its first paper~\cite{Unruh81}, since the editor found it too offensive.}, the sonic analogue of a black hole (see Fig.\ref{sonicBHpic_fig}).

In~\cite{Unruh81}, Unruh made the precise computation of the sound equation in an ideal, inviscid (no dissipation) and irrotational flow. Here, we just present the result without any proof. For detailed computations, we refer to~\cite{Barcelo05,Balbinot06,Visser12}. We first define the velocity potential $\phi$ such that 
\be
\delta \vect v = \vect{\grad}(\phi).
\ee
Using Navier-Stokes and the continuity equation, we show that $\phi$ obeys the massless wave equation of \Sec{fieldprop_Sec} in the geometry
\be
ds^2 = \frac{\rho}{c} \left[ c^2 dt^2 - (dx - v_x dt)^2 - (dy - v_y dt)^2 - (dx - v_z dt)^2 \right].
\ee
Here, $\rho$ is the density of the fluid. $\vect v$, $\rho$ and $c$ are not independent, but related through the continuity equation. Most of the time, we assume a unidimensional flow, and the wave equation then reads 
\be
\left[ \frac1{\rho} (\p_t + \p_x v)\frac\rho{c^2} (\p_t + v\p_x) - \frac1{\rho} \p_x \rho \p_x \right] \phi_\om(x) = 0 .
\label{fluidwavequ}
\ee
Later, this analogy was extended to many condensed matter systems, among which surface waves~\cite{Schutzhold02}, light in optical fibers~\cite{Leonhardt}, liquid helium~\cite{Jacobson98} or Bose-Einstein condensates (BEC)~\cite{Macher09b}\footnote{The suggestion of references here is far from being exhaustive. The interested reader is highly encouraged to consult reference~\cite{Barcelo05} where a historical survey of the work in analog gravity is provided, together with a well-furnished bibliography.}. We shall provide more details about this last analogue in Chapter \ref{mass_Ch}. All this cases share the same equation, \eqref{fluidwavequ}, for the propagation of their perturbations. As a consequence, exactly as was derived in \Sec{eternalHR_Sec}, analogue black holes should radiate a thermal flux of phonons, the quanta of sound waves, at the Hawking temperature $T_H = \kappa/2\pi$. The surface gravity is given by the same formula as \eq{PGsurfgrav}, 
\be
\kappa = \p_x v,
\ee
but now, this has a new interpretation, as the gradient of the flow velocity. 

This analogy is very rich in potential research directions. Not only it gives an opportunity to perform experiments on the subject, it also sheds new light on potential answers to the `transplanckian question'~\cite{Parentani02b}. Of course, an analogy is never a proof, but their study is both stimulating from the astrophysics point of view, or the hydrodynamical one. On the experimental side, an analogue black hole has been produced in BEC~\cite{Lahav09}, but the detection of Hawking radiation is currently under investigations. In a water tank experiment, the classical pendant of Hawking radiation was successfully measured, and the thermal law verified~\cite{Weinfurtner10}. There has also been progressing results in optical fibers~\cite{Belgiorno10}, but the precise interpretation of the data is still under debate~\cite{Schutzhold10b}.\\

We end this section by a small remark concerning the so-called \emph{white hole} flows. A white hole geometry corresponds to the time reverse of a black hole. It is obtained when the gradient of the velocity profile $v$ is of opposite sign, {\it i.e.}, $\p_x v_{|\mathcal H} < 0$. This means that a white hole possesses a negative surface gravity. As we shall see in Chapter \ref{mass_Ch}, the behavior of a white hole is very similar to that of a black hole. From an astrophysical point of view, a white hole geometry is an irrelevant concept. Unlike a black hole, such a solution can never be obtained dynamically, as the final state of a star, or another stellar object. There might still be some speculative conjecture about their possible existence, but up to now, it is very unlikely that these have any physical significance for astrophysics. On the other hand in analog gravity, white holes are easily obtained and provide a very interesting playground for both experimental and theoretical probes of the Hawking effect. As a first example, the well-known `hydraulic jump'~\cite{Volovik05} is an analog white hole. 

\subsubsection{Transplanckian question revisited}
Since the free wave equation is exactly the same in a fluid flow as in a black hole, the same questions concerning the infinite redshift on the horizon might be asked. Already in his first paper on the matter~\cite{Unruh81}, Unruh wrote {\it `The low-energy fluid equations which have led to quantum thermal sonic emission by a transonic background flow break down at high frequencies because of the atomic nature of the fluid. At distances of $10^{-8}$ cm, the assumptions which I use of a smooth background flow are no longer valid just as in gravity one expects the concept of a smooth space-time on which the various relativistic fields propagate to breakdown at scales of $10^{-33}$ cm. Furthermore, the phonons emitted are quantum fluctuations of the fluid flow and thus affect their own propagation in exactly the same way that graviton emission affects the space-time on which the various relativistic fields propagate.'} However, unlike in gravity, when considering atoms we basically know the microscopic quantum theory. It is simply given by a $N$-body Schrödinger equation. Therefore, the transplackian question can now be addressed and possesses a definite answer. Either some mechanism protects the appearance of Hawking radiation, or it is simply cancelled by microscopic effect. As understood by Jacobson, to obtain the answer, it is unnecessary to solve the full $N$-body problem. Indeed, the first deviations due to the atomic structure of matter will manifest themselves as a modification of the dispersion relation for short wavelengths~\cite{Jacobson91,Jacobson93}. 

\subsubsection{Taking dispersive effects into account}
We consider the frequency of the wave in the frame of the atoms, {\it i.e.}, the co-moving frequency 
\be
\Om = \om - v p.
\ee
In the hydrodynamical regime, this frequency is related to the wavelength by the relativistic relation
\be
\Om^2 = c^2 p^2.
\ee
The combination of the last two equations is simply the eikonal approximation of the wave equation \eqref{fluidwavequ}. When considering shorter length, or higher $p$'s, this dispersion relation will be smoothly modified, and becomes 
\be \bal
\Om^2 &= c^2 F^2(p). \\
&= c^2 p^2 \pm c^2 \frac{p^4}{\Lambda^2} + O(p^6). 
\eal \label{analogdisp} \ee
We see that this introduces a new length scale $\Lambda^{-1}$, which is in analog context, essentially the interatomic scale. Therefore, the question of the \emph{robustness} of Hawking radiation in analog models, can be addressed by solving a linear equation, but containing higher order \emph{spatial} derivatives terms. Note also that dispersive effects can be divided into two classes, depending on the choice of sign in \eq{analogdisp}. A + sign leads to a superluminal dispersion, where the group velocity exceeds the seep of light for increasing $p$'s. The - sign gives the opposite result, {\it i.e.} a subluminal dispersion. Before addressing the robustness of Hawking radiation in \Sec{dispHRpaper_Sec}, we will present how one can implement this ideas in general relativity.

\subsection{General relativity with ultraviolet Lorentz violations}
\label{GRwithUTVF}
\subsubsection{How would Lorentz invariance be broken ?}
There are basically two lines of thought concerning the possibility of having an ultraviolet violation of local Lorentz invariance. On one hand, one can consider that there are, microscopically, extra degrees of freedom, which specify a preferred frame. In other words, Lorentz invariance is not exact in the very high energy regime, but rather emerges as an infrared symmetry. This would be the spirit of theories like Einstein-Aether~\cite{Jacobson01} or Ho\u rava-Lifshitz gravity~\cite{Horava09}. On the other hand, one can still assume that Lorentz invariance is a fundamental symmetry, but which would be broken near the horizon of a black hole, due to strong interaction effects~\cite{Parentani02}. As we discussed in the preceding section, the infinite redshift might seem to generate strong gravitational interactions. It is a possibility, that interactions could be effectively described for intermediate energies by a modified dispersion relation. In fact, there is quite a few known mechanisms that lead to a dynamical Lorentz violation, see {\it e.g.}~\cite{Libanov05}. The novelty here would be that such violation occurs in the ultraviolet regime.\\ 

We point out that in the first philosophy, the extra degrees of freedom, which break Lorentz invariance, can either be fundamental or collective degrees of freedom, {\it i.e.} `phonon like'. An interesting example can be found in the concept of `emergent gravity'~\cite{Barcelo01}. Keeping the idea at a naive level, it consists in pushing a step forward the Unruh condensed matter analogy. In this picture, space-time is not a fundamental degree of freedom. Rather it emerges from collective behavior of microscopic degrees of freedom, `atoms of space-time'. In that case, the metric emerges when considering a large number of atoms, as in the preceding section. However, a preferred frame then also emerges, as the frame co-moving with the atoms. Therefore, even though the preferred frame is not necessarily present microscopically, the emergent structure in that is always twofold: a metric together with a preferred frame. Only phonons of long wavelength lose track of that second structure. Moreover, if several types of excitations are present, they will generically feel a different `speed of light', meaning that the preferred frame is still tractable in the long wavelength regime. We refer to~\cite{Carlip12} for a discussion on this kind of difficulties in the emergent approach\footnote{Note that there exist other approaches to quantum gravity inspired by the ideas of emergent gravity, where the emergence of a  preferred frame could be generically avoided. This is the case for {\it e.g.} Group Field Theory~\cite{Oriti11,Carrozza12}.}. Therefore, in this emergent gravity picture, a preferred frame will emerge for \emph{any} geometric background. This is in contrast with the second philosophy, were the preferred frame emerges only in \emph{some} space-times, such as black holes. This would allow to keep Lorentz invariance as an exact symmetry in Minkowski, while being violated around a black hole horizon. \\

In our work though, we shall not address such delicate questions. On the contrary, we will \emph{assume} that Lorentz invariance is broken at short distances, and we study the \emph{consequences} for black hole physics. Of course, in astrophysics perspectives, this is quite speculative. However, for analog gravity purpose, our work has direct implications, since there, Lorentz invariance is never exact. In the following, we describe the main ingredients needed to model a broken Lorentz invariance.

\subsubsection{Einstein-Aether theory}
In order to break local Lorentz invariance, we must necessarily choose a `preferred frame', in which the dispersion relation is imposed. Indeed, in order to mimic what happens in analog gravity, we must define what plays the role of the `co-moving frequency' $\Om$. As exposed in~\cite{Jacobson96}, this can be done by disposing of a vector field $\uf^\mu$, which is time-like everywhere. Without lack of generality, we can assume it as being of unit norm, $\uf^\mu \uf_\mu = 1$. The preferred frequency is then defined as 
\be
\Om = -p_\mu \uf^\mu.
\ee
This new field $\uf$ must be considered as an extra background structure, on equal footing with the metric tensor $\g$. However, in general relativity, such a vector field breaks general covariance, {\it i.e.}, diffeomorphism invariance. This has dramatic consequences, {\it e.g.}, the stress-energy tensor of matter is no longer conserved~\cite{Kostelecky03}. This would make Einstein's equation \eqref{Einsteinequ} as it stands inconsistent, since the Einstein tensor $G_{\mu \nu}$ of \eq{Gmunu_def} is divergenceless by construction. To restore general covariance, the preferred frame must be \emph{dynamical}. In~\cite{Jacobson01}, such a theory, with both a metric and a preferred frame, was developed. Of course, studying Lorentz violation, or the consequences of a preferred frame is much older (see various references in~\cite{Jacobson01}). However, the main new ingredient is that the field $\uf$ is constrained to stay time-like of unit norm. This implies that Lorentz invariance is always broken by the presence of this preferred frame. To consider the most general dynamics, the action is obtained by adding all possible terms that respect the symmetry (here, diffeomorphism invariance) and that contains at most two derivatives~\cite{Jacobson08}. This action reads 
\be \bal
\mathcal S[\g, \uf] = -\frac1{16\pi G} \int \big[R &+ c_1 \nabla^\mu \uf^\nu \nabla_\mu \uf_\nu + c_2 (\nabla_\mu \uf^\mu)^2 + c_3 \nabla^\mu \uf^\nu \nabla_\nu \uf_\mu \\
&+ c_4 (\uf^\mu \nabla_\mu \uf^\nu) (\uf^\rho \nabla_\rho \uf_\nu) + \lam (\uf^\mu \uf_\mu - 1) \big] \sqrt{-g}d^4x.
\eal \label{EAaction} \ee 
The last term is the constraint that $\uf$ is a unit time-like vector field, and $\lam$ is the corresponding Lagrange multiplier. This action can be viewed from two angles. It can either be considered as a modified theory of gravity, where $\uf$ is an extra fundamental field, or, it can correspond to an \emph{effective} theory. In that case, \eqref{EAaction} would be the long wavelength approximation of an underlying theory, which is the motivation for considering only two derivatives in the action. Non zero values of $c_1$, $c_2$, $c_3$, and $c_4$ account for possibly different `speeds of light'~\cite{Jacobson08}, as residual large scale effects from the fundamental theory. Dispersion would be a next-to-leading order effect. This will be implemented in the next paragraph. 

We also point out that an alternative way to describe a preferred frame is to use a scalar field whose gradient is everywhere time-like. This field then plays the role of a time function. This line of thought was used for example in Ho\u rava-Lifshitz gravity~\cite{Horava09}. However the scalar field approach is less general than the unit time-like vector field one. Indeed, the vector field can be viewed as proportional to a gradient if and only if it is hypersurface orthogonal, while a scalar field always produces a vector field through its gradient. For this reason we will prefer the Einstein-Aether approach. Of course, this discussion concerns only the way of parametrizing the preferred frame. If the action differs from \eq{EAaction}, the theory can become drastically different. \\

In analog gravity, the effective action of \eq{EAaction} is always imposed by the microscopic physics. However, we point out that the extra ingredient present in analog context, which we do not have in pure general relativity is not the static laboratory frame. Indeed, in astrophysics, such frame would be given by the Killing field $K$. In both cases, it consists of the class of observers that see a stationary background (the geometry or the fluid flow). In condensed matter, we have an extra privileged class of observers, which is the one co-moving with the atoms. In other words, $\uf^\mu$ is always given by microscopic structure of the fluid. In general relativity, one does not have such field. Rather one has  \emph{several families} of preferred observers, those freely falling (see \eq{framefamily}). Breaking local Lorentz invariance exactly means that we pick up one special member of that family. Then, a physical system experience the violation depending on how it couples to the preferred frame. As we shall see, in our case, the test field couples through higher order derivatives, which alter the dispersion relation in the ultraviolet.

\subsubsection{Radiation field in Einstein-Aether background}
To study the propagation of a dispersive field on a space-time geometry, basically two lines of thought can be followed. On on hand, one can study a particular fluid or theory and derive the equations for linearized perturbations from the known microscopic theory,  {\it e.g.}, of a Bose condensate~\cite{Macher09b}, or perturbations around an Einstein-Aether black hole solution. On the second hand, more abstractly, one can identify the ingredients that must be adopted in order to obtain a well-defined mode equation. We adopt here this second attitude~\cite{Unruh95,Brout95,Jacobson96} as it is more general, and as it reveals what are exactly the choices that should be made.

To this aim, we first need to choose the two background fields, namely $\g$ and $\uf$. In a black hole space-time, the geometry is stationary, {\it i.e.}, there is a Killing field $K$. For the preferred frame, the natural choice is that it is also stationary, that is it commutes with $K$, 
\be
[\uf, K] = 0. \label{uhyp1}
\ee
This hypothesis is crucial. It means that the preferred frame has reached an equilibrium with the black hole. Moreover, we choose the preferred frame as being freely falling. This is realized if it obeys the geodesic equation
\be
\uf^\mu \nabla_\mu \uf^\nu = 0. \label{uhyp2}
\ee
This second assumption is merely an idealization. If one chooses to derive the background configuration from, {\it e.g.}, the Einstein-Aether action \eqref{EAaction}, the $\uf$ field most probably possesses pressure-like terms that give it a non zero acceleration. For the present purpose, what really matters is that this acceleration stays finite\footnote{In particular, this would \emph{not} be the case if one chooses the preferred frame given by the Killing field $K$~\cite{Jacobson91,Jacobson93}.}. In particular, the preferred frame must fall regularly across the black hole horizon. For the sake of simplicity, we additionally assume the problem to be 1+1 dimensional, the studied geometry being described by a Painlevé-Gullstrand metric 
\be
ds^2 = dt^2 - (dx - v(x) dt)^2. \label{metric}
\ee
In this background geometry, with both \eqref{uhyp1} and \eqref{uhyp2}, one is left with a one parameter family of time-like vector fields. This family is the $\uf_{v_\infty}$ described in \Sec{PGgeod_Sec}. At this point, there is \emph{no} preferred choice among the $\uf_{v_\infty}$. As we discussed above, this is reminiscent of the local Lorentz symmetry. To simplify the expressions, we make the choice to use 
\be
\uf = \p_t + v(x) \p_x. \label{UTVF}
\ee
We see that in these coordinates, the function $v(x)$ fully characterizes both the geometry \eqref{metric} and the preferred frame \eqref{UTVF}. Secondly, we must choose a dispersion relation, which we impose in the preferred frame, {\it i.e.}, by using the frequency $\Om = -\uf^\mu p_\mu$, 
\be
\Om^2 = F^2(p) = p^2 \pm \frac{p^4}{\Lambda^2} +O(p^6). \label{disprelCF}
\ee
In 1+1 dimension, we define the space direction with the space-like vector $s^\mu$, which is orthogonal to $\uf^\mu$. $s^\mu$ is of unit norm ($s^\mu s_\mu = -1$) and satisfies $\uf^\mu s_\mu = 0$. It defines the spatial momentum $p = s^\mu p_\mu$. In Painlevé-Gullstrand coordinates, it simply reads $s = \p_x$. Together with the (conserved) Killing frequency, we rewrite the full dispersion relation as 
\be
(\om - v p)^2 = F^2(p) . \label{disp2}
\ee
The dynamics of the field is then obtained by building an action that reduces to \eq{disprelCF} in the eikonal (or WKB) approximation. If we focus on the quartic term, and consider the dispersion relation in \eqref{disp2}, the equation of motion is derived from the action 
\be
\mathcal S_\pm = \frac12 \int \left[g^{\mu\nu}\p_{\mu}\phi \p_{\nu}\phi \pm \frac{(h^{\mu\nu}\nabla_{\mu}\nabla_{\nu}\phi)^2}{\Lambda^2} + \textrm{ higher derivative terms } \right] \sqrt{-g} d^4x, \label{dispCFaction}
\ee
where $h_{\mu \nu} =  \uf_\mu \uf_\nu - g_{\mu \nu} = - s_\mu s_\nu$ is the spatial metric induced on the preferred foliation. The corresponding equation of motion reads, using an arbitrary $F$ function, 
\be
\left[ \left( \partial_t + \partial_x v \right) \left( \partial_t + v \partial_x \right)  +  F^2(i\partial_x) \right] \phi = 0 \label{wavequ},
\ee
When applied to a stationary mode $\phi = e^{- i \om t} \varphi_\om$, this becomes
\be
\left[ \left( \om + i\partial_x v \right) \left( \om + iv \partial_x \right)  -  F^2(i\partial_x) \right] \varphi_\om = 0 \label{modequ}.
\ee
The associated conserved scalar product is
\be
(\phi_1|\phi_2) = i\int \left[ \phi_1^*(\partial_t + v\partial_x)\phi_2 - \phi_2 (\partial_t + v\partial_x) \phi_1^*\right] dx.\label{scalt}
\ee
The stationary (positive norm and real frequency) modes $\phi_\om$ are then normalized by
\be
(\phi_{\om'}|\phi_\om) = \delta(\om' - \om).
\label{scalom}
\ee
In conformity with \Sec{collapseHR_Sec}, negative norm modes are named $(\phi_{-\om})^*$, 
so that $\phi_{-\om}$ are positive norm modes of negative frequency $-\om$. 

When we take the limit $\Lambda \to \infty$, the equation \eqref{wavequ} reduces to that of the D'Alembert equation described in \Sec{fieldprop_Sec}. We stress that this choice for the action is not unique, even when requiring the correct $\Lambda \to \infty$. Indeed, different choices of ordering between $\p_x$'s and $v(x)$ will lead to the same dispersion relation, but with inequivalent field dynamics. As examples, we refer to~\cite{Brout95,Schutzhold08} in which left and right moving modes remain exactly decoupled even when $F(p)$ is non linear. Moreover, in analog models, the theory of the corresponding medium imposes one choice. For example, \eq{wavequ} differs in several respects from the (two dimensional) Bogoliubov-de Gennes equation~\cite{Macher09b}. It first differs in the hydrodynamical regime, {\it i.e.}, in the limit $\Lambda \to \infty$, in that the latter equation is not conformally invariant. As a result, left and right moving modes remain coupled in that case. Moreover, it also differs from \eq{modequ} when taking into account the quartic dispersive effects. The differences arise from different orderings of $\partial_x$'s and $v(x)$. For water waves, the Unruh analogy is valid only in the hydrodynamical regime ($\Lambda \to \infty$). The fully dispersive equation is most often unknown, despite efforts in that direction~\cite{Unruh12}.

Nevertheless, these wave equations share the same characteristics since these are determined by \eqref{disprelCF}. 
What we will show in \Sec{dispHRpaper_Sec}, which is less obvious, is that these models also share, at leading order, the same 
deviations of the spectrum which are due to dispersion. This follows from the fact that these deviations are based on asymptotic expansions that are 
governed by Hamilton-Jacobi actions associated with \eqref{disp2}.

\section{Hawking radiation in the presence of dispersion}
\label{dispHRpaper_Sec}
The question we address now concerns the robustness of the Hawking scenario when one introduces dispersive effects through \eq{disp2}. If we assume that ultraviolet Lorentz violations are induced by quantum gravity effects, it is sensible to suppose that the cut-off energy $\Lambda$ is of the order of Planck mass $M_{\rm P}$\footnote{In fact, this is not a trivial assumption, see discussion in~\cite{Parentani07}.}. If we study a black hole of one solar mass, then 
\be
\frac{\kappa}{\Lambda} \approx 3.10^{-39}.
\ee
This very small number legitimates a perturbative approach in $1/\Lambda$. 
The modifications induced by $\Lambda$ on Hawking radiation have been the subject of many studies. The first quantitative one goes back to 95, when Unruh wrote a dispersive wave equation in an acoustic black hole metric~\cite{Unruh95}. He then numerically observed that the thermal properties of the flux are robust, {\it i.e.}, not significantly affected when $\Lambda \gg \kappa$. In a subsequent numerical analysis~\cite{Corley96}, it was observed that `{\it the radiation is astonishingly close to a perfect thermal spectrum}'. This was confirmed to a higher accuracy in~\cite{Macher09}, and partially explained by analytical treatments~\cite{Jacobson93,Brout95,Corley97,Himemoto00,Unruh04,Balbinot06}. In spite of these works, the origin of this astonishing robustness is not completely understood. This is due to our ignorance of the parameters 
governing the first deviations with respect to the standard thermal spectrum. Indeed, as we shall see in \Sec{validity}, it is \emph{not} governed solely by the ratio $\kappa /\Lambda$. \\

This section contains a detailed presentation of the work~\cite{Coutant11}. In this study, we showed that when $\Lambda \gg \kappa$ the most relevant parameter is the extension of the near horizon region in which second order gradients can be neglected, as in \eq{vNHRapprox}. We obtained these results by studying the validity limits of the connection formula~\cite{Brout95,Corley97,Himemoto00,Unruh04} encoding the scattering across the horizon. While the core of the calculation is based on the mode properties near the horizon (which are universal as they rely on a first order expansion around the horizon) the validity limits are essentially governed by the extension of the region where this expansion is valid. As a result, on the one hand, the leading order expressions are universal and agree with the standard relativistic ones, and on the other hand, the first deviations depend on this extension. Their evaluation is carried out using a superluminal dispersion relation. Interestingly, these results also apply to subluminal dispersion, as we show in establishing a correspondence between these two cases.

\subsection{Modifications at the eikonal approximation}
\label{eikonal_Sec}
\subsubsection{Hamilton-Jacobi actions and turning point}

\begin{figure}[!ht]
\begin{subfigure}[b]{0.5\textwidth}
\includegraphics[scale=0.6]{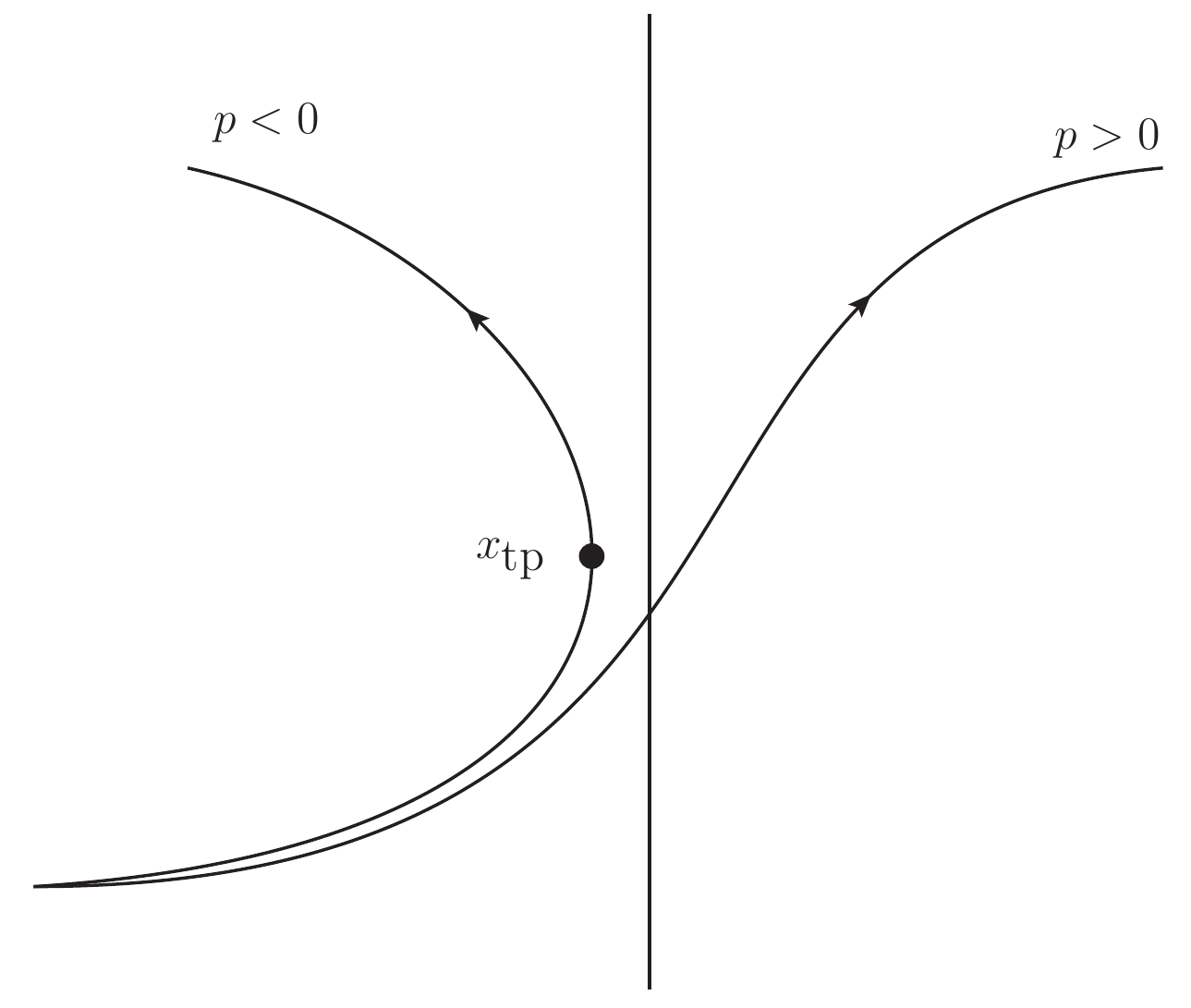}
\caption{Trajectories for superluminal dispersion.}
\end{subfigure}
\begin{subfigure}[b]{0.5\textwidth}
\includegraphics[scale=0.6]{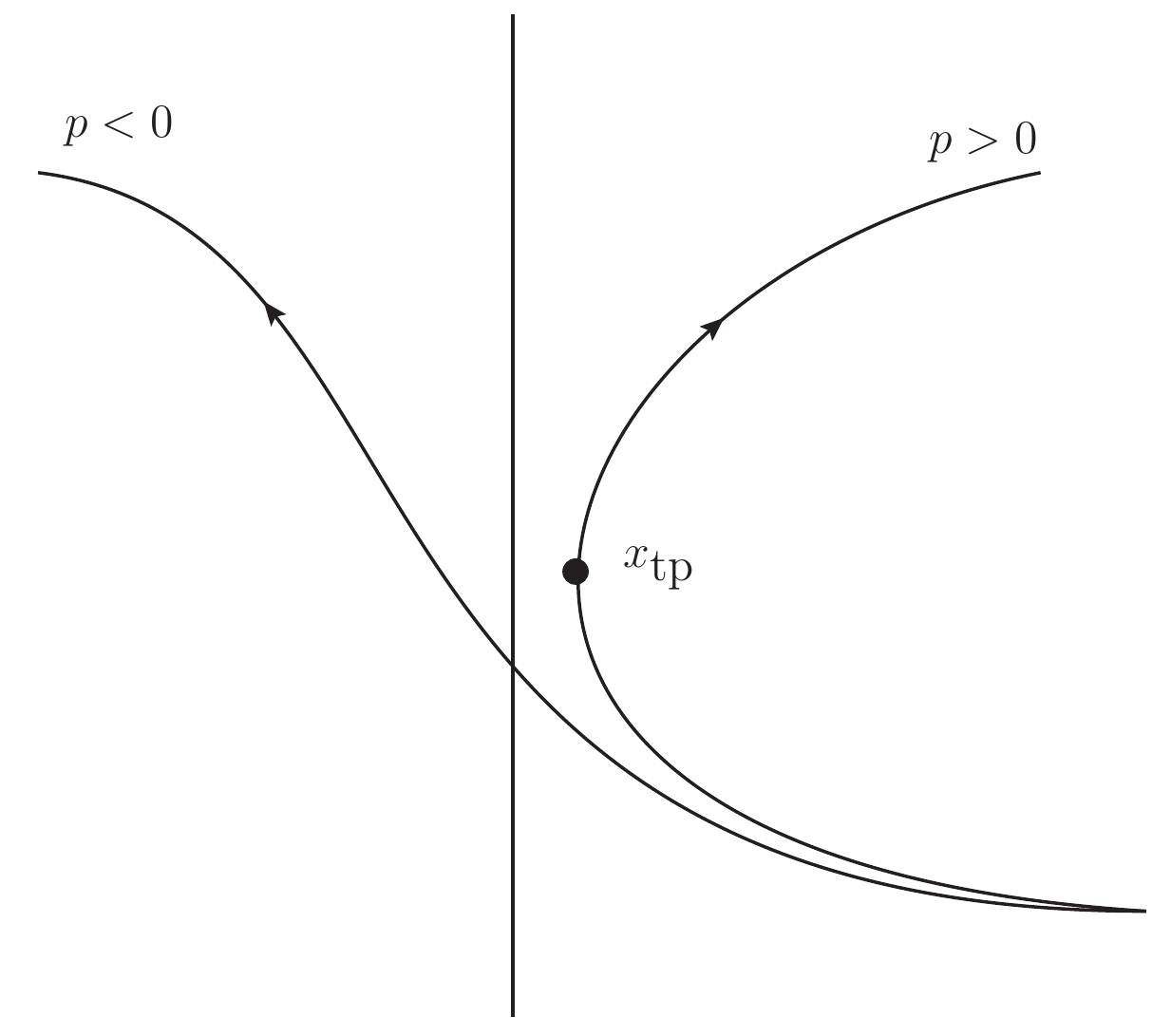}
\caption{Trajectories for subluminal dispersion.}
\end{subfigure}
\caption{Trajectories of massless particles near the horizon, in the presence of dispersion. The behavior depends on the sign of the momentum $p$, or equivalently, the freely falling frequency $\Om$~\cite{Jacobson96}.} 
\label{dispersive_traj_fig} 
\end{figure}

To study the modification introduced by dispersive effects, we shall first focus on the modifications of the geodesics, {\it i.e.}, the characteristics of \eq{wavequ}. This section consists in reviewing \Sec{HRNHR_Sec} in the presence of ultraviolet dispersion. For this, we interpret \eqref{disp2} as the Hamilton-Jacobi equation of a point particle. It suffices to consider the solution for $p$ as a function of $x$ at fixed $\om$, which we call $p_\om(x)$, as $p_\om(x) = \partial_x S_\om$ where the action is decomposed as $S = - \om t + S_\om(x)$. One thus works with the standard expression
\be
S_{\om}(x) =\int^x  p_{\om}(x') dx'.
\ee
In these classical terms, left and right moving solutions with respect to the fluid, 
{\it i.e.} the roots of $\om - v p = \pm  F$, decouple and can be studied separately. 
Restricting attention to the right moving ones, the relevant ones for Hawking radiation (see \Sec{eternalHR_Sec}), 
one deals with 
\be
\om - v(x) p_\om(x) =  F(p_\om(x)).
\label{rDR}
\ee
In the sequel, it will be useful to work in the $p$-representation with 
\be
W_\om(p) = - px + S_\om(x) =  - \int^p  X_{\om}(p') dp' \label{Wom}.
\ee
In this representation, at fixed $\om$, the position $x$ is viewed as a function of $p$. It is given by $X_\om = \partial_p W_\om$ 
and obeys
\be
\om - v(X_\om(p))\ p =  F(p) .
\label{Xomp}
\ee
The usefulness of the $p$-representation can be appreciated when considering 
the trajectories in near the horizon region, where $v \sim - 1 + \kappa x$. 
In this region, irrespectively of $F(p)$, one finds
\be
{dp \over dt} = -\left( \frac{\partial X_\om}{\partial \om} \right)^{-1} = -\kappa p,
\ee
which gives the exponential redshift of \eqref{Carterkappa}, as in the relativistic settings of \Sec{HRNHR_Sec}. Then the trajectory 
is algebraically given by
\be \label{Xp}
x_F(t) = X_\om(p(t)) = \frac{\om}{p \kappa} - \frac{F(p) - p}{\kappa p}.
\ee
Unlike what is found for $p(t)$, using \eq{disprelCF}, $x_F$ completely differs from \eqref{NHRtraj} of \Sec{HRNHR_Sec} for $|p|  \geq \Lambda$. 
To adopt a language appropriate to the study of the modes, we shall work with $\om > 0$ only.
Then negative frequency roots $p_{-\om} > 0$ of \eqref{rDR}
are replaced by the negative roots $p_\om< 0$ associated with $\om > 0$, as explained in~\cite{Jacobson96}. 
Hence, \eqref{Xp} defines two trajectories, one with $p > 0$, and one with $p < 0$. 
At early times, {\it i.e.}, for large $|p| \gg \om $ and for superluminal dispersion, $|F/p| > 1$, both are coming from the supersonic region $x< 0$.  
Then, for $p > 0$ the trajectory crosses the horizon and reaches $x = +\infty$, whereas for $p  < 0$, it is reflected back to $x = -\infty$. On the contrary, for subluminal dispersion, $|F/p| < 1$, both incoming trajectories come from the right region before one is reflected, and the other falls inside the hole. Both types of trajectories are represented on Fig.\ref{dispersive_traj_fig}. 
What is important is that all trajectories stay in the near horizon region a finite time $\Delta t$. 
The integrated red-shift effect $p_{\rm final}/p_{\rm initial} = e^{-\kappa \Delta t}$ is thus finite, unlike what is found for relativistic propagation 
where \eqref{Carterkappa} applies to arbitrary early times. 

From now on, we focus on a specific dispersion relation. To obtain the simplest expression, we shall work with 
\be
F^2(p) = \left(p + \frac{p^3}{2\Lambda^2} \right)^2 = p^2 + \frac{p^4}{\Lambda^2} + \frac{p^6}{4\Lambda^4}. \label{dispr}
\ee
For $p < 0$, the location of the turning point $x_{tp}$ is obtained by solving 
$d x/dt = (\partial_\om p_{\om})^{-1} = 0$. Using Eq.~(\ref{dispr}), 
\eqref{rDR} gives
\be
\om = (1+v) p_\om + \frac{p_\om^3}{2\Lambda^2}. 
\label{xHJ}
\ee
Hence, the momentum and the velocity at the turning point are (exactly) 
\be
- p_{tp}= \left(\om \Lambda^2 \right)^{\frac13},
\label{ptp}
\ee
and 
\be
v(x_{tp})+1 = - \frac{3}{2} \left(\frac{\om}{\Lambda} \right)^{\frac23} \label{vtp}.
\ee
If $\om$ is sufficiently small, {\it i.e.}, $\om < \Lambda \left( D^L_{\rm lin}\right)^{\frac32}$,
the turning point is located in the near horizon region, and it is given by
\be
\kappa x_{tp} = - \frac{3}{2} \left(\frac{\om}{\Lambda} \right)^{\frac23} \label{xtp}.
\ee

In classical terms, the main consequence of dispersion 
is the introduction of this turning point. It introduces a non trivial multiplicity of the real roots $p_\om(x)$, 
solutions of Eq.~(\ref{xHJ}). This multiplicity will play a crucial role when solving the mode equation~(\ref{modequ}). From \eqref{vtp} and \eqref{Das}, 
we see that there is no turning point for $\om$ above 
\be
\om_{\rm max} = \Lambda \left(\frac23 D_{\rm as} \right)^{\frac32} \label{ommax}.
\ee
This threshold frequency corresponds to the limiting case where the turning point $x_{tp}$ is sent to $- \infty$. 
For $\om > \om_{\rm max} $ only the positive root of \eqref{xHJ} and the positive norm mode $\phi_\om$ remain. 
In \Sec{modeanalys}, $\om_{\rm max}$ will determine the dimensionality of the set of modes.

While these results have been obtained with a superluminal dispersion relation, however, they hold when the dispersion is subluminal.
Indeed \eqref{xHJ} is invariant under the three replacements: 
\bsub \label{subsuper} \bea
1+v &\rightarrow& -(1+v), \\
\om &\rightarrow& -\om,  \\
\frac1{\Lambda^2} &\rightarrow& - \frac1{\Lambda^2}. 
\eea \esub 
The first replacement 
exchanges the subsonic region and the supersonic one (for $v< 0$). 
As a result, a black hole horizon is replaced 
by a white one and {\it vice versa}.
The second one amounts to a time reverse symmetry, $t \to - t$. 
At the classical level, it exchanges the roles of positive and negative 
roots of \eqref{disp2}. At the mode level, it changes the sign of their norm, as discussed in Sec.~\ref{modeanalys}. 
The third replaces a superluminal quartic dispersion by a subluminal one. This exchange
applies to all dispersion relations when expressed as $F(p)- p \to -(F(p)-p)$. It replaces 
any dispersion that exhibits a superluminal character when $p$ approaches $\Lambda$ 
by the corresponding subluminal one.
This correspondence thus applies to dispersion relations that pass from super to sub, as it is the case for gravity waves in water
when taking into account capillary waves~\cite{Rousseaux10}.

Under \eqref{subsuper}, the trajectories are mapped into each other, as the function $X_\om(p)$ of \eqref{Xomp} is {\it unchanged}. 
Hence Eqs. (\ref{ptp},\ref{vtp},\ref{xtp}), are also
unaffected. Because we changed the sign of $\om$, for subluminal dispersion,
it is the trajectory of positive frequency that has a turning point. 

\subsubsection{The action $S_\om$ in the near horizon region}

In preparation for the mode analysis we study the behavior of $S_\om$ 
in the near horizon region where $v$ is linear in $x$. Because of this linear character, 
it was appropriate to first solve the equations of motion in the $p$-representation and then go in 
the $x$-representation. This is also true for the action itself. 
Moreover, when solving \eqref{modequ}, $\phi_\om(x)$ will be obtained by inverse Fourier transforming the mode in $p$-space $\tilde \phi_\om(p)$. 
Thus we express the action as $S_\om = xp - W_\om(p)$. 
Imposing $\partial_p S = 0$, we get
\be
S_{\om}(x,p_0) = x p_\om(x) - \int_{p_0}^{p_\om(x)} X_\om(p) dp \label{Somint},
\ee
where $p_\om$ is a solution of \eqref{xHJ} and $p_0$ fixes the integration constant. 
Using \eqref{Xp} and \eqref{dispr}, one gets
\be
S_{\om}(x,p_0) = x p_{\om}(x) - \frac{\om}{\kappa} \ln \left (p_{\om}(x)\right) + \frac{(p_{\om}(x))^3}{6\Lambda^2 \kappa} + \theta_0,
\label{Som}
\ee
where $\theta_0$ is 
\be
\theta_0 = \frac{\om}{\kappa} \ln \left(p_0 \right) - \frac{p_0^3}{6\Lambda^2 \kappa}.
\label{theta0}
\ee
To consider all solutions of \eqref{modequ}, we shall compute the 
action for all roots of \eqref{xHJ}, including the complex ones. 
To this end, we need to define the integral $\int^p_{p_0} dp'/p' = \ln (p/p_0)$, that arises 
from the first term of \eqref{Xp}, for $p$ complex. We shall work with the argument of cut equal to $\pi - \epsilon$,
so that $\ln(-1) = -i\pi$.

\subsection{Hawking radiation as a scattering problem}
\label{profileSec}

\subsubsection{The relevant properties of the profile $v$}

To be able to compute these leading deviations  it is necessary to further discuss the properties of the profile $v(x)$. When using relativistic fields, the 
temperature of HR is completely fixed by $\kappa = \partial_x v$, the gradient of $v$ evaluated at the horizon, 
even though the asymptotic flux generally depends on other properties of $v(x)$ which fix the grey body factors of \Sec{Observables_Sec}. However these describe an elastic scattering between $\phi^{\rm left}_\om$ and $\phi_\om$, and therefore do not affect the temperature, as can be verified by considering the equilibrium state described by the Hartle-Hawking vacuum~\cite{BirrellDavies,Primer}. 
When dealing with dispersive fields, $\kappa$ is no longer the only relevant quantity. Indeed, as we shall show in the sequel, {\it several} properties of $v(x)$ now become relevant. Moreover they govern 
{\it different} types of deviations with respect to the standard flux. For {\it smooth} profiles, there are basically two important properties,
near horizon ones, and asymptotic ones. 

\begin{figure}[!ht]
\begin{center} 
\includegraphics[scale=0.9]{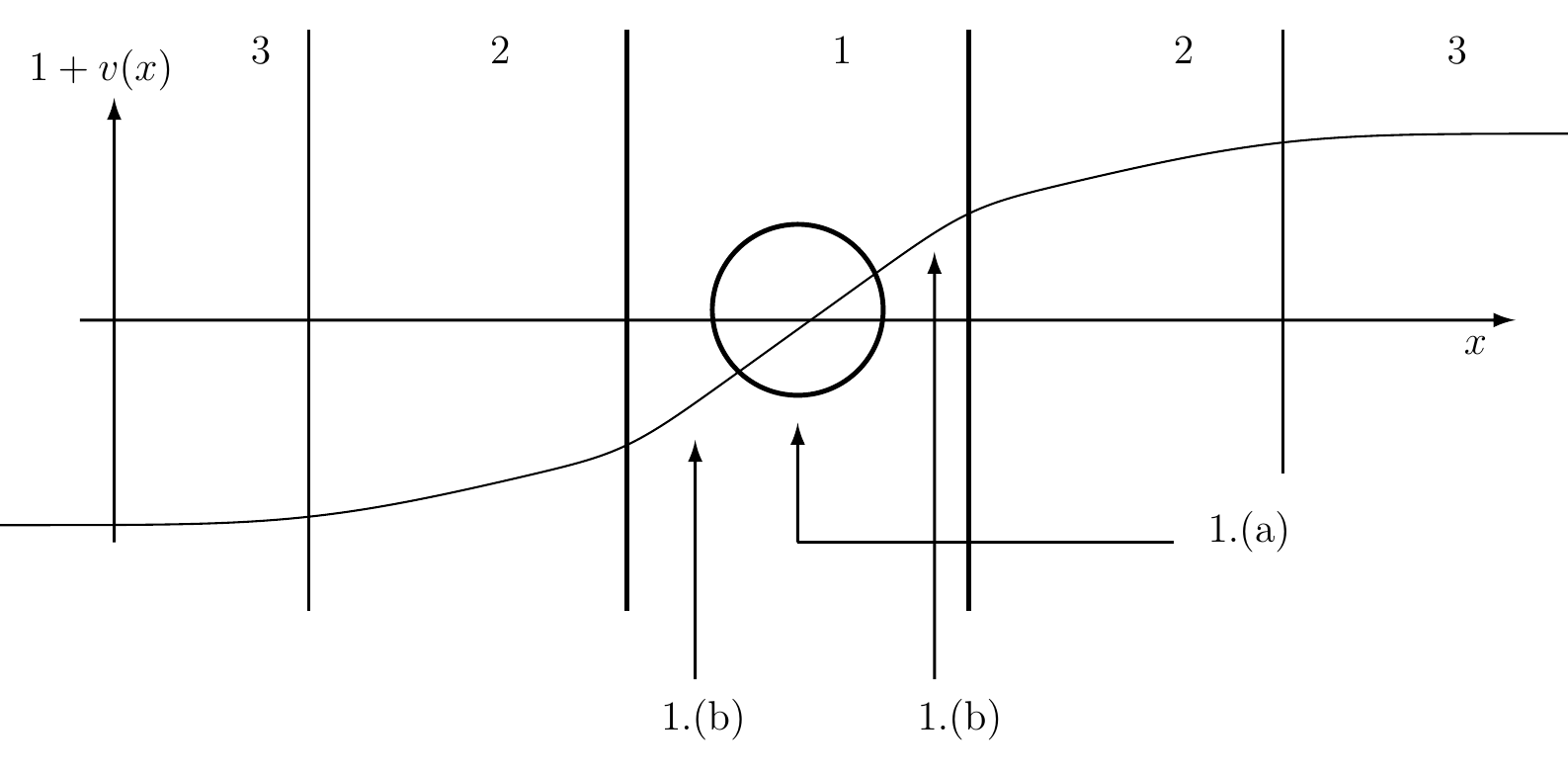}
\end{center}
\caption{Shape of the typical velocity profile $v$, 
together with the extension of the relevant regions. For a given value of $\om$, 
the near horizon region $1$ where $v \sim -1 + \kappa x$ splits into two: 
a region close to the turning point of \eqref{xtp} (1.(a)) 
where the WKB approximation fails, and two intermediate regions (1.(b)) where 
this approximation becomes reliable, and whose sizes
are fixed by $D_{\rm lin}^{L}$ and $D_{\rm lin}^{R}$. 
In the asymptotically flat regions 3, the solutions are superpositions of plane waves. 
The intermediate regions 2 play no significant role when $v$ is sufficiently smooth, 
because the propagation is accurately WKB, {\it i.e.} there is no mode mixing.
As we shall see, mode mixing essentially occurs at the scale of the turning point, {\it i.e.} in 
region 1.(a).}
\label{regions} 
\end{figure}

If there is a regular horizon at $x=0$, $v$ can be expanded as
\be
v(x) =-1 + \kappa x + O(x^2). \label{vlin}
\ee
This near horizon behavior is only valid in a certain interval, not necessarily symmetric about $0$. 
Hence we define $D_{\rm lin}^{L}$ and $D_{\rm lin}^{R}$ such that for
\be
- D_{\rm lin}^{L} \lesssim \kappa x \lesssim D_{\rm lin}^{R}, \label{Dlin}
\ee
$v$ is linear, to a good approximation (see region 1 in Fig.~\ref{regions}). 
As we shall later establish, $D_{\rm lin}^{L}$ and $D_{\rm lin}^{R}$ 
control the leading deviations of the spectrum.
It is worth noticing that in the limit $D_{\rm lin}^{L}, D_{\rm lin}^{R} \to \infty $, 
one effectively works in de Sitter space endowed with an {\it homogeneous} preferred frame\footnote{In that case, the field $\uf$ respects a subgroup of the de Sitter isometry group. Compared to the general case of arbitrary $v$, there is thus an \emph{extra} Killing field for $\g$ and $\uf$, corresponding to the homogeneity of de Sitter. This was noticed in~\cite{Parentani07}, and further developed with Jean Macher during his PhD~\cite{Macherthesis}. It was exploited in~\cite{Parentani10}, and further developed in~\cite{Busch12}.}. In that limiting case, as we shall explain, the relativistic result of \eqref{relats} is found with a higher accuracy.

The other relevant parameter is related to the
asymptotic values of $v$, that we assume to be finite. For superluminal dispersion, 
what matters is 
\be
v(x = - \infty) = -1 - D_{\rm as} < -1 \label{Das}.
\ee
For subluminal instead, it is $v (x = +\infty)$ that matters.
As discussed in \Sec{eikonal_Sec}, $D_{\rm as}$ determines the critical frequency $\om_{\rm max}$ 
(computed below in \eqref{ommax})
above which the flux vanishes~\footnote{In that work~\cite{Macher09b}, the profile was $v= - 1 + D \tanh(\kappa x/D)$. Hence, when looking at the deviations of the spectrum upon changing $D$,  the deviations associated with $D_{\rm lin}$ and $D_{\rm as}$ were confused since both scaled in the same way. In fact, one of the main novelties of the present analysis, and the companion numerical works~\cite{Finazzi10b,Finazzi11}, 
is to remove this confusion by analyzing the deviations due to $D_{\rm lin}^R$ and $D_{\rm lin}^L$ only.}.

For finite values of $x$, $v$ can have a complicated behavior. As we said, we only suppose that the interpolation between the asymptotic regions and around the horizon is {\it smooth} enough, so we can neglect non-adiabatic effects. Indeed, a sharpness in 
$v(x)$ induces non-adiabatic effects~\cite{Corley96,Macher09,Finazzi10b} 
not related to the Hawking effect that both destroy the thermality of the spectrum and induce 
higher couplings between left and right moving modes, see Figs. 12 and 16 in \cite{Macher09}. 
These effects are due to the breakdown of the WKB approximation studied in \Sec{modeanalys}. 

Because of the hypothesis $|v(-\infty)| < \infty$, it seems that our framework is irrelevant for astrophysical black holes. Indeed, if this condition is natural in any analog model, in general relativity, black holes hide a singularity inside where the function $v$ blows up (see Eqs.~\eqref{Schwf} and \eqref{PGvdef}). However, for \emph{subluminal} dispersion, relaxing the hypothesis $|v(-\infty)| < \infty$ does not affect the results of this chapter, since everything happens on the right side and in the near horizon region. On the other hand, the \emph{superluminal} case is more subtle to consider. Indeed, since modes can travel faster than light, the singularity is no longer `future time-like', and information can come out from it. Therefore, new boundary conditions must be imposed, either on physical grounds, or by some deduction of the full theory. This complex discussion will no longer be addressed in the present work.

\subsubsection{{\it in} and {\it out} mode basis, connection formulae, and Hawking radiation}

\begin{figure}[!ht]
\begin{subfigure}[b]{0.5\textwidth}
\includegraphics[trim = 0 0 8.2cm 0, clip]{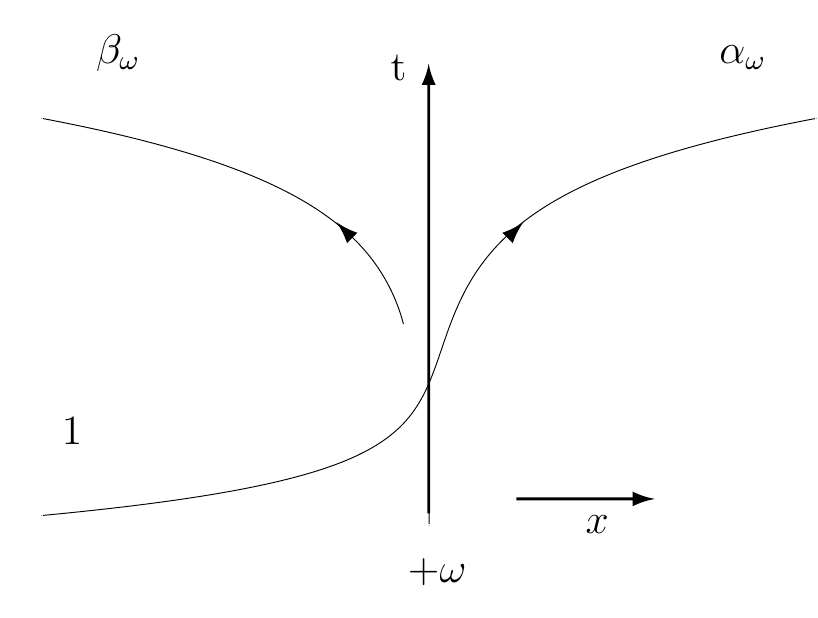}
\caption{Trajectories associated with $\phi_\om^{\rm in}$.}
\end{subfigure}
\quad
\begin{subfigure}[b]{0.5\textwidth}
\includegraphics[trim = 8.3cm 0 0 0, clip]{figs/paper1/HRBogo.pdf}
\caption{Trajectories of the partner mode  $(\phi^{\rm in}_{-\om})^*$.}
\end{subfigure}
\caption{Space-time representation of the near horizon trajectories followed by wave packets made 
with right moving (with respect to the fluid) 
{\it in} modes in the presence of superluminal dispersion, {\it e.g.}, that given in \eqref{dispr},
and in the coordinate system $(t,x)$ of \eqref{metric}.
We have also indicated their asymptotic amplitudes as given in Eq.~(\ref{BogHR}). } 
\label{BogoBH} 
\end{figure}

In this section, the properties of HR 
are approximately determined by making use of connection formulae that relate 
asymptotic solutions of Eq.~(\ref{modequ}).
Before describing this 
procedure in precise terms, let us briefly explain it. 
For the stationary profiles we consider, {\it i.e.}, with $v$ defined on the entire real axis and asymptotically constant, because of dispersion, the Bogoliubov transformation encoding the Hawking effect has the standard form of a scattering matrix. It should be stressed that this is not the case for the relativistic fields. 
In that case indeed, as we saw in \Sec{HRNHR_Sec} and \ref{HRrelat_Sec}, wave packets propagated backwards in time hug onto 
the horizon for arbitrary long time, and thus never transform into waves incoming from an asymptotic region. 
Instead, when there is dispersion, \eqref{NHRtraj} is followed only for a finite
time, and wave packets (propagated backwards in time) leave the near horizon region and reach, for superluminal dispersion,
$x = -\infty$, see Fig.~\ref{BogoBH}. For this reason, the question of the robustness of Hawking radiation against dispersion is non trivial. Indeed, the specification of the {\it in}-vacuum is much different, since in one case it is defined as the local vacuum in a close vicinity of the horizon (see \Sec{eternalHR_Sec}), and in the other one, it is defined asymptotically.

When wave packets reach the asymptotic regions where $v$ is constant, they can be decomposed in terms of stationary plane waves $e^{- i \om t }e^{i p_\om x}$.
Hence, the definition of the {\it in} and {\it out} modes is the standard one, see {\it e.g.}, the scattering in a constant electric field~\cite{Primer}. The {\it in}-modes $\phi_\om^{\rm in}$ are solutions of \eqref{modequ} such that the group velocity $v_{\rm gr} = (\partial_\om p_\om)^{-1}$ of their 
asymptotic branches is oriented toward the horizon for one of them only. Hence when 
forming a wave packet of such modes, it 
initially describes a single packet traveling toward the horizon,
whereas at later times it describes several packets moving away from the horizon. Similarly,
the {\it out} modes $\phi_\om^{\rm out}$ contain only one asymptotic branch with a group velocity oriented 
away from the horizon.

At this point it should be noticed that the dimensionality of these two sets depends 
on the asymptotic values of $v$. 
When $v(x)$ contains one horizon, {\it i.e.}, crosses $-1$ once, the dimensionality is $3$ 
below the threshold frequency $\om_{\rm max}$ of \eq{ommax}: for $ \om_{\rm max} > \om > 0$, 
there is one positive norm left moving mode (with respect to the fluid, but 
not the lab) $\phi^{\rm left}_\om$, and a pair of right moving ones of opposite norm that we shall call 
$\phi_\om$ and $(\phi_{-\om})^*$ according to the sign of their norm. 
Hence the scattering matrix is $3 \times 3$. 
However, when $v$ is smooth enough, 
$\phi^{\rm left}_\om$ essentially decouples. This has been numerically shown in~\cite{Macher09}, and will be mathematically justified in 
\Sec{modeanalys}. Hence to a very good approximation, one recovers a $2 \times 2$ matrix characterizing right moving modes only. From now on we shall work within this approximation. 

Introducing the {\it in} and {\it out} sets of modes, they are related by
\be
\begin{pmatrix} \phi_\om^{\rm in} \\ (\phi_{-\om}^{\rm in})^* \end{pmatrix} =\begin{pmatrix} \alpha_\om & \beta_\om \\ \tilde \beta_\om & \tilde \alpha_\om \end{pmatrix} \cdot \begin{pmatrix} \phi_\om^{\rm out} \\ (\phi_{-\om}^{\rm out})^* \end{pmatrix} \label{BogHR}.
\ee
Because starred modes have a negative norm, the matrix is an element of $U(1,1)$. That is, the coefficients obey:
\bsub \label{scatrel}
\bea
|\alpha_\om|^2 - |\beta_\om|^2 &= 1, \\
\alpha_\om^*\tilde \beta_\om - \beta_\om^*\tilde \alpha_\om &= 0, 
\eea 
\esub
and $|\beta_\om|^2 =  |\tilde \beta_\om|^2 $. Hence, when working in the {\it in} vacuum, the mean number of emitted pairs of quanta of opposite frequency is 
\be
\bar n_\om = \vert \beta_\om \vert^2.
\label{e13}
\ee
In the relativistic limit, {\it i.e.} $\Lambda\to\infty$, one gets the standard result
\be
\bar n_\om^{\rm relativistic} = \frac{1}{e^{2 \pi \om/\kappa} - 1},
\label{relats}
\ee
which was obtained in \Sec{HRrelat_Sec}. We point out that because of dispersion, this is only true for $0 < \om < \om_{\rm max}$. When $\om > \om_{\rm max}$, only one mode subsists, the Bogoliubov transformation \eqref{BogHR} no longer exists, and the flux \eqref{e13} identically vanishes.

Our aim is to compute the coefficients of \eqref{BogHR} using a connection formula. To this end, we  
first identify the various asymptotic solutions, and then we evaluate 
the globally defined solutions which we match to the asymptotic ones.
These techniques have already been used in~\cite{Brout95,Corley97,Himemoto00,Unruh04}. 
The novelty of our treatment concerns a careful control of the various 
approximations involved in this procedure. This will enable us to 
control the leading deviations from \eqref{relats} which are due to dispersion, given $v(x)$.

\subsection{Mode analysis}
\label{modeanalys}
\subsubsection{The $x$-WKB approximation}

Since the WKB approximation fails near a turning point, we cannot compute
the coefficients of \eqref{BogHR} using this approximation. In fact, under this approximation one 
would get a trivial result, namely $\beta_\om = \tilde \beta_\om = 0$, 
$\vert \alpha_\om \vert = \vert \tilde \alpha_\om \vert = 1$. 
To get a non-trivial result, 
 we shall compute these coefficients by inverse Fourier 
transforming the modes in $p$-space and identifying the various terms sufficiently far away 
from the turning point, in a calculation that generalizes that of the Airy function~\cite{Olver,AbramoSteg}.

In this section, we present the calculation of the WKB modes of \eqref{wavequ} 
which generalizes the usual treatment since \eqref{wavequ} is not second order. For this, we write the mode as 
\be
\phi_\om(x) = e^{i\int^x k_\om(x') dx'}\, .
\label{wkbm}
\ee
Injecting this into \eqref{modequ}, we obtain 
\be \bal
(\om - v(x) k_\om)^2 - F^2\left(k_\om \right) =& -i \partial_x \left[\om v(x) - v^2 k_\om + \frac12 \partial_k F^2\left(k_\om\right) \right] \\
& - \frac{4 k_\om \partial_x^2  k_\om + 3 (\partial_x k_\om)^2 - i \partial_x^3 k_\om}{\Lambda^2} .
\eal  \label{rica} \ee 
For definiteness and simplicity, the last term is given for 
$F(k) = k^2 + k^4/\Lambda^2$ but can be generalized to any polynomial dispersion relation. 
So far, this equation is exact. It is known as a Riccati equation \cite{Olver} 
and was already used in the present context in~\cite{Corley97}. 
It is adapted to a perturbative resolution where
the different terms are sorted in order of derivatives~\cite{BirrellDavies,Gottfried}, here
spatial gradients. Hence we write $k_\om$ as
\be
k_\om = k_\om^{(0)} + k_\om^{(1)} + k_\om^{(2)} +...
\label{series}
\ee
where the superscript gives the number of derivatives (one way to 
sort the terms of \eqref{rica} is to make the scale change $x \rightarrow \lambda x$. The superscript then stands for the power of $1/\lambda$).
It is easy to show that $k_\om^{(0)}(x)= p_\om(x)$, 
the classical momentum, solution of the Hamilton-Jacobi equation \eqref{disp2}.
The second equation is not less remarkable: 
it is a total derivative and it has the universal form
\be
k^{(1)}_\om = \frac i2 \partial_x \ln \left[\om v(x) - v^2p_\om +\frac12 \partial_p 
F^2(p_\om) \right] = \frac i2 \partial_x \ln \left[F(p_\om)\, v_{gr}(p_\om) \right],
\label{k1k0}
\ee
where $v_{gr} = 1/\partial_\om p_\om$ is the group velocity. 
Since $k_\om^{(1)}$ is purely imaginary, it governs the mode amplitude. 
\eqref{wkbm} constructed with $k_\om = p_\om + k_\om^{(1)}$ 
 gives the generalized $x$-WKB expression 
\be
\varphi_\om^{\rm WKB}(x) = \sqrt{\frac{\partial p_\om}{\partial \om}} \frac{e^{i\int^x p_\om(x') dx'}}{\sqrt{4\pi  F(p_\om) }}.
\label{xWKB}
\ee
The prefactor, given by \eqref{k1k0}, is such that WKB modes of \eq{xWKB} are exactly normalized with respect to the scalar product of \eq{scalt}. More precisely, when computing \eq{scalt} with \eq{xWKB} at leading order\footnote{It would be interesting to understand if and how to go beyond this approximation.} in $\om - \om'$, we obtain the Dirac normalization of \eq{scalom}. This ensure that tight wave packets made of WKB modes have a conserved norm. In addition, \eqref{rica} also guarantees that the expansion \eqref{series} of $k$ is alternating: even terms  are real, while odd ones are imaginary. 

Taking the Fourier transform of \eq{xWKB} evaluated at the saddle point gives the $p$-WKB mode 
\be
\tilde \varphi_\om^{\rm WKB}(p) = \sqrt{\frac{\partial X_\om(p)}{\partial \om}} \frac{e^{iW_\om(p)}}{\sqrt{4\pi  F(p) }} \label{YYY}.
\ee
These $p$-WKB modes play an important role, because their domain of validity (in parameter space) differs from the $x$-WKB of \eq{xWKB}. As we shall see in \Sec{SecpWKB}, they will provide a much better approximation in the near horizon region. 

All over the study of this chapter, we will compute every error term, in order to explicit the domain of validity of the relativistic result \eqref{relats}. For this, we define the parameter 
\be
\Delta(x) = \frac{\Lambda}{2\kappa}|2(1+v(x))|^{\frac32} \label{Delta}.
\ee
As we shall see, all errors will be characterized by this single parameter.

\subsubsection{The six roots $p_\om$ far away from the turning point}

Since we work in a weak dispersive regime, {\it i.e.} $\Lambda \gg \kappa$, and since HR is related to 
frequencies $\om \sim \kappa$, we have $\om \ll \Lambda$ for relevant frequencies. 
Moreover, since we impose that $D_{\rm as}$ is not too small, we also 
have $\om \ll \om_{\rm max}$, see \eqref{ommax}.
Hence $\om_{\rm max}$ only concerns the high frequency properties of HR, which we no longer study. 
We focus instead on frequencies $\om \sim \kappa$. Even for such frequencies, the expressions of $p_{\om}(x)$, solutions of Eq.~(\ref{xHJ}), are 
complicated. However, their exact expression is not needed.
It is sufficient to estimate them far away from the turning point, in order to build the 
mode basis. 

\begin{figure}[!ht]
\begin{subfigure}[b]{0.5\textwidth}
\includegraphics[scale=0.6]{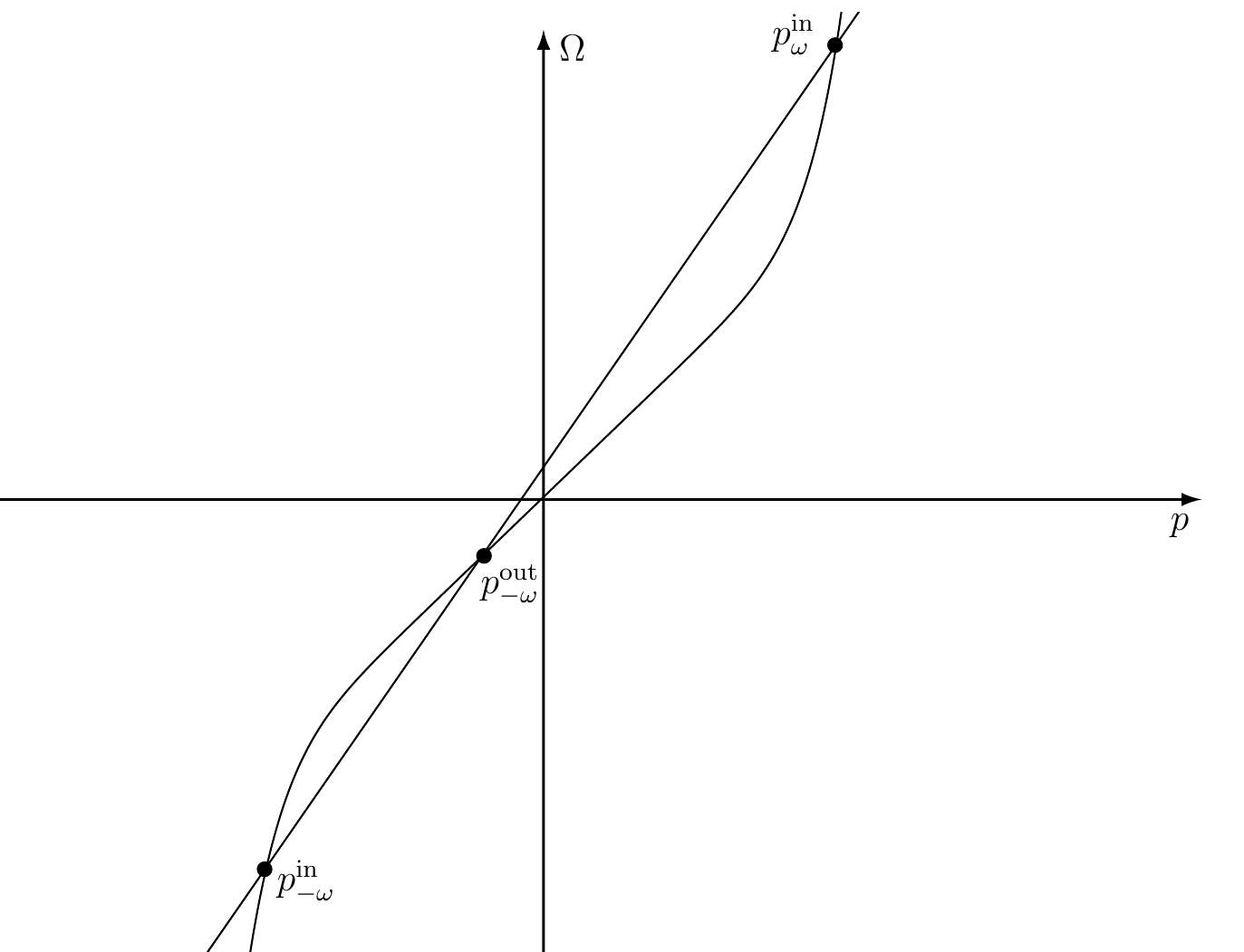}
\caption{Resolution for $1+v<0$.}
\end{subfigure}
\hspace{2mm}
\begin{subfigure}[b]{0.5\textwidth}
\includegraphics[scale=0.6]{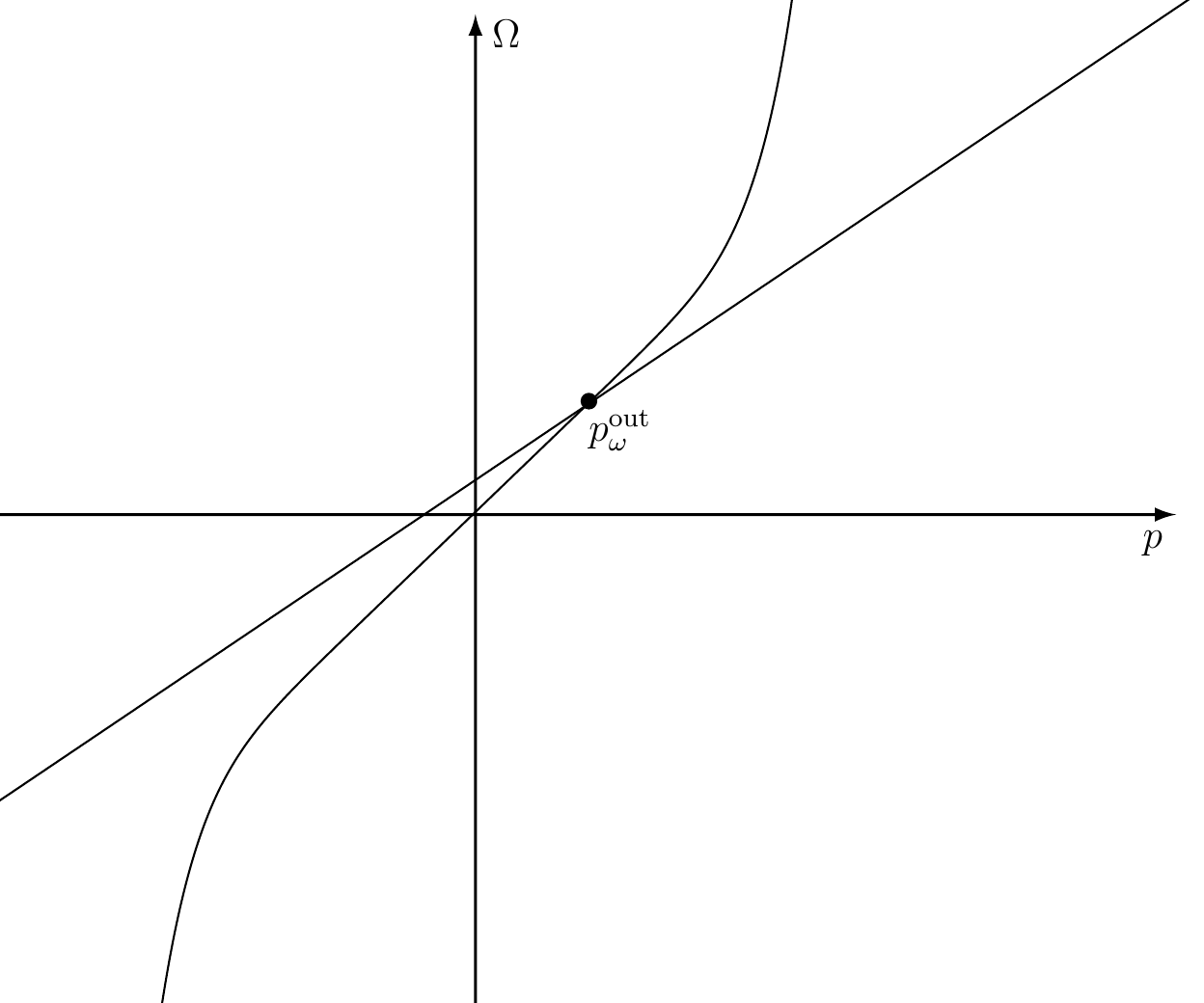}
\caption{Resolution for $1+v>0$.}
\end{subfigure}
\caption{Graphical resolution of \eqref{xHJ} restricted to right moving modes.
The dispersion relation $F = p + p^3/2\Lambda^2$ and 
the straight lines $\Om = \om - v p$ are plotted for a common value of $\om$ and two 
different values of $v$. When $1+v<0$ and we have 3 real roots given by Eqs. (\ref{plusinroot}, \ref{minusinroot}, \ref{minusoutroot}), and when $1+v>0$ and we have the single real root of \eqref{plusoutroot}. The two complex roots of Eqs. (\ref{grroot}, \ref{decroot}), are not represented.} 
\label{graphroots} 
\end{figure}

The denomination of the roots we use is based on that of the corresponding mode,
which is itself based on the sign of the group velocity, as explained in \Sec{profileSec}. 
Moreover, we exploit the fact that for right moving modes 
the sign of the norm is that of the corresponding root $p_\om$ (see {\it e.g.}~\cite{Corley96}). 
Therefore, for $\om > 0$, positive roots shall be called $p_\om$, whereas 
negative roots shall be called $p_{-\om}$ in accord with the fact that 
negative norm modes are called $\left( \phi_{-\om}\right)^*$, 
see Fig.~\ref{graphroots}. 

Using this terminology, on the left of the turning point, one finds 
\bsub \label{left_roots}
\bea
p_\om^{\rm in} &=& \Lambda \sqrt{-2(1+v)} - \frac{\om}{2(1+v)} (1+ O(y)), \label{plusinroot}\\
p_{-\om}^{\rm in} &=& -\Lambda \sqrt{-2(1+v)} - \frac{\om}{2(1+v)} (1+ O(y)), \label{minusinroot}\\
p_{-\om}^{\rm out} &=& -\frac{\om}{1+ v} (1+ O(y^2)), \label{minusoutroot}
\eea \esub  
where the small parameter $y$ is related to $\Delta$ of \eqref{Delta} by 
\be
y = \frac{ \om/\kappa}{\Delta(x)}. 
\label{y}
\ee
Far away from the turning point of \eqref{xtp}, $y \ll 1$ and our expressions are reliable approximations.
In the right (subsonic) region, one has only one real root 
\be
p_\om^{\rm out} = \frac{\om}{1+ v} (1+ O(y^2)).\label{plusoutroot}
\ee
On this side there are also two complex solutions which do not correspond to any classical trajectory. 
However, when looking at the solutions of \eqref{modequ}, 
they govern the growing mode $\phi_\om^{\uparrow}$ and the decaying mode
$\phi_\om^{\downarrow}$ exactly as real roots govern WKB modes. These roots are 
\bsub \label{gr_dec_roots}
\bea
p_\om^{\uparrow} &=& - i\Lambda \sqrt{2(1+v)} - \frac{\om}{2(1+v)} (1+ O(y)),\label{grroot}\\
p_{\om}^{\downarrow} &=& i\Lambda \sqrt{2(1+v)} - \frac{\om}{2(1+v)} (1+ O(y)). \label{decroot}
\eea \esub 
On this side of the horizon, the corrections are again governed by $y$ of Eq.~(\ref{y}). 
Hence, for $\om \sim \kappa$, the corrections to the roots are on both sides 
controlled by $\Delta$ of \eqref{Delta}. It should be also noticed that
the errors for the two {\it out} roots are subdominant with respect 
to the other ones.

\subsubsection{The six WKB modes far away from the turning point}

In order to distinguish globally defined modes from their WKB approximations, 
the former shall be noted $\phi_\om$, and the latter $\varphi_\om$. 
Since we consider only $\om>0$, negative frequency modes (of negative norm) 
shall be written as $\left( \phi _{-\om}\right)^*$ or $\left( \varphi _{-\om}\right)^*$.

Sufficiently far away from the turning point, {\it i.e.} for $\Delta \gg 1$, 
the WKB modes offer reliable solutions of \eqref{modequ}. 
They can thus be used as a basis to decompose globally defined solutions. 
We also assume that $D_{\rm lin}^{R,L}$ of \eqref{Dlin}  
are large enough so that one can be at the same time be far away from the turning point
and still in the region where $v$ is linear. 
Using the expressions for the six roots in this region, 
\eqref{Som}, and neglecting the common phase depending on $\theta_0$ of \eqref{theta0},  
on the left side of the horizon one obtains 
\bsub \bea
\varphi_\om^{\rm in} &\sim& [2\Delta(x)]^{-\frac12} \frac{e^{-i\frac{\om}{\kappa} \ln(\Lambda \sqrt{2\kappa})}}{\sqrt{4\pi \kappa(1+ \kappa |x|)}} |x|^{-i\frac{\om}{2\kappa}} e^{- i \frac23 \Delta(x)}, \label{plusin}\\
\left( \varphi_{-\om}^{\rm in} \right)^* &\sim& [2\Delta(x)]^{-\frac12} \frac{e^{-i\frac{\om}{\kappa} \ln(\Lambda \sqrt{2\kappa})}}{\sqrt{4\pi \kappa(1+ \kappa |x|)}} |x|^{-i\frac{\om}{2\kappa}} e^{i \frac23 \Delta(x)},\label{minusin}\\
\left( \varphi_{-\om}^{\rm out}\right)^* &\sim& e^{i\frac{\om}{\kappa} -i \frac{\om}{\kappa} \ln (\frac{\om}{\kappa}) } \frac{|x|^{i\frac{\om}{\kappa}}}{\sqrt{4\pi \om}}.\label{minusout} 
\eea \esub 
On the right one gets, 
\bsub \bea
\varphi_{\om}^{\rm out} &\sim& e^{i\frac{\om}{\kappa} -i \frac{\om}{\kappa} \ln (\frac{\om}{\kappa}) } \frac{|x|^{i\frac{\om}{\kappa}}}{\sqrt{4\pi \om}},\label{plusout}\\
\varphi_{\om}^{\downarrow} &\sim& \frac{e^{\frac{\om \pi}{2\kappa}}}{[2\Delta(x)]^{\frac12}} \frac{e^{-i\frac{\om}{\kappa} \ln(\Lambda \sqrt{2\kappa})}}{\sqrt{4\pi \kappa(1- \kappa |x|)}} |x|^{-i\frac{\om}{2\kappa}} e^{- \frac23 \Delta(x)}, \label{decmode} \\
\varphi_{\om}^{\uparrow} &\sim& \frac{e^{-\frac{\om \pi}{2\kappa}}}{[2\Delta(x)]^{\frac12}} \frac{e^{-i\frac{\om}{\kappa} \ln(\Lambda \sqrt{2\kappa})}}{\sqrt{4\pi \kappa(1- \kappa |x|)}} |x|^{-i\frac{\om}{2\kappa}} e^{\frac23 \Delta(x)}.\label{grmode} 
\eea \esub 
We now comment these expressions. Firstly, whereas the normalization of the four oscillating 
modes is standard and based on the conserved scalar product of \eqref{scalt}, those of 
the decaying $\varphi_{\om}^{\downarrow}$ and growing $\varphi_{\om}^{\uparrow}$ modes 
follow from \eqref{xWKB} and the fact that $S_\om$ of \eqref{Som} 
is complex since the roots of Eqs. (\ref{grroot}) and (\ref{decroot}) are. 

Secondly, since these two modes do not appear in the `on-shell' Bogoliubov transformation of \eqref{BogHR}, 
we should explain why we are still considering them. First, the forthcoming connection formula 
will be a transfer matrix relating the general solution on each side of the horizon.
Thirdly, when considering problems with several horizons, these modes
could contribute to the `on-shell' S-matrix if they live in a finite (supersonic)  size region between two horizons, see Chapter \ref{mass_Ch} and~\cite{aQuattro}.

Finally, the relative errors of the two {\it out} modes are
\be
O\left(\frac{\om^2/\kappa^2}{\Delta(x)^2} \right), \label{outerror}
\ee
whereas those of the four other modes are 
\be
O\left(\frac{1+\om/\kappa}{\Delta(x)} \right). \label{inerror}
\ee
To get these expressions,  we must take into account two sources of errors: 
 those coming from the approximate roots, see \eq{y}, and those from the $x$-WKB approximation. 
As we shall now see, they all depend only on the parameter $\Delta$.

\subsubsection{WKB deviations due to dispersion} 

In the preceding paragraph, we obtained a mode basis in the limit of weak dispersion, 
that is, for $\Lambda$ large enough.  The aim of this section is to precise the meaning of `$\Lambda$ large enough' 
by computing the scaling of the errors made when building the WKB basis. 

To proceed, we estimate the next order term of \eqref{series} 
in the limit of weak dispersion (WD), {\it i.e.} by dropping terms of order 2 in $1/\Lambda$. 
In this regime, $k^{(2)}$ is solution of
\be
-2\left[ F(p_\om) \, v_{gr}(p_\om) \right] k_{WD}^{(2)}(x) = \left[ -i\partial_x + k_\om^{(1)} \right]\cdot \left[  k_\om^{(1)}\, \partial_k(F v_{gr}) \right] \label{k2}.
\ee
Because we master the modes near the horizon in the $p$-representation (see \Sec{SecpWKB} and \ref{CF}), we are only interested in the error accumulated from infinity, where the $x$-WKB approximation becomes exact, to a certain $x_{\rm pasting}$ at the edge of the near horizon region. Its precise characterization will be obtained in \eqref{Dp}. 
This error is estimated by evaluating the integral of $k_{WD}^{(2)}$ from $x_{\rm pasting}$ till $\infty$. 
Indeed the exact mode $\phi_\om$ can be approximated by 
\be
\phi_\om \simeq \varphi^{\rm WKB}_\om e^{i\int^x k_{WD}^{(2)}(x') dx'} \simeq \varphi_\om^{\rm WKB}
(1 + \epsilon(x)). \label{WKBapprox}
\ee 
To evaluate $\varphi^{\rm WKB}_{\om}$ we use \eqref{xWKB} and the approximate roots of Eqs.~\eqref{left_roots},\eqref{plusoutroot}, and \eqref{gr_dec_roots}. This introduces extra errors 
governed by $y$ of \eqref{y}. Hence, near $x_{\rm pasting}$ the total error is 
\be
\epsilon_T  \simeq \int_{\infty}^{x_{\rm pasting}} k_{WD}^{(2)}(x') dx' + O(y(x_{\rm pasting})). 
\ee
This error term behaves quite differently for {\it in} and {\it out} modes, hence we shall study it 
separately. 
\begin{itemize}
\item For the {\it in} modes, solving \eqref{k2}, we get
\be
k_{WD}^{(2)} = \frac{9 v'^2}{16\Lambda |1+v|^{\frac52}} - \frac{3 v''}{4\Lambda |1+v|^{\frac32}}.
\ee
Since this is not a total derivative, 
the integral depends on what happens all along the way from $\infty$ to $x_{\rm pasting}$. However, when the profile $v$ is smooth enough, the accumulated error is essentially 
\be
\epsilon^{\rm in}_T \sim \left( \frac{v'}{\Lambda |1+v|^{\frac32}} \right)_{x = x_{\rm pasting}} +O(y_p) = \frac1{\Delta_p^L} + \frac{\om/\kappa}{\Delta_p^L}.
\ee

\item For the {\it out} modes, the leading order correction arises from $k^{(1)}$. 
Indeed, in the limit $\Lambda \to \infty$, the {\it out} modes are WKB exact because of conformal invariance 
and $k^{(1)}_{\Lambda \to \infty} = 0$. 
Therefore, for finite $\Lambda$, $k^{(1)}$ will be the dominant contribution to the error term of \eqref{WKBapprox}. 
One finds
\be
k_{WD}^{(1)} = - \frac{6 \om^2 v'}{\Lambda^2 (1+v)^4}.
\ee
This means that
\be
\epsilon^{\rm out}_T \sim \left( \frac{\om^2}{\Lambda^2 |1+v|^3} \right)_{x = x_{\rm pasting}} = \frac{\om^2/\kappa^2}{\Delta_p^2}.
\ee
Here the pasting happens on both sides, hence in the latter expression, one should understand $\Delta_p^L$ for $\left(\varphi_{-\om}^{\rm out}\right)^*$ and $\Delta_p^R$ for $\varphi_\om^{\rm out}$. As expected, the corrections to red-shifted {\it out} modes are subdominant with respect to those of {\it in} modes. 

\end{itemize}

It is interesting to notice that  the validity of the $x$-WKB approximation in a given flow 
and for a given dispersion relation depends on the exact form of the mode equation associated with \eqref{disp2}. 
Indeed, when the mode equation is not conformally invariant in the limit $\Lambda \to \infty$
there is an extra validity condition:
\be
\left| \frac{v'}{\om}\right| \ll 1.
\ee
It is due to the fact that for low frequencies, the left and right moving modes mix
even in the absence of dispersion. This mixing was studied in \cite{Macher09b}, and it was numerically shown that 
these effects stay (in general) subdominant. 

The lesson to retain from this analysis, is that errors to WKB modes are {\it bounded} by the inverse of the dimensionless parameter $\Delta$ of \eq{Delta}. 
Far away from the horizon $\Delta$ is large and the WKB approximation is accurate. More precisely $\Delta$ becomes of order 1 near $x=x_{tp}$ of \eqref{xtp} evaluated for $\om \sim \kappa$. Hence for these frequencies, which are the relevant ones for HR, $\Delta \gg 1$ is reached for $x/x_{tp} \gg 1$. One also sees that at fixed $x$, $\Delta$ grows like $\Lambda/\kappa\to \infty$, hence $\Delta$ also governs the dispersion-less limit. Moreover, we saw that the errors on the {\it out} modes, of low momentum $p$, are subdominant with respect to those of {\it in} modes which have a high momentum. As a result, $D_{\rm lin}^L$ of \eqref{Dlin} will be more relevant than $D_{\rm lin}^R$ to characterize deviations with respect to the relativistic case.

\subsection{Globally defined modes in the near horizon region}
\label{SecpWKB}
\subsubsection*{The $p$-representation}
To accurately describe the behavior of the modes across the horizon one cannot use the $x$-WKB modes  
of the former section. Rather, one should work in $p$-space, and look for solutions of the form:
\be
\phi_{\om}(x) = \int_{\mathcal C} \tilde{\phi}_{\om}(p)\,  e^{ip x} \frac{dp}{\sqrt{2\pi}},
\label{ft}
\ee
where $\mathcal C$ is a contour in the complex $p$-plane. If it is well chosen, {\it i.e.} such that the integral converges and integrations by part can be performed, then it is sufficient 
that the dual mode $\tilde{\phi}_{\om}$ satisfies 
\be
\left(- \om + p \hat{v} \right) \left(- \om + \hat{v}p \right)\tilde{\phi}_{\om} = F^2(p) \tilde{\phi}_{\om} \label{pmodequ},
\ee
where $\hat v = v(\hat x) = v(i\partial_p)$. We should notice that \eqref{ft} is a standard Fourier transform only if 
$\mathcal C$ is on the real line, something we shall not impose. 
The main interest of considering generalized contours
is that it will enable us to compute {\it all} solutions of \eqref{modequ}, including the growing ones.

Because we only need the behavior of the modes near the horizon, the operator
$\hat{v}$ in \eqref{pmodequ} can  be replaced by $\hat v= -1 + i\kappa \partial_p$.
Hence one gets a second order ODE, irrespectively of $F(p)$.
The advantages to work in $p$-space are then clear~\cite{Brout95,Balbinot06}. 
Firstly, the solutions of \eqref{pmodequ} can be (exactly) written 
as a product 
\be
\tilde{\phi}_{\om}(p) = \chi(p) \, e^{-i \frac{p}{\kappa}} \times 
\frac{p^{-i\frac{\om}{\kappa} - 1}}{\sqrt{4\pi \kappa}},
\label{pchi}
\ee
where $\chi$ obeys the $\om$-independent equation:
\be
-\kappa^2 p^2 \partial_p^2 \chi = F^2(p) \chi(p),
\label{chiequ}
\ee
and where ${p^{-i\frac{\om}{\kappa} - 1}}$ is independent of $F$. 
Hence, the deviations due to the dispersion $F$ are entirely encoded in $\chi$.
The origin of this factorization comes from the underlying structure of de Sitter space, as discussed after \eq{Dlin}.
In addition, when considering the limit $\Lambda \to \infty$ in \eqref{modequ}, 
${p^{-i\frac{\om}{\kappa} - 1}}$ is exactly the relativistic (conformaly invariant) mode in $p$ space. 

Secondly, unlike the original equation in $x$, \eqref{chiequ} is perfectly regular. Moreover, 
when dispersion effects are weak, {\it i.e.}, $\Lambda/\kappa \gg 1$, the $p$-WKB approximation,
so called not to confuse it with that used in the former Section, is very good.
It generalizes what is done for the Airy function~\cite{AbramoSteg,Olver} 
where the mode equation in $p$-space is WKB exact. 
At this point, it is worth comparing the expression of the $p$-WKB modes in general and near the horizon. 
Using \eqref{YYY} and \eqref{Wom}, one finds 
\be
\tilde{\varphi}_{\om}(p) = \frac{e^{i \left(- \frac{\om}{\kappa} \ln(p) + \int \frac{F(p') - p'}{\kappa p'}dp' \right)}}{\sqrt{4\pi \kappa\, p F(p)}} \times \left(1+O\left( \frac{\kappa}{\Lambda}\right) \right) 
\label{dualmode}.
\ee
As we saw in \eq{YYY}, $\tilde{\varphi}_{\om}$ is universally governed by $W_\om$ and $X_\om$, irrespectively of $F(p)$ and $v(x)$. 
This equation \eqref{dualmode} now shows how exactly $F(p)$ enters in $\tilde{\varphi}_{\om}$ in the near horizon region. In this region, since the mode equation is second order in $\partial_p$, the corrections are uniformly {\it bounded} by $\kappa/\Lambda$~\cite{Olver}. 
We also notice that using $\chi^*$ instead of $\chi$ in \eqref{pchi} would describe a left moving mode 
of negative norm~\cite{Balbinot06}. This shows that the corrections to the 
$p$-WKB approximation describe creation of pairs of left and right moving modes, as in cosmology~\cite{Macher08}. 
It is also of interest to notice that in models where left and right movers stays completely decoupled~\cite{Brout95,Schutzhold08}, 
the $p$-WKB modes of \eqref{dualmode} are {\it exact} solutions in the near horizon region. 

We finally notice that 
\eqref{dualmode}, as the relativistic mode ${p^{-i\frac{\om}{\kappa} - 1}}$, 
is well defined only when having chosen the 
branch cut of $\ln p$~\cite{Brout95,Corley97,Himemoto00,Unruh04}. As explained below, different possibilities, and different contours ${\cal C}$, lead to different modes.

\subsection{The various modes in the near horizon region} 
\label{CF}
Using $F(p)$ of \eqref{dispr} and \eqref{dualmode}
one gets
\be
\varphi_{\om}(x) = \frac1{\sqrt{4\pi \kappa}} \int_{\mathcal C} \frac{e^{i (px- \frac{\om}{\kappa} \ln(p) + \frac{p^3}{6\Lambda^2 \kappa})}}{(1+\frac{p^2}{2\Lambda^2})^{\frac12}} \frac{dp}{p\sqrt{2\pi}} \label{contourmode}.
\ee
The forthcoming analysis generalizes former treatments for several  
reasons: 
\begin{enumerate}
\item Unlike~\cite{Corley97,Himemoto00,Unruh04}, we shall consider contours that are not homotopic to the real line.
This will allow us to obtain the general connection formula which includes the growing mode.
\item We will make use of mathematical theorems of asymptotics~\cite{Olver} 
 under their exact form. This will lead to the proper identification of the validity conditions in \Sec{validity}.
\item We will compute the {\it phases} of the Bogoliubov coefficients. 
These are essential to compute the correlation pattern (see \eqref{correl}) and 
in the presence of several horizons, see Chapter \ref{laser_Ch}.
\end{enumerate}

To evaluate \eqref{contourmode}, the first thing to take care of is the convergence of the integral. Indeed, for large $p$, the dominant term is of the form $e^{ip^3}$, hence, we should impose that our contour $\mathcal C$ goes to infinity in regions where $\Im(p^3) > 0$. The second step is to perform the integration with a saddle point approximation.  For this, we make a change of variable such that the $p$ and $p^3$ terms are of the same order for all values of 
$\Lambda/\kappa$. We thus write $p = \Lambda \sqrt{2 \kappa |x|}t$ and get:
\be
\varphi_{\om}(x) = \frac{e^{-i\frac{\om}{\kappa} \ln(\Lambda \sqrt{2\kappa |x|})}}{\sqrt{4\pi \kappa}} \int_{\mathcal C} \frac{e^{-i\frac{\om}{\kappa} \ln(t)}}{(1+t^2 \kappa |x|)^{\frac12}} e^{i\Delta(x) \left({\rm sign}(x) t + \frac{t^3}3\right)} \frac{dt}{t\sqrt{2\pi}}.
\label{t}
\ee
Hence we see that the large parameter $\Delta(x)$ defined in \eqref{Delta} can be used to perform a saddle point approximation (see $z$ in \eqref{SP}). Therefore $\Delta$ will govern the deviations due to this approximation.

For completeness we recall the saddle point theorem. If $z$ is some large parameter:
\be
\int_{\mathcal C} A(p) e^{i z f(p)} \frac{dp}{\sqrt{2\pi}} = \sum_j A(p_j) \frac{e^{i z f(p_j)}}{\sqrt{-if''(p_j) z}} \ \ \left(1+ O\left( \frac{E_j(f,A)}z \right)\right) \label{SP},
\ee
where the $p_j$ are saddle points of $f$, {\it i.e.} $f'(p_j) = 0$ of smallest imaginary part, and 
where the square root takes its principal value. 
This formula is valid if and only if one can deform $\mathcal C$ such that $\Im \left( f(p) - f(p_j) \right)$ is always 
positive~\cite{Olver}. The first correction $E_j(f,A)$ involves higher derivatives of $f$ and $A$ evaluated at $p_j$ 
\be
E_j(f,A) = \left(-i \frac{A''}{f''} + i\frac{f'''\, A'}{(f'')^2} -i \frac{5(f''')^2A - 3f''''\, f''\, A}{12(f'')^3} \right).
\ee

\subsubsection{The decaying mode}
\label{decayingSec}
We saw in \Sec{eikonal_Sec} that for negative frequency, the particle is reflected near the horizon, at the turning point of \eqref{Xp}.
Hence we expect that the corresponding mode will decay on the other side, for $x>0$.
This behavior is implemented by imposing that the branch cut is $-i\mathbb R_+$ and that the contour is homotopic to the real line, 
as we now show.

To evaluate \eqref{contourmode} for  $x>0$, we use \eqref{t} and perform a saddle point calculation. The saddle points obey $1+t^2 = 0$. Just like in the Airy case~\cite{AbramoSteg,Olver}, 
only $t = i$ is relevant and its contribution is 
\be
\varphi_{\om}(x) = e^{-i\frac{\pi}2} \frac{e^{\frac{\om \pi}{2\kappa}}}{[2\Delta(x)]^{\frac12}} \frac{e^{-i\frac{\om}{\kappa} \ln(\Lambda \sqrt{2\kappa}) }}{\sqrt{4\pi \kappa(1- \kappa |x|)}} |x|^{-i\frac{\om}{2\kappa}} e^{- \frac23 \Delta(x)} \times \left( 1 + O\left(\frac{1+\frac{\om^2}{\kappa^2}}{\Delta(x)} \right) \right). \label{firstSP}
\ee
As required, the mode decays on this side of the horizon. The error term has been estimated using
\eqref{SP}. (More precisely, when computing $E_j$, we found a bounded function of $\om/\kappa$ and $x$ times 
$(1+ \om^2/\kappa^2)/\Delta$. This justifies our expression.) 
We notice that this expression coincides with $\varphi_{\om}^{\downarrow}$ 
of \eqref{decmode} up to a 
factor. Therefore, this factor defines the scattering coefficient 
and its $x$-dependent correction:  
\be
\varphi_{\om} = \varphi_{\om}^{\downarrow} \times \left( e^{-i\frac{\pi}2} \right)\times 
\left( 1 + O\left(\frac{1+\frac{\om^2}{\kappa^2}}{\Delta(x)} \right) \right). 
\ee
We underline that this identification introduces no new errors because those due to the saddle point
are of the same order as those already present in \eqref{decmode}. 
Since the general method is now understood, we proceed with the same mode on the other side of the horizon, and then apply the same method for two other modes
so as to get the general connexion formula.

\begin{figure}[!ht]
\begin{subfigure}[b]{0.5\textwidth}
\includegraphics[scale=0.55]{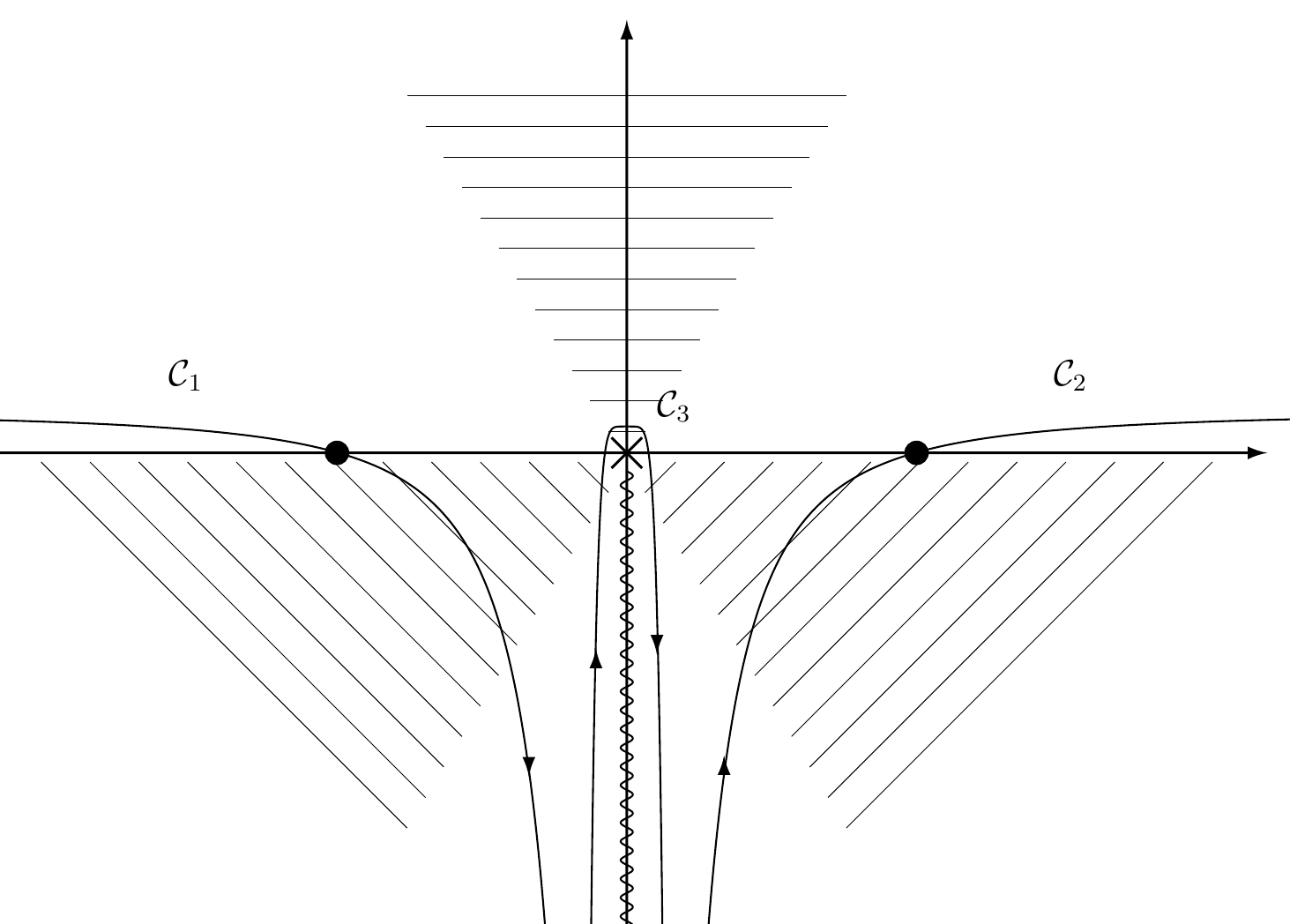}
\caption{$x$ negative}
\end{subfigure}
\hspace{2mm}
\begin{subfigure}[b]{0.5\textwidth}
\includegraphics[scale=0.55]{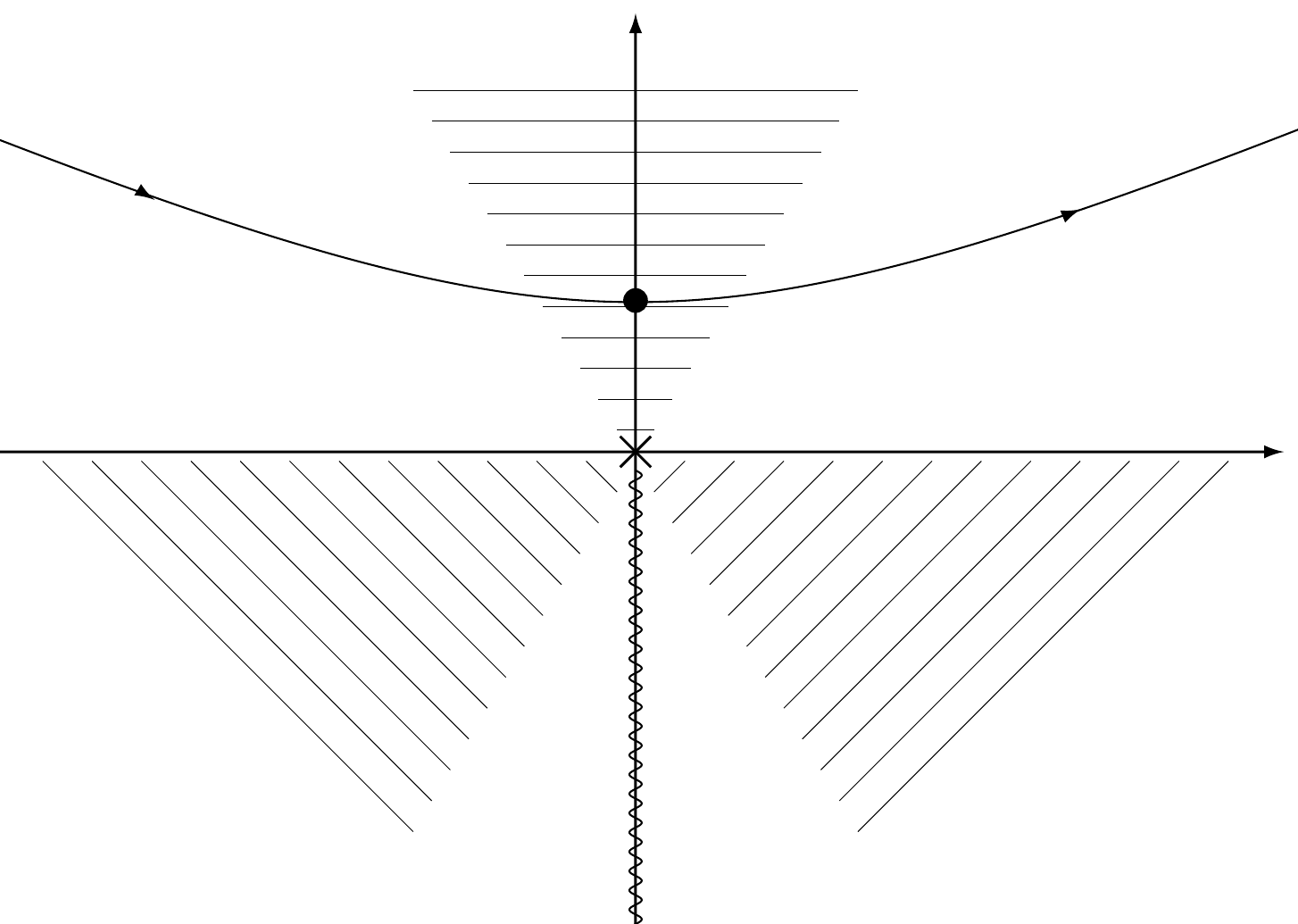}
\caption{$x$ positive}
\end{subfigure}
\caption{Representation of the contours in the complex $p$-plane determining the decaying mode, for both signs of $x$. 
The hatched regions are the asymptotically forbidden ones, and the dots indicate the saddle points
that contribute to the integral. }
\label{contours} 
\end{figure}

For $x < 0$, the saddle points are now solutions of $1-t^2 = 0$. However, because of the branch cut, one must 
cut the contour into three separate branches, as shown in Fig.\ref{contours}. ${\mathcal C_1}$ and ${\mathcal C_2}$
 go from $\pm \infty + \epsilon$ and dive toward $-i\infty$ on each side of the branch cut. 
${\mathcal C_3}$ encircles the branch cut, and is necessary for the union of the $3$ new contours to be homotopic to the original one. Separating the three contributions, 
$\varphi_{\om} = \varphi^{\mathcal C_1} + \varphi^{\mathcal C_2} + \varphi^{\mathcal C_3}$,
$ \varphi^{\mathcal C_1}$ and  $\varphi^{\mathcal C_2}$ are evaluated by the saddle point method and, after identification with the WKB modes of \eqref{plusin} and  \eqref{minusin}, respectively give, 
\bsub \bea
\varphi^{\mathcal C_1} &=& \left(\varphi_{-\om}^{\rm in}\right)^* \times ( e^{\frac{\om \pi}{\kappa}} e^{i\frac{3\pi}4} )\times \left( 1 + O\left(\frac{1+\frac{\om^2}{\kappa^2}}{\Delta(x)} \right) \right)
\label{C1},\\
\varphi^{\mathcal C_2} &=& \varphi_\om^{\rm in} \times e^{i\frac{\pi}4}
\times  \left( 1 + O\left(\frac{1+\frac{\om^2}{\kappa^2}}{\Delta(x)} \right) \right).\label{C2}
\eea \esub 
To properly evaluate $\varphi^{\mathcal C_3}$, one cannot use the saddle point method. 
However, because the factor $e^{ipx}$ decays along 
$\mathcal C_3$, one can use a `dominated convergence theorem', 
 {\it i.e.} take the limit $\Lambda \rightarrow \infty$ in the integrand of \eqref{contourmode}. 
Using the Euler function, and $\varphi_{-\om}^{\rm out}$ of \eqref{minusout} we get
\bsub \bea
\varphi^{\mathcal C_3} &=& \left( \varphi_{-\om}^{\rm out} \right)^* \times  \left(- \sinh \left(\frac{\om \pi}{\kappa} \right) \sqrt{\frac{2\om}{\pi \kappa}} \Gam\left( -i \frac{\om}{\kappa}\right)e^{\frac{\om \pi}{2\kappa}} e^{-i\frac{\om}{\kappa} +i\frac{\om}{\kappa} \ln(\frac{\om}{\kappa})} \right) \\
&&\times \left( 1 + O\left( \frac{\kappa |x|(1+\om^3/\kappa^3)}{\Delta(x)^2} \right) \right). \label{gammalike}
\eea \esub 
The correction term has been calculated by expanding the integrand in \eqref{contourmode} to first order in $\kappa/\Lambda$ and computing the integral again with the use of the $\Gam$ function.

\subsubsection{The transmitted mode}
To get another mode, we keep the same contour but 
the branch cut is now taken to be $i\mathbb R_+$. 
As we shall see it corresponds to a transmitted mode. For $x< 0$ the saddle points still obey $1-t^2 = 0$, but we can now use the saddle point approximation 
because the branch cut is no longer in the way. 
Taking into account that on the negative real $t$-axis $\ln t = \ln|t| -i\pi$, we get: 
\be
\varphi_{\om} = \left[  \varphi_\om^{\rm in}\times e^{i\frac{\pi}4}
+ \left( \varphi_{-\om}^{\rm in}\right)^* 
\times  e^{-\frac{\om \pi}{\kappa}} e^{i\frac{3\pi}4} 
\right] \times \left( 1 + O\left(\frac{1+\frac{\om^2}{\kappa^2}}{\Delta(x)} \right) \right).
\ee
On the other side, for $x>0$, the saddle points obey $1+t^2 = 0$, as previously. 
Because of the branch cut one cannot pick the contribution of the decaying saddle point. 
We must instead deform the contour to a region where the `dominated convergence theorem' can be used and then stick it to the branch cut. 
With a computation similar to what was done for \eqref{gammalike}, we find 
\bsub \bea
\varphi_{\om} &=& \varphi_{\om}^{\rm out} \times \left(\sinh \left(\frac{\om \pi}{\kappa}\right) \sqrt{\frac{2\om}{\pi \kappa}} \Gam \left( -i \frac{\om}{\kappa}\right) e^{-\frac{\om \pi}{2\kappa}} e^{-i\frac{\om}{\kappa} +i\frac{\om}{\kappa} \ln(\frac{\om}{\kappa}) } \right) \\
&& \times \left( 1 + O\left( \frac{\kappa |x|(1+\om^3/\kappa^3)}{\Delta(x)^2}\right) \right).
\eea \esub 

\subsubsection{The growing mode}
To get a third linearly independent solution of \eqref{modequ}, we must construct the growing mode.
To get it, we re-use the above defined contours $\mathcal C_1$ and $\mathcal C_2$,
and we choose the branch cut to be $i\mathbb R_+$. 
For $x<0$, the relevant saddle points are $-1$ for $\mathcal C_1$ and $+1$ for $\mathcal C_2$, 
and respectively give \eqref{C1} and \eqref{C2}. 
For $x>0$ instead, for both contours, the relevant saddle point is $t=-i$, and this gives 
\be
\varphi^{\mathcal C_2}_{\om} = - e^{- \frac{2\om \pi}{\kappa}} \, \varphi^{\mathcal C_1}_{\om} = 
\varphi_\om^{\uparrow} \times \left( 1 + O\left(\frac{1+\frac{\om^2}{\kappa^2}}{\Delta(x)} \right) \right) .
\ee 
Since all combinations of $\varphi^{\mathcal C_1}_{\om}$ and $\varphi^{\mathcal C_2}_{\om}$
behave as $ \varphi_\om^{\uparrow}$  when $\Delta \to \infty$,
there is an ambiguity in choosing the third 
mode we shall use. We appeal to the conserved Wronskian to fix the choice. 
For a forth order differential equation, the Wronskian is a $4\times 4$ determinant, but because we neglected the $v$-modes, it becomes $3\times 3$. 
Once we have chosen 2 propagating modes, there is a {\it unique} choice of growing mode such that the basis has 
a unit Wronskian. The connection matrix is then an element of the group $SL_3(\mathbb C)$, {\it i.e.} of unit determinant (up to an overall gauge phase).

Therefore, our third mode is $\varphi_{\om}^g= \varphi_{\om}^{\mathcal C_2} - e^{-\frac{2\om \pi}{\kappa}} \varphi_{\om}^{\mathcal C_1}$. 
On the right and on the left we respectively get: 
\bsub \bea
\varphi_{\om}^g &\underset{x>0}{=}& 2 \varphi_{\om}^{\uparrow}\times \left( 1 + O\left(\frac{1+\frac{\om^2}{\kappa^2}}{\Delta(x)} \right) \right), \\
& \underset{x<0}{=}& \left[ e^{i\frac{\pi}4} \varphi_{\om}^{\rm in} + e^{-\frac{\om \pi}{\kappa}} e^{-i\frac{\pi}4} \left( \varphi_{-\om}^{\rm in} \right)^* \right] \times \left( 1 + O\left(\frac{1+\frac{\om^2}{\kappa^2}}{\Delta(x)} \right) \right).\label{Bilike}
\eea \esub 

\subsection{Connection matrix and on-shell Bogoliubov transformation} 
\label{transfermatrix_Sec}
\subsubsection{The connection formula} 
The results of the former subsection can be synthesized 
by the following $3 \times 3$ `off-shell transfer matrix' that connects  
the WKB modes defined on either side of the horizon
\be
\begin{pmatrix}\varphi_{\om}^{\rm out} \\ \varphi_{\om}^{\downarrow} \\ \varphi_{\om}^{\uparrow} \end{pmatrix} = (U_{\rm BH})^T \cdot \begin{pmatrix}\varphi_{\om}^{\rm in}  \\ \left( \varphi_{-\om}^{\rm in} \right)^* \\ \left( \varphi_{-\om}^{\rm out} \right)^* \end{pmatrix} \label{Udef}.
\ee
We define $U_{\rm BH}$ through its transpose so that it relates the  three amplitudes of {\it any} mode decomposed on the left and right side basis of \Sec{modeanalys}. 
Ignoring for the moment the correction terms, the matrix is
\be
U_{\rm BH} = \begin{pmatrix}  \tilde{\Gam} \left(\frac{\om}{\kappa} \right)^{-1} & e^{i\frac{3\pi}4} & \frac{e^{i\frac{\pi}4}}2 \\ e^{-\frac{\om \pi}{\kappa}} e^{i\frac{\pi}2} \tilde{\Gam} \left(\frac{\om}{\kappa} \right)^{-1} & -e^{i\frac{\pi}4} e^{\frac{\om \pi}{\kappa}}  & e^{-\frac{\om \pi}{\kappa}} \frac{e^{-i\frac{\pi}4}}2 \\ 0 & e^{\frac{\om \pi}{\kappa}} e^{-i\frac{\pi}4} \tilde{\Gam} \left(\frac{\om}{\kappa} \right) & 0  \end{pmatrix}.
\label{U}\ee
To simplify the above, we defined the `normalized' $\Gam$ function:
\be
\tilde{\Gam}(z) =  \Gam \left( -iz \right)\, \sqrt{\frac{2z}{\pi}} \sinh(\pi z) e^{-\frac{\pi z}2} e^{i z \ln(z) - i z} e^{-i \frac{\pi}4},
\label{Gr0}
\ee
which obeys for large $z$ (see App.\ref{Specialfunction_App})
\bsub \label{Gr} \bea
\left| \tilde{\Gam}(z) \right|^2 &=& 1 - e^{-2\pi z},\\
{\rm Arg}\left(\tilde{\Gam}(z) \right) &=& 0 + \frac{1}{12z} + O \left(\frac{1}{z^2}\right).
\eea \esub 
As expected from our choice of modes, the determinant is a pure phase:
\be
\det (U_{\rm BH}) = e^{i \frac{\pi}2}.
\ee

\subsubsection{Robustness of black hole radiation}
Using \eqref{U} we can now easily extract the Bogoliubov coefficients of \eqref{BogHR}. Let us start with $\phi_\om^{\rm in}$. Being a physical mode, it is 
asymptotically bounded, and therefore the 
amplitude multiplying the growing mode should vanish. Moreover, on the left it asymptotes to 
$\varphi_\om^{\rm in}$ of \eqref{plusin}. Hence its six amplitudes obey
\be
\begin{pmatrix} 1 \\ 0 \\ \beta_\om \end{pmatrix}_{x<0} = U_{\rm BH} \cdot \begin{pmatrix} \alpha_\om \\ d_\om \\ 0 \end{pmatrix}_{x>0},
\ee
where $d_\om$ is the amplitude of the decaying mode. 
From these equations, and the corresponding ones for 
the negative frequency mode $\left(\phi_{-\om}^{\rm in} \right)^*$,
the coefficients of \eqref{BogHR} are
\bsub \label{bogocoef}
\bea
\alpha_\om &=& \frac{\tilde{\Gam} \left(\frac{\om}{\kappa}\right)}{1 - e^{-\frac{2\om \pi}{\kappa}}}= e^{ -i\frac{\pi}2}\, \tilde{\alpha}_\om ,\\
\frac{\beta_\om}{\alpha_\om} &=&  e^{-\frac{\om \pi}{\kappa}} = {\tilde \beta_\om \over \tilde \alpha_\om } .
\eea \esub 
The amplitudes of the decaying modes in $\phi_\om^{\rm in}$ and $\left(\phi_{-\om}^{\rm in} \right)^*$ are also fixed and given by 
\bsub \bea
d_\om &=& e^{i\frac{\pi}4} \frac{e^{-\frac{\om \pi}{\kappa}}}{2 \sinh(\frac{\om \pi}{\kappa})}, \\
d_{-\om} &=& e^{i\frac{3\pi}4} \frac1{2 \sinh(\frac{\om \pi}{\kappa})}.
\eea \esub 

For $\om \ll \om_{\rm max}$ of \eqref{ommax},
and when ignoring the $x$-dependent corrections of the former subsection
({\it i.e.} to leading order in $\kappa/\Lambda$),
the mean occupation number \eqref{e13} 
is exactly the relativistic result of \eqref{relats}. 
This is in agreement with what was found in~\cite{Brout95,Corley97,Himemoto00,Unruh04}, although the conditions are now stated more precisely.
This result implies that the spectral deviations due to dispersion
are to be found by examining the various approximations that have been used. 

\subsubsection{The correlation pattern}
In addition, it should be noticed that sufficiently far away from the horizon, 
{\it i.e.} in a region where $\Delta(x)$ of \eqref{Delta} is much larger than 1,
the space-time correlation pattern of the Hawking particles of positive frequency and their inside
partners of negative frequency are also unaffected by dispersion. 
This second aspect of the robustness of HR can be established
by either forming wave packets of {\it in}-modes~\cite{Brout95}, {\it i.e.} 
considering non-vacuum states described by coherent states, see App.C. in~\cite{Macher09b},
or by computing the 2-point correlation function $G(t,x; \, t'x')= \langle \phi(t,x)\, \phi(t',x') \rangle$~\cite{Carusotto08,Balbinot08,Schutzhold10}.
For a comparison of the two approaches, see~\cite{Parentani10}. 
In the {\it in}-vacuum, at equal time, the $\om$ contribution of $G$ 
for $x> 0$ and $x' < 0$
is given by
\be
G_\om(x,x') = \alpha_\om \beta_\om^* \, 
\varphi_\om^{\rm out}(x) \varphi_{-\om}^{\rm out}(x') + \tilde \beta_\om \tilde \alpha_\om^* \left( \varphi_\om^{\rm out}(x)\right)^* \left( \varphi_{-\om}^{\rm out}(x')\right)^* \label{correl},
\ee
see the ${\cal B}_\om$ term in Sec.IV.F. in~\cite{Macher09b}. 
Using the expressions of \Sec{modeanalys} together with \eqref{bogocoef}, far away from the turning point of \eqref{xtp} but still in the near horizon region, we get 
\be
G_\om(x,x') = {1 \over \sinh(\frac{\om \pi}{\kappa})} \frac{\Re |x/x'|^{i\frac{\om}{\kappa}}}{4\pi \om} \times \left(1+ O \left(\frac{1+\om^2/\kappa^2}{\Delta(x)}\right)\right). \label{BHcorrel}
\ee
At leading order in $\kappa/\Lambda$, this is exactly the relativistic result. Hence, 
the long distance correlations are also robust when 
introducing short distance dispersion. This follows from the fact that 
the {\it phase} of $\beta_\om/\alpha_\om$ (and not just its norm) is not modified by dispersion. 
At this point we should say that this phase actually depends 
on those of the {\it in} and {\it out} modes that can be arbitrarily chosen. 
Thus, as such it is not an observable quantity. However, the phase of $\alpha_\om \beta_\om^* \varphi_\om^{\rm out} \varphi_{-\om}^{\rm out}$ is an observable,  
independent of these choices. We have chosen to work with {\it in} and {\it out} bases 
where all modes have a common phase at a given $p$, see \Sec{eikonal_Sec},
as this ensures that $\arg(\beta_\om/\alpha_\om)$ is unaffected by this arbitrary phase. 

Using these bases, the phases of $\alpha_\om$ and $\tilde \alpha_\om$ of \eqref{bogocoef} also have a clear meaning as can be seen by 
considering the scattering  of classical waves. They characterize the phase shifts which are not taken 
into account by the WKB modes of Sec.~\ref{modeanalys}.
In fact, using \eqref{Gr0} one verifies that in the limit $\om/\kappa \to \infty$ 
one recovers the standard WKB results, {\it i.e.} $\arg(\alpha_\om) =\arg(\tilde\Gam(\om/\kappa)) \to 0$ 
for the transmitted mode, and
$\arg(\tilde \alpha_\om)  \to \pi/2$ for the reflected one. 
For smaller values of $\om/\kappa$, 
$\arg(\tilde \Gam(\om/\kappa))$ thus accounts for the non-trivial phase shift.
In Chapter \ref{laser_Ch} we show that in the presence of two horizons, this shift affects the
spectrum of trapped modes.

\subsection{Validity of the connection formula}
\label{validity}
Our computation is based on two approximations. The first one is the $p$-WKB approximation introduced 
when solving \eqref{chiequ} in the near horizon region. Its validity requires
\be
\frac{\Lambda}{\kappa} \gg 1.\label{uvmixvalcond}
\ee
This condition is the expected one. 
It involves neither $\om$ nor the parameters $D_{\rm lin}$ of \eqref{Dlin}.
For this reason, as we shall see below, it will {\it not} be the most relevant one in the general case. This is a non-trivial result. 
In addition, the corrections to this approximation encode a mixing between left and right moving modes~\cite{Balbinot06}. Therefore, 
at leading order, the spectral deviations we shall describe below will be the same in our model as in models where the decoupling between these modes is exact~\cite{Brout95,Schutzhold08}. It would be interesting to validate this prediction by numerical analysis.

The second approximations are controlled by 
$\Delta$ of \eqref{Delta}. This quantity governs both the validity of the saddle point approximation, 
as in \eqref{firstSP}, and that of the WKB  
modes of Eq. (\ref{plusin}-\ref{grmode}).
Since these corrections decrease when $\Delta$ increases, 
the pasting of the near horizon modes on the 
WKB ones should be done at the edges of the near horizon region. 
One could imagine pasting the modes further away, but this would require to control 
$\tilde \phi_\om(p)$ outside the region where $v$ is linear in $x$, {\it i.e.} to deal 
with ODE in $\partial_p$ of order higher than $2$. This is perhaps possible
but it requires other techniques than those we used, see~\cite{Scottthesis} for recent developments.  
In any case, what we do is sufficient to control the error terms in the relativistic limit, and more precisely to find an upper bound.

\begin{figure}[!ht]
\begin{center} 
\includegraphics[scale=2]{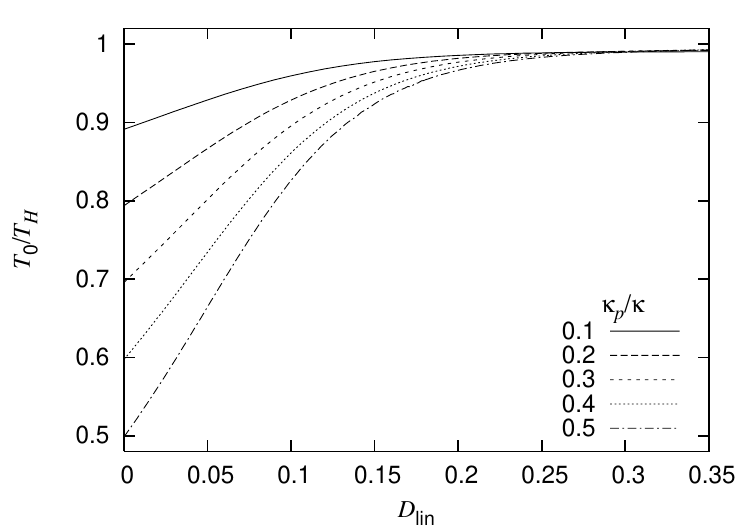}
\end{center}
\caption{Deviation of the temperature found using 
the code of~\cite{Finazzi10b} for various flows $v(x)$ which all
have the same surface gravity $\kappa$.
$T_H = \kappa/2\pi$ is the usual Hawking temperature. 
$T_0$ is the actual temperature given quartic superluminal dispersion
computed for $\om \ll \kappa$.
The parameter $\kappa_p$ (see~\cite{Finazzi10b} for its precise definition) 
characterizes the slope of $v(x)$ {\it outside} the near horizon region, {\it i.e.} in regions 2 of Fig.\ref{regions}. 
At fixed $\Lambda/\kappa=15$, the critical value of $D_{\rm lin}^c \sim 0.2$ 
below which $T_0$ deviates from $T_H$ does not depend on $\kappa_p$, in agreement with \eqref{Dp}. Moreover, an analysis for various values of $\Lambda/\kappa$ in~\cite{Finazzi10b} has shown that the scaling law for $D_{\rm lin}^c$ is consistent with \eq{Dp}.}
\label{Xlinfig} 
\end{figure}

Being confined to stay within the near horizon region, the validity of the pasting procedure requires that 
\be
\Delta_{\rm p} \equiv
\Delta(x_{\rm pasting}) =\frac{\Lambda}{\kappa} (\kappa x_{\rm pasting})^{\frac32} 
 \sim \frac{\Lambda}{\kappa}\left(D_{\rm lin}\right)^{\frac32} \gg 1,
\label{Dp}
\ee
where $D_{\rm lin}$ characterizes the extension of this region on either side of the horizon.
As we see in Fig.\ref{Xlinfig}, at fixed $\Lambda/\kappa$, the spectrum is very close to the relativistic one of \eqref{relats} if 
 $D_{\rm lin}$ is large enough. Instead, below a certain threshold, the deviations become non negligible. Our criterion in \eqref{Dp}  
indicates that this will happen when $\Delta_{\rm p}$ is of order 1. 
Hence the threshold value for $D_{\rm lin}$ should scale as $({\kappa}/{\Lambda})^{2/3}$. This prediction 
is confirmed by the numerical analysis of~\cite{Finazzi10b}.

We can be more precise. 
Indeed, as discussed after \eqref{vlin},  
the near horizon region is not necessarily symmetric. Hence, the values of $x_{\rm pasting}$ 
on the right and on the left will be different. 
This is important because error terms coming from {\it in} modes 
are dominant compared to those from {\it out} modes,
see \Sec{modeanalys}. Since the former lives on the left side, 
the validity condition is
\be
\Delta_{\rm p}^L \sim \frac{\Lambda}{\kappa}\left(D_{\rm lin}^L\right)^{\frac32} \gg 1,
\label{DpL}
\ee
The higher sensitivity of the spectrum to perturbations of $v$ localized on the left hand side 
was clearly observed in~\cite{Finazzi10b}, see Fig. 8 right panel. This sensitivity has been
recently exploited in~\cite{Zapata11} 
to produce resonant effects. We can now estimate the deviations on the spectrum. 
Considering Eqs. (\ref{outerror},\ref{inerror},\ref{firstSP},\ref{gammalike},\ref{DpL}), 
we obtain
\be
\left|\frac{\bar n_\om - \bar n_\om^{\rm relativistic}}{\bar n_\om^{\rm relativistic}}\right| = O\left( \frac{P(\om/\kappa)}{\Delta_p^L} \right)\label{leadcorr}, 
\ee
where $P$ is a polynomial function of degree 2. Therefore, at fixed $\om \lesssim \kappa$, {\it i.e.} in the relevant regime for Hawking radiation, 
the leading deviations are {\it bounded} by a quantity scaling as $1/\Delta_{\rm p}^L$. 
In particular, \eqref{leadcorr} demonstrates that the spectrum is thermal for arbitrary small frequencies,
contrary to what has been suspected in~\cite{Jacobson93,Corley97,Jacobson99}. 
Moreover, our analysis indicates that the deviations 
should grow with $\om$. This is compatible with the fact that $\bar n$ vanishes for 
$\om > \om_{\rm max}$ defined in \eqref{ommax}, irrespectively of the value of the ratio $\Lambda/\kappa$.
However, we have not yet been able to confirm this precisely with the code of~\cite{Finazzi10b}. 
One of the reasons is that the spectrum is radically modified when $\om $ approaches $\om_{\rm max}$. 
So far the $p$-WKB approximation has been a subdominant effect. However, 
if one takes the de Sitter limit, {\it i.e.} $D_{\rm lin} \to \infty$, the correction term in \eqref{Dp} 
vanishes and $p$-WKB becomes the only source of deviations.

To conclude, we recall that the deviations have been computed using \eqref{modequ}.
In other mode equations, like that for phonons in a BEC,
the corrections to the $x$-WKB approximation will be in general larger. 
However these corrections are not due to dispersion but rather
to the fact that the conformal invariance of  \eqref{modequ} in the dispersionless limit
will be lost. These corrections can thus be studied without introducing dispersion.
This is also true when introducing an infrared modification of \eqref{modequ} associated with a 
mass or with a non-vanishing perpendicular momentum. 
In Chapter \ref{mass_Ch}, we show that the spectral properties are still robust in that \eqref{DpL} is sufficient to guarantee 
that to leading order the Bogoliubov coefficients are unaffected by short distance dispersion. 

\subsection{General superluminal dispersion}
\label{generalization}
Instead of \eqref{dispr}, we now consider
\be
F^2(p) = \left(p + \frac{p^{2n+1}}{\Lambda^{2n}} \right)^2 \label{generalF},
\ee
taken again, for simplicity, to be a perfect square. 
In \Sec{SecpWKB} we computed p-WKB modes for any dispersion in \eqref{dualmode}, hence the globally defined 
modes for $F$ satisfying \eqref{generalF} are, see \Sec{CF}, 
\be
\phi_{\om}(x) = \frac1{\sqrt{4\pi \kappa}} \int_{\mathcal C} \frac{e^{i (px- \frac{\om}{\kappa} \ln(p) + \frac{p^{2n+1}}{(2n+1)\Lambda^{2n} \kappa})}}{(1+\frac{p^{2n}}{\Lambda^{2n}})^{\frac12}} 
\frac{dp}{p\sqrt{2\pi}}.
\ee
There now exist $2n+1$ linearly independent modes. 
Hence in terms of contours, there are $2n+1$ sectors (Stokes lines) 
toward $\infty$ in the complex $p$ plane. 
By using a contour homotopic to the real line, and the same two possible branch cuts of $\ln p$
on $\pm i\mathbb R^+$, we can compute the $2$ `on-shell' modes 
that are asymptotically bounded. Even though, there  
exist $n$ pairs of growing and decaying modes on the subsonic side, and $n-1$ 
pairs on the other side, only one pair 
in the subsonic sector will be relevant in the `off-shell' connection formula of \eqref{U}. 
Indeed, all the other pairs do not mix with propagating modes. 
Therefore, the different contours giving rise to the relevant modes 
will be quite similar to those of \Sec{CF}. 

To perform a saddle point approximation, we introduce $t= p/ \Lambda|\kappa x|^{\frac{1}{2n}}$ 
and get:
\be
\varphi_{\om}(x) = \frac{e^{-i\frac{\om}{\kappa} \ln(\Lambda|\kappa x|^{\frac{1}{2n}})}}{\sqrt{4\pi \kappa}} \int_{\mathcal C} \frac{e^{-i\frac{\om}{\kappa} \ln(t)}}{(1+t^{2n} \kappa |x|)^{\frac12}} e^{i\Delta_n(x) \left({\rm sign}(x) t + \frac{t^{2n+1}}{2n+1}\right)} \frac{dt}{t\sqrt{2\pi}} \label{modifdual}.
\ee
By a computation similar to that of \Sec{CF}, 
at leading order in $\kappa/\Lambda$,
we recover the Bogoliubov coefficients of \eqref{bogocoef}, thereby establishing their robustness for arbitrary integer values of $n$.
Moreover, the deviations from this result are now governed by
\be
\Delta_{p, n}^L = \frac{\Lambda}{\kappa} \left( D^L_{\rm lin}
\right)^{\frac{2n+1}{2n}}.
\ee
Hence, the error on the mean number of emitted quanta satisfies 
\be
\left|\frac{\bar n_\om - \bar n_\om^{\rm relativistic}}{\bar n_\om^{\rm relativistic}}\right| = O\left( \frac{\kappa}{\Lambda (D_{\rm lin}^L)^{\frac{2n+1}{2n}}}\right) \label{corrgeneral}.
\ee

\subsection{Relating subluminal dispersion relations to superluminal ones}
\label{subsuper_Sec}

So far we analyzed only superluminal dispersion relations. We should thus inquire 
how subluminal dispersions would affect the spectrum. At the classical level, as noticed in Sec.~\ref{eikonal_Sec}, there is an exact correspondence between these two cases. 
At the level of the modes \eqref{subsuper} does not leave \eqref{modequ} invariant 
as it does not apply to the left moving solutions
governed by $\om - vp = - F$. However, it becomes a symmetry 
when neglecting the mode mixing 
between left and right movers. Therefore, in models where the decoupling between these is exact~\cite{Brout95,Schutzhold08}, 
\eqref{subsuper} is an {\it exact} symmetry. 
Moreover, since  the mode mixing between left and right movers is subdominant  
for general mode equations, the discrepancy of the spectral deviations between superluminal and subluminal dispersion 
will not show up at leading order. This is {\it precisely} what has been observed in Sec.VI.2 of~\cite{Macher09}.

At the level of the connection formula of \eqref{U}, the three exchanges of \eqref{subsuper}
still are an exact symmetry since the $U_{\rm BH}$-matrix 
is based on the right moving mode of \eqref{dualmode}
which is determined by the action $W_\om(p)$. 
Therefore,  at leading  order in $\kappa/\Lambda$, without any further calculation, 
this symmetry implies that the spectrum of HR is equally robust for subluminal dispersion.
Moreover, the {\it leading order deviations} from the thermal spectrum will be 
governed by the {\it same} expression as \eqref{corrgeneral}. 
It should be noticed that when applying \eqref{subsuper}, 
$D_{\rm lin}^L$ characterizes the extension on the sub-sonic region. 

The above symmetry should not be confused with the one that relates black and white hole geometries without changing the dispersion relation, see \eqref{BHWHsym} in \Sec{BHWHcorresp_Sec}. Indeed, the symmetry \eqref{subsuper} presented here exchanges the role of left and right moving modes, while that of \eq{BHWHsym} applies only to the right moving sector. Instead, these two symmetries can be \emph{composed} with each other. This allows to compare black hole spectra for both types of dispersion 
without referring to white holes. 

To conclude the discussion, we mention that this approximate symmetry not only allows to \emph{predict} several effects, but also to predict how the observables will quantitatively behave. In this discussion, we anticipate the analysis of Chapters \ref{mass_Ch} and \ref{laser_Ch} concerning the undulation~\cite{aQuattro,Coutant12} and black hole laser~\cite{Coutant10}.
 \begin{itemize}
\item
A laser effect will be found for subluminal dispersion in a flow possessing two horizons that passes from super to sub and then back to a supersonic 
flow, and this exactly for the same reasons that the laser effect was found in the `reversed' flow in the case of superluminal dispersion. This laser effect will be the subject of Chapter \ref{laser_Ch}. The frequencies and the growth rates of this subsonic laser effect will be governed by the same expressions as those of Chapter \ref{laser_Ch} 
(when neglecting the coupling to the left moving modes).
\item
Subsonic phonon propagation in a non-homogeneous flow that remains everywhere subsonic,
{\it i.e.}, without a sonic horizon, will be governed by a $4\times 4$ $S$-matrix that encodes new pair creation
channels with respect to those found in the presence of a sonic horizon~\cite{Scottthesis} 
exactly for the same reasons that a supersonic phonon propagation in a non-homogeneous flow that remains everywhere supersonic does so, as mentioned in~\cite{Finazzi11}. 
\item
The behavior of the Bogoliubov coefficients in the two cases will behave {\it quantitatively} 
in the same way.
\item
The undulation observed in white hole flow for subluminal gravity waves in the experiment of~\cite{Weinfurtner10} is generated for the 
same reasons as that found in white holes for Bose condensates using the Bogoliubov-de~Gennes 
equation~\cite{Mayoral11}. As we discussed in Chapter \ref{mass_Ch}, this is obtained by composing the symmetries of \eqref{BHWHsym} and \eqref{subsuper}.
\item
When the dispersion relations and the profiles $v(x)+c(x)$ 
(where $c(x)$ is the speed of sound as in \eq{fluidwavequ}) obey \eqref{subsuper}
up to a possible rescaling of $\Lambda$, the momentum $p$, and distances, 
these undulations should have the same spatial profile.
\end{itemize}

\section{Main conclusions of this study}
\label{Cl_LIV_Sec}
In this chapter (and in~\cite{Coutant10}), we studied the modification of the Hawking flux when Lorentz invariance is no longer exact, but only emergent in the infrared. We confirmed that the dispersionless limit is regular, in that both the spectrum (\eq{bogocoef}) and the correlation pattern (\eq{BHcorrel}) becomes exactly relativistic in the limit $\Lambda \to \infty$. Moreover, we identified the role of the size of the near horizon region, the parameter $D_{\rm lin}$, in this limit, as explained by Eqs. \eqref{Dp} and \eqref{DpL}. It is important to note that a small $D_{\rm lin}$ can \emph{invalidate} the relativistic result even though $\kappa/\Lambda$ is quite small. This might be the case {\it e.g.} in optical fibers. 

As can be understood by combining our results with the numerical ones of~\cite{Finazzi10b}, dispersive effects introduce a `resolution length' $\sim \kappa^{-1/3} \Lambda^{-2/3}$. If the gradient of the geometry varies too much on this distance, the field cannot identify the surface gravity of the horizon, and the flux starts deviating from Hawking's result. Moreover, unlike what could be suspected~\cite{Jacobson93,Corley97,Jacobson99}, dispersion generates no deviations in the infrared, see \eq{leadcorr}. New infrared effects can appear, but they are essentially insensitive to dispersion, we refer to Chapter \ref{mass_Ch} for more details, and in particular \Sec{massiveBogo_Sec}. In addition, thanks to \eq{subsuper}, we established that our results do not depend on the nature of the dispersion relation, {\it i.e.} sub- or superluminal. 

This work not only obtained interesting results concerning the robustness of Hawking radiation, it also presents a whole formalism to compute the Bogoliubov transformation in the weak dispersive regime. This has many interesting applications.In the following chapters, we shall see some a few examples. In particular, it will be used to deal with configurations with several horizons. By computing properly the phases of the coefficients, we can characterize precisely interference effects arising in the black hole laser effect (see Chapter \ref{laser_Ch}). Moreover, because our transfer matrix of \eq{U} is valid `off-shell', we are also able to incorporate effects due to the growing and decaying modes. As we show in Chapter \ref{mass_Ch}, they contribute when they live in a finite size region.

This work seems to indicate that adding ultraviolet dispersion barely affects Hawking radiation, if the cut-off scale $\Lambda$ is high enough. This is true when restricting our attention to the emitted flux of a black hole. However, as we shall see, new stability issues must be addressed. In fact, in some configurations, dispersion might strongly alter the relativistic picture (see {\it e.g.} the discussion in \Sec{Sma}).

\chapter{Infrared instabilities}
\label{mass_Ch}

\minitoc

\section{Infrared divergences in Hawking radiation}
\label{IRdiv_Sec}
In the preceeding chapter, we saw that the Hawking effect is robust against the introduction of ultraviolet dispersion, when $\kappa/(\Lambda D_{\rm lin}^{3/2}) \ll 1$, see \eq{Dp}. In the following (Chapters \ref{mass_Ch} and \ref{laser_Ch}), we analyze more general geometries than the simple black hole case. In contrast with Chapter \ref{LIV_Ch}, we will observe new features, which appear through dispersive effects and that were absent in the relativistic theory. To start with, we analyze the nature of the infrared divergence of Hawking radiation. Indeed, the Bogoliubov coefficient or equivalently, the occupation number of Hawking quanta (see {\it e.g.}, \eq{HRrelat_flux}) diverges in the infrared regime, {\it i.e.}
\be
n_\om \sim \frac{T_H}\om \quad {\rm for} \quad \om \to 0.\label{HRdivIR}
\ee
It is well known though that there are no physical infrared divergences in Hawking radiation~\cite{Primer}, exactly as there is no infrared divergence for any object radiating as a black body. This is simply because when computing observables, as the energy density, $n_\om$ is multiplied by powers of $\om$ which regulate the divergence \eqref{HRdivIR}. However, when introducing dispersion, this statement is no longer clear. Indeed, as we saw in Fig.\ref{graphroots}, for dispersive theories, there might exist non-zero values of the momentum at zero frequency, which could spoil the preceding argument. In fact, the infrared divergence of \eq{HRdivIR} becomes physically relevant if three conditions are met.
\ben
\item The Bogoliubov coefficient  $\beta_\om$ must diverge, or at least be much larger than 1 for $\om \to 0$ (see {\it e.g.}, \eq{bogocoef}). 
\item There must exist a zero frequency root $p_{U} = p_{\om = 0} \neq 0$ of the Hamilton-Jacobi equation (see {\it e.g.}, Figs.\ref{graphroots} and \ref{WHdisprel_fig}).
\item The group velocity of the corresponding mode must be oriented away from the sonic horizon. 
\een
In that case, we will observe a contribution from the zero mode to the physical observables. Such phenomenon will be called `undulation'. Indeed, it corresponds to some well known solutions in hydrodynamics~\cite{Chanson95}\footnote{Even though these are well known, they are not completely understood. In the Introduction of Ref.~\cite{Wolsthesis}, one finds the statement:  `Still, the characteristics and the formation of an undular hydraulic jump are not fully understood'.} produced by hydraulic jumps. Moreover, they also share some features with `bow waves' in that both phenomena concerns the excitation of the zero-frequency mode associated with $p_U$. However, the important difference is that for bow waves, the excitation is due to a source term coming from a defect in the medium ({\it e.g.}, a boat in water)~\cite{Carusotto12}, while in our case, the undulation is produced by the large amplification of infrared modes due to the horizon. As we shall see, this effect should be conceived as an important {\it prediction} of the linearized treatment, 
and could be validated for a BEC using numerical techniques similar to those of~\cite{Mayoral11},
and perhaps also in future experiments.

To begin with, we shall assume that the three conditions are met, and we derive very general features of the undulation, and how it can affect observables. To this aim, we consider the two-point correlation function evaluated in the {\it in}-vacuum
\be
G(t,t';x,x') = \vev{\rm in}{\hat \phi(t,x) \hat \phi(t',x')}.
\ee
Using the decomposition of the field in Fourier modes \eqref{phidecomp} we find for equal times that
\be
G(t,t;x,x') = \int_0^\infty G_\om(x,x') d\om.
\label{Gint}
\ee
In the infrared sector, for low enough frequencies, the relevant term in the integrand is
\be
G_\om(x,x') = \phi_\om^{\rm in}(x) \left(\phi_\om^{\rm in}(x')\right)^* + \phi_{-\om}^{\rm in}(x) \left(\phi_{-\om}^{\rm in}(x')\right)^* . \label{Gom}
\ee
In general, $\phi_\om^{\rm in}$ and $\phi_{-\om}^{\rm in}$ are two different functions of $x$, and $G_\om$ is complex and cannot be factorized. To see that, under some conditions, it factorizes, we use \eq{BogHR} to work with the {\it out}-modes. Then the first term in Eq.~(\ref{Gom}) becomes 
\bsub \label{phiphi} 
\bea
\phi_\om^{\rm in}(x) \left(\phi_\om^{\rm in}(x')\right)^* &=& |\alpha_\om|^2\,  \phi_\om^{\rm out}(x) \left(\phi_\om^{\rm out}(x')\right)^* + |\tilde \beta_\om|^2 \, \left(\phi_{-\om}^{\rm out}(x)\right)^* \phi_{-\om}^{\rm out}(x') \\
&&+ \alpha_\om \tilde \beta_\om^* \phi_\om^{\rm out}(x) \phi_{-\om}^{\rm out}(x') + \alpha_\om^* \tilde \beta_\om \left(\phi_{-\om}^{\rm out}(x)\right)^* \left(\phi_{\om}^{\rm out}(x')\right)^*.
\eea \esub
In the limit $\om \to 0$, two effects are combined.
Firstly $\phi^{\rm out}_{\om}$ and $\phi^{\rm out}_{-\om}$ become the same function of $x$,
$\phi^{\rm out}_0(x)$. This is true in general, but not
in the particular case of the two dimensional massless field in a black hole metric
because in that case $\phi^{\rm out}_{\om}$ and $\phi^{\rm out}_{-\om}$
vanish on the L and R quadrant respectively (see~\cite{Coutant11}).

Secondly, when assuming
that $\vert \beta_\om|^2  \gg 1$ for $\om \to 0$, since the $S$-matrix of \eq{BogHR} is an element of $U(1,1)$,
one has
\bsub \label{alphabetarelation}
\bea
|\alpha_\om|^2 &\sim& |\tilde \alpha_\om|^2 \sim |\beta_\om|^2 \sim |\tilde \beta_\om|^2 , \\
\alpha_\om \tilde \beta_\om^* &\sim& \tilde \alpha_\om^* \beta_\om \sim e^{2i\theta}|\beta_\om|^2,
\eea \esub
where $e^{2i\theta}$ is a phase. These two facts guarantee that $G_\om$ becomes real and factorizes as
\be
G_\om(x,x') \sim 8 |\beta_\om|^2 \times \Phi_{\rm U}(x)\,  \Phi_{\rm U}(x') ,
\label{omundul}
\ee
where the real wave
\be
\Phi_{\rm U}(x) \doteq \Re\left\{e^{i\theta} \phi^{\rm out}_0(x) \right\}, \label{Undulprofile}
\ee
gives the profile of the undulation. It should first be noticed that its phase is locked. Indeed, if one modifies
the arbitrary phase of the {\it out}-mode $\phi^{\rm out}_0$, the modified phase $\theta$
would exactly compensate this change so that $\Phi_{\rm U} $ would remain unchanged.
This can be understood from the fact that $\Phi_{\rm U} $ oscillates on one side of the horizon and decays on the other. 
One should point out that this factorization means that the undulation contributes to observables in the same way that a coherent state does, see App.\ref{QHO_App}, or the appendix of reference~\cite{Macher09b}.
It thus behaves as a classical wave in that its profile and its phase are not random.
However, in the present linearized treatment,
its amplitude is still a random variable, {\it i.e.}, its mean value
is identically zero, and $|\beta_\om|^2$
gives the ($\om$-contribution of its) standard deviation.

So far we have worked at fixed $\om$. We now consider the integral over low frequencies in \eq{Gint}. As was understood in~\cite{Mayoral11,Coutant11}, the divergence of $|\beta_\om|^2$ for $\om \to 0$ accounts for a growth in time of the root mean square (r.m.s.) amplitude of the undulation. When considering an observable evaluated
at a finite time $t$ after the formation of the horizon, the stationary setting of \eq{Gint} with a dense set of frequencies should be used with care.
Indeed, after such a lapse, one cannot resolve frequencies separated by less than $2\pi/t$, as in the Golden Rule. This effectively introduces an infrared cut-off in the integral over $\om$. Taking this into account  gives the growing rate that depends on the power of the divergence of $|\beta_\om|^2$.
For example, when the Bogoliubov coefficients take their values of \eq{WHcoef}, in the {\it in}-vacuum, the infrared contribution of $G$ grows as
\bsub \label{General_undul} 
\bea
G_{\rm IR}(t;x,x') &\sim& 8 \int_{2\pi/t} |\beta_\om|^2 d\om \, \times   \Phi_{\rm U}(x)\,  \Phi_{\rm U}(x') ,\\
&\sim& 8 f(t) \times  \Phi_{\rm U}(x)\,  \Phi_{\rm U}(x') ,
\eea \esub
where $f$ is a growing in time function. More precisely, $f$ grows in $t$ as fast as $|\beta_\om|^2$ diverges in $1/\om$. Therefore, the infrared divergence manifests itself as a growing in time zero-mode. Note also that in a black hole, $\Phi_{\rm U}(x)$ is a constant, therefore its contribution is an artifact and disappears from all observables. Indeed, they all consists of derivatives of the 2-point function $G$ ($T_{\mu \nu}$ in gravity, density fluctuations in BEC, height variations for surface waves~\cite{Schutzhold08}, {\it etc.}). This argument is equivalent to the first one we provided after \eq{HRdivIR}. 

In the next section, we first discuss the relevance of the undulation for Bose-Einstein condensates and then present the structure of the present chapter.

\section{The Bose-Einstein condensate context}
\label{BECmass_Sec}

As we discussed in the preceding section, the analogue Hawking radiation of white hole flows emit a standing zero-frequency wave, {\it i.e.} an undulation~\cite{Mayoral11,Coutant11}. This wave possesses a macroscopic amplitude and a short wavelength
fixed by the dispersive properties of the medium. As we saw in \Sec{IRdiv_Sec}, in hydrodynamic, this wave corresponds to an undular hydraulic jump. They also seem closely related to the appearance of `layered structures' in$~^4$He~\cite{Pitaevskii84}, or in BEC~\cite{Baym12}.
In addition, undulations have been recently observed in water tank experiments~\cite{Rousseaux07,Weinfurtner10}
aiming at detecting the analogue Hawking radiation, but their relation with the Hawking effect was
not pointed out. This relation was understood in the context of atomic Bose-Einstein condensates (BEC), where
the emission of an undulation was explained in terms of the infrared divergence of the mean occupation number, as in \eq{HRdivIR}.
More precisely, the undulation arises from the combination of the three conditions mentioned earlier.
Firstly, $n_\om$, the spectrum of massless phonons spontaneously produced {\it \`a la Hawking}
from the sonic horizon diverges like $1/\om$ for $\om \to 0$, as in the Planck distribution. Secondly, $p_U = p_{\om = 0}$,
the wave number of the undulation is a non-trivial solution of the dispersion relation, and thirdly, its group velocity is oriented away from the horizon.

\begin{figure}[!ht]
\begin{center}
\includegraphics[scale=0.65]{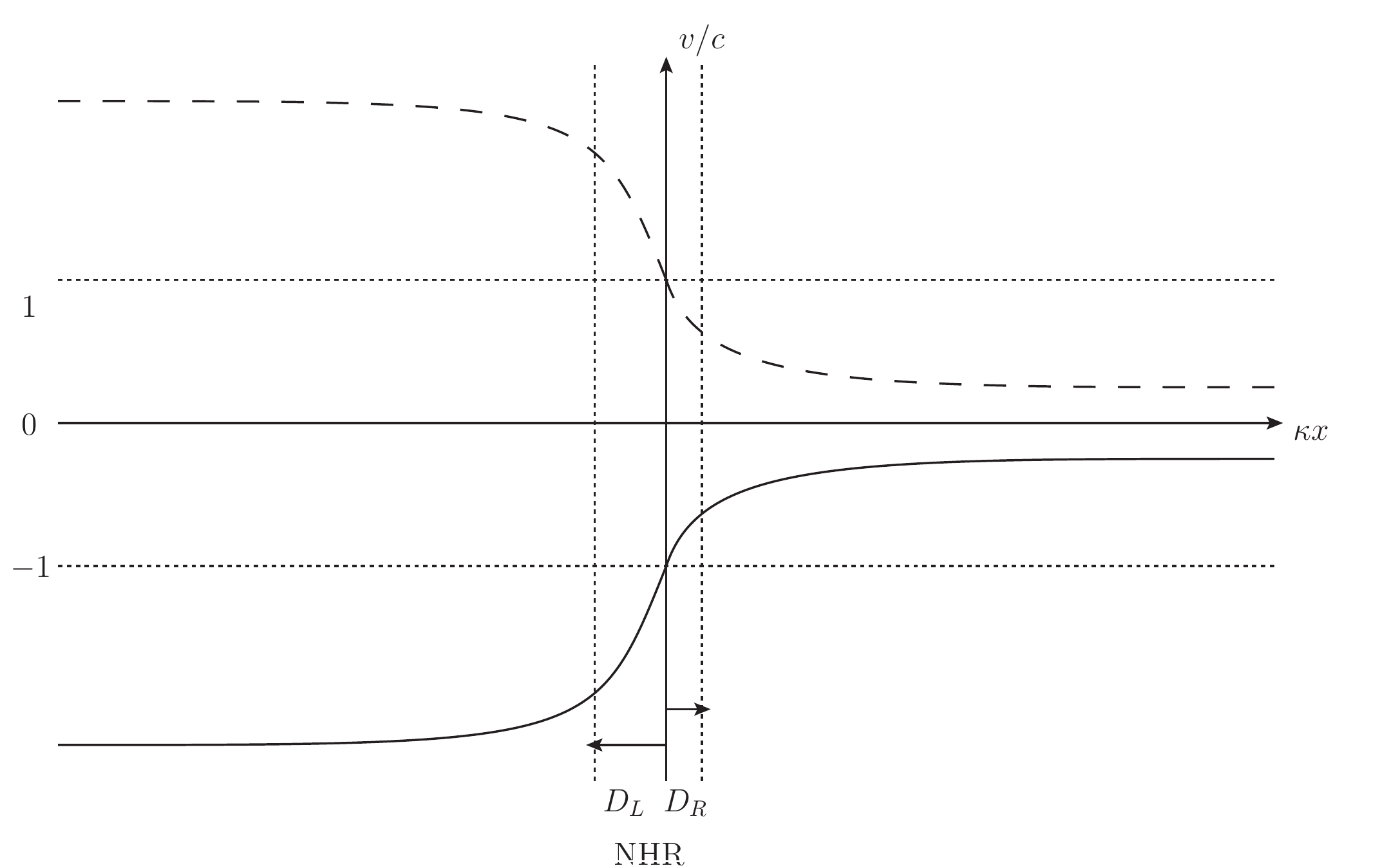}
\end{center}
\caption{Examples of one dimensional black hole flow (solid line) and white hole flow (dashed line) with regular asymptotic properties. They are related to each other
by reversing the sign of the velocity $v(x) \to -v(x)$. In both cases, the subsonic R region $\vert v \vert < c = 1$ is on the right of the horizon, while the supersonic L region is on the left.
The near horizon region (NHR), where $v \sim -1 + \kappa x$  is a good approximation, has a width in units of $\kappa$ of $D_L$ on the left and of $D_R$ on the right. }
\label{vprofile_fig}
\end{figure}

\begin{figure}[!ht]
\begin{center}
\includegraphics[scale=0.8]{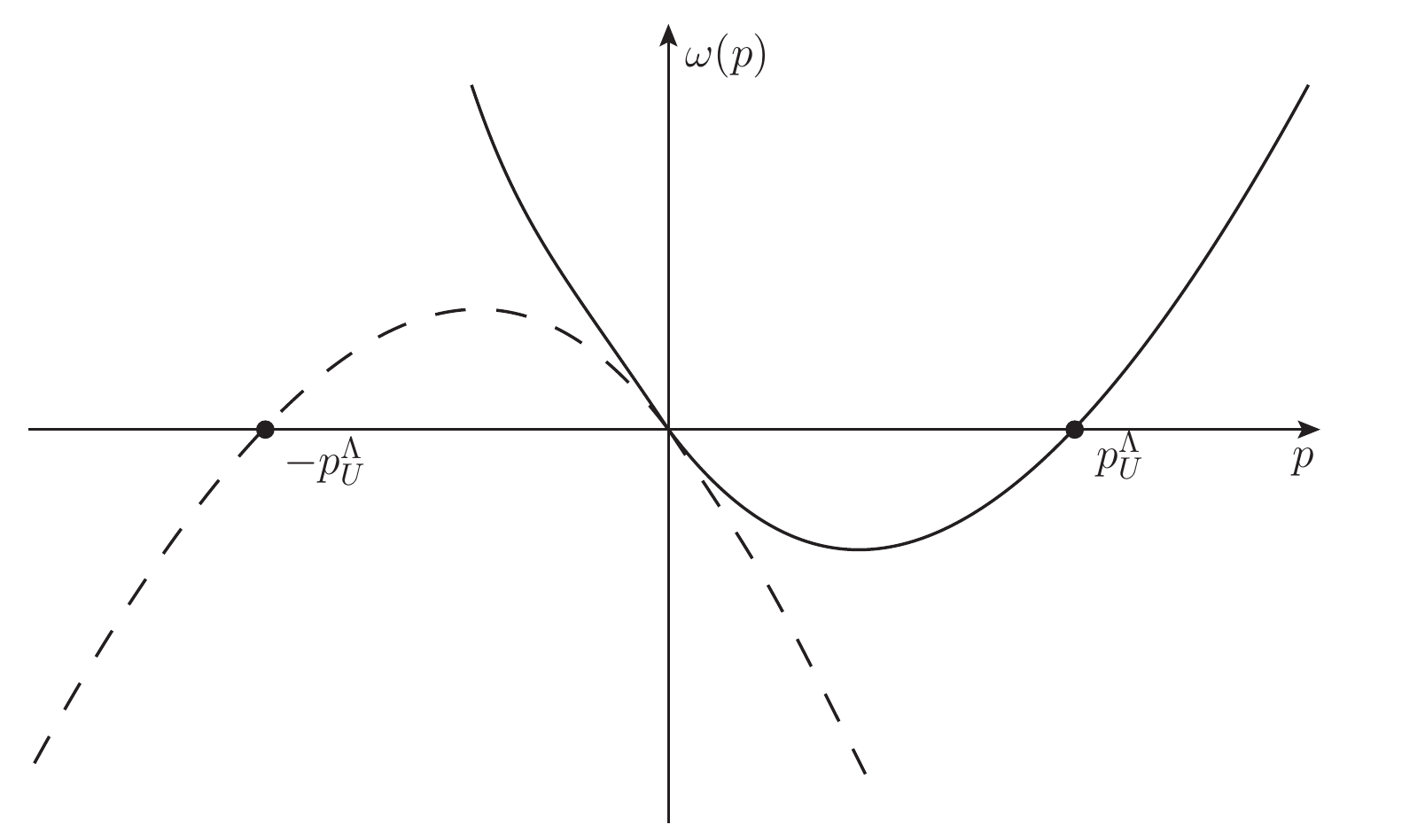}
\end{center}
\caption{The solid line represents the positive branch of the comoving frequency $\Om$, while the
dashed line represents the negative branch. One clearly sees that the superluminal Bogoliubov dispersion
is responsible for the two zero-frequency roots $\pm p^\Lambda_U$. The sign of the group velocity
can be seen from the slope of the solid line at the corresponding root.}
\label{WHdisprel_fig}
\end{figure}

To see this in more detail, we first note that the Bogoliubov dispersion relation in a one dimensional stationary flow, and for a longitudinal wave vector $p$, is
\be
\Om = \om - v p = \pm \sqrt{c^2 p^2 (1 + \xi^2 p^2)} ,
\label{Bogdr}
\ee
where $\om$ is the conserved frequency, $v$ is the flow velocity, $c$ is the speed of sound, and $\xi = \hbar/2m_{\rm at} c$ is the healing length, given in terms of the mass of the atoms $m_{\rm at}$.  
The $\pm$ sign refers to positive and negative norm branches, see {\it e.g.}~\cite{Mayoral11} for details. The flow profiles giving rise to a black hole (BH) and a white hole (WH) sonic horizon are represented in Fig.~\ref{vprofile_fig} (this is an alternative way of solving \eqref{Bogdr} equivalent to Fig.\ref{graphroots}).
The dispersion relation of \eq{Bogdr} evaluated in the asymptotic supersonic region is plotted in Fig.~\ref{WHdisprel_fig}. The zero-frequency roots are $\pm p^\Lambda_U$, where $p^\Lambda_U$ is
\be
p^\Lambda_U = \Lambda\,  \sqrt{v_L^2 - c_L^2} .
\label{pU}
\ee
In this equation, $v_L$ and $c_L$ are the asymptotic values of the velocity and speed of sound in the
supersonic region L, and $\Lambda = 1/c_L \xi_L =2m_{\rm at}/\hbar$ characterizes the short distance dispersion.
This root only exists in a supersonic flow, and its associated group velocity $v_{\rm g} = 1/\partial_\om p$ is directed {\rm against} the flow. Hence in a BH flow, it is oriented toward the horizon, whereas
for a WH one it is oriented away from it. This explains why the zero-frequency mode only appears in WH flows, where it is generated  at the sonic horizon.
At this point it should be mentioned that these solutions are \emph{not} restricted to superluminal dispersion.
A completely similar phenomenon exists in fluids characterized by a subluminal dispersion relation, such as that obtained by replacing $\xi^2 \to -\xi^2$ in  \eq{Bogdr}.
This time however, the zero frequency root, and the corresponding undulation,
live in the subsonic R region of the WH flow. This can be understood because of the (approximate) symmetry of the mode equation expose in \Sec{subsuper_Sec} and~\cite{Coutant11}, which replaces a superluminal dispersion by a subluminal one.

\begin{figure}[!ht]
\begin{center}
\includegraphics[scale=0.8]{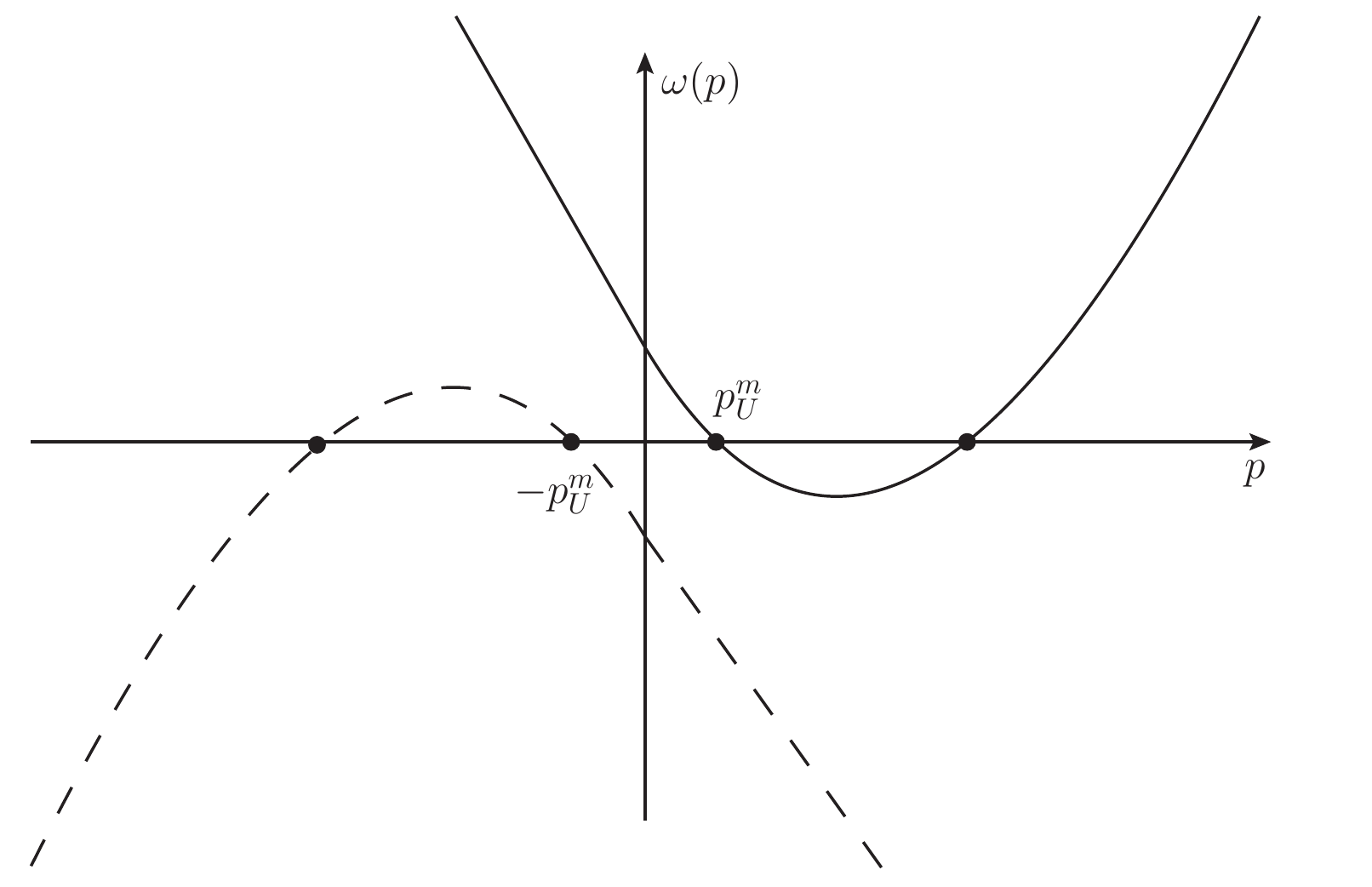}
\end{center}
\caption{As in Fig.~\ref{WHdisprel_fig}, the solid line represents the positive branch of the comoving frequency $\Om$, while the
dashed line represents the negative branch.
One sees that $p_\perp^2$, which acts as a mass, is responsible for new zero-frequency roots $\pm p^m_U$ which occur in the phonon part of
the dispersion relation. }
\label{WHdisprelm_fig}
\end{figure}

When considering elongated quasi one dimensional systems, but relaxing the assumption that the phonon excitations are purely longitudinal, the
phonon modes are now characterized by  their transverse wave number $p_\perp$, which takes discrete values $2 \pi n/L_\perp$, where $n$
is an integer and $L_\perp$ is the characteristic size of the perpendicular dimensions. When
$p_\perp^2 \neq 0$, the modified dispersion relation replacing \eq{Bogdr} is
\be
\Om = \om - v p = \pm \sqrt{c^2 (p^2 +p_\perp^2)(1 + \xi^2 (p^2 +p_\perp^2))} \, .
\label{BogdrNl}
\ee
It is represented in Fig.~\ref{WHdisprelm_fig}.
When $p_\perp^2 \xi^2 \ll 1$, $c^2 p_\perp^2$ acts as a mass squared.
In this regime, there are two new zero-frequency
roots $\pm p^m_U$. They live in the hydrodynamical regime, characterized by a relativistic
linear dispersion relation.
Indeed, in the limit $\xi^2 p_\perp^2 \to 0$, and
if $v_L^2/c_L^2$ is not too close to 1, $ p^m_U$ is independent of $\xi$ and given by
\be
p^m_U = \frac{c_L p_\perp}{\sqrt{v_L^2- c_L^2 }} .
\label{pUm}
\ee
In addition we note that the group velocity of this new solution has a sign
opposite to that of \eq{pU}. Hence, this new solution
will be emitted in BH flows but not in WH ones.\\

In the following, we shall first present the analysis of the massless undulation in white holes (\Sec{WH_Sec}). Then, we study and solve the scattering problem of massive modes in a BH geometry (\Sec{massfields_Sec}) and we exposed the properties of massive undulations in both BH (\Sec{BHundul_Sec}) and WH flows (\Sec{WHdispundul_Sec}). Finally, we apply our framework in a geometry containing both a BH and a WH horizon (\Sec{WD_Sec}).

\section{White hole undulations}
\label{WH_Sec}
In this section, we study in more detail the \emph{white hole} geometry. As mentioned in \Sec{analogmodel_Sec}, a white hole corresponds to the time reverse of a black hole. Its surface gravity is thus of opposite sign, {\it i.e.}, 
\be
\p_x v_{|\mathcal H} < 0.
\ee
For more transparency, we will define the surface gravity as $\kappa = - \p_x v_{|\mathcal H}$, so that the parameter $\kappa$ stays positive. When considering the wave equation \eq{fullPGwavequ}, or its 1+1 reduction, 
\be
(\p_t + \p_x v) (\p_t + v \p_x) \phi - \p_x^2 \phi = 0, \label{WHmodequ}
\ee
a simple way to pass from the black hole to the white hole case is to make the change $v \to -v$. As we see from the above equation \eqref{WHmodequ}, this is equivalent to the change $t \to -t$. 

\begin{figure}[h]
\begin{subfigure}[b]{0.5\textwidth}
\includegraphics[scale=0.7]{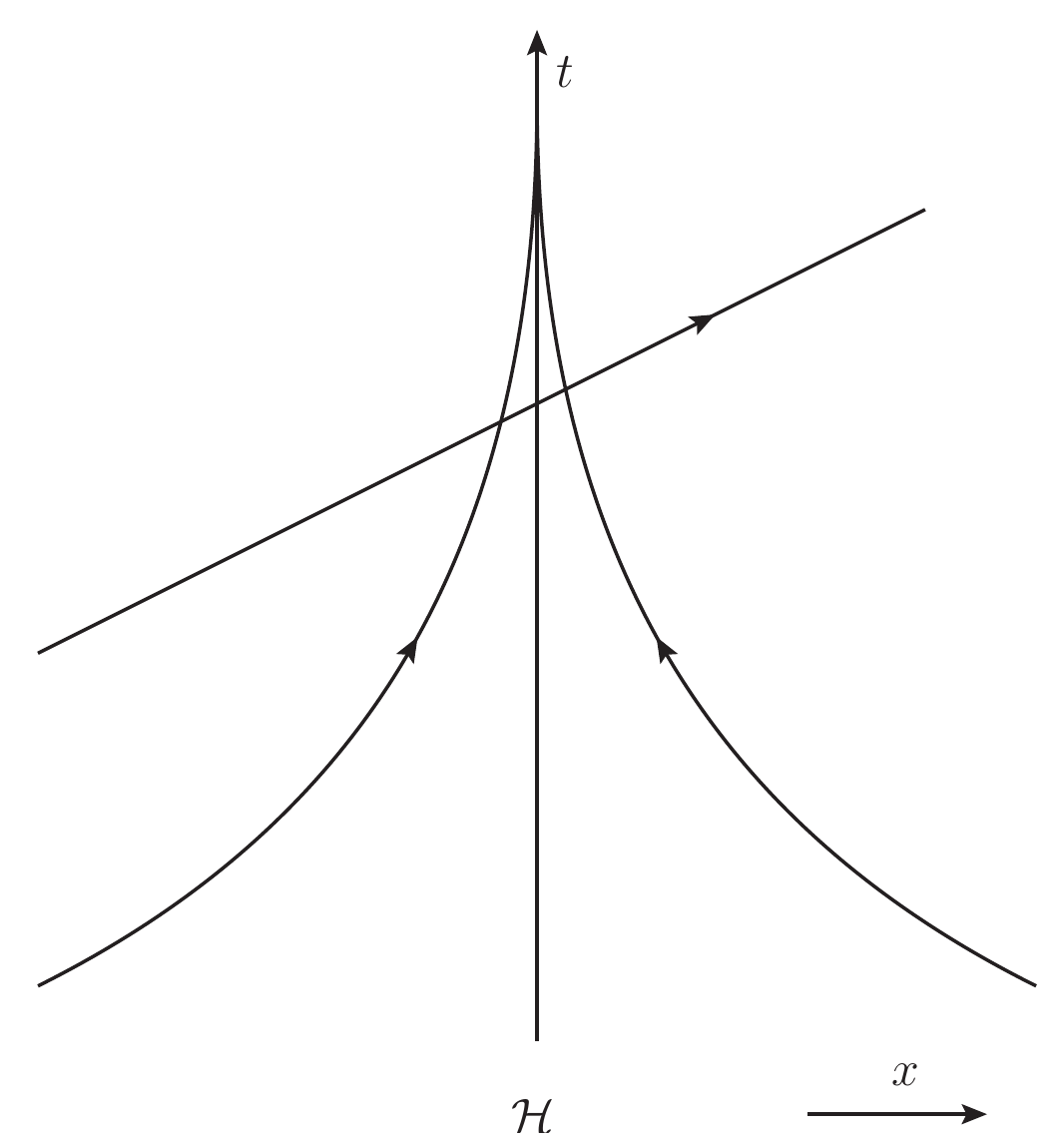}
\caption{Relativistic case.}
\end{subfigure}
\begin{subfigure}[b]{0.5\textwidth}
\includegraphics[scale=0.7]{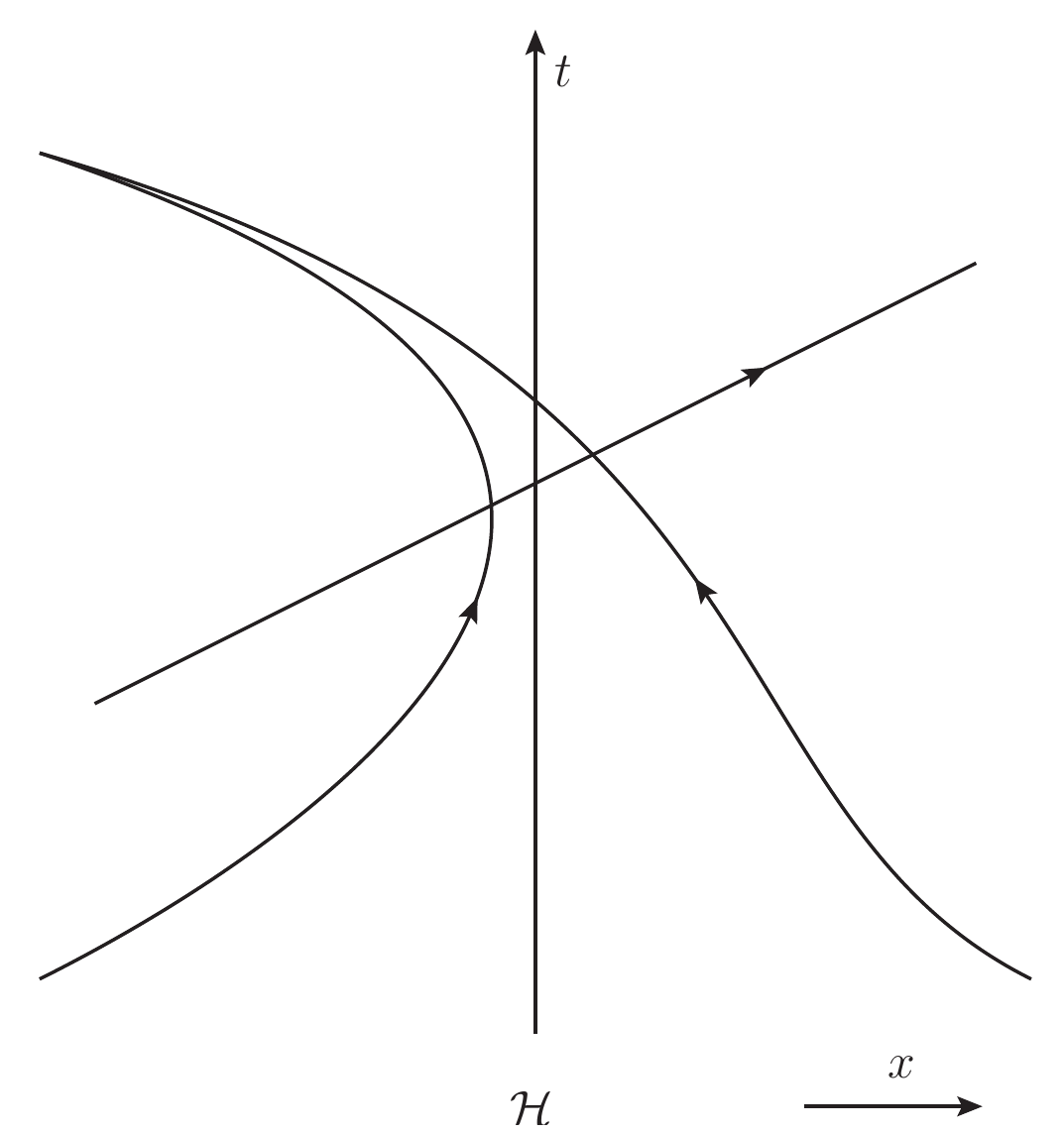}
\caption{Superluminal dispersion.}
\end{subfigure}
\caption{Geodesics around a white hole horizon, for relativistic or dispersive fields. $v$-modes focus and are being blueshifted, while $u$-modes come out regularly. This has to be compared to \Sec{HRNHR_Sec} and \Sec{eikonal_Sec}.}
\label{WHfocus_fig}
\end{figure}

However, inverting the flow of time is not free of consequences. In particular, the infinite redshift is now an infinite blue shift. Modes come in and focus on the white hole horizon, with an increasing co-moving frequency $\Om = \Om_0 e^{\kappa t}$ (see Fig.\ref{WHfocus_fig}). This raises new concerns about the stability of the white hole. Indeed, every noise in the infrared sector will be blueshifted and become highly energetic, with possible dramatic consequences. The question of stability of white holes is quite old~\cite{Frolov}. In general relativity ({\it i.e.}, no dispersion), not only white holes cannot be formed by a physical process, but their horizons are also unstable because of the infinite blueshift and focusing on the horizon, which is now future directed (see Fig.\ref{WHfocus_fig}). Indeed, this makes the renormalized stress-energy tensor infinite on the horizon. In analog gravity or when Lorentz invariance is broken at short distances, this question must be readdressed. In~\cite{Leonhardt02}, it was first believed that these are drastically unstable, because exponentially growing modes are present in the spectrum. In fact, these modes are \emph{not} included in the spectrum, because they do not satisfy the ABM requirement (see \Sec{setofmodes} in Chapter \ref{laser_Ch}). However, (sonic) white holes do display a mild instability, due to the divergence of the Bogoliubov coefficients for $\om \to 0$, as explained in \Sec{IRdiv_Sec}.

We conclude by a very small semantic remark. It is not clear whether an undulation should be called `instability' or not. Indeed, it is non standard in the sense that the growing rate is not exponential, but rather logarithmic of polynomial (unlike what we shall study in Chapter \ref{laser_Ch}). However, it does produce an ever growing contribution, that will ultimately backreact and modify the background structure (irrespectively of the underlying theory, Navier-Stokes, Gross-Pitaevskii or General Relativity). We compromise by calling it `mild instability'. In practice, {\it e.g.} in~\cite{Weinfurtner10}, it was observed that the undulation does not completely destroys the set up, but rather deforms it, because non linearities saturate it quite rapidly. 

\subsection{The black hole-white hole correspondence}
\label{BHWHcorresp_Sec}

As far as the stationary mode equation \eqref{modequ} is concerned, a white hole geometry is very similar to that of a black hole. Indeed, all the formalism developed in Chapter \ref{LIV_Ch} equally applies. In fact, there is a simple correspondence between the two cases, when solving the stationary problem, {\it i.e.}, \eq{wavequ}. In terms of stationary modes, the correspondence is made by exchanging {\it in} and {\it out} and performing a complex conjugation:
\be
\phi_\om^{\rm in, WH} = \left( \phi_\om^{\rm out, BH}\right)^*.
\ee
For more details, we refer to the App.D of~\cite{Macher09b}. This mapping follows from a symmetry of the mode equation \eqref{modequ}. Indeed, the latter is invariant under
\be
\om \to -\om \quad {\rm and} \quad v \to -v \label{BHWHsym}.
\ee
In the language of the transfer matrix of \Sec{transfermatrix_Sec}, it implies 
\be
U_{\rm WH} = \left(U_{\rm BH}\right)^* \label{BHtoWH},
\ee
where $U_{\rm WH}$ is defined through the same equation as \eqref{Udef}. As a corollary, the Bogoliubov coefficients of a white hole posses the same norm as those in the corresponding black hole setup. We define the Bogoliubov coefficients of a white hole by the same relation as \eqref{BogHR} and distinguish them with the superscript $W$. We get
\bsub \label{WHcoef}
\bea
\alpha_\om^{W} &=& \frac{\tilde{\Gam} \left(\frac{\om}{\kappa}\right)}{1 - e^{-\frac{2\om \pi}{\kappa}}} = e^{ i\frac{\pi}2} \tilde{\alpha}_\om^{W} ,\\
\tilde \beta_\om^{W} &=&  \frac{\tilde{\Gam} \left(\frac{\om}{\kappa}\right) e^{-\frac{\om \pi}{\kappa}}}{1 - e^{-\frac{2\om \pi}{\kappa}}} = e^{-i\frac{\pi}2} \beta_\om^{W}, 
\eea \esub 
where the function $\tilde \Gamma$ is defined in \eq{Gr0}.

\subsection{Undulation in a white hole}
\label{WHundul_Sec}
We now consider the existence of undulation in a white hole geometry with dispersion. Being motivated mainly by BEC, we assume a superluminal dispersion, however, as pointed out in \Sec{BECmass_Sec}, the results are the same for subluminal, except the fact that the undulation will live on the other side of the horizon. The Hamilton-Jacobi equation is 
\be
(\om - v p_\om)^2 = p_\om^2 + \frac{p_\om^4}{\Lambda^2}. \label{WH_HJ}
\ee
See Figs.\ref{graphroots} and \ref{WHdisprel_fig} for their resolution. Moreover, we take the profile $v$ positive, as in Fig.\ref{vprofile_fig}. On the left side, $v>1$, and on the right $1>v>0$. We anticipate the notations of \Sec{massfields_Sec}, and define the asymptotic value of $v$ by 
\be
D_L = \underset{x\to -\infty}{\lim}(v^2-1),
\ee
as on Fig.\ref{vprofile_fig}. If we work in a regime $\Lambda D_L^{3/2} \gg \kappa$, we can apply the results of Chapter \ref{LIV_Ch}. We now look at the necessary conditions. First, in the limit $\om \to 0$, \eqref{WHcoef} gives 
\be
|\alpha_\om|^2 \sim |\tilde \beta_\om|^2 \sim \frac{\kappa}{2\pi \om} \ 
\text{ and }\ - \alpha_\om^* \tilde \beta_\om \sim e^{-i\frac\pi2}\frac{\kappa}{2\pi \om}. 
\ee
Moreover, when $\om \to 0$, there is 2 non zero roots of \eq{WH_HJ}.  They are opposite to each other (which is necessary to have $\phi_0^{\rm out} = \phi_{-0}^{\rm out}$, see discussion after \eqref{phiphi}). The positive one is given by 
\be
p_U^\Lambda = \Lambda D_L^{1/2}.
\ee
In addition, the corresponding group velocity is, in a white hole, {\it out}-going. In fact, this last condition is what makes the difference between black and white holes. Therefore, all the conditions are met for an undulation to be spontaneously produced. In the 2-point function, using \eq{General_undul}, we obtain 
\bsub \label{Gundul} 
\bea
G_{\rm IR}(t;x,x') &\sim& 8 \int_{2\pi/t} \frac{ d\om}{\om} \, \frac{ \kappa}{2\pi} \times   \Phi_{\rm U}(x)\,  \Phi_{\rm U}(x') ,\\
&\sim& \frac{4\kappa}{\pi} \ln(t/2\pi)\times  \Phi_{\rm U}(x)\,  \Phi_{\rm U}(x') .
\eea \esub
This logarithmic growth was already observed in~\cite{Mayoral11}. Of course, in a medium, this growth would saturate because of the non linearities, as was also observed in~\cite{Mayoral11}. In the experiments of~\cite{Rousseaux07,Weinfurtner10}, only a constant (saturated) amplitude was observed. The saturated value of the amplitude could be obtained by using a treatment similar to~\cite{Pitaevskii84,Baym12}. Moreover, it would be very interesting to understand under which conditions the randomness of the amplitude in the linearized treatment is replaced by a deterministic nonlinear behavior, or if there is some residual randomness when non linearities are included. 

\begin{figure}[!ht]
\begin{center}
\includegraphics[scale=0.8]{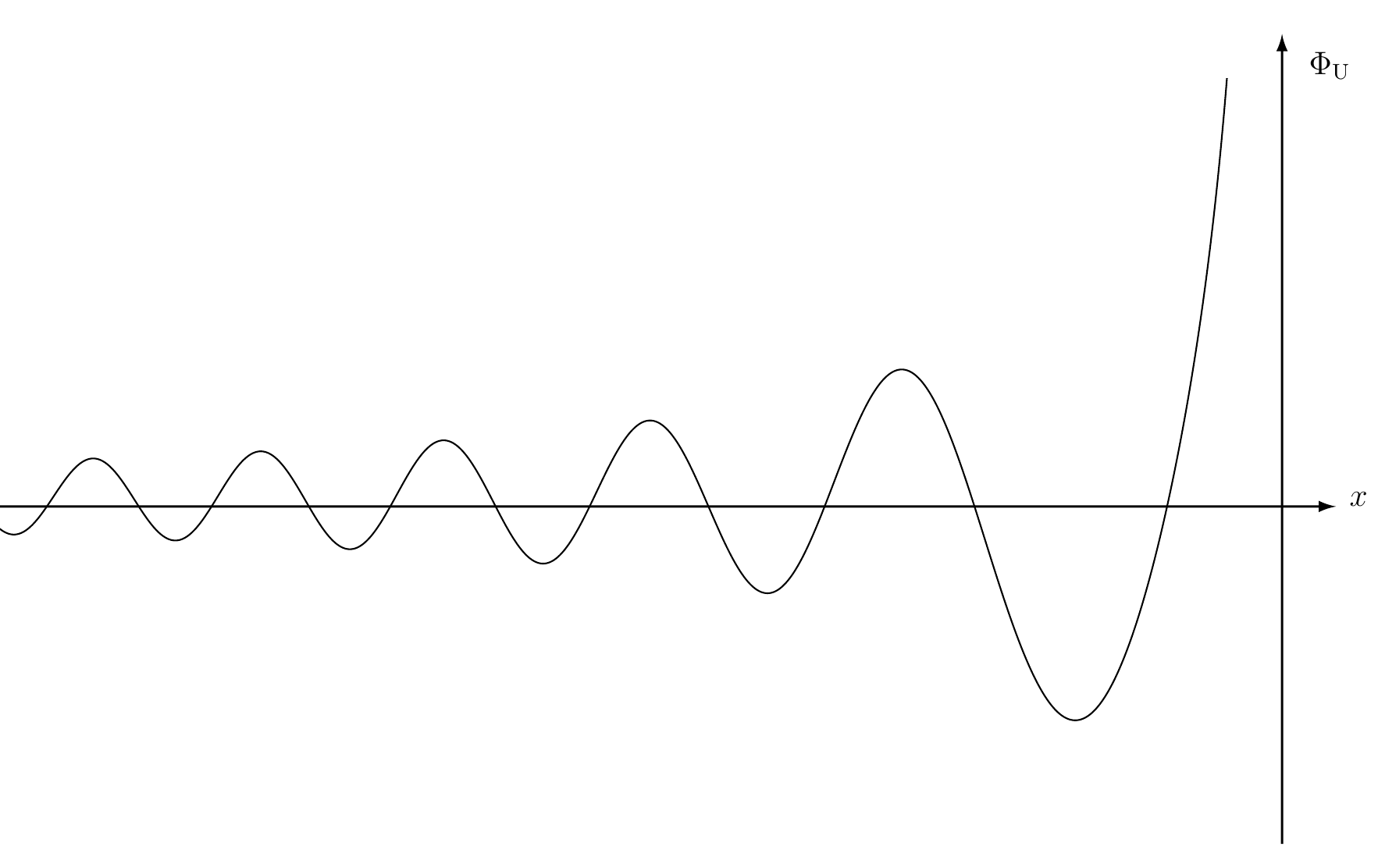}
\end{center}
\caption{Shape of the undulation profile obtained in \eq{undulpi}, on the left of the horizon. This expression is valid in the near horizon, {\it i.e.} $|\kappa x| \lesssim D_L$ and not too close to the horizon ($x=0$), {\it i.e.}, $|\kappa x| \gg (\kappa/\Lambda)^{2/3}$. On this plot, $\Lambda/\kappa \approx 10$. On the right side, the profile decays exponentially, as the $Ai$ function.}
\label{Undul_plot_fig}
\end{figure}

Moreover, using \Sec{modeanalys}, one can compute the profile of the undulation (in the WKB approximation) in the near horizon region (more precisely, region 1.(b) of Fig.\ref{regions}). On the right side, it decays according to the zero frequency limit of \eqref{firstSP}, whereas on the left side, one has
\be
\Re\left\{e^{-i\frac\pi4}\, \varphi_0^{\rm in}(x)\right\} \sim \frac1{\sqrt{8\pi \kappa}} \frac{\cos \left(\frac23\Delta(x)+\frac\pi4 \right)} {\sqrt{\Delta(x)(1+\kappa |x|)}} ,
\label{undulpi}
\ee
where $\Delta$ was defined in \eq{Delta}. The left side profile is represented on Fig.\ref{Undul_plot_fig}. On the right it behaves very much like the decaying Airy function $Ai$~\cite{Olver,AbramoSteg}, and on the left it oscillates in a similar manner but, quite surprisingly, the phase shift $\pi/4$ has the opposite sign. The origin of this flip is to be found in the extra factor of $1/p$ in the integrand of \eqref{contourmode} that is associated with the relativistic (non-positive) norm of \eqref{scalt}. It would be very interesting to observe the profile of \eqref{undulpi} and its unusual phase shift in future experiments. Further away from the horizon, the undulation profile can be obtained from the zero frequency limit of \eqref{xWKB}. Explicitly, we find 
\be
\Re\left\{e^{-i\frac\pi4}\, \varphi_0^{\rm in}(x)\right\} \sim \frac{\cos \left(p_U^\Lambda (x + \frac{D_L}{6\kappa}) + \frac\pi4 \right)}{\sqrt{4\pi \Lambda D_L^{3/2}}} .
\label{WHasundul}
\ee

In brief, this analysis \emph{predicts} that white hole flows should emit a zero frequency wave
which has a large amplitude and which behaves classically. 
Moreover, given the fact that the low frequency Bogoliubov coefficient $\beta_\om$
contributes to the undulation amplitude, we think that studying and observing these waves should be conceived as part of the enterprise
to observe the analogue Hawking radiation.
As a last remark, we wish to stress that the linearized treatment predicts that, when starting from the vacuum the undulation amplitude is described by a Gaussian stochastic ensemble with a vanishing mean amplitude. However, the same result will also hold if the field is in a thermal state. 
Indeed, the undulation acts as an amplifier, which generates a random noise by amplifying fluctuations~\cite{Unruh11,Leonhardt}, exactly as for primordial density fluctuations in the inflationary scenario~\cite{Campo05}. But these fluctuations need not be quantum, and therefore, in a thermal state the same phenomenon will also produce an undulation. As was discussed in~\cite{Mayoral11}, in that case, the only modification with respect to the previous discussion, is the growing rate, which is linear in time instead of logarithmic. 
Moreover, the outcome can also be a deterministic signal if the initial state is a large enough classical wave packet like a coherent state~\cite{Coutant13}. \\

It is also interesting to compare the expression \eqref{Gundul} with the corresponding one obtained in a black hole geometry. In that case, when 
working with the black hole {\it in} vacuum, the $\om$ contribution of the two-point function is, see \eqref{Gom},
\be
G^{\rm BH}_\om = \phi^{\rm in}_\om(x)\left( \phi^{\rm in}_\om(x')\right)^* + \phi^{\rm in}_{-\om}(x) \left(\phi^{\rm in}_{-\om}(x')\right)^* .
\ee 
Hence in the limit $\om \to 0$ it gives (see Eq.~32 in~\cite{Schutzhold10})  
\be
G^{\rm BH}_0 = 2\phi^{\rm in}_0(x) \left( \phi^{\rm in}_0(x')\right)^*,
\ee
which behaves very differently from $G_{\rm IR}$ of \eqref{Gundul}. It does not diverge as $1/\om$ and it is not the product of two {\it real} waves. 
To get an expression that might correspond to that of \eqref{Gundul}
one should express the {\it in} modes in terms of the {\it out} black hole modes given in \eqref{minusout} and (\ref{plusout}), as done in \eqref{correl}. 
Doing so, the prefactor of \eqref{Gundul} is recovered but the spatial behavior of $G^{\rm BH}$
is completely different because the {\it out} modes of \eqref{minusout} and (\ref{plusout})
are defined on opposite sides of the horizon, and become constant in their domain. Hence, unlike $G_{\rm IR}$,
$G^{\rm BH}$ cannot be written in the limit $\om \to 0$ as a product of twice the same real wave.

\section{Massive fields}
\label{massfields_Sec}

In \Sec{BECmass_Sec}, we saw that the introduction of a perpendicular momentum opens the
possibility of finding `massive' undulations in BH flows, which are well-described in the hydrodynamical approximation of the underlying condensed matter system.
To verify if this is the case, one should see how the mass affects the spectrum, and in particular
if it acts as an infrared regulator that saturates the growth of the undulation amplitude found in the massless case in \Sec{WHundul_Sec}.
In this section, these issues will be investigated in a simplified context where the phonon modes
obey a second order differential equation, {\it i.e.} for $\Lambda \to \infty$ in \eq{wavequ}. 
Our results should work not only for BEC but for other condensed matters systems
where the quasi-particle dispersion relation is linear at low frequency. 

As in preceding chapters, we study the propagation in a 1+1 Painlevé-Gullstrand metric
\be
ds^2 = dt^2 - (dx - v(x) dt)^2 . \label{PGds}
\ee
In a condensed matter context, it means that we assume a constant sound velocity (put to 1) and density, see \Sec{analogmodel_Sec} for details. The work presented in the following was obtained in~\cite{Coutant12}. Even though the main motivation concerns the appearance of undulations in analog models and especially BEC, it also provides a complete understanding of the emission of massive modes by black holes. In particular, analytical results are obtained for the scattering in the so-called CGHS black hole~\cite{Witten91,Callan92}. In \Sec{Sett}, we study the solutions of the Klein-Gordon equation in a stationary BH metric. We explain how the $in/out$
scattering matrix can be decomposed into three blocks that each encodes some aspect of mode mixing of massive fields. In \Sec{quattro}, we study three preparatory cases
which are then combined so as to obtain the $S$-matrix in a black hole flow similar to that represented in Fig.~\ref{vprofile_fig}.
In \Sec{BHundul_Sec} and \ref{WHdispundul_Sec}, we study the properties of massive undulations in black and white holes.

\subsection{Settings, mode mixing, structure of the $S$-matrix}
\label{Sett}

In the considered geometry, the definition of the surface gravity of the horizon we shall use is
\be
\kappa = \frac12 \partial_x (1 - v^2)_{|0}. \label{kappadef}
\ee
We adopt this local definition, which no longer refers to the norm of $K$ at infinity, because it
allows us to compare various geometries starting from the near horizon region (NHR). Unless specified otherwise, we shall only consider black holes, {\it i.e.} $\kappa >0$.
Notice also that we shall work with flow velocities that are either asymptotically bounded or unbounded; in the latter case, there will be a singularity.

The field will be studied at fixed Killing frequency $\om = - K^{\mu} p_\mu$, using a decomposition into stationary modes
\be
\phi = \int \phi_\om(x) e^{-i \om t} d\om,
\ee
exactly as in preceding chapters. At fixed $\om$, the Klein-Gordon equation \eqref{gmunuwavequ} in \eqref{PGds} gives
\be
[ (\om + i\partial_x v)(\om + iv\partial_x) + \partial_x^2 - m^2] \phi_\om(x) = 0 \label{mass_modequ}.
\ee
Similar equations are obtained when studying acoustic perturbations on a fluid flow with a velocity profile $v(x)$, see \Sec{analogmodel_Sec} or~\cite{Unruh95,Balbinot06}.
In these cases, a non zero transverse momentum $p_{\perp}$ plays the role of the mass $m$. Moreover, \eq{mass_modequ} is also relevant for standard Hawking radiation, when considering either massive quanta, but also those with high angular momentum. Indeed, from \eq{fullPGwavequ}, we see that in the near horizon region, there is an effective mass 
\be
m_{\rm eff}^2 = m^2 + \frac{\ell(\ell+1)}{r_{\mathcal H}^2} + \frac{2\kappa}{r_{\mathcal H}}. 
\label{m_eff}
\ee
However, this effective mass correctly models the full potential only if all the relevant physics, {\it i.e.} the scattering, occurs in the near horizon region. As we shall see in \Sec{quattro}, this is the case when $m_{\rm eff}$ is large compared to $\kappa$. In what follows, we consider only \eq{mass_modequ}, for profiles $v(x)$ that give rise to analytically soluble equations.
Yet, we aim to extract generic features. When studying numerically the phonon mode equation in
a Bose condensate and with a varying sound speed~\cite{to_appear}, 
we recovered the features found for solutions to \eq{mass_modequ}.

\subsubsection{Classical trajectories}
\label{clatrajSec}

To understand the consequences of the mass on black hole radiation, it is useful to first consider
the corresponding classical problem where $p = (\partial_x)^\mu p_\mu$ is the momentum of the massive particle at fixed $\om$.
In that case, the Hamilton-Jacobi equation associated with \eq{mass_modequ} is
\be
\Om^2 = (\om - v(x) p)^2 = p^2 + m^2 \label{HJ},
\ee
where $\Om = \om - v p$ is the comoving frequency. \eq{HJ} admits two roots
\begin{align}
p_\pm &= \frac{-\om v \pm \sqrt{\om^2 - m^2(1-v^2)}}{1-v^2}, \label{roots}
\end{align}
The classical trajectories obey Hamilton's equations $dx/dt = 1/\partial_\om p$
and $dp/dt = - 1/\partial_\om x$. We summarize here their main features with $\om > 0$,
see \cite{Jacobson07b,Jannes11,Jannes11b} for more details.
\bi
\item Close to the horizon, at first order in $1-v^2 \sim 2\kappa x \ll 1$, one has
\begin{align}
p_+ &= \frac{\om}{\kappa x},\\
p_- &= \frac{m^2 - \om^2}{2\om} .
\end{align}
We see that $p_+$ diverges for $x\to 0$ whereas $p_-$ hardly varies.
The corresponding geodesics follow
\begin{align}
x_+ (t) &= x_+^0 \, e^{\kappa t},\\
x_- (t) &= x_-^0 - \frac{2\om^2}{\om^2 + m^2} t. 
\end{align}
The second trajectory is regularly falling
across the horizon, while the first undergoes an infinite focusing in the past, in a mass independent manner.

\item Far away from the horizon,  in the left region (L) for $1-v^2 < 0$,
both solutions are moving to the left (since $v< 0$) even though $p_+<0$ and $p_- >0$.

\begin{figure}[!ht]
\begin{center}
\includegraphics[scale=1]{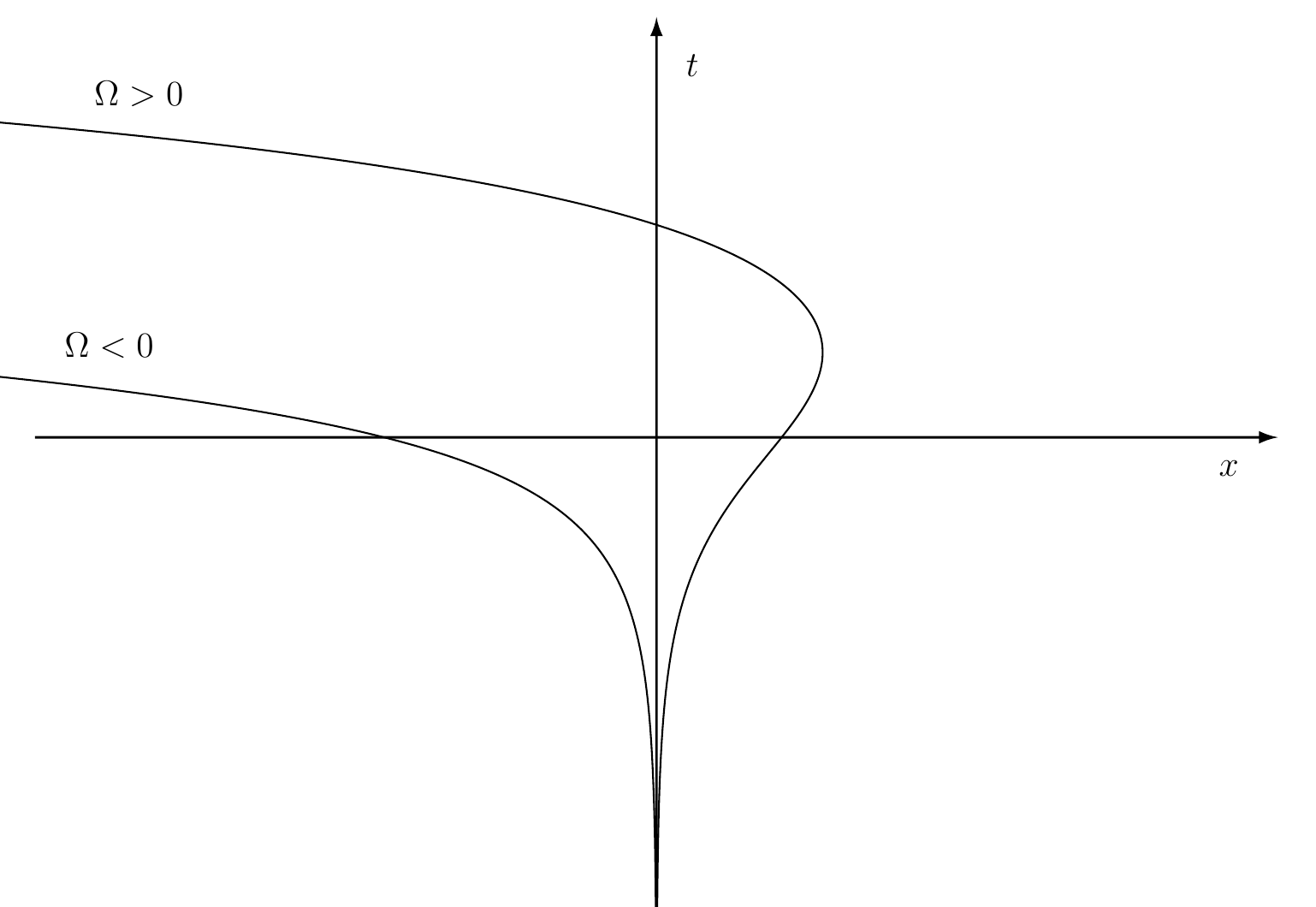}
\end{center}
\caption{In this figure, trajectories of massive particles for a fixed frequency $\om > 0$ below the threshold of \eq{thres} are shown.
The positive momentum trajectory is reflected in the outside region at the turning point $x_{\rm tp}(\om)$
whereas the negative momentum one (or equivalently the positive momentum one with negative $\om$~\cite{Jacobson96})
propagates in the inside region
where the Killing field is space-like. The negative momentum particle
has a negative comoving frequency $\Om = \om - vp$, and corresponds to a negative norm mode,
as explained in \Sec{eternalHR_Sec}.}
\label{mass_traj_fig}
\end{figure}

\item For $1-v^2 >0$, in the right region (R), as long as $(1-v^2)m^2 < \om^2$, there are two real roots.
At some point $x_{\rm tp}$ we reach $(1-v^2)m^2 = \om^2$ where they
become complex. This means that the trajectory is
reflected and falls back across the horizon.
Hence, the asymptotic value of $1 - v^2 > 0$ determines
the threshold frequency
\be
\om_R = m\sqrt{1 - v_{\rm as}^2}, \label{thres}
\ee
above which the trajectory is not reflected.
When $\om < \om_R$, there is a single trajectory with $p_\om > 0$
that starts from the horizon to the right and bounces back across the horizon,
see Fig.\ref{mass_traj_fig}. For $\om > \om_R$ instead, there are 2 disconnected
trajectories, one is moving outwards from the horizon, while the other falls in from $x = \infty$.
As we shall see,
the dimensionality of asymptotic modes will be different above and below $\om_R$.
\ei

\subsubsection{Mode mixing}

As in chapter \ref{HR_Ch}, we decompose the field operator in a basis of stationary modes
\be
\hat \phi(t,x) = \int_0^{+\infty} \sum_j \left[ \hat a_\om^j \phi_\om^j(x) + \hat a_\om^{j \dagger} (\phi_{-\om}^j(x))^* \right] e^{-i\om t} d\om + h.c.\, , \label{massphidecomp}
\ee
where the discrete index $j$ takes into account the dimensionality of mode basis at fixed $\om$. The basis is orthonormal in the sense of the Klein-Gordon scalar product
\begin{align}
(\phi_\om^j|\phi_{\om'}^{j'}) &= \int_{\mathbb R} \left[ \phi_\om^{j*}(\om' + i v\partial_x)\phi_{\om'}^{j'} + \phi_{\om'}^{j'} (\om - i v\partial_x) \phi_\om^{j*}\right] dx \label{scalprod}, \nonumber
\\
&= \pm \delta(\om - \om') \delta_{jj'}.
\end{align}
As before, we name the negative norm modes $(\phi_{-\om})^*$
so that $e^{i \om t}\phi_{-\om}$ is a positive norm mode with negative frequency.

To obtain the dimensionality of the mode basis, one must identify
the solutions of \eq{mass_modequ} that are asymptotically bounded modes (ABM).
This requirement univocally picks out a \emph{complete} basis over which the canonical field $\hat \phi$ must be decomposed.
Asymptotically, solving the mode equation \eqref{mass_modequ} is equivalent to solving the Hamilton-Jacobi equation \eqref{HJ}.
Hence, the dimensionality of the ABM can be found
by considering the real roots of \eq{HJ}. Moreover, the sign of the norm of an asymptotic mode
is given by the sign of the corresponding comoving frequency $\Om(p_i)= \om - v p_i$, as can be directly seen from \eq{scalprod}.

In addition, because the situation is non-homogeneous, modes mix and the basis is not unique. 
As usual, we introduce {\it in} modes $\phi_\om^{\rm in}$ and {\it out} modes $\phi_\om^{\rm out}$ by examining the mode behavior at early and late times. The $S$-matrix then relates the {\it in} and {\it out} bases. When there is an horizon, these two basis are inequivalent because positive and negative norm modes coexist and mix.
To further study the mixing, one should consider separately the frequencies below and above $\om_R$
in \eq{thres}.

For $0 < \om < \om_R$, there are two ABM.
One has a negative norm and propagates behind the horizon.
 The other has a positive norm, comes out from the horizon, and bounces back
across the horizon, see Fig.~\ref{mass_traj_fig}. The $S$-matrix thus has the form
\be
\bmat \phi_\om^{\rm in} \\ \left(\phi_{-\om}^{\rm in}\right)^* \emat = S^T \cdot \bmat \phi_\om^{\rm out} \\ \left(\phi_{-\om}^{\rm out}\right)^* \emat .
\label{Sdef22}
\ee
To follow the standard definition of the $S$-matrix~\cite{Weinberg1,Gottfried},
we use its transpose here.

For $\om > \om_R$, there are three ABM. The negative norm one still propagates behind the horizon.
The second one has a positive norm, comes out from the horizon, and reaches infinity. The third one comes from infinity and falls into the hole.
We denote the first two with the superscript $u$ and the last one with $v$,
because at high momentum, when the mass is negligible, they follow
retarded ($u$) and advanced ($v$) null geodesics.
We then define $S$ by
\be
\bmat \phi_\om^{\rm in, u} \\ \left(\phi_{-\om}^{\rm in, u}\right)^* \\ \phi_\om^{\rm in, v} \emat = S^T \cdot \bmat \phi_\om^{\rm out, u} \\ \left(\phi_{-\om}^{\rm out, u}\right)^* \\ \phi_\om^{\rm out, v} \emat.
 \label{Sdef33}
\ee
In this regime,  when starting from vacuum,
the three {\it out} occupation numbers $n^{\rm u}_\om$, $n^{\rm v}_\om$
and $n^{\rm u}_{-\om}$ obey $n^{\rm u}_\om + n^{\rm v}_\om = n^{\rm u}_{-\om}$
because of the stationarity of the settings. The first two are given by the square of the overlaps
\be
n^{\rm u}_\om = |(\phi_\om^{\rm out, u} | \phi_{-\om}^{\rm in, u*})|^2, \qquad
n^{\rm v}_\om = |(\phi_\om^{\rm out, v} | \phi_{-\om}^{\rm in, u*})|^2. \label{occupnum+}
\ee
As we shall see, in both cases, it is useful to decompose the total $S$-matrix as
 \be
S = S_{\rm far} \cdot S_{\rm ext} \cdot S_{\rm NHR}, \label{Sfactogene}
\ee
where each $S$-matrix describes one step of the {\it in/out} scattering\footnote{Note that this decomposition is very similar to that of~\cite{tHooft96}, but our is \emph{exact}, since we do not consider interactions or dispersive effects.}. The first one, $S_{\rm NHR}$, describes the mode mixing which arises for high momenta $p \gg m$,
near the horizon where the modes are effectively massless. The second matrix $S_{\rm ext}$ encodes the elastic scattering which occurs in the external region R.
Below the threshold, it describes the total reflection, while above it governs the
grey body factors encoding the partial transmission.
The last matrix $S_{\rm far}$ describes the mixing occurring in the left region
between the two modes that are propagating toward $x = - \infty$.

It should be mentioned that this decomposition is not unique, as only the {\it in/out} $S$-matrix
is univocally defined. However, in the absence of dispersion, each $S$-matrix is solution of a well defined and independent scattering problem. Therefore, this decomposition is very useful as it allows us to compute $S$, and to understand its properties. 

\subsubsection{Near horizon scattering}
\label{NHRSec}

We start with $S_{\rm NHR}$ because its properties are valid for all metrics possessing
an horizon and because they are determined for momenta much higher than the mass
and in the immediate vicinity of the horizon. To simplify the mode equation (\ref{mass_modequ}), we introduce the
auxiliary mode $\varphi_\om$
\be
\phi_\om(x) = \frac{e^{-i \om \int^x \frac{v(x')}{1-v^2(x')}dx'}}{\sqrt{|1-v^2|}} \varphi_\om(x). \label{def_varphi}
\ee
\eq{mass_modequ} is then cast in a canonical form, without the term linear in $\partial_x$,
\be
\left[ - \partial_x^2 + \left(\frac{\partial_x^2\sqrt{|1-v^2|}}{\sqrt{|1-v^2|}} +\frac{m^2}{1-v^2} -\frac{\om^2}{(1-v^2)^2}\right) \right] \varphi_\om(x) = 0 . \label{canmodequ}
\ee
We notice that the norm of $\varphi_\om$ is, up to a sign,
given by the Wronskian
\be
W(\varphi) = 2i\pi \left(\varphi_\om^* \partial_x \varphi_\om - \varphi_\om \partial_x\varphi_\om^* \right) \label{Wronskian}.
\ee
Using \eq{def_varphi}, one verifies that unit Wronskian $\varphi_\om$ modes,
give rise to $\phi_\om$ modes that have a unit norm with respect to the scalar product of \eq{scalprod}.
The relative sign is given by that of the comoving frequency $\Om$ of \eq{HJ}.

In the close vicinity of the horizon,  the mass term becomes negligible in \eq{canmodequ}.
More precisely, in the near horizon region where $1-v^2 \sim 2\kappa x$,
keeping only the leading term for $\kappa x \ll 1$, one obtains
\be
\left[ - \partial_x^2 -\left( \frac14 + \frac{\om^2}{4\kappa^2}\right)\frac1{x^2} \right] \varphi_\om(x) = 0 .
\ee
Therefore the leading behavior of $\varphi_\om$ is
\be \bal
\varphi_\om \underset{x \to 0}{\sim} \, &\underbrace{\Theta(-x)\, A |2\kappa x|^{i \frac{\om}{2\kappa} +\frac12}}_{\text{focusing on the left}} + \underbrace{\Theta(x)\ A' \, |2\kappa x|^{i \frac{\om}{2\kappa} +\frac12}}_{\text{focusing on the right}} \\
&+ \underbrace{\Theta(-x)\, B  |2\kappa x|^{-i \frac{\om}{2\kappa}+\frac12}  +\Theta(x) \, B'  |2\kappa x|^{-i \frac{\om}{2\kappa} +\frac12}}_{\text{regularly falling in}}. 
\eal \label{NHRbehav} \ee
When re-expressing this in terms of the original mode $\phi_\om$ using \eq{def_varphi},
we see that the $B$ weighted terms are regular, and are in fact constant.
Therefore, they account for the regularity of the left moving mode as it crosses the horizon. Hence, we impose
\be
B=B'. \label{regularcondition}
\ee
On the other hand, the $A$ parts in \eq{NHRbehav} oscillate infinitely around $x=0$ and account for
high momentum modes living on either side of the horizon, which are singular on it.
As understood by Unruh~\cite{Unruh76}, it is appropriate to combine them in superpositions that are analytic either in the upper, or lower half complex $x$-plane, as was exposed in details in \Sec{eternalHR_Sec}. This characterization applies to the $\phi_\om$ modes which are solutions to \eq{mass_modequ}.
Hence, the modes $\varphi_\om$ of \eq{def_varphi} are products of a non analytic function and an analytic one, which is an Unruh mode:
\bsub \bea 
\varphi_\om^{\rm in} &\sim& \Gamma \left(i\frac{\om}{\kappa}\right) \frac{e^{\frac{\om \pi}{2\kappa}}}{\sqrt{8\pi^2 \kappa}} |x|^{-i\frac{\om}{2\kappa}+\frac12}\times \left( x+i\epsilon \right)^{i\frac{\om}{\kappa}}, \label{inplus}\\
\left(\varphi_{-\om}^{\rm in}\right)^*
&\sim& \Gamma \left(i\frac{\om}{\kappa}\right) \frac{e^{\frac{\om \pi}{2\kappa}}}{\sqrt{8\pi^2 \kappa}} |x|^{-i\frac{\om}{2\kappa}+\frac12} \times \left( x-i\epsilon \right)^{i\frac{\om}{\kappa}}. \label{inminus}
\eea \esub
We have used \eq{Wronskian} to normalize these modes, and their phases have been chosen in order to obtain simple expressions.
When there are turning points, as is the case for dispersive fields~\cite{Coutant11} and for massive fields, one should pay attention to these phases.

The normalized modes that propagate on either side of the horizon and vanish on the other side are
\bsub \label{NHRoutmassmodes} 
\bea 
\varphi_\om^{\rm Right} &\sim& \Theta(x) \frac{|2\kappa x|^{i\frac{\om}{2\kappa} +\frac12}}{\sqrt{4\pi \om}},\label{rightm}\\
\left(\varphi_{-\om}^{\rm Left}\right)^* &\sim& \Theta(-x) \frac{|2\kappa x|^{i\frac{\om}{2\kappa} +\frac12}}{\sqrt{4\pi \om}}.
\eea \esub
They correspond to the massless {\it out} modes of \Sec{eternalHR_Sec}, but here they are not the massive {\it out} modes, since they undergo extra scatterings. The near horizon $S$-matrix $S_{\rm NHR}$  is then defined by
\be
\bmat \phi_\om^{\rm in} \\ \left(\phi_{-\om}^{\rm in}\right)^* \emat = \bmat \alpha_\om^{\rm NHR} & \tilde \beta_\om^{\rm NHR} \\ \beta_\om^{\rm NHR} & \tilde \alpha_\om^{\rm NHR} \emat \cdot \bmat \phi_\om^{\rm Right} \\ \left(\phi_{-\om}^{\rm Left}\right)^* \emat.
\label{SNHR_1}
\ee
Using the analytic properties of the {\it in} modes of Eqs. \eqref{inplus}, \eqref{inminus}, we immediately obtain
\be
S_{\rm NHR} = \sqrt{\frac{\om}{2\pi \kappa}} \Gamma \left(i\frac{\om}{\kappa}\right) \bmat e^{\frac{\om \pi}{2\kappa}} & e^{-\frac{\om \pi}{2\kappa}} \\ e^{-\frac{\om \pi}{2\kappa}} & e^{\frac{\om \pi}{2\kappa}} \emat. \label{SNHR}
\ee
Unlike the other matrices in \eq{Sfactogene}, $S_{\rm NHR}$
is \emph{universal} in that it only depends on $\kappa$ in \eq{kappadef}.
It is independent of the other properties of the profile $v(x)$, and also of the mass $m$.
In fact, when considering a two dimensional massless field, which obeys \eq{mass_modequ} with $m = 0$,
the left moving v-modes decouple, $n^{\rm v}_\om$ in \eq{occupnum+} vanishes, and
the total $S$-matrix reduces to the above $S_{\rm NHR}$
(when the asymptotic flow velocity $v$ is such that {\it out} modes are well defined).
In that case, on both sides of the horizon, the flux of $u$-quanta is Planckian, and
at the standard Hawking temperature $T_H$ of \eq{TH}, since $n^{\rm u}_\om= n^{\rm u}_{-\om}$ and
\be
\frac{n^{\rm u}_\om}{n^{\rm u}_\om + 1} = \left| \frac{\beta_\om^{\rm NHR}}{\alpha_\om^{\rm NHR}}\right|^2 = e^{- 2\pi \om/\kappa} .\label{SNHR+1}
\ee

\subsubsection{Exterior and interior scatterings}
\label{StrucSec}

As mentioned above, $S_{\rm ext}$ and $S_{\rm far}$ both depend on other properties of the profile $v(x)$ than $\kappa$.
However their structure can be analyzed in general terms,
and the meaning of their coefficients can be identified.
Before computing these coefficients in specific flows,
it is of value to present their general features.

Below $\om_R$ of \eq{thres}, the positive norm mode is totally reflected
while the negative norm mode propagates inside the horizon.
The exterior scattering matrix $S_{\rm ext}$
is thus fully characterized by the phase accumulated by the positive norm mode
in the right region
\be
\bmat \phi_\om^{\rm Right} \\ \left(\phi_{-\om}^{\rm Left}\right)^* \emat = \bmat e^{i\delta_\om} & 0 \\ 0 & 1 \emat \cdot \bmat \phi_\om^{\rm Refl} \\ \left(\phi_{-\om}^{\rm Left}\right)^* \emat. \label{Sextbelow}
\ee
Since this mode is totally reflected, it is the unique ABM of \eq{canmodequ} in R.
For small $x\to 0^+$, using \eq{rightm}, its behavior will be
\be
\varphi_\om(x) \underset{0^+}
{\sim} \, C \left[  |2\kappa x|^{i \frac{\om}{2\kappa} +\frac12} + e^{i\delta_\om} \times
|2\kappa x|^{-i \frac{\om}{2\kappa} +\frac12} \right], \label{Boom}
\ee
which will allow us in the next sections to extract the phase $e^{i\delta_\om}$.

In the interior region, there is some extra mode mixing which is described by $S_{\rm far}$.
Because the norms of the two modes are of opposite sign, this mixing
introduces new Bogoliubov coefficients:
\be
\bmat \phi_\om^{\rm Left} \\ \left(\phi_{-\om}^{\rm Left}\right)^* \emat = \bmat \alpha_\om^{\rm far} & \beta_\om^{\rm far*} \\ \beta_\om^{\rm far} & \alpha_\om^{\rm far*} \emat \cdot \bmat \phi_\om^{\rm out} \\ \left(\phi_{-\om}^{\rm out}\right)^* \emat \label{Sfar}.
\ee
This scattering is entirely governed by \eq{canmodequ} in the interior region L.
When working in an appropriate basis, namely when positive and negative frequency modes are complex conjugated,
the real character of \eq{canmodequ} guarantees that the matrix $S_{\rm far}$ is an element of $SU(1,1)$.

\begin{figure}[!ht]
\begin{center}
\includegraphics[scale=1]{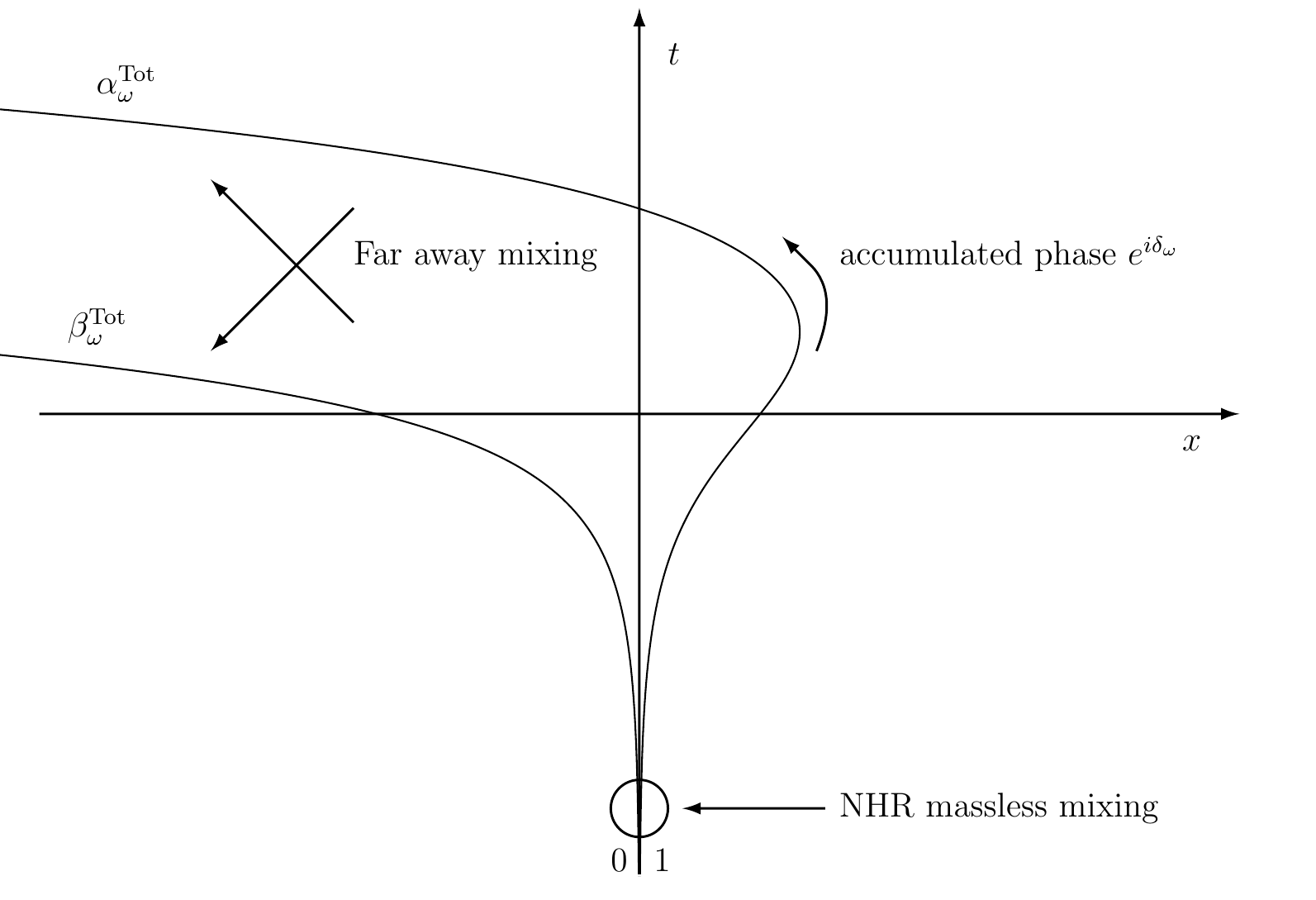}
\end{center}
\caption{In this figure a schematic representation of the $S$-matrix decomposition in \eq{Sfacto22} is shown.
The asymptotic {\it in} and {\it out} amplitudes of the positive
frequency mode $\phi_\om^{\rm in}$ are indicated.}
\label{Sfacto_fig}
\end{figure}

We see that the regularity condition of \eq{regularcondition} for the mode crossing the horizon plus the ABM requirement
reduces the dimensionality of the four unknown coefficients of \eq{NHRbehav} to two.
The $S$-matrix in \eq{Sfactogene} is then
\be
S = \bmat \alpha_\om^{\rm Tot} & \beta_\om^{\rm Tot} \\ \tilde \beta_\om^{\rm Tot} & \tilde \alpha_\om^{\rm Tot} \emat = \bmat \alpha_\om^{\rm far} & \beta_\om^{\rm far} \\ \beta_\om^{\rm far*} & \alpha_\om^{\rm far*} \emat
\cdot \bmat e^{i\delta_\om} & 0 \\ 0 & 1 \emat \cdot  \bmat \alpha_\om^{\rm NHR} & \beta_\om^{\rm NHR} \\ \tilde \beta_\om^{\rm NHR} & \tilde \alpha_\om^{\rm NHR} \emat.\label{Sfacto22}
\ee
This decomposition is depicted in Fig.\ref{Sfacto_fig}.
Notice that  it is the transposed version of
the intermediate $S$ matrices of Eqs.~\eqref{SNHR_1}, \eqref{Sextbelow}, 
\eqref{Sfar} that appear in \eq{Sfacto22}, as explained after \eq{Sdef22}.

For frequencies larger than $\om_R$, the $u$ mode and the $v$ mode, both of positive norm, mix in the exterior region R.
The matrix $S_{\rm ext}$ now is $3\times 3$ and reads
\be
\bmat \phi_\om^{\rm Right} \\ \left(\phi_{-\om}^{\rm Left}\right)^* \\ \phi_\om^{\rm in, v} \emat = \bmat T_\om & 0 & \tilde R_\om \\ 0 & 1 & 0 \\ R_\om & 0 & \tilde T_\om \emat \cdot \bmat \phi_\om^{\rm out, u} \\ \left(\phi_{-\om}^{\rm Left}\right)^* \\ \phi_\om^{\rm NHR, v} \emat.
\ee
The $2\times 2$ non trivial sector of this matrix is an element of $U(2)$,
and it describes an \emph{elastic} scattering.
The interior scattering shares the same properties that occur for frequencies below $\om_R$, namely the positive and negative norm modes propagating in the region L mix.
The outgoing $u$ mode is not affected by this scattering, and is thus left unchanged by $S_{\rm far}$. Therefore, the structure of the total $S$-matrix is
\be
S = \bmat 1 & 0 & 0 \\ 0 & \alpha_\om^{\rm far*} & \beta_\om^{\rm far*} \\ 0 & \beta_\om^{\rm far} & \alpha_\om^{\rm far} \emat \cdot \bmat T_\om & 0 & R_\om \\ 0 & 1 & 0 \\ \tilde R_\om & 0 & \tilde T_\om \emat \cdot \bmat \alpha_\om^{\rm NHR} & \beta_\om^{\rm NHR} & 0 \\ \tilde \beta_\om^{\rm NHR} & \tilde \alpha_\om^{\rm NHR} & 0 \\ 0 & 0 & 1 \emat .\label{Sfacto33}
\ee

\subsection{Exactly solvable models}
\label{quattro}

We shall first compute the above $S$ matrices in three preparatory cases
in order to understand various aspects regarding the
scattering of massive modes on a stationary horizon.
The results of these three cases will then be put together so as
to obtain the $S$-matrix in a background flow
relevant for analogue gravity models,
and similar to that presented in Fig.~\ref{vprofile_fig}.

In what follows, the various geometries will be characterized by a single function, given by the conformal factor of \eq{PGds}
\be
C(x) = 1-v^2(x). \label{conformalF}
\ee
The reason to refer only to this function is double. First, as we see from \eq{canmodequ}, the mode equation for $\varphi_\om$ only depends on $C(x)$.
Secondly, it will allow us to consider ranges of $x$ where $C(x) >1$.
In such regions, the function $v(x)$ of \eq{PGds}
is complex. However, neither the geometry nor the wave equation \eqref{canmodequ} are ill defined,
as can be seen by making the change of time coordinate $t_S = t + \int v dx/(1-v^2)$, which gives
\be
ds^2 = (1-v^2) dt_S^2 - \frac{dx^2}{1-v^2}.
\ee
Note that this change is nothing else than the return to a Schwarzschild time, as in \Sec{PGgeod_Sec}. This line element obviously depends only on $C(x)$, and thus can be extended to $C>1$. In fact, the reason why the mode equation for $\varphi_\om$ depends only on $C$ is simply that the prefactor in \eq{def_varphi} accounts for the change $t \to t_S$. This gives a physical interpretation of the auxiliary field $\varphi_\om$. It is simply the stationary mode as expressed Schwarzschild-like coordinates. Moreover, near an horizon, since $C \approx 0$, $v$ is always well defined in its vicinity. Therefore, we will always be able to use $\phi_\om$ around an horizon, and $\varphi_\om$ in any other regions. The latter obeys an equation that is simpler to solve, while the former allows us to impose regularity conditions across the horizon.

\subsubsection{Rindler horizon}
\label{RindlerSec}

It is instructive to first study a Rindler (future) horizon in the above formalism.
To do so, we use the profile defined by
\be
C(x)= 2\kappa x .
\label{Rindler}
\ee
It is straightforward to check that this metric has a vanishing scalar curvature, since 
\be
R = - \partial_x^2 C/2,
\ee 
and thus \eq{Rindler} describes flat space. In \eq{Rindler} $\kappa$ is the `surface gravity'
as defined by \eq{kappadef}.
The fact that it depends on the arbitrary normalization of the Killing field $K$
is free of physical consequence, because the $S$-matrix depends only
on the ratio $\om/\kappa$, see {\it e.g.}, \eq{SNHR}.

In this geometry, \eq{canmodequ} reads
\be
\left[ - \partial_x^2 +\frac{m^2}{2\kappa x} -\left( \frac14 + \frac{\om^2}{4\kappa^2}\right)\frac1{x^2} \right] \varphi_\om(x) = 0 .
\label{modequRindler}
\ee
The interesting aspect of Rindler space
is that we know the result in advance.
Indeed, since there is no pair creation in flat space, the total Bogoliubov transformation of \eq{Sfactogene} must be trivial, {\it i.e.}, $\beta_\om^{\rm Tot} = 0$. However, from \eq{modequRindler} we see that close to the horizon
the modes behave as in \eq{NHRbehav},
and thus are subjected to the near horizon region mixing described in \Sec{NHRSec}.
Therefore, the extra scattering described by $S_{\rm far}$ and $S_{\rm ext}$ exactly compensates the near horizon one, so that the total $S$-matrix is trivial. 
To show that this is the case, we solve \eq{modequRindler}
following the steps of \Sec{NHRSec}.

\eq{modequRindler} should thus be solved separately for $x>0$ and $x<0$.
On both sides, its solutions can be expressed in terms of Bessel functions. We start by studying the
exterior R region. For $x > 0$, the only ABM is
\be
\varphi_\om(x) = C \frac{2i}{\pi} K_{i \om/\kappa}\left(2\sqrt{\frac{m^2 x}{2\kappa}}\right),
\ee
where $K_{\nu}(z)$ is the Mac-Donald function~\cite{Olver} and $C$ a constant.
At large values of $x$ 
\be
\varphi_\om(x) \underset{+\infty}{\sim} 2i C \left(\frac{\kappa x}{8\pi^2 m^2}\right)^{\frac14} e^{-2\sqrt{\frac{m^2 x}{2\kappa}}}.
\ee
This exponential decrease is expected since $\om_R$ of \eq{thres} is infinite. Near the horizon, for $x \to 0^+$, the ABM behaves as
\be
\varphi_\om(x) \underset{0^+}{\sim} - Ce^{i\frac{\om}{\kappa} \ln(\frac m{2\kappa})} \left(\frac{e^{i\delta_{\rm Rindler}}\times |2\kappa x|^{-i\frac\om{2\kappa} +\frac12} + |2\kappa x|^{i\frac\om{2\kappa} +\frac12}}{\Gamma(1+i\om/\kappa) \sinh(\frac{\om\pi}{\kappa})}\right),
\ee
where
\be
e^{i\delta_{\rm Rindler}} = \frac{\Gamma(i\om/\kappa)}{\Gamma(-i\om/\kappa)} e^{-2i\frac\om\kappa \ln(\frac m{2\kappa})}.
\label{deltaR}
\ee
This is the phase shift that enters in \eq{Sextbelow}.
It will play a crucial role in what follows.

In the interior region L, the general solution reads
\be
\varphi_\om(x) = A \sqrt{-x} J_{-i\om/\kappa}\left(2\sqrt{-\frac{m^2 x}{2\kappa}}\right) + B \sqrt{-x} J_{i\om/\kappa}\left(2\sqrt{-\frac{m^2 x}{2\kappa}}\right).
\ee
Interestingly, these solutions have been considered before, but in a different context, since the interior of the Rindler horizon corresponds to the so-called `Milne universe'~\cite{BirrellDavies}. Near the horizon, for $x \to 0^-$, one finds
\be
\varphi_\om(x) \underset{0^-}{\sim} A_\om \frac{e^{-i\frac\om{\kappa} \ln(\frac m{2\kappa})}}{\sqrt{2\kappa}\Gamma(1-i\om/\kappa)} |2\kappa x|^{-i\frac\om{2\kappa} + \frac12} + B_\om \frac{e^{i\frac\om{\kappa} \ln(\frac m{2\kappa})}}{\sqrt{2\kappa}\Gamma(1+i\om/\kappa)} |2\kappa x|^{i\frac\om{2\kappa} + \frac12}.
\ee
In order to build the normalized positive frequency left mode $\phi_\om^{\rm Left}$ appearing in \eq{Sfar},
we choose
\be
A_\om = -i\sqrt{\frac{\om}{2\pi \kappa}} \Gamma(-i\om/\kappa) e^{i\frac\om{2\kappa} \ln(\frac m{2\kappa})},
\ee
and $B_\om=0$, so as to get
\be
\varphi_\om^{\rm Left}(x) \underset{0^-}{\sim} \frac{ |2\kappa x|^{-i\frac\om{2\kappa} + \frac12}}{\sqrt{4\pi \om}}.
\ee
For $x \to - \infty$,  the asymptotic behavior of this mode is~\cite{Olver,AbramoSteg},
\begin{align}
\varphi_\om^{\rm Left} \,  \underset{-\infty}{\sim} & \sqrt{\frac{\om}{2\pi \kappa}} \Gamma(-i\om/\kappa)\, e^{\frac{\om\pi}{2\kappa}} \, e^{i\frac\om{\kappa} \ln(\frac m{2\kappa}) -i\frac\pi4} \times \left\{ \frac{e^{-i\sqrt{\frac{-2m^2 x}{\kappa}}}}{\left(\frac{8\pi^2 m^2}{- \kappa x}\right)^{\frac14}}
+ e^{-\frac{\om\pi}{\kappa}} e^{ i\frac{\pi}2} \times \frac{e^{i\sqrt{\frac{-2m^2 x}{\kappa}}}} {\left(\frac{8\pi^2 m^2}{- \kappa x}\right)^{\frac14}} \right\} .
\end{align}
In the parenthesis, the first term is the asymptotic positive norm {\it out} mode,
whereas the last factor of the second term gives the negative norm one. Therefore, the coefficients of \eq{Sfar} are
\bsub \label{farR} 
\bea
\alpha_\om^{\rm far} &=& \sqrt{\frac{\om}{2\pi \kappa}} \Gamma(-i\om/\kappa)\, e^{\frac{\om\pi}{2\kappa}} \, e^{i\frac\om{\kappa} \ln(\frac m{2\kappa}) -i\frac\pi4},\\
\beta_\om^{\rm far*} &=& \alpha_\om^{\rm far} \times e^{-\frac{\om\pi}{\kappa}} \, e^{i \frac{\pi}2}.
\eea \esub
Making a similar computation for the negative left mode $(\phi_\om^{\rm Left})^*$, we obtain $\alpha_\om^{\rm far*}$ and $\beta_\om^{\rm far}$ and verify that
$S_{\rm far}$ is an element of $SU(1,1)$.
From \eq{farR} we see that $\vert \beta_\om^{\rm far}/\alpha_\om^{\rm far}\vert^2 = e^{- 2 \pi \om/\kappa}$, which is exactly the ratio of the near horizon coefficients of \eq{SNHR}.
This is a necessary condition for having $\beta_\om^{\rm Tot} = 0$. However it is not sufficient, as one also needs the
phases to match each other, since
\be
\beta_\om^{\rm Tot} =  \tilde \alpha_\om^{\rm NHR} \beta_\om^{\rm far} + \alpha_\om^{\rm far} \beta_\om^{\rm NHR} e^{i \delta_{\om}}. \label{betaTotRind}
\ee
From this equation,
one clearly sees the crucial role played by $e^{i\delta_\om}$ of \eq{deltaR}.
An explicit calculation shows that the total $S$-matrix of \eq{Sfacto22} is
\be
S = \frac{\Gamma(i\om/\kappa)}{\Gamma(-i\om/\kappa)} e^{-i\frac\om\kappa \ln(\frac m{2\kappa})} \bmat e^{-i\frac\pi4} & 0 \\ 0 & e^{i\frac\pi4} \emat.
\ee
We see that the two $in/out$ coefficients $\beta_\om^{\rm Tot}$ vanish for all values of $\om$ and $m$.
Hence the scattering away from the horizon {\it exactly} compensates the near horizon mixing
and there is no pair creation.
Of course, this exact cancellation was expected in the present case.
However, in more general space-times, as we shall see below,
a partial cancellation between $S_{\rm far}$ and $S_{\rm NHR}$
will be obtained for similar reasons.

\subsubsection{Totally reflecting model}
\label{SandroSec}

\begin{figure}[!ht]
\begin{subfigure}[b]{0.3\textwidth}
\includegraphics[scale=0.55]{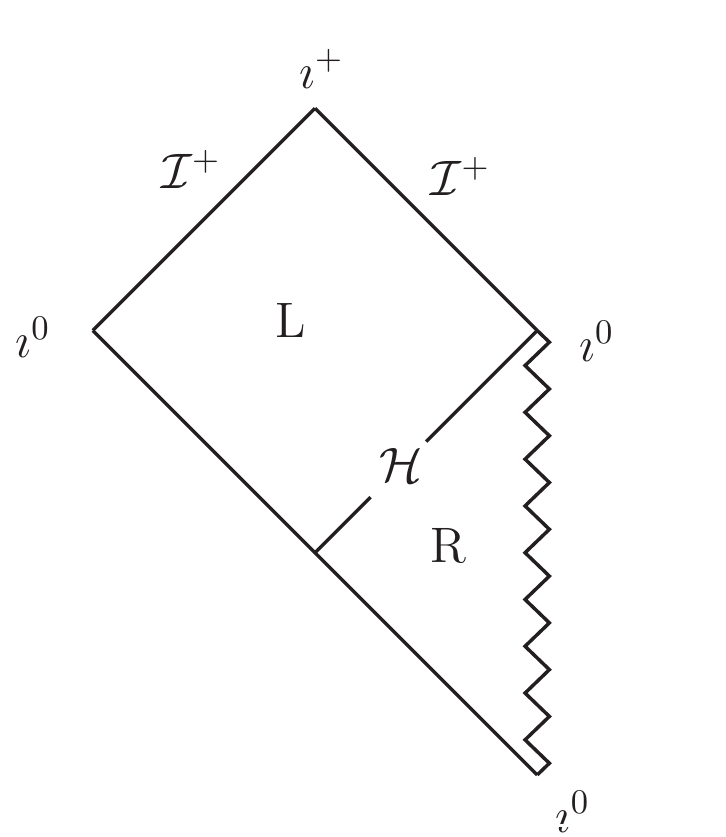}
\caption{Totally reflecting model.}
\end{subfigure}
\begin{subfigure}[b]{0.3\textwidth}
\includegraphics[scale=0.55]{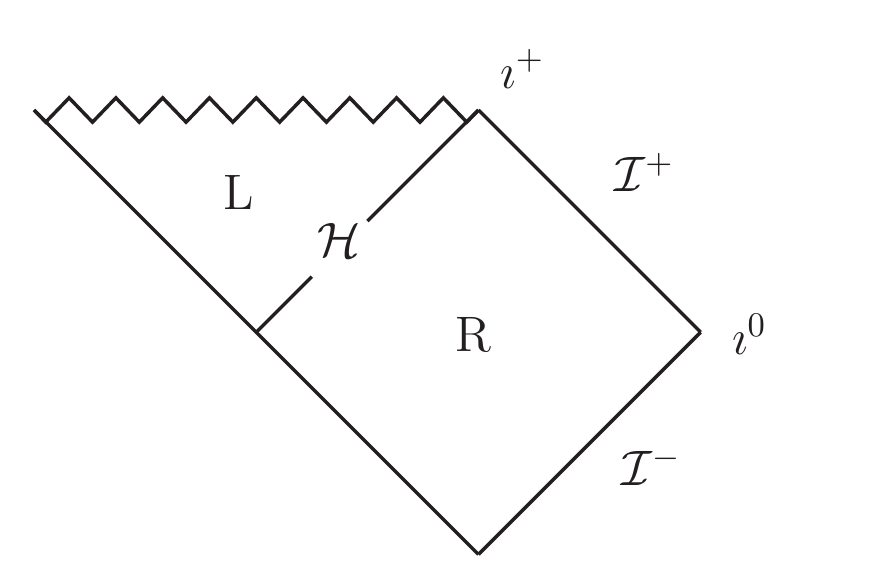}
\caption{CGHS model.}
\end{subfigure}
\begin{subfigure}[b]{0.3\textwidth}
\includegraphics[scale=0.55]{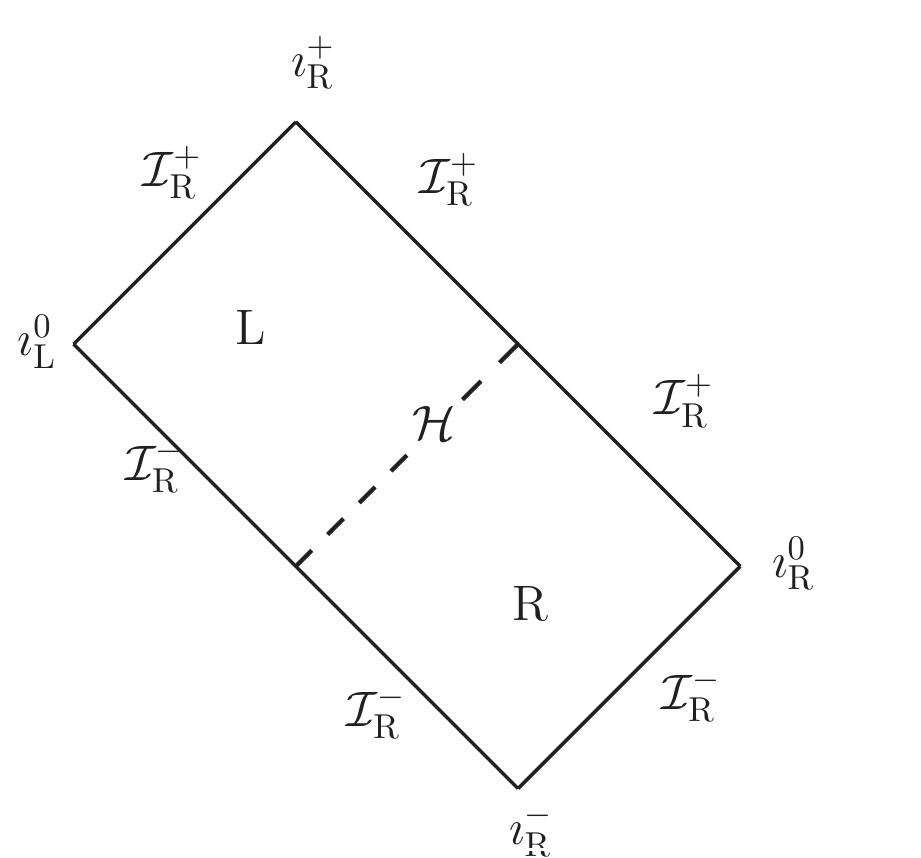}
\caption{Analog model.}
\end{subfigure}
\caption{In this figure, we show the Penrose-Carter diagrams of the geometries we shall use. The totally reflecting model of \eq{Sandrov} is singular in the exterior region. The CGHS model of \eq{CGHSv} has its singularity in the interior. The analog model of \eq{analogv} is everywhere regular, and can be obtained by pasting the L quadrant of the first model with the R quadrant of the second.
These diagrams do not represent the full analytic extension of each space-time, but only the quadrants that are relevant for our $S$-matrix,
namely, the left (L) and Right (R) regions on either side of the (future) horizon $\mathcal H$.
Precise definitions of the various types of infinities along with more details about the 
last diagram, are given in Ref.~\cite{Barcelo04}.}
\label{3Penrose_fig}
\end{figure}

Our second example generalizes the former Rindler case in that
there is still a total reflection,
but the profile $v(x)$ now possesses an asymptotically flat interior region.
As a result, the asymptotic flux of left going particles is well defined,
since the emitted quanta are asymptotically described by plane waves.
The profile which generalizes \eq{Rindler} is
\be
C(x) = D(-1 + e^{2\kappa x\over D}). \label{Sandrov}
\ee
The parameter $D$ characterizes the transition from the Near Horizon Region to the asymptotic one.
In the limit $D\to \infty$, $C(x)$ of \eq{Sandrov}
becomes that of \eq{Rindler}, which describes Rindler space.
In the above metric, \eq{canmodequ} is analytically solvable in terms of hypergeometric functions~\cite{Olver,AbramoSteg}.
The full expression of the general solution is given in App.~\ref{Hypergeo_App}.
To compute the total $S$-matrix, we follow exactly the same procedure as in \Sec{RindlerSec}. On Fig.\ref{3Penrose_fig}, we have drawn the Penrose diagrams of the three models we will solve.  

The first important quantity is the phase shift of \eq{Sextbelow}.
To simplify its  expression, we introduce the dimensionless quantities
\be
\varpi \doteq \frac{\om}{2\kappa},\quad
\bar \Om_+  \doteq \frac{\sqrt{m^2D + \om^2}}{2\kappa}. 
\label{notations} 
\ee
The (exact) phase shift then reads
\be
e^{i\delta_{\rm Refl}} = \frac{\Gamma(2i\varpi) \, \Gamma(1 - i\varpi + i \bar \Om_+ )
\, \Gamma(1 - i\varpi - i \bar \Om_+ )}{\Gamma(-2i\varpi) \, \Gamma(1 + i\varpi - i \bar \Om_+ )\,
\Gamma(1 + i\varpi + i \bar \Om_+)} \,  e^{2i\varpi \ln(D)} .\label{Sandrophase}
\ee
In the interior L region, the scattering coefficients in $S_{\rm far}$ are
\bsub \bea 
\alpha_\om^{\rm far} &=& \left(\frac{\bar \Om_+^2}{\varpi^2}\right)^{\frac14} \frac{\Gamma(1-2i\varpi)\Gamma \left(-2i\bar \Om_+\right)}{\Gamma \left(-i\bar \Om_+ - i\varpi \right) \Gamma \left(1 - i\varpi - i\bar \Om_+\right)} e^{i\varpi \ln(D)}  ,\\
\beta_\om^{\rm far} &=& \alpha_{-\om}^{\rm far} \times e^{-2i\varpi \ln(D)}. 
\eea \esub
The total beta coefficient is then given by
\be
\beta_\om^{\rm Tot} = \tilde \alpha_\om^{\rm NHR} \beta_\om^{\rm far} + \alpha_\om^{\rm far} \beta_\om^{\rm NHR} e^{i \delta_{\rm Refl}}. \label{betaTot}
\ee
Its full expression is rather complicated, and not very transparent. It is more interesting
to study its behavior in different regimes of the parameter space $(\om/\kappa,m/\kappa,D)$. \\

{\it Low frequency regime} - An interesting phenomenon happens in the deep infrared regime, $\om\to 0$. In this regime, we find
\bsub \label{cancel_eq} 
\bea
e^{i\delta_{\rm Refl}} &\sim& -1,\\
\beta_\om^{\rm far} \sim \alpha_\om^{\rm far} &\sim& \sqrt{\frac{m D^{\frac12}}\om} \frac{\Gamma \left(-i\frac{mD^{\frac12}}\kappa \right)}{\Gamma \left(-i\frac{mD^{\frac12}}{2\kappa} \right)\Gamma \left(1-i\frac{mD^{\frac12}}{2\kappa} \right)},\\
\beta_\om^{\rm NHR} \sim \alpha_\om^{\rm NHR} &\sim& -i \sqrt{\frac\kappa{2\pi\om}} .
\eea \esub
This equations show that, even though both $\beta_\om^{\rm NHR}$ and $\beta_\om^{\rm far}$ diverge as $1/\om^{1/2}$,
the total coefficient in \eq{betaTot} stays finite. Moreover, since the $\Gam$ functions are analytic, it is not difficult to see that the next order is $\om^0$, and thus 
\be
\beta_{\om}^{\rm Tot} \underset{\om\to 0}\sim f(\kappa/mD^{1/2}) \label{betazero}, 
\ee
which is finite for all $m > 0$. This completely differs from the massless case where $\beta_\om^{\rm Tot}$ diverges since
$\beta_\om^{\rm Tot}\sim \beta_\om^{\rm NHR} \sim 1/\om^{1/2}$. 
Moreover, when we take the massless limit of \eq{betaTot} at fixed $\om$,
we obtain the massless result,
\be
\beta_\om^{\rm Tot} \underset{m \to 0}{\sim} \beta_\om^{\rm NHR}, \label{betaHawk}
\ee
for all $\om$, and thus in particular we recover the diverging behavior for $\om\to 0^+$.
Before addressing the apparent contradiction between \eq{betazero} and \eq{betaHawk},
it is of value to make a pause and to discuss the lesson from \eq{betaHawk}.
This equation shows that when a massless conformally coupled field is scattered
on a Killing horizon of a stationary metric which is asymptotically singular in the
external region, since the curvature $R= -\partial_x^2 C/2 \to \infty$ for $x\to \infty$ (see Fig.~\ref{3Penrose_fig})
the particle flux is nevertheless well defined in the {\it interior}
 region because it is asymptotically flat,
so that the $out$ modes of {\it negative} Killing frequency are unambiguously defined.
In this case, using \eq{SNHR+1}, one gets a Planck spectrum emitted toward asymptotic left infinity.  This is rather unusual since the
Killing frequency is negative and has in that L region the physical meaning of a momentum
since the Killing field is space-like.

\begin{figure}[!ht]
\begin{center}
\includegraphics[scale=1]{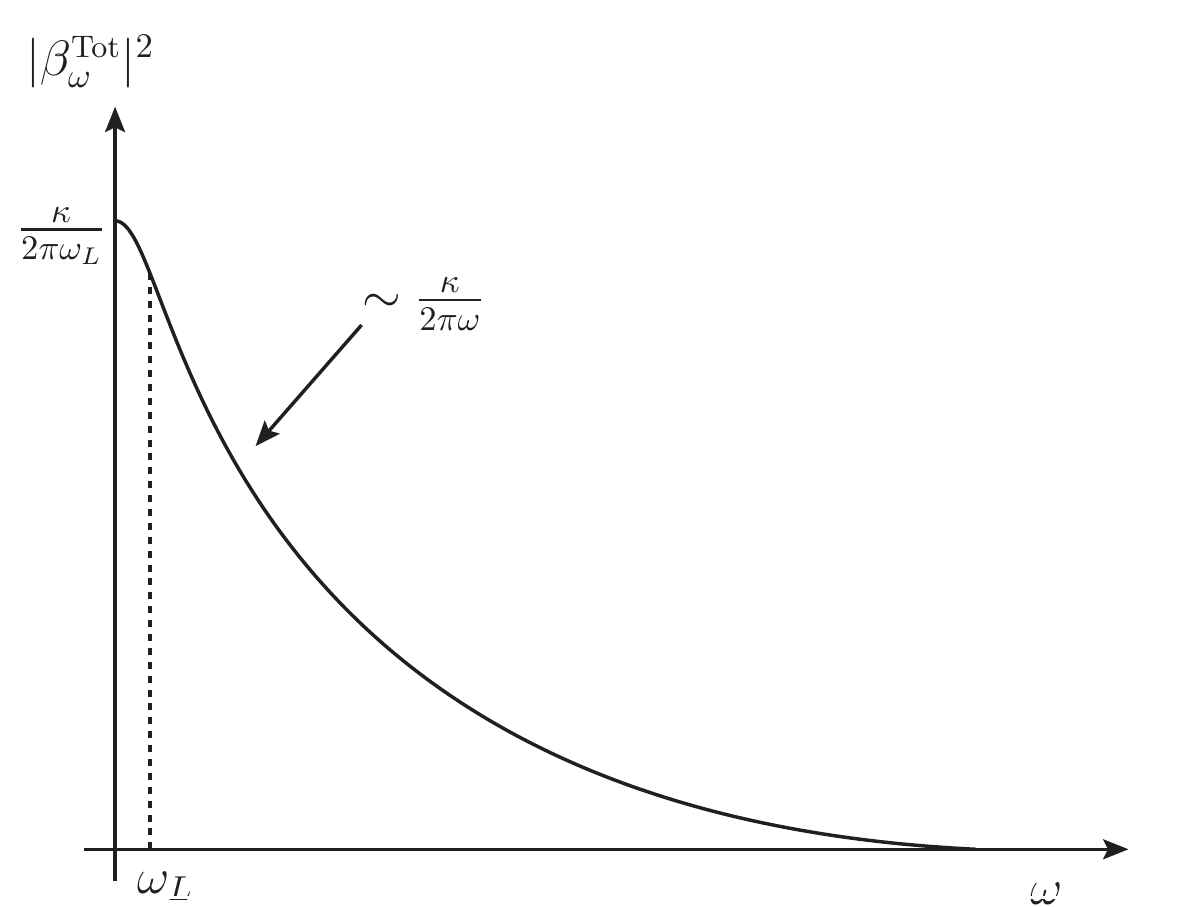}
\end{center}
\caption{In this figure $\vert \beta_\om^{\rm Tot}\vert^2$ is plotted as a function of $\om$ in the regime
of low mass and frequencies, {\it i.e.}, $\om, m \ll \kappa$. For frequencies above the threshold of \eq{omL}, $\vert \beta_\om^{\rm Tot}\vert^2$
behaves as for a massless field, and grows as $\kappa/\om$. For frequencies $\om < \om_L $,
 $\vert \beta_\om^{\rm Tot}\vert^2$ saturates at $\sim \kappa/\om_L \sim  \kappa/m D^{1/2}$.
In the opposite regime, where the mass is larger than $\kappa/2\pi$, the suppression arises at a frequency larger than the
temperature, and $\vert \beta_\om^{\rm Tot}\vert^2$ remains smaller than 1.}
\label{Sat_fig}
\end{figure}

The compatibility between \eq{betazero} and \eq{betaHawk} is understood
when realizing that the saturated value $\beta_0^{\rm Tot}$ of \eq{betazero}
diverges when $m\to 0$. To see this more precisely, we focus on the regime of small mass $m\ll \kappa$ and small frequencies $\om \ll \kappa$, for
arbitrary ratios $\om/m$. In this regime, we get
\be
\beta_\om^{\rm Tot} \sim -i \left(\frac{\kappa^2}{4\pi^2 (\om^2 + m^2 D)}\right)^{\frac14}. \label{betatransit}
\ee
This expression reveals that there is a change of regime near
\be
\om_L = m D^{\frac12}. \label{omL}
\ee
When $\kappa \gg \om \gg \om_L$, $\beta_\om^{\rm Tot}$ is growing as in the massless case,
 whereas for $\om \ll \om_L$ it saturates at a high but finite value, as can be seen Fig.~\ref{Sat_fig}.
As in the case of ultraviolet dispersion~\cite{Macher09,Coutant11}, we observe that the effective
frequency that governs the spectrum depends, as expected, on the dispersive frequency, here the mass $m$,
there the ultraviolet scale $\Lambda$, but also depends in a non-trivial manner on the parameter $D$ that governs the
extension of the near horizon region, see discussions in \Sec{validity}. In the present case, the power of $D$ is $1/2$, whereas for ultraviolet
dispersion, the power is $(n+1)/n$ when
the dispersion relation that replaces \eq{HJ} is $\Om^2 = p^2 + p^{n+2}/\Lambda^{n}$, as we showed in \Sec{generalization}. \\

{\it Large mass regime} - When the mass is large, {\it i.e.}, $m\gg \kappa, \om$, we find
that the coefficients of $S_{\rm NHR}$ and $S_{\rm far}$
go to their Rindler values in a well controlled manner, {\it e.g.},
\be
e^{i\delta_{\rm Refl}} \underset{m \to \infty}\sim e^{i\delta_{\rm Rindler}} \left(1+O(\kappa/mD^{\frac12}) \right).\label{SandrophaselimRindler}
\ee
This implies that $\beta_\om^{\rm Tot} \to 0$ for $m \to \infty$ as
\be
\beta_\om^{\rm Tot} = O(\kappa/mD^{\frac12}) = O(\kappa/\om_L).\label{SandrolimRindler}
\ee
This can be understood by considering the Bessel functions of \Sec{RindlerSec}.
Their behavior reveals that the scattering away from the horizon in region $L$, which compensates the near horizon mixing, occurs on a
distance from the horizon $\sim \kappa/m^2$. Hence, for large masses, the entire scattering
occurs in a close vicinity of the horizon. Therefore, in the large mass limit, the scattering in the present geometry
is indistinguishable from that occurring in Rindler space.
Using WKB techniques~\cite{Coutant11} similar to those of Chapter \ref{LIV_Ch}, 
which furnish reliable approximations in the large mass limit,
one can demonstrate that the residual scattering outside the near horizon region is negligible.
This means that for $m\gg \kappa, \om$,
irrespectively of the properties of the (smooth) profile $v(x)$,
the net $in/out$ Bogoliubov coefficient $\beta_\om^{\rm Tot}$ is
suppressed by the mass. This behavior radically differs from that of the
massless case given in \eq{betaHawk}, even though both cases share the same $S_{\rm NHR}$.

\subsubsection{CGHS model}
\label{CGHSSec}

We now study another exactly soluble example, which is given by the CGHS black hole~\cite{Witten91,Callan92},
except for the definition of the surface gravity $\kappa$ which is here given by \eq{kappadef}.
 In Painlev\'e-Gullstrand coordinates, the conformal factor reads
\be
C(x) = D(1 - e^{-\frac{2\kappa x}{D}}). \label{CGHSv}
\ee
Even though this geometry is very different from that of \eq{Sandrov} as it is
singular in the interior region, see Fig.~\ref{3Penrose_fig},
at the level of the mode equation, it gives something very close since
the discrete interchange $C \to -C$ and $x\to -x$ maps one problem into the other.
For this reason, the solutions of \eq{canmodequ} will also be
hypergeometric functions, see App.~\ref{Hypergeo_App}.
As in \Sec{SandroSec}, $\kappa$ is the surface gravity and $D$ characterizes the transition from the near horizon region to the asymptotic one.
However, here $D$ also controls the value of the threshold frequency $\om_R$ in \eq{thres} since
\be
\label{omR}
\om_R = m D^{\frac12}.
\ee

When $\om < \om_R$, the positive mode is totally reflected, and  the accumulated phase shift
characterizes $S_{\rm ext}$. As in the preceding section, to obtain simpler expressions,
we introduce
\bsub \bea
\bar \Om_< &=& \frac{\sqrt{ \om_R^2 - \om^2}}{2\kappa},\\
\bar \Om_> &=& \frac{\sqrt{\om^2 - \om_R^2}}{2\kappa},
\eea \esub
which are modified versions of \eq{notations}.
The exterior phase shift is then
\be
e^{i\delta_{\rm CGHS}} = \frac{\Gamma(2i\varpi)\Gamma \left(1 - i\varpi + \bar \Om_< \right)\Gamma \left(- i\varpi + \bar \Om_< \right)}{\Gamma(-2i\varpi)\Gamma \left(1 + i\varpi + \bar \Om_< \right)\Gamma \left(i\varpi + \bar \Om_< \right)}e^{2i\varpi \ln(D)} \label{deltaCGHS}.
\ee
From this, conclusions similar to that of \Sec{SandroSec} can be drawn.
For instance, when $\om \to 0$, we recover
\be
e^{i\delta_{\rm CGHS}} \underset{\om \to 0}\sim -1 ,
\ee
which is the main ingredient needed to obtain a canceling effect as in \eq{betazero}, and to have $\beta^{\rm Tot}_{\om \to 0}$ be regular in the limit $\om \rightarrow 0$.
If the mass is large then the behavior is essentially that found for Rindler spacetime in \eq{SandrophaselimRindler},
as can be seen by an explicit calculation.

When $\om > \om_R$, we are in the configuration where there exist three ABM, as in \eq{Sfacto33}. The greybody factors in the external region R are analytically obtained from the hypergeometric functions. The
transmission and reflection coefficients are
\bsub \label{CGHSgf}
\bea
T_\om = \tilde T_\om &=& \left(\frac{\varpi^2}{\bar \Om_>^2}\right)^{\frac14} \frac{\Gamma \left(-i\bar \Om_> - i\varpi \right) \Gamma \left(1 - i\varpi - i\bar \Om_> \right)}{\Gamma(1-2i\varpi)\Gamma \left(-2i\bar \Om_> \right)} e^{-i\varpi \ln(D)} . \\
\tilde R_\om &=& - \frac{ \Gamma(1+ 2i\varpi) \Gamma \left(-i\bar \Om_> - i\varpi \right) \Gamma \left(1 - i\varpi - i\bar \Om_> \right)}{\Gamma(1- 2i\varpi )\Gamma \left(-i\bar \Om_> + i\varpi \right) \Gamma \left(1 + i\varpi - i\bar \Om_> \right)} e^{-2i\varpi \ln(D)} ,\\
R_\om &=& \frac{ \Gamma \left(2i\bar \Om_>\right) \Gamma \left(-i\bar \Om_> - i\varpi\right) \Gamma \left(1 - i\varpi - i\bar \Om_>\right)}{ \Gamma \left(-2i\bar \Om_>\right) \Gamma \left(i\bar \Om_> - i\varpi\right) \Gamma \left(1 - i\varpi + i\bar \Om_>\right)} .
\eea \esub
Using \eq{Sfacto33}, the asymptotic out-going flux of \eq{occupnum+} is
\be
n_\om^{\rm u} = \vev{\rm in}{(a_\om^{{\rm out, u}})^\dagger a_\om^{{\rm out, u}}} = |\beta_\om^{\rm NHR} T_\om|^2
\ee
At $\om =\om_R$, $T_\om$ vanishes, and below $\om_R$ it is trivially 0. Note that the coefficients in \eqref{CGHSgf} are greybody factors as we discussed in \Sec{Observables_Sec}. 
In the next section, the transition shall be analyzed in more detail.

\subsubsection{Analog model}
\label{acousticSec}

We now consider a profile that combines the regular interior region of \Sec{SandroSec}
with  the regular exterior region of the above CGHS model so as to get a flow
similar to that of Fig.~\ref{vprofile_fig}.
The resulting geometry is relevant for analog models where the velocity profile is everywhere bounded.
We thus consider
\be
\label{analogv}
C(x) = 1-v^2(x) = \left\{ \begin{aligned}
&D_L(-1 + e^{\frac{2\kappa x}{D_L}}) \text{ for } (x<0),\\
&D_R(1 - e^{-\frac{2\kappa x}{D_R}}) \text{ for } (x>0) .\\
\end{aligned} \right.
\ee
This profile is $C^1$, {\it i.e.}, it is continuous and its first derivative is continuous. This ensures that the global geometry obtained is {\it regular}, in the sense that the
curvature does not contain a distributional contribution~\cite{Poisson}.
Since the scattering matrices $S_{\rm ext}$ and  $S_{\rm far}$ have been
already studied, both in the  exterior and interior regions,
all we need to do here is combine them to get the total $S$-matrix
\be
S =\underbrace{S_{\rm far}}_{\text{\Sec{SandroSec}}}\cdot \underbrace{S_{\rm ext}}_{{\text{\Sec{CGHSSec}}}} \cdot  \,\underbrace{ S_{\rm NHR} }_{{\text{\Sec{NHRSec}}}}\, . \label{Stotanalo}
\ee
The two threshold frequencies of \eq{omR} and \eq{omL} are now
\be
\label{2critfr}
\om_R = mD_R^{\frac12}, \quad {\rm and} \quad \om_L = mD_L^{\frac12}.
\ee
Having different values for $D_R$ and $D_L$ will allow us to distinguish their roles. \\

We first consider the totally reflecting regime, $\om<\om_R$.
Interestingly, we recover the transition seen in \Sec{SandroSec} and in Fig.~\ref{Sat_fig}.
Indeed, for $\om \ll \om_R$
\be
e^{i\delta_{\rm CGHS}} \sim -1.
\label{deltaCG}
\ee
Together with the coefficients of $S_{\rm far}$, this ensures that $\beta_\om^{\rm Tot}$ has a finite value in the limit $\om \to 0$. More precisely, in the high $\kappa$ regime, for $m, \om \ll \kappa$, we have
\be
\vert \beta_\om^{\rm Tot} \vert^2 \sim \frac{\kappa\, (D_L+D_R)}{2\pi D_R\, \sqrt{\om^2 + \om_L^2}}. \label{betalowkappa}
\ee
To observe a divergent regime $|\beta^{\rm Tot}_\om|^2 \propto \kappa/\om$,
one needs to assume that $\om_L \ll \om < \om_R$,
where the last inequality is required in order to be
below the threshold $\om_R$. This implies $D_L \ll D_R$, hence
\be
\vert \beta_\om^{\rm Tot} \vert^2 \sim \frac{\kappa}{2\pi}\frac{1}{ \sqrt{\om^2 + \om_L^2 }}. \label{betatransitL}
\ee
This expression shows the transition between the diverging regime
at the standard temperature, which is independent of $D_R$ and $D_L$,
and a saturating regime governed by $\om_L$.
For large masses $m \gg \kappa, \om$, the results are the same as 
for the totally reflecting and CGHS models, 
namely the various scattering coefficients asymptote to their Rindler values.
In numerical simulations~\cite{to_appear} all of these results have been recovered using rather different settings where the sound speed $c$ varies with $x$ and the velocity $v$ is a constant. This demonstrates that the low frequency behavior of \eq{deltaCG} applies to a much wider class of situations
than the one considered here.

We now have all the ingredients necessary to study the effects of a massive field
on the outgoing fluxes when starting from vacuum.
On the right side, the outgoing particle flux is that of
\Sec{CGHSSec}:  as expected, it vanishes below $\om_R$
and above it is given by
\be
n_\om^{\rm u} = |\beta_\om^{\rm NHR}\,  T_\om|^2.
\ee
To observe the transition, we work in the high $\kappa$ regime, {\it i.e.}, $\om, m \ll \kappa$, and obtain
\be
n_\om^{\rm u} \simeq \frac{\kappa}{2\pi} \Theta(\om - \om_R) \frac{ 4\sqrt{\om^2 - \om_R^2 }}{(\sqrt{\om^2 -\om_R^2 } + \om)^2}. \label{outfluxR}
\ee
The flux is thus continuous when crossing $\om_R$.

\begin{figure}[!ht]
\begin{center}
\includegraphics[scale=0.6]{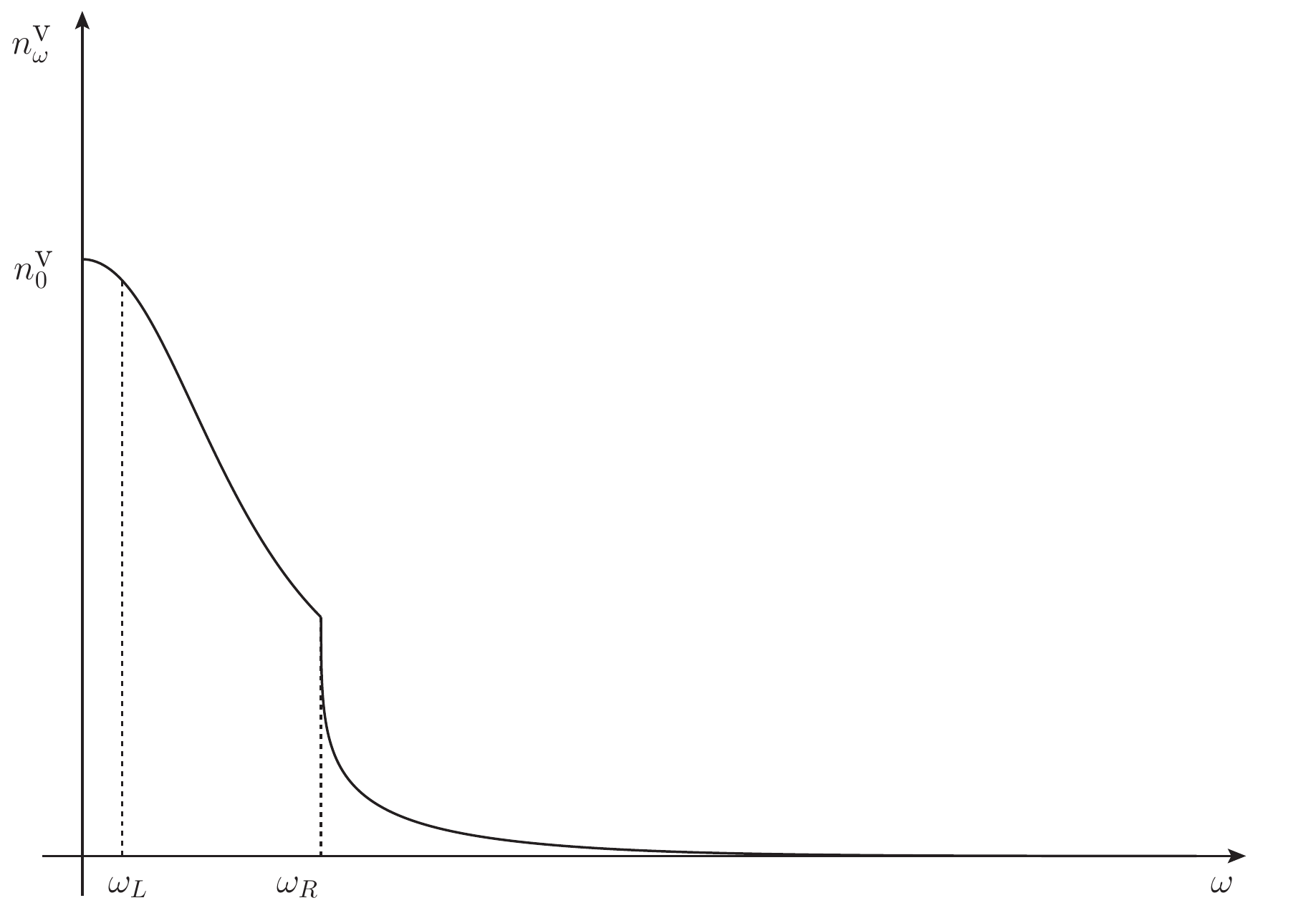}
\end{center}
\caption{Particle flux of positive frequency quanta spontaneously
emitted from the horizon toward $x \to - \infty$ in the high $\kappa$ regime.
Above $\om_R$, one finds a small contribution
which is due to the reflection (the backscattering) of Hawking quanta emitted toward $x = \infty$.
Below the threshold, the entire thermal flux is reflected and grows like $\kappa/\om$ for decreasing
values of $\om$, until one reaches $\om_L$ where it saturates, as explained in \Sec{SandroSec}.}
\label{fluxes_fig}
\end{figure}

On the left side, the particle fluxes are
more complicated since two contributions are present, see \eq{occupnum+};
$n_\om^{\rm v}$ is composed of positive frequency particles
and $n_{-\om}^{\rm u}  = n_\om^{\rm u} +  n_\om^{\rm v}$ is composed of 
the negative frequency partners. We first notice that both of these are
well defined since the profile of \eq{analogv} is asymptotically flat in L.
Using \eq{Sfacto33}, in full generality, $n_\om^{\rm v}$ reads
\be \bal
n_\om^{\rm v} = \, &\Theta(\om - \om_R) \, |\tilde R_\om \, \alpha_\om^{\rm far}  \beta_\om^{\rm NHR} + \tilde \alpha_\om^{\rm NHR} \beta_\om^{\rm far}|^2 \\
 & +\Theta(\om_R - \om) \, | e^{i \delta_{\rm CGHS}}\, \alpha_\om^{\rm far}  \beta_\om^{\rm NHR} + \tilde \alpha_\om^{\rm NHR} \beta_\om^{\rm far} |^2.
\eal \ee
The first term in the first line, which is proportional to $\tilde R_\om$,
describes the stimulated production in the L region due to the reflected Hawking quanta.
The other terms describe the interference between the mixing in the near horizon region and the scattering in the $L$ region away from the horizon.
Just as for $e^{i \delta_{\rm CGHS}}$ below the threshold, the phase of $\tilde R_\om$ is crucial since there is interference between these two terms.
The behavior of $n_\om^{\rm v}$ near the threshold frequency $\om_R$ is particularly interesting.
In the regime of large surface gravity,  $\kappa \gg \om, m$,
for $\om>\om_R > \om_L$, one finds
\be
n_\om^{\rm v} \simeq \frac{\kappa}{2\pi}\frac{1}{ \sqrt{\om^2+\om^2_L}} \left(\frac{\sqrt{\om^2 + \om^2_L} - \sqrt{\om^2 - \om^2_R}}{\sqrt{\om^2 - \om^2_R} + \om}\right)^2,
\ee
whereas for $\om<\om_R$
\be
n_\om^{\rm v} \simeq \frac{\kappa }{2\pi}\frac{(1+ D_L/D_R)}{ \sqrt{\om^2 + \om^2_L}}.
\ee
These two equations describe the 
effects on the spectrum in the $L$ region
which are due to a small mass. We first notice that
$n_\om^{\rm v}$ is continuous across $\om_R$, but with a cusp, see Fig.~\ref{fluxes_fig}.
From \eq{outfluxR}, we see that this is also true for $n_\om^{\rm u}$ and hence for $n_{-\om}^{\rm u}$ as well.

We also see that
the spectrum depends on the mass $m$
only through the two critical frequencies of \eq{2critfr}. It is thus through them
that the profile properties, namely $D_R$ and $D_L$ which govern the extension of the near horizon region on the right and on the left, see Fig.~\ref{vprofile_fig}, affect the spectrum. The lesson here is that when dealing with a conformally invariant massless field, the surface gravity is the
only background quantity that affects the spectrum. When breaking conformal invariance,
by a mass, or a non conformal coupling as in 3+1 dimensions, or by adding some
ultraviolet dispersion, other properties of the background flow affect the spectrum.
From the above analysis, and that of~\cite{Coutant11,Macher09,Finazzi10b,Finazzi12},
the most important ones are the extensions of the near horizon region, on both sides.

\section{Massive undulations in black holes}
\label{BHundul_Sec}

We now study the undulation in the analog black hole metric of \eq{analogv}. In preceding sections, we saw that the low frequency massive modes end up in the inside L region for both signs of $\om$, see Fig.~\ref{mass_traj_fig}. Moreover, in the limit $\om \to 0$, their momentum (solution of \eq{HJ}, see also Fig.\ref{WHdisprelm_fig}), is finite and given by
\be
p_U^m = p_{\om \to 0} = m D_L^{-\frac12} \left(1+O\left(\frac{\om}{\om_{\rm U}}\right)\right).
\label{omU}
\ee
This means that the zero frequency mode $\phi^{\rm out}_0(x)$ is a non trivial function of $x$,
opening the possibility of finding a behavior similar to that of \eq{Gundul}.
When $\om_L \ll \kappa$, an explicit calculation of
$ \Re\left\{e^{i\theta} \phi^{\rm out}_0(x) \right\}$,
similar to that made in \cite{Coutant11},
tells us that the asymptotic profile of $\Phi_{\rm U}^m$ is
\be
\Phi_{\rm U}^m(x) = \frac{1}{\sqrt{4\pi \om_L}}\cos\left( p_U^m \, x \right) .
\ee

The next important aspect concerns the
calculation of the net contribution of low frequency modes to $G$.
In this respect, two aspects should be discussed.
The first one concerns the fact that
$|\beta_\om^{\rm Tot}|^2$ no longer diverges for $\om\to 0$.
However, the criterion for the factorization of $G$ is only that $|\beta_\om^{\rm Tot}|^2 \gg 1$.
When $\om_L \ll \kappa$, as shown in \eq{betatransitL},
this is the case for frequencies $\om \ll \kappa/2\pi$.
The second aspect concerns the frequency interval $0\leqslant \om < \om_{\rm U}$ such that the momenta $p_\om$ are close
enough to the undulation momentum $p_U$ so that the $out$ modes $\phi_\om^{\rm out}$
contribute coherently to $\phi^{\rm out}_0$.
Using \eq{HJ} in the asymptotic interior region, we get
\be
\om_{\rm U} \simeq \frac{\om_L} {\sqrt{1+D_L}} \ll \kappa. \label{omUm}
\ee
Therefore,  at time $t$ after the formation of the horizon, the
contribution of the low frequency modes to the two-point function is
\be
\label{Gir}
G_{\rm IR}(t;x,x') = 8\int_{2\pi/t}^{\om_{\rm U}}  |\beta_\om^{\rm Tot}|^2 d\om \times \Phi_{\rm U}^m(x)\,  \Phi_{\rm U}^m(x') .
\ee
When assuming $\om_{\rm U} \ll \kappa$, which is the case for a small enough mass,
using \eq{betatransitL} we get
\be
\int_{2\pi/t}^{\om_{\rm U}}  |\beta_\om^{\rm Tot}|^2 d\om = \frac{\kappa}{2\pi} \left[ {\rm sinh}^{-1}\left(\frac{\om_{\rm U}}{\om_L} \right) - {\rm sinh}^{-1}\left(\frac{2\pi}{\om_L t}
\right) \right].
\ee
Hence, for short times, the amplitude grows as $ln(t)$, as in the massless case. However, when $t > 2\pi/\om_L$, the amplitude saturates and stays constant afterwards. This is an important prediction of this analysis. It shows how the transition in $\om$ near $\om_L$ with respect to the massless spectrum (see Fig.~\ref{Sat_fig}) produces here a change in time of the growth rate.

To conclude this section, we wish to provide a qualitative evaluation of the importance of
this infrared contribution to $G$. To do so, we need to consider some
observables, such as the stress energy tensor.
In particular, its trace accounts for the mass density of the field
\be
{\rm Tr}(\hat T) = \langle \hat T^\mu_{\ \mu} \rangle = m^2 \langle \phi^2(x) \rangle.
\ee
At late times, {\it i.e.}, $t\gg 2\pi/\om_L$,
the contribution of the undulation to the trace is
\be
{\rm Tr}(\hat T_{\rm IR}) = \frac{\kappa m}{\pi^2 D_L^\frac12} \underbrace{\sinh^{-1}\left(\frac{\om_{\rm U}}{\om_L}\right)}_{\lesssim 1} \times
\cos^2\left({p_U^m\, x} \label{BHundul} \right).
\ee
Since we work with $\kappa \gg \om_L = m D_L^{1/2}$,
this contribution to the trace is much smaller than $\kappa^2$, the typical energy density
contained in the Hawking flux for a light field.
Hence we expect that the undulation will not be easily
visible in this case. Moreover, when $D_L\to \infty$, which is the Rindler limit, the amplitude goes to 0 as $1/D_L$,
confirming the stability of Minkowski space. It is also interesting to notice that the parameter $D_R$ plays no role (as long as $\om_{\rm} \ll \om_R$, so that \eq{betatransitL} stands), confirming that the undulation is controlled by the interior geometry.

\section{Massive dispersive undulations in white holes}
\label{WHdispundul_Sec}

For white holes, undulations can be found when the dispersion relation is non-relativistic in the ultraviolet sector,
as discussed in \Sec{WHundul_Sec}. In this section, we first provide an argument of why the introduction of a dispersive cut-off $\Lambda$ does not alter the massive propagation described in \Sec{massfields_Sec} when $m\ll \Lambda$. To this aim, we generalize the treatment of Chapter \ref{LIV_Ch} in the presence of a mass.

For simplicity purposes, we consider here the superluminal relation 
\be
\om^2 = F^2(p) = m^2 + p^2 + p^4/\Lambda^2,
\ee
which is obtained from \eq{BogdrNl} in the limit $\xi \, p_\perp \to 0$. As we said in \Sec{BECmass_Sec}, both superluminal and subluminal dispersion relations
give rise to undulations in white hole flows, in virtue of the symmetry which relates them 
when interchanging at the same time the R and L regions, see Sec.~III.E in~\cite{Coutant11}.

\subsection{The scattering}
\label{massiveBogo_Sec}
Because of dispersion, the factorization of the $S$-matrix in \eq{Sfactogene} is no longer exact. However, as we shall see, when the two scales are well separated, {\it i.e.} $ m \ll \Lambda$, this factorization is recovered at first order. The reason is very similar to that of \Sec{dispHRpaper_Sec}, and therefore our computation will be performed using the same formalism. In the regime $m\ll \Lambda$, one has 
\be
F(p) = \sqrt{m^2 + p^2} +\frac{p^4}{2\Lambda^2\sqrt{m^2 + p^2}}.
\ee
Moreover, since the last term becomes non negligible only when $p\sim \Lambda \gg m$, 
we get
\be
F(p) = \sqrt{m^2 + p^2} +\frac{p^3}{2\Lambda^2}.
\ee
Hence, to first order in $m/\Lambda$, $F_m$ is a sum of the relativistic massive dispersion plus a dispersive term. 
We shall work in the near horizon region, and therefore, our aim is to compute the scattering coefficients between {\it in}-modes, which are now given by Eqs.~\eqref{plusin}, \eqref{minusin} rather than Eqs.~\eqref{inplus}, \eqref{inminus}, and near horizon outgoing modes of \eq{NHRoutmassmodes}. In particular, we shall establish that $S_{\rm NHR}$ is still given by \eq{SNHR} (in fact \eq{bogocoef}, since the {\it in}-modes are now dispersive) when the mass is low compared to the dispersive scale. Following \Sec{CF}, in the near horizon region, the mode in $p$-space is still given by \eq{dualmode}, and the various modes in $x$-space are given by contour integrals
\be
\phi_{\om,m}^{\mathcal C}(x) = \frac1{\sqrt{4\pi \kappa}} \int_{\mathcal C} \left(\frac{p}{F(p)}\right)^{\frac12} e^{i (px- \frac{\om}{\kappa} \ln(p) + G(p) +\frac{p^3}{6\Lambda^2 \kappa} )} \frac{dp}{p\sqrt{2\pi}} \label{massmode},
\ee
where
\be
G(p) = -\int_p^{\infty} \frac{\sqrt{m^2+p'^2} - p'}{\kappa p'}dp',
\ee
encodes the modification of the phase due to the mass. 
As before, the choice of the contour $\mathcal C$ dictates which mode one is considering. 

\begin{figure}[!ht]
\begin{subfigure}[b]{0.5\textwidth}
\includegraphics[scale=0.5]{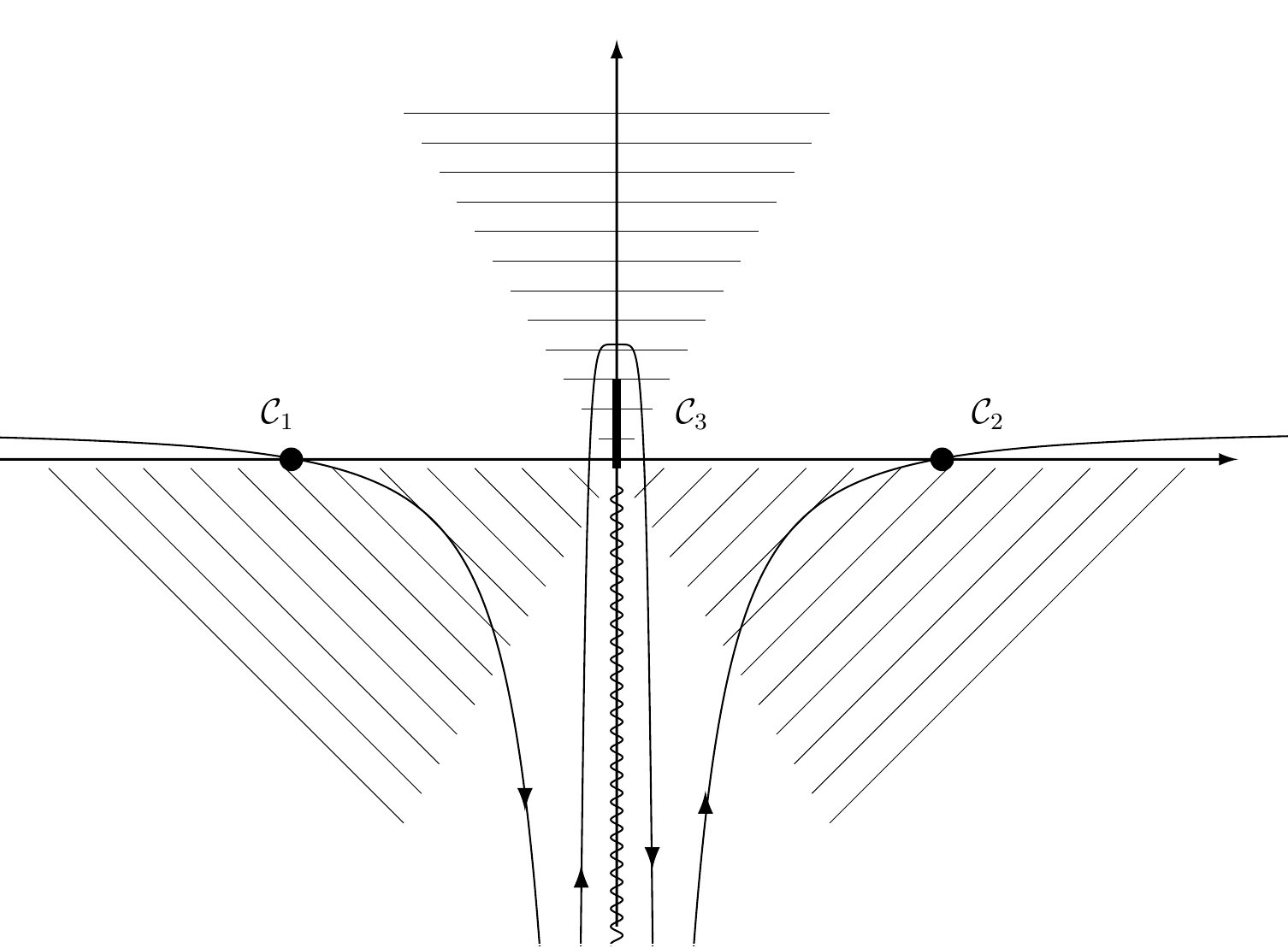}
\caption{$x<0$.}
\end{subfigure}
\begin{subfigure}[b]{0.5\textwidth}
\includegraphics[scale=0.5]{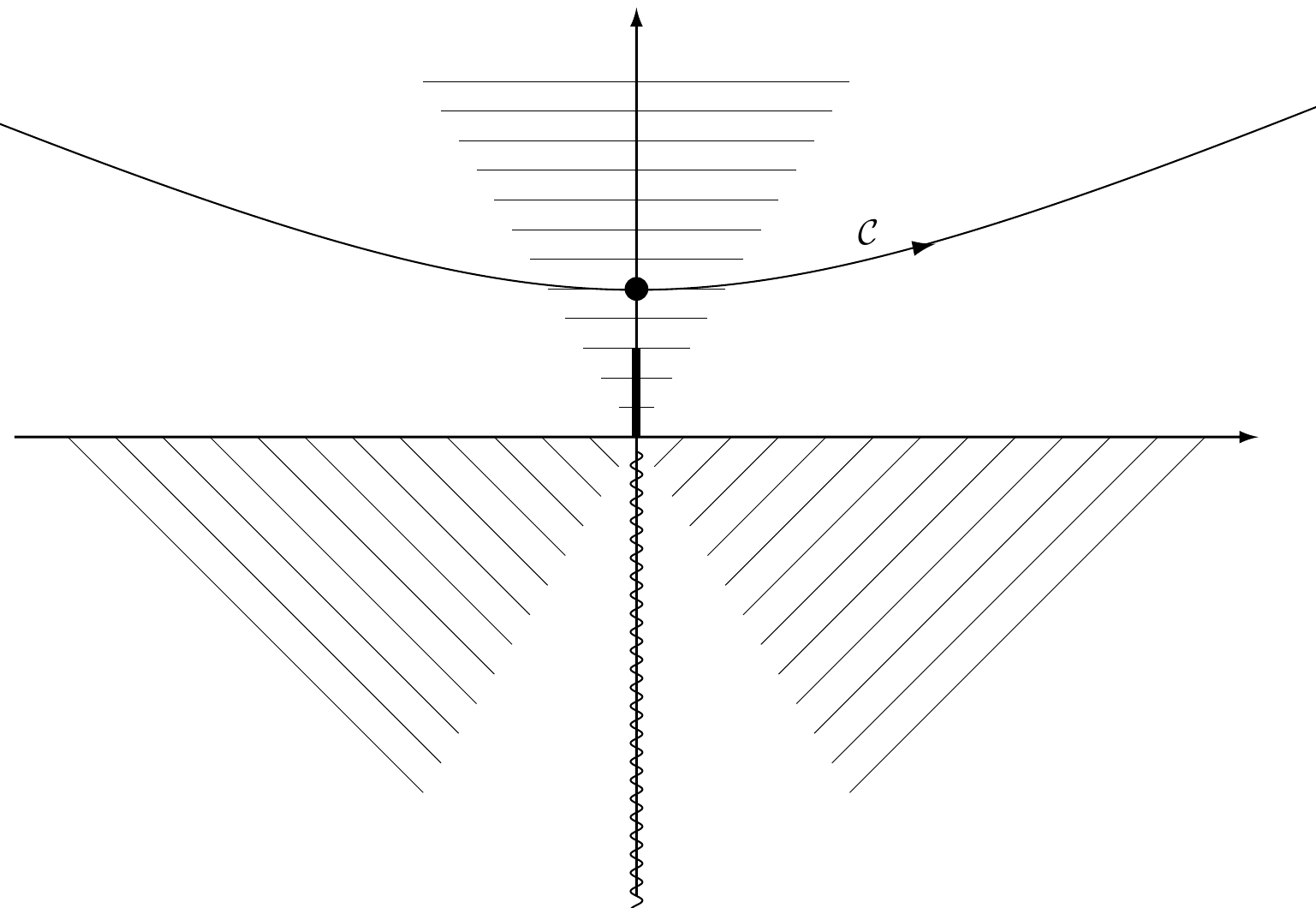}
\caption{$x<0$.}
\end{subfigure}
\caption{Representation of the contours in the $p$-plane to get $(\phi_{-\om}^{\rm Left})^*$,  the {\it out} mode of negative norm.  
The hatched regions are the asymptotically forbidden ones. The wavy line is the branch cut of $\ln(p)$, and the bold line is what must be added because of the mass.}
\label{masscontours} 
\end{figure}

In the following, we construct the generalization of the decaying mode of \Sec{decayingSec} 
because this is enough to demonstrate our claim. To this end, we must choose a branch cut to define both the  
$\ln(p)$ and 
$\sqrt{m^2+p^2}$ appearing in $G$. These functions introduce three branching points, $p=0$, and $p=\pm im$.
Here, we take the line $-i\mathbb R^+$ extended until $im$ to be the branch cut, 
as shown on Fig.~\ref{masscontours}. Hence in the limit $m \to 0$ we recover what we did in \Sec{CF}, see Fig.~\ref{contours}. 
To compute  \eq{massmode}, we proceed as in \Sec{decayingSec}.

When $x>0$, the introduction of a mass 
does not alter the discussion of \Sec{decayingSec} since the saddle 
at $p_s = i\Lambda \sqrt{2\kappa x}$ is well 
above the singularity at $im$ for $x$ sufficiently far away from the horizon. Hence
\be
\phi_{\om,m}^{\mathcal C} = e^{iG(p_s)} \times  \phi_{\om, {\rm as}}^{\downarrow} \times \left( e^{-i\frac{\pi}2} \right),
\ee
where $\phi_{\om, {\rm as}}^{\downarrow}$ is the {\it massless dispersive} decaying mode of \eq{decmode}\footnote{In the notations of Chapter \ref{LIV_Ch}, this mode should be referred with a $\varphi$ instead of a $\phi$. Here, in order not to confuse it with the auxiliary mode \eqref{def_varphi}, we add the subscript $as$, to specify that it corresponds to an asymptotic WKB mode. These modes generalize the mode basis of \Sec{modeanalys} in the presence of a mass.}. 
Moreover, we have $G(p_s) \approx 0$ because $p_s \gg m$ in the region of interest. For  $x>0$, the mode is thus rapidly decaying, on a scale $\kappa x\sim (\kappa/\Lambda)^{\frac23}$, and in the relativistic limit ({\it i.e.} $\Lambda \to \infty$) it vanishes. 
Therefore, as in the massless case, this mode is proportional to $(\phi_{-\om}^{\rm Left})^*$, the negative norm {\it out}-going mode \emph{in the near horizon region}. 
Indeed, if it were containing a small amount of the positive norm {\it out}-going mode 
$\phi_{\om}^{\rm Right}$, it would oscillate on the right side of the horizon 
until $\kappa x\sim (\om/m)^2 \gg (\kappa/\Lambda)^{\frac23}$ which gives the location of the turning point where
the mode is reflected due to its mass, as we saw in Sec.\ref{clatrajSec}. 

For $x<0$, using the analytic properties of $\tilde \phi_\om (p)$, we  deform the contour $\mathcal C$ into the union of $\mathcal C_1$, $\mathcal C_2$ and $\mathcal C_3$ shown in Fig~\ref{masscontours}. 
On $\mathcal C_1$ and $\mathcal C_2$, there are two saddle points 
at $p = \pm \Lambda \sqrt{\kappa |x|}$ that describe the high momentum 
incoming modes, as in \Sec{decayingSec}. Their contribution is 
\be
\label{alpbet}
\phi^{\mathcal C_1 \cup \mathcal C_2}_{\om,m} = ( e^{\frac{\om \pi}{\kappa}} e^{i\frac{3\pi}4} ) \times\left(\phi_{-\om, {\rm as}}^{\rm in}\right)^* 
+  e^{i\frac{\pi}4}\times\phi_{\om, {\rm as}}^{\rm in}, 
\ee
where $\phi_{\pm \om, {\rm as}}^{\rm in}$ are the {massless} dispersive {\it in} modes of \eq{plusin} and \eq{minusin}. 
Moreover, the saddle point approximation is controlled by the parameter $\Delta(x)$ of \eq{Delta}, irrespectively of the mass $m \ll \Lambda$. 
Hence, as expected, the high momentum contributions of the {\it out} mode are mass independent. 
Along $\mathcal C_3$ instead, 
we perform a strong limit $\Lambda \to \infty$ as in \eq{gammalike},
and we get
\be
\phi^{\mathcal C_3}_{\om,m}(x) = \frac1{\sqrt{4\pi \kappa}} \int_{\mathcal C_3} 
p^{-i\frac{\om}{\kappa}-1} e^{i (x p + G(p) )} \left(\frac{p}{\sqrt{p^2 + m^2}}\right)^{\frac12}
\frac{dp}{\sqrt{2\pi}},
\ee
which is a {\it massive relativistic} mode of negative norm. Up to a complex amplitude $A_\om$, it gives $(\phi_{-\om}^{\rm Left})^*$,
the low momentum out branch of the globally defined mode $(\phi_{-\om}^{\rm Left})^*$.
In brief, the mode obtained with the contour $\mathcal C$ is $A_\om \left(\phi_{-\om}^{\rm Left}\right)^*$: For all $\om$, it decays for $x > 0$, 
and on the left side, it contains three WKB branches 
\be
\label{concl}
A_\om \times \left(\phi_{-\om}^{\rm Left}\right)^* = \left( \frac{\alpha_{\om}^{{\rm NHR}*}}{\beta_{\om}^{{\rm NHR}*}} e^{i\frac{\pi}4} \right) \times\left(\phi_{-\om, {\rm as}}^{\rm in}\right)^* +  e^{i\frac{\pi}4}\times \phi_{\om, {\rm as}}^{\rm in} + A_\om \times \left(\phi_{-\om, {\rm as}}^{\rm Left}\right)^*.
\ee
Since $\left(\phi_{-\om}^{\rm in}\right)^*$ and $\phi_\om^{\rm in}$ are normalized
and have opposite norms, their relative coefficient furnishes the ratio of the {\it near horizon} Bogoliubov coefficients 
$| \beta_{\om}^{\rm NHR}/\alpha_{\om}^{\rm NHR}|$. From \eq{alpbet} and \eq{concl}, we obtain  
$| \beta_{\om}^{\rm NHR}/\alpha_{\om}^{\rm NHR}| = e^{- \pi \om/\kappa}$. 
It is independent of $m$ and has the standard relativistic value. \\

By studying only the mode $\left(\phi_{-\om}^{\rm Left}\right)^*$, we have demonstrated that the introduction of dispersion does not affect the near horizon Bogoliubov coefficients of $S_{\rm NHR}$. However, as we understood in \Sec{Sett}, what matters is the total $S$-matrix. However, the extra scattering, governed by $S_{\rm far}$ and $S_{\rm ext}$ concerns effects in the infrared, where the modes hardly feel dispersion. This shows that we can directly apply the results of \Sec{massfields_Sec} for the dispersive white hole, as it will be correct up to subleading terms in $O(m/\Lambda)$.

\subsection{The undulation}
\label{WHdispmassundul_Sec}
For superluminal quartic dispersion, the outgoing momentum at zero frequency is found in the supersonic region and is given by
\be
p_U^\Lambda = p_{\om \to 0} = \Lambda D_L^{1/2} \left( 1+O\left(\frac{\om}{\om_{\rm U}^\Lambda}\right)\right),
\ee
and the asymptotic behavior of the undulation is~\cite{Coutant11}
\be
\Phi_{\rm U}^\Lambda (x)=\frac{\cos(p_U^\Lambda \, x + \theta_U)}{\sqrt{4\pi p_U^\Lambda D_L}}.
\ee
The phase $\theta_U$ cannot be obtained from the preceeding equations, because it
is mainly determined by the dispersive properties of the modes.
In Chapter \ref{LIV_Ch}, or~\cite{Coutant11},  when $m=0$, it was established that $\theta_U = (\Lambda D_L^{3/2})/(6\kappa) + \pi/4$, see \eq{WHasundul}.

In the presence of ultraviolet dispersion, the width of frequencies that contribute coherently to $G$ is
\be
\om^\Lambda_{\rm U} = \Lambda D_L^{\frac32}. \label{omUL}
\ee
Since $\Lambda$ can be much larger than $\kappa$, $\om^\Lambda_{\rm U}$ can be either smaller
or larger than the Hawking temperature $\kappa/2\pi$.
In what follows, we work with $\om^\Lambda_{\rm U} \gg \kappa$
where the Bogoliubov coefficients are
well approximated~\cite{Coutant11} by their relativistic values computed in preceding sections. Therefore,
the contribution of the low frequency dispersive modes is given by
\be
\label{GirUV}
G_{\rm IR}(t;x,x') = 8\int_{2\pi/t}^{\om^\Lambda_{\rm U}}  |\beta_\om^{\rm Tot}|^2 d\om \times
\Phi_{\rm U}^\Lambda(x)\,  \Phi_{\rm U}^\Lambda(x') ,
\ee
which is \eq{Gir} with $\Phi_{\rm U}^m$ and $\om_{\rm U}$ replaced by $\Phi_{\rm U}^\Lambda$ and $\om^\Lambda_{\rm U}$.
The exact expression of $\beta_\om^{\rm Tot}$ in \eq{betaTot} is quite complicated.
To get an undulation, we assume $\om_L \ll \kappa \ll \om_{\rm U}^\Lambda$.
In that regime, $\beta_\om^{\rm Tot}$ is large for $\om < T_H$, but for $\om > T_H$ it becomes exponentially small. 
Thus
\be
\int_{2\pi/t}^{\om_{\rm U}^\Lambda}  |\beta_\om^{\rm Tot}|^2 d\om \simeq \int_{2\pi/t}^{\kappa/2\pi}  |\beta_\om^{\rm Tot}|^2 d\om.
\ee
In that range of frequencies, $\beta_\om^{\rm Tot}$ is well approximated by \eq{betatransitL}, therefore
\be
G_{\rm IR}(t;x,x') = \frac{4\kappa}\pi \left[ \text{sinh}^{-1}\left(\frac{\kappa}{2\pi \om_L} \right) - \text{sinh}^{-1}\left(\frac{2\pi}{\om_L t} \right) \right] \Phi_{\rm U}^\Lambda(x)\,  \Phi_{\rm U}^\Lambda(x').
\ee
Hence, at late times and for $\om^\Lambda_{\rm U} \gg \kappa$, 
we obtain
\be
G_{\rm IR}(t;x,x') = \frac{4\kappa}\pi \ln\left(\frac{\kappa}{\pi \om_L} \right)  \Phi_{\rm U}^\Lambda(x)\,  \Phi_{\rm U}^\Lambda(x').
\ee
When considering a BEC, the relationship between the scalar field $\phi$ and the density fluctuation
$\delta\rho$ is $\delta\rho \propto \partial_x \phi$~\cite{Balbinot07,Balbinot08}. Hence the mean value of the 
equal-time density-density two-point function is
\be
\langle \partial_x \phi(x)\,  \partial_{x'} \phi(x') \rangle =
\frac{\kappa p_U^\Lambda}{\pi^2 D_L}  \ln\left(\frac{\kappa }{\pi \om_L}\right)\times
\sin \left(p_U^\Lambda x + \theta_U \right) \, \sin \left(p_U^\Lambda x' + \theta_U \right). \label{WHundul}
\ee
This generalizes what was found in~\cite{Mayoral11} in that, in the supersonic region, one still finds
a short distance checker board pattern in the $x, \, x'$ plane,
and the undulation amplitude still grows initially as $\ln(t)$. However, when there is a mass term,
it grows only for a finite amount of time $\sim 2\pi/\om_L$, after which it saturates.
The mass therefore provides a saturation mechanism that can occur before nonlinearities take place.
Moreover, because $p_U^\Lambda \propto \Lambda \gg \kappa$,
the r.m.s. amplitude of the undulation is large.\\

So far, we have considered only the case where the initial state is the vacuum. 
As discussed in \Sec{WH_Sec} and~\cite{Mayoral11}, when dealing with a thermal state the initial
growth rate is no longer logarithmic but linear in time. However, the mass term
acts again as an infrared regulator because the initial distribution of phonons
is expressed in terms of $\Om > m$, and not in terms of the constant frequency $\om$~\cite{Macher09b}. 
Hence no divergence is found when integrating over $\om$ when computing the two-point function.
In addition, the random character of the undulation amplitude is fully preserved
when taking into account some initial thermal noise.
What is modified is the  undulation r.m.s. amplitude. When the initial
temperature $T_{\rm in}$ is much larger than $T_H = \kappa/2\pi$, the above two-point function
is, roughly speaking, multiplied by $T_{\rm in}/T_H$.

\section{Double horizon undulation}
\label{WD_Sec}
In this section we study a flow configuration that contains two horizons.

\subsection{Warp-drive analogy}
Warp-drives~\cite{Alcubierre94} allow, at least theoretically, to travel at arbitrary high superluminal speeds and consequently to travel in time~\cite{Everett95}.
However, besides the fact that they require matter distributions violating positive energy conditions~\cite{Pfenning97,VanDenBroeck99,Lobo04,Lobo04b}, 
they are quantum mechanically unstable because they possess a white hole horizon and a Cauchy horizon on which the renormalized stress-energy tensor blows up exponentially~\cite{Finazzi09}. In this section, we re-examine the question of their stability when postulating that Lorenz invariance is broken at ultra-high energy.
One of our motivations comes from the fact that nonlinear dispersion relations remove Cauchy horizons and regulate the fluxes emitted by white holes~\cite{Macher09}.

A superluminal warp-drive metric describes a bubble containing an almost flat region, moving at some constant speed $v_0 > 1$ within an asymptotically flat spacetime:
\be
\label{eq:3Dalcubierre}
 ds^2= dt^2-\left[dX-V(r)dt\right]^2 - dY^2 - dZ^2,
\ee
where $r\equiv \sqrt{(X-v_0 t)^2+Y^2+Z^2}$ is the distance from the center of the bubble.
Here $V=v_0 f(r)$, with $f$ a smooth function satisfying $f(0)=1$ and $f(r) \to 0$ for $r \to \infty$.
Along the direction of motion, the backward and forward locii where $V(r) = 1$ behave respectively as a future (black) and past (white) event horizon~\cite{Hiscock97}.
In fact, the Hawking flux emitted by the black horizon accumulates on the white horizon while being unboundlessly blueshifted.
However, since the whole analysis rests on relativistic quantum field theory, one should examine whether the warp drives instability is peculiar to the local Lorentz symmetry.
Although current observations constrain to ultra high energy a possible breaking of that symmetry~\cite{Liberati09}, one cannot exclude this possibility which has  been suggested by theoretical  investigations~\cite{Gambini98,Jacobson01,Horava09}.

\begin{figure}[!ht]
\begin{center}
\includegraphics[scale=1.3]{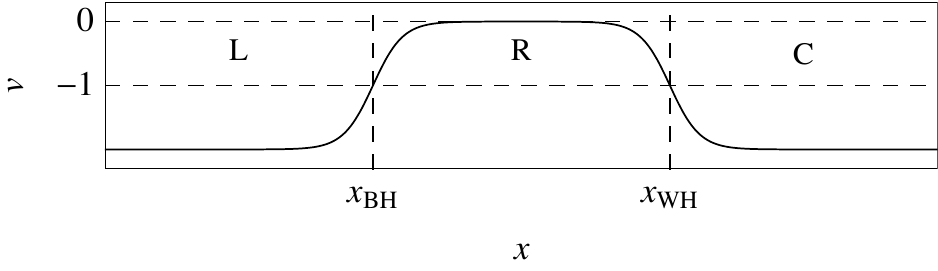}
\end{center}
\vspace{-0.3cm}
\caption{Velocity profile for a right-going warp drive in the Painlev\'e--Gullstrand~\cite{Barcelo05} coordinates of Eq.\eqref{PGds}. Two {\it superluminal} asymptotic regions $L$ and $R$ are separated by a black and a white horizon from a compact internal {\it subluminal} region $C$.
The Killing field $\partial_t$ is space-like in L and R, light-like on both horizons, and time-like in C.}
 \label{fig:velocity}
\end{figure}

For the sake of simplicity we work in $1+1$ dimensions and ignore the transverse directions $Y$ and $Z$.
Defining a new spatial coordinate $x=X - v_0 t$, the metric becomes the Painlevé-Gullstrand of \eq{PGds} with the profile 
\be
v(x) = v_0(f(x) - 1),
\ee
which is negative. In this space-time, $\partial_t$ is a globally defined Killing vector field whose norm is given by $1-v^2$: it is time-like within the bubble, its norm vanishes on the two horizons, and it is space-like outside. One thus gets three regions {L, C, and R (see Fig.~\ref{fig:velocity})}, separated by two horizons $x_{\rm BH}<0<x_{\rm WH}$ which are respectively the black and the white one.

\subsection{Propagation on a double horizon metric}
\subsubsection{Settings}
\begin{figure}[!ht]
\begin{subfigure}[b]{0.5\textwidth}
\begin{center}
\includegraphics[scale=1.3]{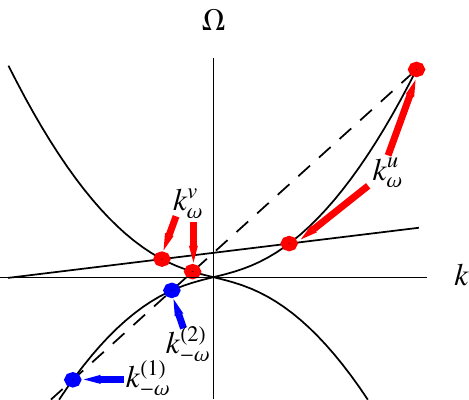}
\end{center}
\caption{Superluminal dispersion}
\end{subfigure}
\begin{subfigure}[b]{0.5\textwidth}
\begin{center}
\includegraphics[scale=1.3]{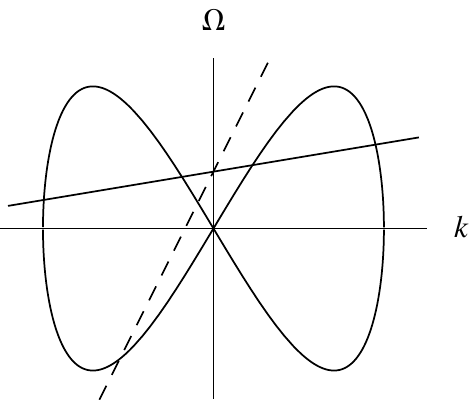}
\end{center}
\caption{Subluminal dispersion}
\end{subfigure}
 \caption{Graphical solution of \eq{eq:dispersion} for super (left panel), and subluminal dispersion (right panel). In both panels, the straight lines represent $\om-v k$ for $|v|<1$ (solid) and $|v|>1$ (dashed). The curved lines represents $\Om^2(k)$.}
 \label{fig:dispersion}
\end{figure}

We now consider a massless scalar field with a quartic dispersion relation.
In covariant terms, its action reads 
\be
\mathcal S_\pm=\frac{1}{2}\int \left[g^{\mu\nu}\partial_{\mu}\phi \partial_{\nu}\phi \pm \frac{(h^{\mu\nu}\partial_{\mu}\partial_{\nu}\phi)^2}{\Lambda^2}\right] \sqrt{-g} d^2 x,
\label{action}
\ee
which is the action described in \Sec{GRwithUTVF} without any higher derivative term.
We recall the expression of the preferred frame $\uf$ in Painlevé-Gullstrand coordinates 
\be
\uf = \p_t + v\p_x.
\ee
Then the {\it aether} flow is geodesic and it is asymptotically at rest in the $t,X$ frame of \eq{eq:3Dalcubierre}.
The sign $\pm$ in Eq.~(\ref{action}) holds for superluminal and subluminal dispersion, respectively.
In these settings, the wave equation has the usual form of \eq{WHmodequ}. Again, thanks to stationarity, the field can be decomposed in stationary modes $\phi = \int e^{- i \om t }\phi_\om d\om$, where $\om$ is the conserved (Killing) frequency. We also recall the Hamilton-Jacobi equation 
\be\label{eq:dispersion}
(\om - v k_\om)^2=k_\om^2\pm\frac{k_\om^4}{\Lambda^2},
\ee
where the solutions are graphically given in Fig.~\ref{fig:dispersion}.

The mode analysis is very similar to that of \Sec{modeanalys}, and we expose here the results for our configuration. For superluminal dispersion and $|v|<1$, there are two real roots ($k^v_\om$, $k^u_\om$) describing left- and right-going waves ($\phi^{v}_{\om}$, $\phi^{u}_{\om}$), and two complex ones ($k^\uparrow_\om$, $k^\downarrow_\om$) describing a spatially growing and decaying mode ($\phi^\uparrow_{\om}$, $\phi^\downarrow_{\om}$).
For $|v|>1$, there is a cut-off frequency $\om_{\rm max}$ below which the complex roots turn into real ones ($k^{(1)}_\om$, $k^{(2)}_\om$) with negative $\om$.
Correspondingly there exist two additional propagating waves $(\phi^{(1)}_{-\om})^*$, $(\phi^{(2)}_{-\om})^*$ with negative norm.
When the dispersion relation is subluminal, the negative norm modes are trapped in the region with $|v|<1$. \\

In the following, we assume a superluminal dispersion relation. As we shall analyze, in that case, undulation types of effect are present, as those studied in preceding sections. Note that the case of subluminal dispersion, up to the symmetry \eq{subsuper}, will be studied in detail in Chapter \ref{laser_Ch}. In that case, the instability of the warp-drive is maintained because exponentially growing mode are present in the spectrum of the field.

\subsubsection{The scattering}
\begin{figure}[!ht]
\begin{center}
\includegraphics[scale=0.8]{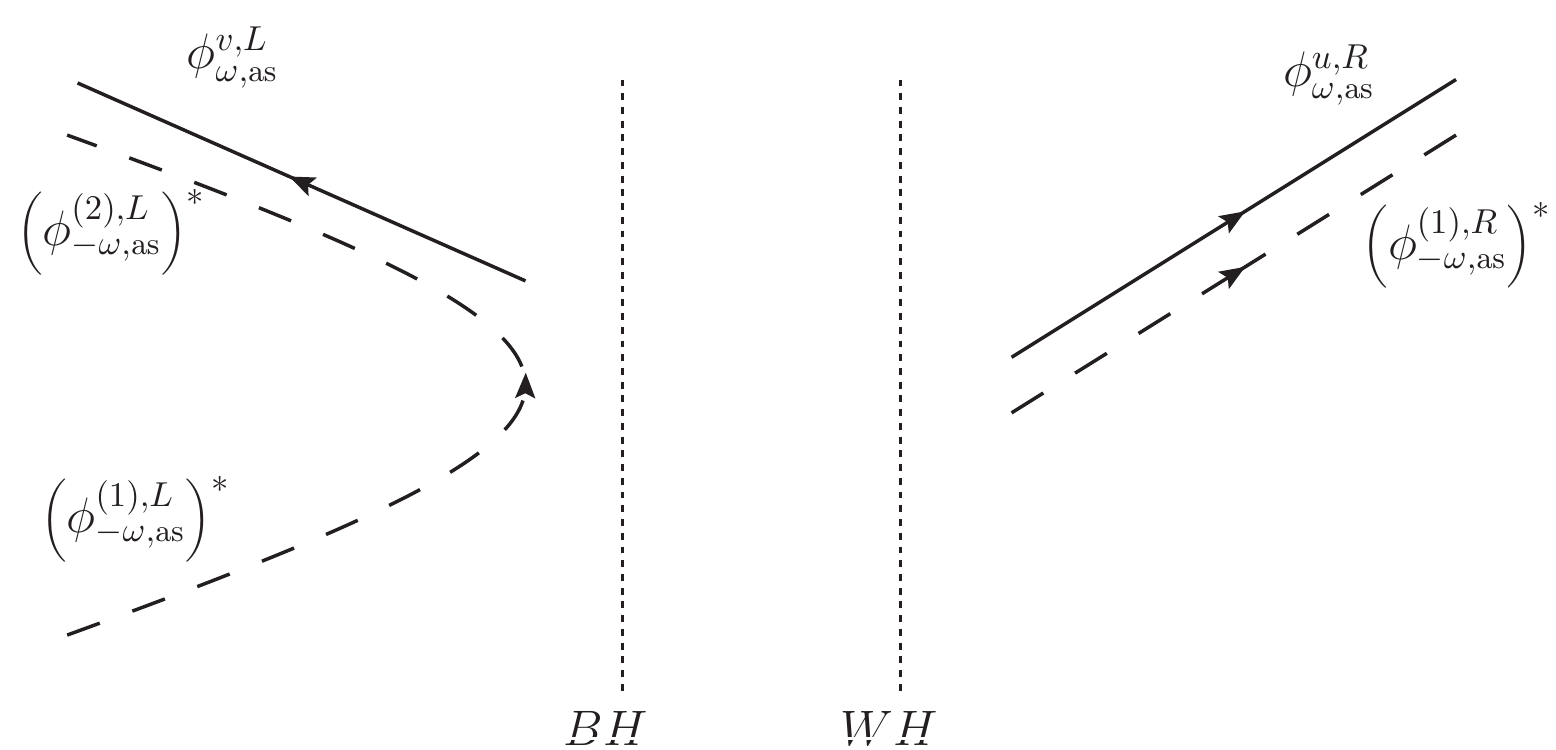}
\end{center}
\caption{Asymptotic decomposition in plane waves $\phi^{(i),  L/R}_{\om, {\rm as}}$ of the {\it in} mode $(\phi^{(1),\rm in}_{-\om})^*$. Note that only $\phi^{(1),  L}_{-\om, {\rm as}}$ has group velocity directed toward the horizons, with wavevector $k_\om^{(1)}$.}
\label{fig:mode}
\end{figure}

We underline that, up to the symmetry of \Sec{subsuper}, this configuration is exactly the one realized in the experiment of~\cite{Weinfurtner10}. In a geometry with two infinite asymptotic `superluminal' regions, for each $\om<\om_{\rm max}$, 4 asymptotically bounded modes can be defined. Moreover, by examining their asymptotic behavior, an {\it in} and an {\it out} bases can be defined by the standard procedure: each \emph{in} mode $\phi^{(i),\rm in}_{\om}$ (\emph{out} mode $\phi^{(i),\rm out}_{\om}$) possesses a single asymptotic branch $\phi^{(i),  L/R}_{\om, {\rm as}}$ carrying unit current and with group velocity directed toward region C (from C to $\infty$). This is exemplified in Fig.~\ref{fig:mode} using the \emph{in} mode $\phi^{(1),\rm in}_{-\om}$.

When the dispersive scale and the horizon surface gravity $\kappa$ are well separated, {\it i.e.}, $\kappa \ll \Lambda D_{\rm lin}^{3/2}$ as in \Sec{validity}, the left-going mode does not significantly mix with the other three modes, all defined on the right-going branch of \eq{eq:dispersion}~\cite{Coutant11}. 
Thus, the {\it in}-{\it out} scattering matrix is effectively a $3\times3$ matrix
\be
\bmat  \phi^{u,\rm in }_{\om}\\
 \left(\phi^{(1),\rm in }_{-\om}\right)^*\\
 \left(\phi^{(2),\rm in }_{-\om}\right)^*\\
\emat
=\bmat \alpha_{\om}^u & \beta^{(1)}_{-\om} & \beta^{(2)}_{-\om}\\ 
 \beta^{(1)}_{\om} & \alpha_{-\om}^{(1)} & A_{-\om}\\ 
 \beta^{(2)}_{\om} & \tilde A_{-\om} & \alpha_{-\om}^{(2)} 
\emat
\bmat
 \phi^{u,\rm out}_{\om}\\
 \left(\phi^{(1),\rm out}_{-\om}\right)^*\\
 \left(\phi^{(2),\rm out}_{-\om}\right)^*\\
\emat.
\label{eq:bogo4x4}
\ee
Given that the two $(\phi^{(i)}_{-\om})^*$ have negative norms, the matrix coefficients satisfy normalizations conditions such as
\be 
| \alpha^{u}_{\om} |^2-| \beta^{(1)}_{\om}|^2 - | \beta^{(2)}_{\om} |^2=1.\label{current}
\ee
When working in the $in$-vacuum, the state without incoming particle, the mean occupation numbers of outgoing particles with negative here frequency are $\bar n^{(i)}_{-\om}= |\beta^{(i)}_{-\om} |^2$, whereas that  with positive frequency is $\bar n^{u}_{\om}=\bar n^{(1)}_{-\om}+ \bar n^{(2)}_{-\om}$, by energy conservation.
That is, pair production occurs here through a two-channel Hawking-like mechanism.
To approximatively compute the coefficients of \eq{eq:bogo4x4} in the regime $\kappa\ll \Lambda$ we use connection formula techniques developed in Chapter \ref{LIV_Ch}. We first decompose $\phi_\om$ in both asymptotic regions L and R as a sum of plane waves:
\bsub \bea
\phi_\om &=& L_\om^u \, \phi_{\om, {\rm as}}^{u,L}+ L^{(1)}_{\om}  (\phi^{(1),L}_{-\om, {\rm as}})^* + L^{(2)}_{\om} (\phi^{(2),L}_{-\om, {\rm as}})^*, \\
\phi_\om &=& R_\om^u \, \phi_{\om, {\rm as}}^{u,R} + R^{(1)}_{\om} (\phi^{(1),R}_{-\om, {\rm as}})^* + R^{(2)}_{\om} (\phi^{(2),R}_{-\om, {\rm as}})^*.
\eea \esub
The coefficients are connected by
\be\label{eq:connection}
 \bmat R_\om^u \\ R_\om^{(1)} \\ R_\om^{(2)} \emat = U_{\rm WH} \cdot U_{\rm HJ} \cdot U_{\rm BH}^{-1} \cdot \bmat L_\om^u \\ L_\om^{(1)} \\ L_\om^{(2)} \emat,
\ee
where $U_{\rm BH}$ and $U_{\rm WH}$ respectively describe the {\it off-shell} scattering on the two horizons~\cite{Coutant11} and $U_{\rm HJ}$ describes the WKB propagation from one horizon to the other, {\it i.e.}\/ it is diagonal and contains the exponential of $i S_\om^a= i\int k^a_\om(x) dx$, where $k^a_\om$ is $k^u_\om$, $k^\uparrow_\om$ or $k^\downarrow_\om$. The gauge fixing of the phases are obtained by the same conventions as in \Sec{eikonal_Sec}. 
In Chapter \ref{LIV_Ch}, we derived {\it off-shell} transfer matrices. By this we mean that these three matrices are not restricted to the two modes that govern the asymptotic scattering on each horizon considered separately, that is the growing mode is here kept in the mode mixing. In fact, since $k^\uparrow$ has negative imaginary part, $e^{iS^\uparrow_\om}$ is exponentially large. Simple WKB algebra shows that it grows as $e^{\Lambda \Delta}$ where $\Delta$ is the distance between the two horizons.
Concomitantly, since $k^\downarrow=k^{\uparrow*}$, $e^{iS^\downarrow_\om}$ is exponentially small. This is very useful in the present case, because growing and decaying modes live in a finite size region. Therefore, they cannot be discarded from the propagation, they will contribute to the scattering coefficients. In usual quantum mechanical problems, the decaying mode leads to exponentially small transmission coefficient across classically forbidden zones, {\it i.e.} tunnel effects. The growing mode it always absent, in the sense that there is no exponentially large coefficients, proportional to $e^{\Lambda \Delta}$. This is guaranteed by the current conservation $|R|^2+|T|^2 = 1$. However, in our case, the current conservation leads to \eq{current}, and hence, coefficients are not in principle bounded. Hence, there is no guarantee that the growing mode will not amplify some scattering. As we shall see, this is in fact never the case. 

We now pick an example to show how to determine the coefficients of \eq{eq:bogo4x4}. 
The globally defined $\phi^{(1),\rm in}_{-\om}$ is constructed by imposing that the asymptotic amplitudes of the two incoming branches $\phi^{u, L}_{\om, {\rm as}}, \phi^{(2), R}_{-\om, {\rm as}}$ both vanish, see Fig.~\ref{fig:mode}.
Therefore the three outgoing amplitudes are given by the second row of the matrix of \eq{eq:bogo4x4}. Moreover, using \eq{eq:connection}, these coefficients correspond to $(R_\om^u,R_\om^{(1)},R_\om^{(2)})=(\beta_\om^{(1)} , \alpha_{-\om}^{(1)} , 0) $ and $(L_\om^u , L_\om^{(1)} , L_\om^{(2)}) =(0 , 1 , A_{-\om})$. 
Solving the resulting system, we obtain
\bsub \label{3Bogo}
\bea
\beta^{(1)}_{\om} &=& \tilde\beta_\om^{\rm BH} \times e^{iS^u_\om} \times \alpha_\om^{\rm WH}+O(e^{iS_\om^\downarrow}), \\
\alpha_{-\om}^{(1)} &=& - \tilde \beta_\om^{\rm BH} \times e^{iS^u_\om} \times \beta_\om^{\rm WH} +O(e^{iS_\om^\downarrow}) ,\\
A_{-\om} &=& \tilde\alpha_\om^{\rm BH},
\eea \esub
where the $\alpha$'s and $\beta$'s in the above are the standard Bogoliubov coefficients for black and white holes that encode the thermal Hawking radiation~\cite{Primer}.
By a similar analysis of other modes, all coefficients of \eq{eq:bogo4x4} can be computed.
Although the non-positive definite conservation law of  Eq.~(\ref{current}) does not bind these coefficients, the exponentially large factor in $e^{\Lambda \Delta}$ cancels out from all of them.
As a consequence, as can be seen in \eq{3Bogo}, the leading term is, up to some phase coming from the WKB propagation in region C, given by the Bogoliubov coefficients of $S_{\rm BH}$ and $S_{\rm WH}$. In the first two lines, one finds a product of two coefficients because the associated semi-classical trajectory passes through both horizons. Instead, in the third line only one coefficient is found because there is only a reflection on the black horizon. 
\begin{figure}[!ht]
\begin{center}
\includegraphics[scale=0.8]{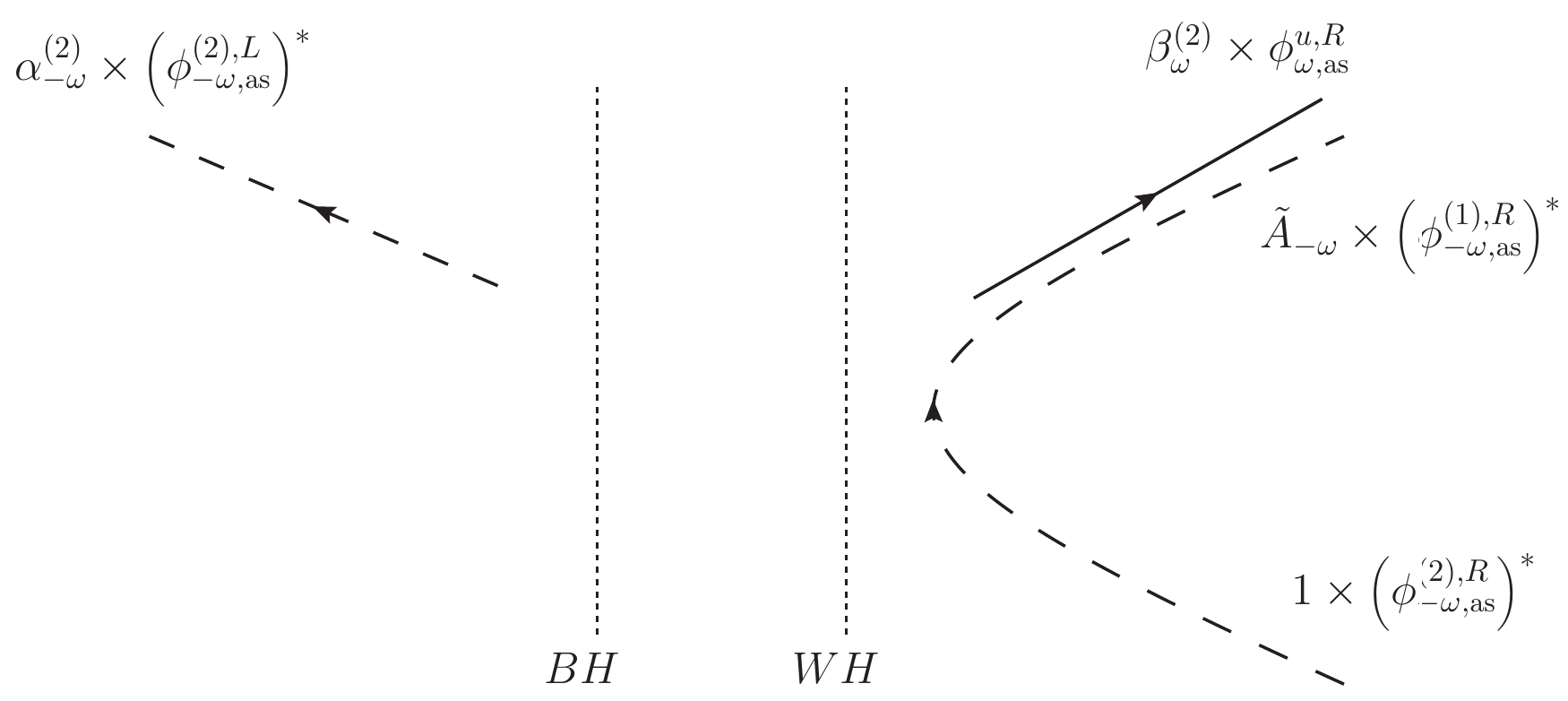}
\end{center}
\caption{Asymptotic decomposition in plane waves $\phi^{(i),  L/R}_{\om, {\rm as}}$ of the {\it in} mode $(\phi^{(2),\rm in}_{-\om})^*$.}
\label{WDtunnel_fig}
\end{figure}

We also analyze the scattering of the mode $\left(\phi_{-\om}^{(2), {\rm in}}\right)^*$ exposed in Fig.\ref{WDtunnel_fig}. Using the same technics, we derive the scattering coefficients 
\bsub \label{WDtunnel_eq}
\bea
\tilde A_{-\om} &=& -\tilde \alpha_{\rm WH} ,\\
\beta_\om^{(2)} &=& - \tilde \beta_{\rm WH},\\
\alpha_\om^{(2)} &=& - e^{-i S_\om^{\uparrow}} \tilde \beta_{\rm WH} \beta_{\rm BH}. 
\eea \esub 
The first two lines are easily interpreted, since they are the Bogoliubov coefficients of the white hole horizon. The last one is more interesting. Indeed, in the central region C, no modes are propagating toward the left side. Therefore, it is only through a decaying or growing mode that the coefficient $\alpha_{-\om}^{(2)}$ can be turned on\footnote{This is true because coupling to $v$-modes are negligible, see \Sec{modeanalys}. A complete characterization of $\alpha_{-\om}^{(2)}$ would require to take the $u$-$v$ mixing into account.}. We see that our transfer matrix technic predicts a precise value for this effect. Moreover, if the presence of $e^{-i S_\om^{\uparrow}}$, it is the usual tunnel amplitude at the WKB approximation, the contributions of $\beta_{\rm WH}$ and $\beta_{\rm BH}$ could not have been guest from usual methods. Therefore, \eq{WDtunnel_eq} is an nontrivial prediction of the formalism developed in Chapter \ref{LIV_Ch}. It would be interesting if this expression could be validated in future experiments or numerical simulations.

\subsubsection{The stress-energy tensor}
To identify the physical consequences of the pair creation encoded in Eqs.~\eqref{eq:bogo4x4} and \eqref{WDtunnel_eq}, we compute the expectation value of the stress energy tensor
\be
 T_{\mu\nu} \equiv \frac{2}{\sqrt{-g}}\frac{\delta S_+}{\delta g^{\mu\nu}}=T_{\mu\nu}^{(0)}+T_{\mu\nu}^{(\Lambda)},
\ee
where $T_{\mu\nu}^{(0)}$ is the standard relativistic expression and $T_{\mu\nu}^{(\Lambda)}$ arises from the Lorentz violating term of \eq{action}: 
\be
 T_{\mu\nu}^{(\Lambda)} =\frac1{\Lambda^2} \left[h^{\alpha\beta}\left(\phi_{,\alpha\beta}\phi_{,\mu\nu} + \phi_{,\mu\nu} \phi_{,\alpha\beta} \right) -\frac12 \left(h^{\alpha\beta}\phi_{,\alpha\beta}\right)^2g_{\mu\nu}\right].
\ee
In the asymptotic region on the right of the white horizon, the field can be expanded as the superposition of the two right-going modes $\phi_\om^{u,\rm out}$ and $\phi_{-\om}^{(1),\rm out}$, see Fig.~\ref{fig:mode},
\be
 \phi=\int \left[\phi_\om^{u,\rm out} \hat a_\om^{u,\rm out}+\phi_{-\om}^{(1),\rm out} \hat a_{-\om}^{(1),\rm out}\right] d\om +{\rm h.c.}
\ee
In this region, the geometry is stationary and homogeneous. Hence the renormalized tensor $T_{\mu\nu}^{\rm ren}$ is obtained by normal ordering the above creation and destruction {\it out}\/ operators. The fact that there exist negative frequency asymptotic particles causes no problem in this respect.
In fact all asymptotic excitations have a positive comoving frequency $\om$ of \eq{eq:dispersion}. 
Imposing that the initial state is vacuum, using \eq{eq:bogo4x4}, it is straightforward to compute $\vev{\rm in}{T_{\mu\nu}^{\rm ren}}$.
The final expression contains an integral over $\om$ of a sum of terms, each being the product of two modes $\phi_\om^{u,\rm out}, \phi_{-\om}^{(1),\rm out}$ and two coefficients of \eq{eq:bogo4x4}.
We do not need the exact expression because we only consider possible divergences.
When $\om>\om_{\rm max}$ there are no negative frequency modes. Hence the $\beta$ coefficients of \eq{eq:bogo4x4} vanish for $\om > \om_{\rm max}$, and the stress-energy tensor cannot have ultraviolet divergences.
Therefore the only possible divergence can be found for $\om\to0$.

In each term of $T_{\mu\nu}^{(0)}$, there are two derivatives with respect to $t$ or $x$, leading to two powers of $\om$, $k_\om^{(u)}$ or $k_\om^{(1)}$. 
Analogously in $T_{\mu\nu}^{(\Lambda)}$, there are 4 powers of these.
Now, from Fig.~\ref{fig:dispersion} we see that the wavenumbers $k_\om^{(u)}, k_\om^{(1)}$ do not vanish for $\om \to 0$ in L. Rather they go to constant opposite values, that we call $k_0$ and $-k_0$ respectively. These modes contribute to an undulation, exactly as explained in \Sec{WHundul_Sec}. 
Finally, the terms containing only spatial derivatives will not be suppressed for $\om \to 0$.
The leading terms in $\langle 0_{\rm in}|T_{\mu\nu}^{(0), \, \rm ren}|0_{\rm in}\rangle$ are thus proportional to
\be
\frac{k_0^2}{{4\pi\om(k_0)}\, v_{g0}}\, \int \left[\bar n^{(u)}_\om +\bar n^{(1)}_{-\om}\right] d\om.
\label{leading}
\ee
The above integral gives the integrated mean occupation number of the two {\it out}\/ species, and $v_{g0}$ is their asymptotic group velocity in the $t,X$ frame.
The leading terms of $\langle 0_{\rm in}|T_{\mu\nu}^{(\Lambda), \, \rm ren}|0_{\rm in}\rangle$ are proportional to \eq{leading} up to an extra factor of $k_0^2/\Lambda^2$.
Since $k_0 = \Lambda\sqrt{v_0^2-1} $, one finds that the $(0)$ and $(\Lambda)$ components of the stress energy yield typically the same contribution.\\

The key result comes from the fact that $|\beta^{(1)}_\om|^2$
diverges as $1/\om^2$ for $\om\to 0$, being the product of $|\beta^{\rm BH}_\om|^2\sim 1/\om$ and $|\beta^{\rm WH}_\om|^2\sim 1/\om$.
This infrared behavior has been validated by numerical analysis. Moreover, as explained in detail in \Sec{IRdiv_Sec}, this divergence accounts for a growing in time behavior. Because it diverges as $1/\om$ instead of $1/\om^{1/2}$, the growth is \emph{linear} in time, unlike in a single white hole, where it was logarithmic, see \Sec{WHundul_Sec}.
Thus, the energy density scales as 
\be
{\cal E} \propto \Lambda \int_{1/t} \left[\bar n^{(u)}_\om +\bar n^{(1)}_{-\om}\right] d\om \propto \Lambda \kappa^2 t.
\ee
That is, there in an infrared divergence that leads to a linear growth of the energy density.
This result can be understood from the findings of~\cite{Mayoral11} and the discussion at the end of \Sec{WHdispmassundul_Sec}. The BH radiation emitted toward the WH horizon stimulates the latter as if a thermal distribution was initially present. In that case, it was also found that the observable (the density correlation function) increased {linearly} in $t$. 

\subsubsection{Warp-drive stability}

In the case of a warp-drive metric, this undulation is very large, and thus truly destabilizes the system. Indeed, using quantum inequalities~\cite{Roman05}, it was argued~\cite{Finazzi09} that $\kappa$ must be of the order of the Planck scale, which implies that the growth rate is also of that order (unless $\Lambda$ is very different from that scale). In the presence of superluminal dispersion, warp drives are therefore unstable on a short time scale. To conclude, we note that in~\cite{Everett95} it was shown that close timelike curves can be obtained by combining several warp drives. Our results, together with those of~\cite{Finazzi09}, weaken that possibility because isolated warp drives are unstable irrespectively of the features of the dispersion relation in the ultraviolet regime. As a consequence, whereas former attempts to tackle the issue of chronology protection deeply relied on local Lorentz invariance~\cite{Kay96}, the present result suggests that this conjecture may be valid also for quantum field theories violating Lorentz invariance in the ultraviolet sector.

\section{Conclusions about undulations}
\label{Cl_mass_Sec}
In this chapter, we have detailed how the infrared growth of the beta coefficient of Hawking radiation can lead to a dynamical process, namely, the growth in time of a zero frequency mode. This mode consists of a classical real wave with a definite profile, see \eq{Undulprofile}. On the contrary, the amplitude of this wave is a fluctuating variable, of vanishing mean value but large spread, which is governed by the two-point function, see \eq{General_undul}. The appearance of this undulation is due to the amplification of initial fluctuations, that can be from quantum vacuum or a thermal state.

This phenomenon has first been studied in the case of dispersive white hole (\Sec{WHundul_Sec}). There, a field initially in its vacuum state develops an undulation growing logarithmically with time. We then considered massive fields. In that case, the near horizon mode mixing is not altered, 
as can be seen in \eq{SNHR}. However, the mass does regularize the infrared behavior of the net $in/out$ Bogoliubov transformation
in \eq{Stotanalo}. The reason for this is the extra mode mixing,
described by $S_{\rm far}$ of \eq{Sfar}, which occurs in the supersonic inside
region, and which interferes with the near horizon scattering
so as to cancel out the divergence in $1/\om$ of the $|\beta_\om|^2$ coefficient.
Indeed, the squared norm of the
total Bogoliubov coefficient saturates as $| \beta_\om^{\rm Tot}|^2 \sim \kappa/2\pi \om_L$ for $\om \to 0$,
where $\om_L$ is the threshold frequency of \eq{2critfr}.
As a consequence, the undulation r.m.s. amplitudes
now saturate after a lapse of time $\sim 2\pi /\om_L$ and then stay constant.
This has to be contrasted with the massless case where the saturation
of the amplitude can only occur because of non linearities, or dissipation, in the system.

The presence of a mass induces a new type of undulation in the supersonic region
that exists in black hole flows. Unlike the undulations occurring in white hole flows
which are due to some ultraviolet dispersion, this new type occurs in the hydrodynamical regime if the mass term is small enough.
It will thus appear both in superluminal and subluminal media. However, as shown by \eq{BHundul}, the typical energy
density carried by an undulation is small, and thus this new type should be
difficult to detect.

Although the $S$-matrix coefficients governing black hole and white hole flows are the same, 
the frequency ranges that contribute to the massive and the dispersive undulations are very different as can be seen by
comparing \eq{omUm} with \eq{omUL}. As a result, the white hole undulations
possesses larger amplitudes. In addition, since the wave length of the undulation
is smaller in the white hole case, it gives rise to even larger amplitudes for a BEC,
 as can be seen from \eq{WHundul}. These results might also be relevant for surface waves where `transversal instabilities' have been observed~\cite{Chaline12}.

In addition, the properties of the spectrum and the undulations
depend on the mass $m$ essentially through the effective frequencies $\om_L$
and $\om_R$ of \eq{2critfr}. These frequencies are both proportional to $m$
but also depend in a non trivial way on $D_L$ and $D_R$ which determine the
spatial extension of the near horizon region, on the inside and on the outside
respectively. Therefore, as in the case of ultraviolet dispersion, these two quantities
should be conceived of as the most relevant geometrical properties, after the surface gravity $\kappa$.

Finally, we analyzed the propagation of a field on a flow containing two horizons. The scattering coefficients are obtained with help of the connection formula developed in Chapter \ref{LIV_Ch} and~\cite{Coutant11}. As illustrated by \eq{WDtunnel_eq}, some features are new predictions of our formalism, that cannot be obtain only by knowing the Bogoliubov transformation of a black and a white hole. Moreover, in this configuration, undulation type of phenomena are obtained. Since there are several channels, several diverging terms are present. The dominant one corresponds to an undulation growing linearly in time. This can be interpreted as a white hole undulation like in \eq{Gundul}, but where the state is thermal instead of being the vacuum. Indeed, by Hawking radiating, the black hole horizon feed the white hole with a thermal flux.

\chapter{Dynamical instability}
\label{laser_Ch}

\minitoc

\section{Presentation of the black hole laser effect}

In the preceding chapters, we have established that in the presence of dispersion, even though the emitted spectrum of black holes is very close to the Hawking one, the dynamical properties of certain configurations can be strongly affected. In particular, the stability properties of horizons must be readressed. In this chapter, we study an example of configuration that becomes highly unstable because of dispersion. Indeed, in 98, Corley and Jacobson noticed that the propagation of a superluminal dispersive field in a stationary geometry containing two horizons, leads to a self amplified Hawking radiation of bosonic fields~\cite{Corley98}. The origin of this laser effect can be attributed to the closed trajectories followed by the negative Killing frequency partners of Hawking quanta. This closed trajectory exists because dispersion opens new scattering channels (see Fig.\ref{dispersive_traj_fig}). Moreover, the superluminal character of the dispersion makes Hawking partners bounce from one horizon to the other. 

The original analysis~\cite{Corley98} and that of~\cite{Leonhardt08} were  both carried out using wave packets. 
In~\cite{Coutant10}, we showed that there exists a more fundamental description based on frequency eigenmodes which are spatially bounded. This chapter is devoted to a detailed description of that work. When $v(x)$ is constant and subsonic ({\it i.e.}, $|v|<1$) for both $x\to \pm  \infty$, 
there is a discrete set of complex frequency modes and a continuous family of real frequency modes\footnote{If instead the subsonic region is finite and periodic conditions imposed, the situation is more complicated~\cite{Jain07,Garay00} and will not be considered here.}.
In our case, the real frequency modes are asymptotically oscillating and normalized by a delta of Dirac. Moreover, they are all of positive norm (for positive frequency), unlike what is found in the case of Hawking radiation. Therefore, they are not subject to any Bogoliubov transformation. The scattering matrix at fixed $\om$ only
contains reflection and transmission coefficients mixing right and left moving (positive norm) modes.
This was not {\it a priori} expected since the matrix associated with a single BH (or a WH)
is $3 \times 3$ and mixes positive and negative norm modes in a nontrivial way, as we saw in chapters \ref{LIV_Ch} and \ref{mass_Ch}.

The discrete set is composed of modes that vanish for $x \to \pm \infty$. 
They form two modes subsets of complex conjugated frequencies, each containing a growing and a decaying mode. 
The time dependence of the coefficients in each subset corresponds to that 
of a complex upside down and rotating harmonic oscillator. 
Both the real and the imaginary part of the frequency play important
roles in determining the asymptotic properties of the fluxes. We notice such modes were encountered
in several other situations~\cite{Fulling,Damour76b,Damour78,Kang97,Cardoso04,Greiner}.
We also mention that a stability analysis of BH-WH flows in Bose Einstein condensates
was presented in \cite{Barcelo06}. We reach different conclusions because we use different boundary conditions.

In Sec.~\ref{settingsBHL}, we present our settings. In Sec.~\ref{setofmodes}, we demonstrate that the  
set of spatially bounded modes contains a continuous part and a discrete part 
composed of complex frequency modes. 
In Sec.~\ref{modeprop} and Sec.~\ref{predic2}, 
we study the properties of the modes, and show how the complex frequency modes determine the fluxes.
We also relate our approach to that
based on wave packets~\cite{Corley98,Leonhardt08},
and explain why the predictions differ in general, and in particular 
when the number of discrete modes is small. In Sec.~\ref{condi} we give the conditions to get 
complex frequency modes in general terms. 

\section{The settings}
\label{settingsBHL}

\begin{figure} 
\includegraphics[height=80mm]{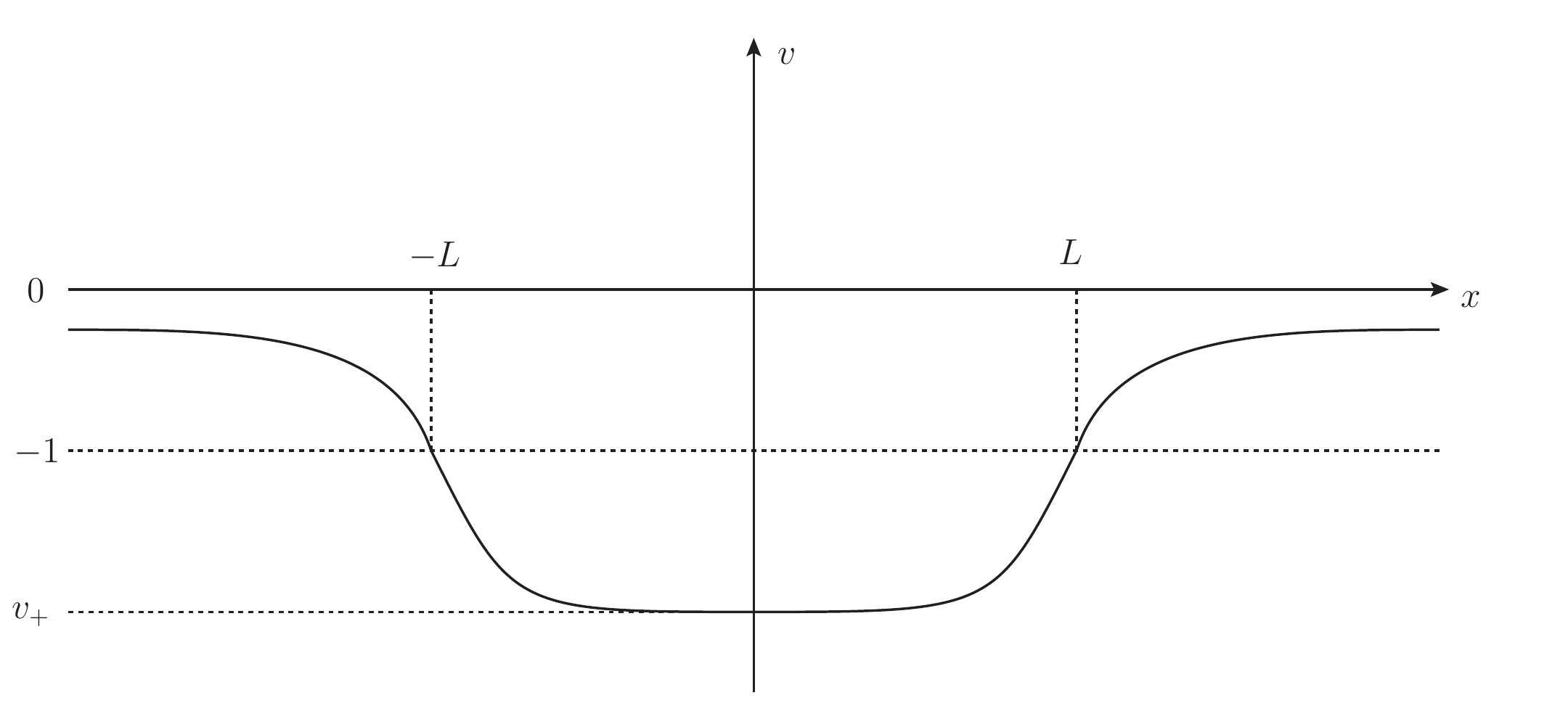}
\caption{Velocity profile $v(x)$ as a function of $x$, for $\kappa_B = \kappa_W$. The horizons are located at $x = \pm L$, where $v(x) + 1 = 0$.}
\label{vlaserprofile_fig} 
\end{figure}

As in the previous chapters, we work in the 2 dimensional stationary Painlevé-Gullstrand metric of \eq{metric}. In addition, we assume that this geometry contains both a black hole and a white hole horizon. We restrict ourselves to flows that are asymptotically constant, {\it i.e.}, we consider velocity profiles such as 
\be
v(x) = -1 + D \tanh\left[\frac{\kappa_B (x - L) }D \right] \tanh\left[\frac{\kappa_W (x + L) }{D}\right] ,
\label{vdparam}
\ee
see Fig.~\ref{vlaserprofile_fig}. The BH (WH) horizon is situated at $x=L$ ($x=-L$).
We suppose that the inequality $\kappa L/D \gg 1$ is satisfied for both values of $\kappa$, where $D\in \, ]0,1]$.
In this case, the two near horizon regions of width $\Delta x \sim D/\kappa$
are well separated, and the surface gravities of the BH and
the WH are, respectively, given by 
$\kappa_B= \p_x v|_{x=L}$ and $\kappa_W = -\p_x v|_{x=-L}$, as explained in \Sec{HRNHR_Sec}.
The minimal speed $| v_-| < 1$ is reached for $x \to \pm \infty$, whereas 
the maximal speed $| v_+ | > 1$ 
is found at $x=0$ between the two horizons.
When $\kappa L/D \gg 1$, their values are
\be v_\pm = -1 \mp D.
\label{v+}
\ee
As explained in \Sec{eikonal_Sec}, when using nonlinear dispersion relations, $D$ fixes the critical frequency $\om_{\rm max}$ above which no radiation is emitted by a single black hole, or white hole. Similarly here, there will be no unstable mode above $\om_{\rm max}$. As we discussed in \Sec{profileSec}, the above assumptions concerning the flow $v$ are very convenient, but not necessary. Our analysis of the unstable modes equally applies to {\it e.g.}, a Reissner-Nordstrom geometry, which is described by the profile 
\be
v(r) = - \sqrt{\frac{2GM}{r} - \frac{Q^2}{r^2}}.
\ee

As in \cite{Corley98} we work with a real field $\phi$ obeying a quartic superluminal dispersion relation
\be
\Om^2 = k^2 + k^4/\Lambda^2, 
\label{disprela}
\ee
where $\Om = -\uf^\mu p_\mu$ is the freely falling frequency and $k = s^\mu p_\mu$ the spatial momentum, exactly as in Chapter \ref{LIV_Ch}. Most of the results we shall derive also apply to higher order superluminal dispersion relations, but also to subluminal dispersion when the locus of BH and WH horizons are exchanged, see the discussion in \Sec{subsuper_Sec}. They also apply to the Bogoliubov theory of phonons in Bose condensates. 
We recall the action of $\phi$ in the metric \eqref{metric} 
\be
S = \frac12 \int\left[ \left(\p_t \phi + v\p_x \phi\right)^2 - (\p_x \phi)^2- \frac{1}{\Lambda^2}(\p_x^2 \phi)^2 \right] dt dx , 
\label{laseraction}
\ee
and the wave equation is 
\be
\left[(\p_t+ \p_x v)(\p_t + v \p_x) - \p_x^2 + \frac1{\Lambda^2}\p^4_x \right]\phi = 0.
\label{twaveeq}
\ee
When the flow is stationary, 
one can look for solutions with a fixed frequency $\lam = i \p_t$. 
Inserting $\phi = e^{-i\lam t}\, \phi_{\lam}(x)$ in \eq{twaveeq} yields
\be
\left[(-i \lam + \p_x v)( -i \lam + v \p_x) - \p_x^2 + \frac{1}{\Lambda^2}\p^4_x \right]\phi_\lam = 0.
\label{waveeq}
\ee
Because of the quartic dispersion, the number of linearly independent solutions
is four. It would have been $n$ if the dispersion relation weas $\Om^2 = k^2 + k^n/\Lambda^{n-2}$ rather
than \eq{disprela}. However, when imposing that the modes $\phi_\lam$
be asymptotically bounded for $x\to \pm \infty$, the dimensionality is reduced to $2$ or $1$
depending on whether $\lam$ is real or complex, but irrespectively of the value of the power $n$. 
(To avoid confusion about the real or complex character of $\lam$
we shall write it as $\lam = \om + i \Gamma$,  with $\om$ and $\Gamma$ both real and positive.
The other cases can be reached by complex conjugation and by multiplication by $-1$.)

The necessity of considering only asymptotically bounded modes (ABM) comes from the requirement 
that the observables, such as the energy of \eq{Hlindec}, be well defined. Returning to 
\eq{laseraction}, the conjugate momentum is 
\be
\pi = \p_t \phi + v\p_x \phi,
\label{pi}
\ee
the scalar product is 
\be
(\phi_1|\phi_2) = i\int_{-\infty}^{\infty} \left[\phi_1^* \pi_2 - \phi_2 \pi_1^* \right] dx , 
\label{KGnorm}
\ee
and the Hamiltonian is given by 
\be
H = \frac{1}{2}\int dx \left[(\p_t \phi)^2 + (1 - v^2) (\p_x \phi )^2 + \frac{1}{\Lambda^2}(\p_x^2 \phi)^2 \right] .
\label{Hphi}
\ee
In the subspace of ABM, the Hamiltonian is hermitian, {\it i.e.}, $(\phi_1| H \phi_2 ) = (H \phi_1  | \phi_2)$.

We conclude with some remarks.
First, when written in the form \eq{KGnorm}, the scalar product is 
conserved in virtue of Hamilton's equations, and the hermiticity of $H$. 
Second, from \eq{Hphi}, one sees that the Hamiltonian density is negative where the flow is supersonic: $v^2 > 1$. 
We shall later see that
the supersonic region should be `deep' enough so that it can sustain at least a bound mode,
thereby engendering a laser effect.
Third, when considering ABM, eigenmodes characterized by different frequencies are orthogonal in
virtue of the identity~\cite{Fulling,Kang97}
\be
(\lam' - \lam^*)\, (\phi_\lam |\phi_{\lam'}) = 0,
\label{ortho}
\ee
which follows from the hermiticity of $H$. 
Finally, we notice that complex frequency ABM can exist  
in the present case because 
neither the scalar product \eqref{KGnorm}, nor the Hamiltonian \eqref{Hphi} are positive definite, see \cite{Fulling} page 228.
On the contrary, since fermionic fields are endowed with a positive scalar product~\cite{Itzykson}, 
no complex frequency ABM could possibly be found in their spectrum~\cite{LevyBruhl,Reed}.

\section{The set of asymptotically bounded modes}
\label{setofmodes}

\subsection{Main results}

In the BH-WH flows of \eq{vdparam}, the set of ABM contains a continuous spectrum of dimensionality $2$ labeled by 
a positive real frequency $\om$, and a discrete spectrum of $N < \infty$ pairs of complex frequencies eigenmodes. 
Moreover, this set is complete. That is, any solution of \eq{twaveeq}, with \emph{finite Hilbert norm}\footnote{In our problem, the precise definition of a Hilbert norm is delicate because no positive scalar product is conserved. Here, we shall use
\be ||\phi||^2 = \int_{-\infty}^{\infty} \left[ |\phi(x)|^2 + |\pi(x)|^2 \right] dx. \ee 
This norm is \emph{not} conserved by the dynamics, but it guarantees that \eqref{KGnorm} is well defined. This definition is close to the positive frequency part $(\phi^+|\phi^+)$ in~\cite{WaldQ}, but it avoids using the sign of the frequency, since complex ones are present here. Unfortunately, our choice is not the unique one, and it is a subtle problem to define a Hilbert norm on a Krein space (defined by \eqref{KGnorm}) that in addition makes the dynamics self-adjoint~\cite{Langer82,Bognar}. Even though the spectral theory in Krein space is still a developing domain, in our case, the completeness of the basis can be shown rigorously, because the Hamiltonian \eqref{Hphi} posses the property of being `Pontryagin', see~\cite{Langer82} for details and~\cite{Gerard11} where the black hole laser problem is explicitly mentioned.}, can be decomposed as
\bsub \label{lindec}
\bea
\phi(t,x) &=& \int_0^\infty \left( e^{- i \om t}\, [ a_{\om, \,u}\, \phi_{\om}^u(x)  + a_{\om, \,v}\, \phi_{\om}^v(x)] + h.c.\right) d\om \\
&& + \sum_{a=1,N} \left( e^{- i \lam_a t} \,  b_a \,\varphi_a(x) +  e^{- i \lam^*_a t}\, c_a \psi_a(x) +  h.c.\right).
\eea 
\esub
For flows that are asymptotically constant 
on both sides of the BH-WH pair, 
we shall show that the real frequency modes can be normalized according to
\be
(\phi^i_{\om'}|\phi^j_\om) = \delta^{ij} \, \delta(\om - \om') ,\quad (\phi^{i\, *}_{\om'}|\phi^j_\om) = 0,
\label{delta_om}
\ee
where the discrete index $i$ takes two values $u,v$, and where $\om , \om' >0$.
The index $u,v$ characterizes modes which are asymptotically left ($v$) or right moving modes ($u$)
with respect to stationary frame.

When $\lam$ is complex, the situation is unusual. Yet, it closely corresponds to that 
described in the Appendix of \cite{Fulling}. In fact whenever a hermitian Hamiltonian
possesses complex frequency ABM, one obtains 
a discrete set of two-modes $(\varphi_a, \psi_a)$ of complex conjugated frequency $\lam_a, \lam_a^*$.
Their `normalization' can be chosen to be
\be 
(\varphi_{a'} | \varphi_a) =0 ,\quad (\psi_{a'} | \varphi_a) = i\,  \delta_{a, a'}, 
\label{psivar}
\ee
with all the other (independent) products vanishing in virtue of \eq{ortho}. 
Since the overlap between modes belonging to the continuous and discrete sectors, such as $(\phi_{\om}|\varphi_a)$, also vanish,
eigenmodes of different frequency never mix. Moreover, 
since the positive norm modes $\phi^i_\om$ all have $\om > 0$, one cannot obtain
Bogoliubov transformations as those characterizing the 
Hawking radiation associated with 
a single BH (or WH). This implies that the (late time) radiation emitted by BH-WH pairs 
{\it entirely} comes from the discrete set of modes.

Using the above equations, the energy carried by $\phi$ of \eq{lindec} is given by 
\be
E = (\phi| H \phi)= \int_0^\infty \om \left( | a_{\om, \,u}|^2 + | a_{\om, \,u}|^2 \right) d\om + \sum_{a=1,N} \left( -i \lam_a \,  b_a c_a^* +  h.c.\right) .
\label{Hlindec}
\ee
This resembles very much \eq{MinkQHam}, with the additional contribution of the complex frequency modes. Because of them, the energy is unbounded from below. Notice also that the absence of terms such as $| b_a|^2$ is necessary to have at the same time complex frequency eigenmodes and real energies. 

\subsection{Asymptotic behavior and roots $k_\lam$}
\label{lasrootsbehav_Sec}

The material presented below closely follows that of \cite{Macher09}. 
In fact the lengthy presentation of that work was written having in mind its applicability
to the present case. The novelties are related to the fact that, for the metrics of \eq{vdparam}, 
the supersonic region is compact (from $-L$ to $L$), and the velocity is subsonic for $x\to \pm \infty$.

Since the velocity $v$ is asymptotically constant for $|x/L| \gg 1$, in both asymptotic regions,
the solutions of Eq.~(\ref{waveeq}) are superpositions of four exponentials  
$e^{i k_\lam x}$ weighted by constant amplitudes. 
To characterize a solution, one thus needs to know (on one side, say on the left) 
the four amplitudes $A_k$ 
associated to the corresponding
asymptotic wave vectors $k(\lam)$. 
These are the roots of 
\be
(\lam - v({x}) k )^2  = k^2 +  \frac{k^{4}}{ \Lambda^{2 }} = \Omega^2(k),
\label{reldisp}
\ee
evaluated for $v(x \to - \infty)= v_-$. We shall not assume {\it a priori} that $\lam$ is real. 
Rather we shall look for all ABM. 
Notice that when considering complex frequencies $\lam = \om + i \Gamma$, the roots of \eq{reldisp} are 
continuous functions of $\Gamma$. In addition, when the scales are well separated, {\it i.e.}, 
when $\kappa/\Lambda \ll 1$, the relevant values of $\Gamma$ will obey $\Gamma/\Lambda\ll 1$.
It is therefore appropriate to start the analysis with $\lam= \om$ real, and then to study how the roots migrate when $\Gamma$ increases.

When $\om > 0$,  since the flow is subsonic for $x \to \pm\infty$, there exist two real asymptotic roots:  
$k^u_\om > 0$ and $k^v_\om <  0$ which correspond to a right and 
a left mover respectively. There also exists a pair of complex conjugated roots, since \eq{reldisp} is real.
Thus, on each side of the BH-WH pair there is a growing and a decaying mode. As in \cite{Macher09} we define 
them according to the behavior of the mode when moving away from the BH-WH horizons.

\begin{figure}[!ht]
\begin{subfigure}[b]{0.5\textwidth}
\includegraphics[scale=0.6,trim = 0 0 12cm 0, clip]{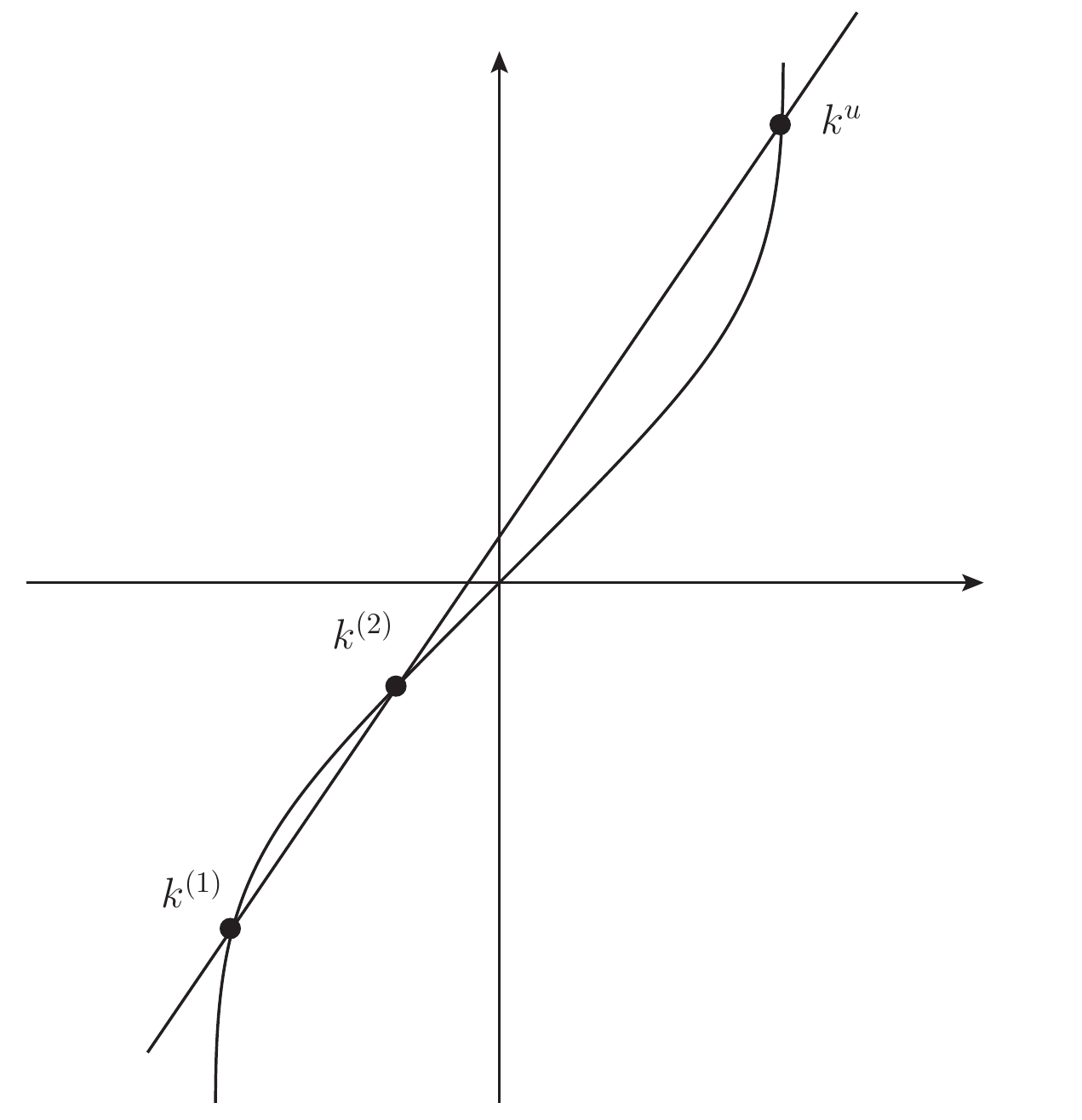}
\caption{Resolution between the horizons ($v_+ < -1$).}
\end{subfigure}
\begin{subfigure}[b]{0.5\textwidth}
\includegraphics[scale=0.6,trim = 13cm 0 0 0, clip]{figs/paper2/dispersionrelation.pdf}
\caption{Resolution outside the horizons.}
\end{subfigure}
\caption{Graphical resolution of \eq{reldisp}. This figure is the same as Fig.\ref{graphroots}, with names for the roots adapted to this chapter.}
\label{figdisp} 
\end{figure}

In preparation for what follows, we study
the roots in the supersonic region between the horizons.
For $\om$ smaller than the critical frequency $\om_{\rm max}$ discussed in \Sec{eikonal_Sec}, the four  roots are real. 
For flows to the left, $v < 0$, the two new real roots correspond to two right movers (with respect to the fluid).
Indeed, they live on the $u$ branch of the dispersion relation \eq{reldisp}, 
that with $\p_k \Omega > 0$, see Fig.\ref{figdisp}. 
When $\om$ increases at fixed $| v |> 1$, these roots approach to each other.
Thus for flows characterized by a maximal velocity $v_+$, 
there is a frequency $\om_{\rm max}$ above which they no longer exist as real roots. It is given
by the value of $\om$ where they merge for $v = v_+$. When the two horizons are well separated, $v_+$ is given in \eq{v+},
up to exponentially small terms.
As shown in \Sec{eikonal_Sec}, $\om_{\rm max}$ is of the form $\om_{\rm max} = \Lambda f(D)$, where $D$ is defined in \eq{vdparam}.
For $D\ll 1$, one finds $f(D)\propto D^{3/2}$. 
Thus, for a given dispersion scale $\Lambda$, 
$\om_{\rm max}$ can be arbitrarily small. This will be important when considering the appearance
of the laser effect in parameter space. 

\subsection{The continuous spectrum}

We now have all the elements to show that the continuous part 
of the spectrum of ABM is labeled by positive real frequencies $\om$, 
and that, for a fixed $\om$, its dimensionality is $2$. 
The general solution of \eq{waveeq} with $\om$ real can be characterized by the $4$ amplitudes, which multiply
the four exponentials evaluated in the asymptotic left region. When imposing that the growing mode is absent on that side,
only three independent solutions remain. However, when propagating these solutions in the asymptotic
region on the right side of the BH-WH pair, the growing mode will be generally present on that side. 
Thus when requiring that it also be absent imposes to take particular combinations, 
and this reduces the dimensionality of ABM to two. 

One can then take appropriate linear combinations to construct the $in$ ($out$) modes describing the left and right movers
propagating toward (escaping from) the BH-WH pair. The $in$ right moving solution $\phi_\om^{u, {\rm in}}$ is the combination which on the left 
is asymptotically proportional to $e^{i k^u_\om x}$ where $k^u_\om$ is the asymptotic
real positive root of \eq{reldisp}. Similarly one can identify the $in$ left moving mode $\phi_\om^{v, {\rm in}}$,
and the two $out$ modes $\phi_\om^{u, {\rm out}}$ and $\phi_\om^{v, {\rm out}}$. 

More precisely, because of the infinite and flat character of the space on either side of the BH-WH pair, the
two $in$ and the two $out$ modes can be normalized as in a constant velocity flow 
(by considering a series of broad wave packets localized in one asymptotic region 
and whose spread in frequency progressively vanishes).
Considering $\phi_\om^{u, {\rm in}}$ for ${x \to -\infty}$, \eq{delta_om} and \eq{KGnorm} 
imply that it asymptotes to 
\be
\phi^{u, {\rm in}}_{\om}(x) \to \sqrt{\frac{dk_\om^u}{d \om}}\, \frac{\exp{i k^u_\om x}}{\sqrt{4 \pi  \Omega(k^u_\om)}}.
\label{normmodes}
\ee
A similar expression valid for $x\to \infty$ gives $\phi^{v\, in}_{\om}$.
These two $in$ modes are orthogonal to each other, establishing the Kronecker $\delta^{ij}$ in \eq{delta_om}. 
For $\om > 0$, these modes have a positive KG norm.
For $\om < 0$, they  have a negative norm. Therefore negative norm modes can all be described 
as superpositions of complex conjugated positive norm modes. 

When propagated across the BH-WH geometry, the $in$ modes are scattered by the
gradients of $v(x)$. When the variation of $v$ is slow, {\it i.e.}, $\p_x \ln v \ll \p_x \ln k_\om$,
the exact solutions are globally well approximated by the WKB solutions of \eq{waveeq}. This was obtained in details in \Sec{modeanalys}. For the $u$-mode, one finds 
\be
\varphi^{u}_{\om}(x) = \sqrt{\frac{\p k_\om^u(x)}{\p \om}}\, \frac{\exp \left(i \int^x dx' k^u_\om(x')\right)}{\sqrt{4 \pi  \Omega(k^u_\om(x))}}.
\label{WKBmodes}
\ee
We use the symbol $\varphi^u_\om$ (resp. $\varphi^v_\om$) to differentiate the WKB solution from the exact one $\phi^u_\om$ (resp. $\phi^v_\om$). At high frequency, $\om/\kappa \gg 1$, the inequality $\p_x \ln v \ll \p_x \ln k_\om$
is satisfied, and $u$ and $v$ modes do not mix, as we obtained in \Sec{modeanalys} and~\cite{Coutant11}. 
At lower frequency, they do. Nevertheless, far away from the BH-WH pair, since $v(x)$ is asymptotically constant, exact solutions decompose into superpositions of $\varphi^u_\om$ and $\varphi^v_\om$ with constant amplitudes.
This applies for both the real and the complex frequency modes of \eq{lindec}.
We shall use this fact several times to characterize the properties of the exact solutions.

Introducing the $out$ modes as in \eq{normmodes}, this scattering is described by 
\be
\begin{split}
\phi^{u, in }_{\om} &= T_{\om} \, \phi^{u, out}_{\om} 
+ R_{\om} \,  \phi^{v, out}_{\om},\\
\phi^{v, in }_{\om} &= \tilde T_{\om}\,  \phi^{v, out}_{\om} + \tilde R_{\om}\,  \phi^{u, out}_{\om} .  
\end{split}
\label{RT}
\ee
Unitarity imposes $| T_{\om}|^2 +|R_{\om}|^2 = 1 = | \tilde T_{\om}|^2 +|\tilde R_{\om}|^2 $, 
and $R_\om \tilde T_\om^* + T_\om \tilde R_\om^* = 0$. 
For all values of $\om$, one thus has an elastic scattering, without spontaneous pair creation. For frequencies $\om > \om_{\rm max}$,
 \eq{RT} coincides with what is found in single horizon scatterings.
Instead, for frequencies $0 < \om < \om_{\rm max}$,
this radically differs from the scattering on a single because, in that case, 
the matrix was $3 \times 3$ and mixed $\phi^u_\om$, $\phi^v_\om$
with the negative frequency $u$-mode $(\phi_{-\om}^{u})^*$. 
The presence of the second horizon therefore `removes' these modes. As we shall later see, they shall be `replaced' by a finite and discrete set of complex frequency modes.
This is not so surprising since the classical trajectories associated with the 
negative frequency modes are closed (hence the discretization), and since these trapped
modes mix with the continuous spectrum through each horizon (hence the imaginary part of the frequency).
In fact this is reminiscent, but not identical, to quasinormal modes~\cite{Frolov,Berti09} or resonances. 
The main difference is that quasinormal modes are not asymptotically bounded.
Thus, they should not be used in the mode expansion of \eq{lindec}.

It should be also mentioned that both $in$ and $out$ modes contain a trapped component  
in the supersonic region\footnote{Because of this trapped wave, 
the $in$ (and $out$) modes are {\it not} asymptotic
modes in a strong sense since a wave packet made with $\phi_\om^{u,\,  in}$
will have a double spatial support for $t\to -\infty$:
the standard incoming packet coming from $x= -\infty$, and the unusual trapped piece. 
This additional component, see \eq{phiuin}, ensures that $\phi_\om^{u,\,  in}$ 
is orthogonal to the complex frequency modes of \eq{lindec}.}
which plays no role as far as their normalization is concerned
since the modes are everywhere regular
and the supersonic domain is finite. 
The case where the subsonic domain is also finite should be analyzed
separately. 
If periodic conditions are imposed at the edges of the condensate, 
one obtains discrete frequencies and resonances effects~\cite{Garay00,Jain07}.
If instead absorptive conditions are used, the frequencies are continuous
and the situation is closer to the case we are studying.

\subsection{The discrete spectrum}
\label{discrete}

On general grounds we explain why, when considering \eq{waveeq} in BH-WH flows of 
\eq{vdparam}, there exists a discrete and finite set of complex frequency modes. 
To this end, we first show that for a generic complex frequency $\lam = \om + i \Gamma$,
there exists no ABM. When $\Gamma\ll \Lambda$, the four roots have not crossed each other with respect to the case where $\Gamma = 0$ 
since the imaginary part of the two complex roots $k^\pm_\om$ 
with $\om$ real is proportional to $\Lambda$. 
Thus, in this regime, one can still meaningfully
talk about the two 'propagating' roots $k^u_\lam, k^v_\lam$, and the growing and decaying roots $k^\pm_\lam$. 
Then, as in the former subsection, when imposing that the growing mode be asymptotically absent on both sides of the BH-WH pair,
the space of solutions is still two. However, when $\lam$ leaves the real axis, developing a small imaginary part $\Gamma > 0$, we have 
\be
k_\lam \sim k_\om^{(0)} + i \Gamma /v_g, \label{k_complex}
\ee
with $k_\om^{(0)}$ is the real solution in the $\Gamma \to 0$ limit, and $v_g^{-1} = \p k / \p \om$ the group velocity of the corresponding mode. In particular, when $\Gamma > 0$ the $u$-root $k^u$ acquires a positive imaginary contribution,
which means that the $u$-mode of \eq{normmodes} diverges for $x \to -\infty$. To get a bounded mode, its amplitude should be set to $0$.
For similar reasons, on the right, the incoming $v$ mode diverges for $x \to \infty$. Hence, for a general value of $\lam$,
the set of ABM is empty.
The above reasoning applies to flat backgrounds with $v$ constant, and establishes that in that case, complex frequency modes should not be considered in \eq{lindec}. 

We should now explain why, when $v$ is supersonic in a finite region, 
some complex frequency ABM exist. The basic reason is the same as that which gives rise to a discrete set of bound modes 
when considering the Schröedinger equation in a potential well (or the propagation of light through a cavity).
In the supersonic region, (in the well), there exist two additional real roots $k_\om$ when $\lam = \om$ real.
The classical trajectories associated with them are closed, and, 
as in a Bohr-Sommerfeld treatment, the discrete set of modes is related to the requirement that the mode be single valued and bounded. 
The complex character of the frequency $\lam$ is due to 
the finite tunneling amplitude across the horizons. Indeed, as we shall later see, the imaginary part of $\lam_a$ is proportional to some quadratic expression in
the $\beta$ Bogoliubov coefficients characterizing the scattering through the horizons. Were these coefficients equal to zero,  $\lam_a$ would have been real.
When `turning on' these coefficients, the frequency $\lam_a$ migrates in the complex plane, 
and the bound modes are continuously deformed.  

These ABM appear in pairs with complex conjugated frequencies. 
This stems from the hermiticity of $H$ which guarantees that there exists an ABM 
of frequency $\lam_a^*$ whenever there is one of frequency $\lam_a$.
At this point it should be re-emphasized that the existence of these complex frequency ABM
is due to the fact that the scalar product is not definite positive.
Indeed, these modes all have a vanishing norm
in virtue of \eq{ortho}. We can also conclude that the discrete set of complex frequencies ABM is 
finite. In fact there are no closed orbits for  
$\om < -\om_{\rm max}$, since the extra real roots $k_\om$
no longer exist  and since there is a gap between the  eigenfrequencies.

\subsection{The quantization}

The canonical quantization of the field $\phi$ is straightforward since each eigenfrequency sector evolves 
independently from the others. 
Indeed when decomposing the field as in \eq{lindec}, with the coefficients $a, b, c$ promoted operators,
the equal time commutation given in \eq{ETC} and the orthonormality conditions
\eq{delta_om} and \eq{psivar} entirely fix their commutation relations. The procedure is very close to that of \Sec{canonicalquantize_Sec}, but the role of the operators must be reconsidered due to the presence of complex frequencies. More precisely, for real frequency modes, the operators $a_{\om,\, i}= (\phi_\om^i| \phi)$ obey the standard commutation relations
\be
[a_{\om,\, i},\, a_{\om,\, j}^\dagger ]= \delta_{ij} \, \delta(\om - \om').
\ee
Instead, for complex frequency modes, one gets 
\be
\big[b_{a}^{\phantom{\dagger}},\, c_{a'}^\dagger \big]= i \delta_{a\, a'}.
\label{combc}
\ee
All the other commutators vanish.

Because of this disconnection, the ground state of the real frequency modes is
stable and in fact subject to no evolution. Hence the number of quanta of these modes
is constant. The evolution of the states associated with the complex frequency modes is also rather simple
and described in App.\ref{QHO_App}. However, asymptotically, an observer cannot distinguish a flux coming from real frequency quanta or from the unstable mode sector, as we shall detail in \Sec{QMsettings_Sec}. What remains to be done is to determine the properties of the asymptotic fluxes. To this end we need 
a better understanding of the modes. 

\section{The properties of the modes}
\label{modeprop}

From \eq{waveeq}, it is not easy to determine the complex frequencies $\lam_a$ and the properties of the modes.
Several routes can be considered. One can adopt numerical techniques. This was done in~\cite{Finazzi10}, in parallel to our work~\cite{Coutant10}, and in \Sec{zsmall}, we briefly compare our results with the numerical ones. One can also bypass the calculation of the eigenmodes and directly compute the propagation of coherent states, or the density-density correlation function~\cite{Balbinot08}, using the techniques of~\cite{Carusotto08}. 
One can also envisage to use analytical methods by choosing the flow $v(x)$ as in~\cite{Barcelo06,Mayoral10}. 
This method is also currently under study. 

In what follows, we use an approximative treatment which is valid when the two horizons are well separated. 
Doing so, we make contact with the original treatment~\cite{Corley98,Leonhardt08} based on wave packets.
More importantly, we determine algebraic relations 
which do not rely on the validity of our approximations. 
In particular, we establish that the real frequency modes $\phi_\om^u$ are intimately related to 
the complex frequency modes even though their overlap vanishes.

\subsection{The limit of thin near horizon regions}

To simplify the mode propagation, 
we assume that the near horizon regions are thin and well separated, {\it i.e.}, $L \kappa \gg D$ in \eq{vdparam}. 
In this case, the propagation through the BH-WH geometry resembles very much to that through a cavity. 
Indeed, the following applies. First, the nontrivial propagation across the two thin horizon regions
can be described by matrices that connect a solution evaluated on one side to that on the other side.
Second, the modes can be analyzed separately in three regions: in $L$, the external left region, for $-(x + L) \gg  D/\kappa_W$;
 in $R$, the  external right region, for $(x - L) \gg  D/\kappa_B$; 
and in the inside region $I$, 
for $L - | x | \gg D/\kappa$. 
Within each region, the gradient of $v$ is small. Hence, any solution is well approximated 
by a superposition of WKB waves \eq{WKBmodes} with constant amplitudes.

To further simplify the analysis, we use the fact that 
the $u$-$v$ mode mixing coefficients are generally much smaller than those mixing the negative frequency
modes to the positive $u$ ones, as exposed in \Sec{modeanalys}. Hence, it is a reliable (and consistent) 
approximation to assume that the $v$ modes completely decouple. After having analyzed this case,
we shall briefly present the modifications introduced by relaxing this hypothesis.
Adopting the hypothesis that the $u-v$ mixing can be neglected, 
for each $\om$ real, one has the following situation. In the left region $L$, 
one only has the WKB mode $\varphi^u_\om$ of \eq{WKBmodes}. Thus, the only solution is $\phi^{u,\, in}_\om$ (up to an overall irrelevant phase we take to be $1$).
In the inside region $I$, one has three modes: 
\be
\phi_\om^{u, {\rm in}} = {\cal A_\om} \, \varphi^u_\om + 
{\cal B}^{(1)}_\om (\varphi^{(1)}_{-\om})^* + 
{\cal B}^{(2)}_\om (\varphi^{(2)}_{-\om})^* ,
\label{phiuin}
\ee
since in supersonic flows, there exist two extra real roots in \eq{reldisp}. 
The superscripts $u, (1), (2)$ characterize the coefficients and the WKB  modes associated with 
the three roots shown in Fig. \ref{figdisp}. 
Since we are considering a solution with $i \p_t = \om > 0$, the (positive norm) negative frequency modes $\varphi_{-\om}^{(i)}$
appear complex conjugated in \eq{phiuin}.

In the external $R$ region the solution must be again proportional to 
the WKB mode $\varphi^u_\om$. By unitarity, the solution must be of the form $\phi_\om^{u, {\rm in}} = e^{i \theta_\om}\varphi_\om^{u} $.
Thus, a full characterization of $\phi^{u,\, in}_\om$
requires to compute the phase $\theta_\om$ and the above three coefficients. 
At this point, it should be noticed that $\lam = \om$ is {\it a priori} real. 
However, the $S$ matrix, and therefore the three coefficients, are holomorphic
functions in $\lam$. Hence nothing prevents to leave the real axis. In fact we shall show
that the complex frequencies correspond to poles associated with a coefficient of the $S$ matrix.

\subsection{An $S$ matrix approach}
\label{Sma}

\subsubsection{The $S$ matrix}

The simplest way we found to compute the above coefficients is
to follow the  approach  of \cite{Leonhardt08}, up to a certain point.
In this treatment, a solution of frequency $\om > 0$ 
is represented by a two component vector $(\varphi^u_\om, \varphi^*_{-\om})$.
The time evolution of a wave packet of such solutions is then considered 
in the thin horizon limit. Since the frequency content of the wave packet plays no role, 
we do not need to introduce a new notation to differentiate it from an eigenmode. 
In this language, the $S$-matrix characterizing a bounce of the trapped mode  $\varphi^*_{-\om}$
can be decomposed as  
\be
S = S_4 \cdot S_{\rm BH} \cdot S_2 \cdot S_{\rm WH} .
\label{Utot}
\ee
The first matrix describes the scattering across the WH horizon. In full generality 
we parameterize it by
\be
S_{\rm WH} =  \bmat \alpha_\om &\alpha_\om\, z_\om \\ \tilde \alpha_\om \,  z_\om^*  & \tilde \alpha_\om \emat .
\ee
Unitarity imposes that $| \alpha_\om|^2 = | \tilde \alpha_\om|^2$ and
$| \alpha_\om|^2\, (1 - | z_\om|^2 ) = 1$.
The second matrix describes the free propagation ({\it i.e.}, without backscattering) from the WH to the BH horizon 
\be
S_2 = \bmat e^{iS^u_\om} & 0  \\ 0 & e^{-iS^{(1)}_{-\om}} \emat .
\label{U2}\ee
In a WKB approximation, the two phases are respectively given by the actions
\bsub \bea
S^u_\om &=& \int  k^u_\om(x) dx, \\
S^{(1)}_{-\om} &=&  \int \left[ - k^{(1)}_{\om}(x)\right] dx.
\eea \esub
In $S_2$, $S^{(1)}_{-\om}$ is multiplied by $-i$ since it governs the evolution of $\varphi_{-\om}^*$. Its momentum is $k^u_{-\om} = - k^{(1)}_{\om} > 0$, where $k^{(1)}_{\om}$ is the most negative root of \eq{reldisp} found in supersonic flows. The end points of these integrals are unfortunately hard to determine. Of course, there is a gauge choice to make here. But any choice for these phases will affect the phases in $S_{\rm BH}$ and $S_{\rm WH}$ so that this choice always disappear from physical observables. To fix this choice, we define these actions through their construction in momentum space, as was done in \Sec{eikonal_Sec}, see \eq{Somint}. In fact, for the negative frequency mode, a canonical choice is to end the integrals at the turning points $x_{tp}^{WH}$, $x_{tp}^{BH}$, see \eq{xtp} for a definition. For the positive frequency mode, there is no clever choice. In fact, as we shall see, it will be more appropriate to stay in momentum space. Unlike for $S_{\rm BH}$ and $S_{\rm WH}$, unitarity brings no conditions on these phases.

The third matrix describes the scattering across the BH horizon and we write it as 
\be
S_{\rm BH} = \bmat \gamma_\om &\gamma_\om \, w_\om \\ \tilde \gamma_\om \, w_\om^*  & \tilde \gamma_\om \emat .
\label{U3}
\ee
Unitarity imposes that $| \gamma_\om|^2 = | \tilde \gamma_\om|^2$
and $| \gamma_\om|^2 \, (1 - | w_\om|^2 ) = 1$.
The fourth matrix describes the return of the negative frequency partner toward the WH horizon, 
whereas the positive frequency mode propagates away in the $R$ region. This is described by 
\be
S_4 = \bmat 1 & 0  \\ 0 & e^{iS^{(2)}_{-\om}} \emat .
\label{U4}
\ee
In the WKB approximation, this backwards movement (hence the $+i$ in the front of $S^{(2)}$) 
is governed by 
\be
S^{(2)}_{-\om} = \int [-k^{(2)}_{\om}(x)] dx ,
\ee
where the momentum $k^{(2)}_{\om}$ is the least negative $u$-root of \eq{reldisp}. 
Since the positive frequency mode further propagates to the right, 
there is no meaning to attribute it a phase in $S_4$. In any case this phase would drop from all physical quantities.

The matrix $S$ of \eq{Utot} is unitary since its 4 constituents are.
Hence $| S_{22} |^2 = | S_{11} |^2 = 1+  | S_{12} |^2 $. 
The components $S_{22}$ and $S_{21}$ we shall later use are given by
\bsub \label{S22}
\bea 
S_{22} &=& \tilde \gamma_\om  \tilde \alpha_\om \, e^{-i(S^{(1)}_{-\om} -S^{(2)}_{-\om})} \left(1 +  z_\om w^*_\om  \frac{\alpha_\om}{\tilde \alpha_\om }\, e^{i(S^{u}_{\om}+S^{(1)}_{-\om})} \right), \label{S22a} \\ 
S_{21} &=& \tilde \gamma_\om  \tilde \alpha_\om \, e^{-i(S^{(1)}_{-\om} -S^{(2)}_{-\om})}
\left( z_\om^* + w^*_\om  \frac{\alpha_\om}{\tilde \alpha_\om }\, e^{i(S^{u}_{\om}+S^{(1)}_{-\om})} \right) .
\eea 
\esub
Hitherto we followed the method of \cite{Leonhardt08}.
Henceforth, we proceed differently by adding a key element. 
We require that the mode propagated by $S$ be single valued. 
For real $\om$, this unequivocally defines $\phi^u_\om$ of \eq{phiuin}. 
Moreover, when looking for complex frequency bounded modes, this will give us the modes $\varphi_a,\psi_a$ we are seeking. 

\subsubsection{The real frequency modes}

Imposing that the trapped mode of negative frequency 
is single valued translates into 
\be
\left(\begin{array}{c} e^{i \theta_\om}\\ b_\om \end{array}\right)
= 
\begin{array}{cc} S \end{array}
\left(\begin{array}{c} 1 \\ b_\om \end{array}\right).
\ee
The phase $\theta_\om$ is that mentioned after \eq{phiuin}. It
should not be constrained since the positive frequency component keeps
propagating to the right. The matricial equation gives
\bsub \label{Unotyetpole}
\bea
b_\om &=&{\frac{ S_{21} }{1-S_{22}}}  , \nonumber \\
e^{i \theta_\om} &=& - \frac{S_{11}}{S_{22}^* } \frac{ 1 - S_{22}^* }{1-S_{22}}.
\eea 
\esub
These equations constitute the first important result of this section. 
They do not rest on the WKB approximation. 
Of course, this approximation can be used to estimate the 4 elements of $S$.
But once these are known, {\it e.g.}, using a numerical treatment, these equations apply. 
What is needed to get these equations is the neglect of the $u-v$ mixing,
and no significant frequency mixing in the inside region in order to obtain well-defined 
amplitudes in \eq{phiuin}.

\begin{figure}  
\centering
\includegraphics[width=0.9\textwidth]{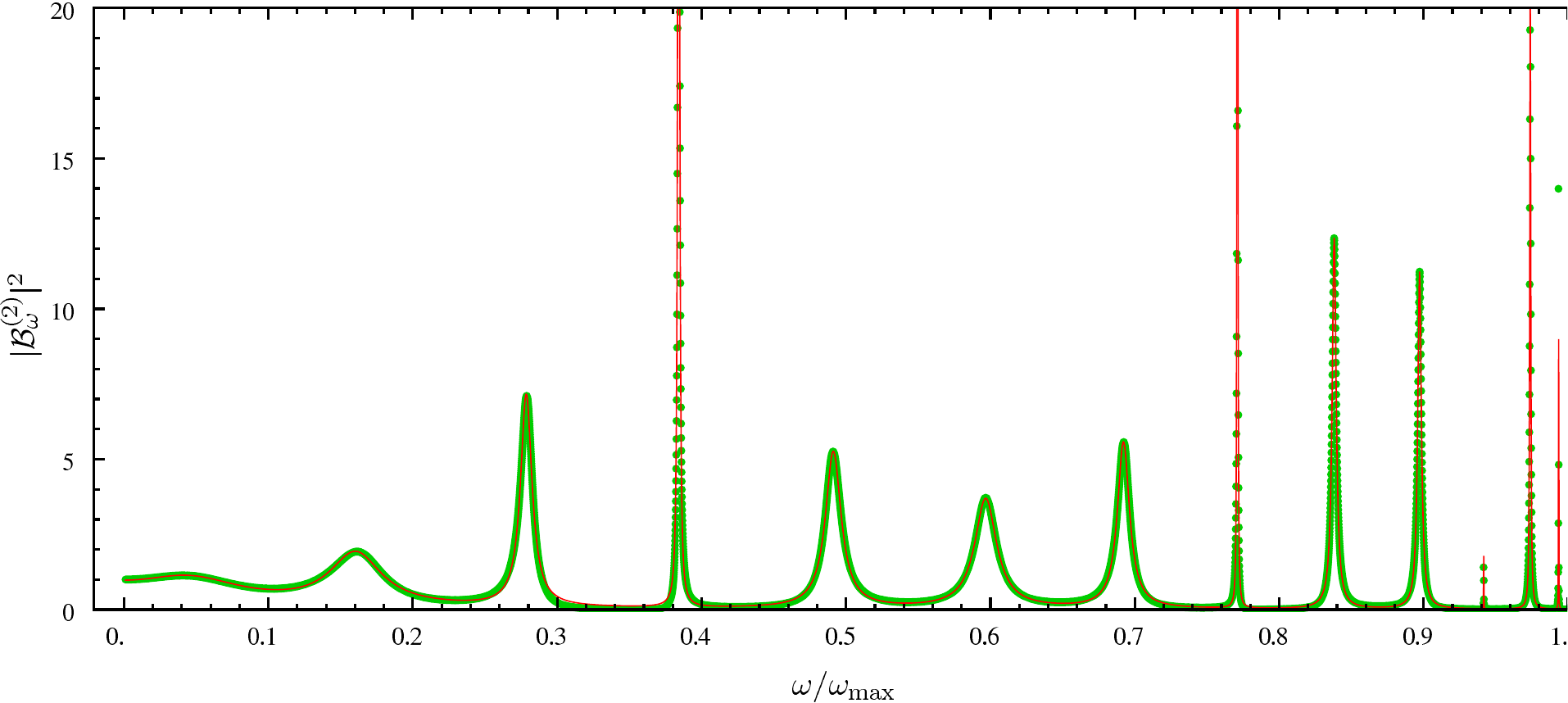}
\caption{By making use of numerical results borrowed from~\cite{Finazzi10}, 
we have represented the square norm of ${\cal B}^{(2)}_\om$, the 
amplitude of the  negative frequency trapped mode of \eq{phiuin}, as a function of $\om$ real.
Near a complex frequency $\lam_a = \omega_a - i\Gamma_a$ which solves $S_{22}=1$, see \eq{Upole}, 
$| {\cal B}^{(2)}_\om |^2 $ behaves as a Lorentzian: $\sim \left| {\omega - \omega_a - i\Gamma_a} \right|^{-2} $, see Eqs. (\ref{Unotyetpole}, \ref{bB}). 
The dots are the numerical values, whereas the continuous line is a fitted sum of Lorentzians. The remarkable agreement establishes that 
the complex frequencies $\lam_a$ 
can be deduced from the analysis of the 
real frequency modes. The frequency $\om$ has been expressed in the units of 
$\om_{\rm max}$, see Fig. 2, so that there is no resonance above $\om/\om_{\rm max} = 1$. In the present case
there are 13 resonances. The narrow peaks, $\Gamma_a \sim 0$, are due to the fact that 
the surface gravities are equal, $\kappa_B= \kappa_W$,  
which leads to $z_\om= w_\om$ in \eq{Gddr}.} 
\label{BHLres_fig}
\end{figure}

Using $S^\dagger = S$, one verifies that the norm of the right hand side of 
the second equation is unity. 
This is as it must be, since in the absence of $u-v$ mixing,
the $u$-component only acquires a phase, here measured with respect to 
the WKB wave. Up to an overall irrelevant phase, the coefficients of \eq{phiuin} are
\bsub \label{bB}
\bea 
{\cal A_\om} &=&  \alpha_\om (1 + z_\om b_\om) ,
\nonumber \\
{\cal B}^{(1)}_\om &=& \tilde \alpha_\om (z^*_\om + b_\om), 
\quad \, {\cal B}^{(2)}_\om =  b_\om .
\eea 
\esub
These amplitudes are governed by $b_\om$, which can {\it a priori} 
be larger or smaller than unity. In particular it diverges if $S_{22} \to 1$ for some $\om$,  thereby approaching a resonance,
see Fig.\ref{BHLres_fig}. We now show that the complex frequency modes correspond to these resonances. 

\subsubsection{The pairs of complex frequency modes} 

Following the discussion of \Sec{discrete}, 
we impose that the amplitude of the incoming $u$ branch be zero, and, as above, that the trapped mode of negative frequency is single valued.
This  gives
\be
\left(\begin{array}{c} \beta_a \\ 1 \end{array}\right) = \begin{array}{cc} S \end{array}
\left(\begin{array}{c} 0 \\ 1 \end{array}\right),
\label{Uc}
\ee
which implies 
\be
\beta_a = S_{12}, \quad 1 =  S_{22}.
\label{Upole}
\ee
Two important lessons are obtained.
First, to get the complex frequencies $\lam_a$ with a positive imaginary part, 
it suffices to solve the roots of $S_{22}=1$. 
Second, as mentioned, these correspond to the poles 
characterizing the propagating modes, see \eq{Unotyetpole}. 

Before computing these frequencies, we explain how to get 
the decaying modes $\psi_a$, the `partners' of the $\varphi_a$ in \eq{lindec}. 
Since these bound modes have a negative imaginary frequency, 
the amplitude of the escaping mode must be zero, {\it i.e.}, 
\be
\left(\begin{array}{c} 0 \\ 1 \end{array}\right)
= \begin{array}{cc} S \end{array}
\left(\begin{array}{c} \tilde \beta_a \\ 1 \end{array}\right).
\ee
Using the hermitian conjugated $S^\dagger$, and the unitarity relation $S^\dagger S = 1$, this condition gives
\be
\bmat \tilde \beta_a\\ 1 \emat
= S^\dagger \cdot \bmat 0 \\ 1 \emat,
\label{eqfpsi}
\ee
from which we get $\tilde \beta_a = (S^\dagger)_{12}$ and $1 =  (S^\dagger)_{22}$. 
These can be reexpressed in terms of the elements of $S$. By imposing the ABM requirement for $\Gam<0$, in particular through \eq{k_complex}, we obtain 
\be
\tilde \beta_a = - [S_{21}(\lam^*)]^* , \quad 1 =  [S_{22}(\lam^*)]^*.
\label{Ugpole}
\ee
Thus the solutions of $1 =  S_{22}^*$ are the complex conjugated of those that solve \eq{Upole}, 
thereby establishing the partnership between $\varphi_a$ and $\psi_a$.
In addition, one has $\tilde \beta_a = -(\det S)^*\beta_a$.

\subsection{The set of complex frequencies $\lam_a$}
\label{zsmall}
\subsubsection{Analytic expressions}
 
To compute the roots of \eq{Upole} we need to know $S_{22}$  as a function of $\lam= \om + i \Gamma$. 
In what follows we shall relate them to the quantities which enter in \eq{S22}. 
To this end, we suppose that the `tunneling' across the horizons is small,
{\it i.e.}, the $\beta$-Bogoliubov coefficients associated with each horizon are small. 
This is true for $\om/\kappa$ sufficiently large, see \cite{Macher09} where it was shown that $z_\om$ and $w_\om$, which are related to the Bogoliubov coefficients 
 by $z_\om  = \beta_\om/\alpha_\om$, behave as $\sim e^{-\pi \om/\kappa}\, (1 - \om / \om_{\rm max})^{1/4}$ where $\om_{\rm max}$ is defined in \Sec{eikonal_Sec}. To proceed we suppose that the norms  of $z_\om^2$, $w_\om^2$ and $z_\om w_\om$ are much smaller than 1. 
In this case, one can expand \eq{S22} in these three products, and in $\Gamma$, since  the roots of \eq{Upole} are real when $z_\om = w_\om = 0$.

To zeroth order in these quantities, $\tilde \alpha_\om \tilde \gamma_\om$ is a pure phase, since the norm is trivially constrained by unitarity. However, when computing it in the weak dispersive regime $\Lambda \gg \kappa$ (or better, under the condition \eqref{Dp}), one does not find the usual phase shift $\pi$ that accounts for the two reflections when dealing with Schrödinger type problems, where the modes near the turning point can be well approximated by Airy functions. Instead, using the results of Chapter \ref{LIV_Ch}, we find 
\bsub \label{epsphase}
\bea
\arg(\tilde \alpha_\om \tilde \gamma_\om) &=& -\pi + \eps(\om), \\
&=& \arg(\tilde \Gam(\om/\kappa_W)) + \arg(\tilde \Gam(\om/\kappa_B)), 
\eea \esub
where the function $\tilde \Gam$ was defined in \eq{Gr0}. Taking this into account, $S_{22}= 1$ gives 
\be
 S^{(1)}_{-\om} - S^{(2)}_{-\om}  - \arg(\tilde \alpha_\om \tilde \gamma_\om) = \int_{x_{tp}^{WH}}^{x_{tp}^{BH}} [- k^{(1)}_{\om}(x) + k^{(2)}_{\om}(x)] dx + \pi - \eps(\om) = 2 \pi\,  n,
\label{BS}
\ee
with  $n \in \mathbb N$. This is a Bohr-Sommerfeld like condition applied to the negative frequency mode $\varphi_{-\om}$.
(In fact, subtracting from both $k^{(1)}$ and $k^{(2)}$ the value of $k$ at the turning points, $k_\om^{tp}$, the differences $k^{(1)}_\om- k_\om^{tp}$, $k^{(2)}_\om- k_\om^{tp}$ have opposite sign.
Hence, \eq{BS} contains a sum of two positive contributions, as in the Bohr-Sommerfeld condition.) Note however that one recovers the standard Bohr-Sommerfeld condition, {\it i.e.}, $\eps = 0$, only in the limit $\om/\kappa \to \infty$. For smaller values of $\om$, $\eps$ accounts for the non trivial phase shift due to the reflections on the two horizons. This corrections come from the fact that the reflected modes cannot be well approximated by Airy functions, something not discussed in~\cite{Nardin09}. We call $\om_a$, $a= 1, 2, ... , N$ the discrete set of frequencies solutions of \eq{BS}, which is finite because no solution exists above $\om_{\rm max}$. 

To first order in $z_\om$ and $w_\om$, for each $\om_a$, 
one gets a complex phase shift $\delta \lam_a = \delta \om_a+ i \Gamma_a$. The imaginary shift is 
\bsub \label{Gddr}
\bea
 2 \Gamma_a T^b_{\om_a} &=& | S_{22}(\om_a) |^2 - 1 , \label{Gddra} \\
\text{thus} \qquad 2 \Gamma_a T^b_{\om_a} &=& | S_{12}(\om_a) |^2 = | z_{\om_a} |^2 +  | w_{\om_a} |^2 + 2 | z_{\om_a} w_{\om_a} |\,  \cos(\Psi_a). \label{Gddrb}
\eea 
\esub
The phase in the cosine is 
\be
\Psi_a = S^{u}_{\om_a} + S^{(1)}_{-\om_a} + \arg\left(z_{\om_a} w_{\om_a}^* \alpha_{\om_a}/\tilde \alpha_{\om_a}\right),
\ee
and $T^b_{\om_a } > 0$ is the
time for the negative frequency 
partner to make a bounce. It is  given by 
\be
T^b_\om =  \frac{\p}{\p{\om}} \left( S^{(2)}_{-\om} - S^{(1)}_{-\om} + \arg (\tilde \alpha_{\om}
\tilde \gamma_{\om}) \right) ,
\label{bouncet}
\ee
evaluated for $\om= \om_a$. The first two terms give the classical (Hamilton-Jacobi) time, whereas the last one gives the contribution from the scattering coefficients. 
The phase shift $\Psi$ can be precisely computed using the results of Chapter \ref{LIV_Ch}. However, we postpone its expression to the next paragraph (see \eq{psiphase}), and first discuss some qualitative features of our expression of $\Gamma_a$. \eq{Gddrb} tells us that $\Gamma_a$ is linearly related to the norm of the effective $\beta$-Bogoliubov coefficient
of the pair, which obeys $| S_{22} |^2 = 1+  | S_{12} |^2 $. It is also worth noticing that $S_{12}$ fixes the amplitude $\beta_a$ of the leaking mode in \eq{Upole}.

As seen from \eq{Gddrb}, $\Gamma_a$ is positive, thereby implying that $\varphi_a$, the solution of \eq{Uc}, is a growing mode in time, 
and an ABM in space. What distinguishes the present case
from usual resonances characterized by a decay rate ($\Gamma < 0$)
is the fact that the norm of the trapped mode $\varphi_{-\om}^*$
is opposite to that of the leaking wave $\varphi^u_\om$. Even though unitarity in both cases 
implies a decrease of the norm of the trapped mode, in the present case it becomes more negative
whereas in the standard case it tends to zero since it has the same sign as that of $\varphi^u_\om$.
As a corollary of this, the fact that resonances have an opposite sign of $\Gamma$ while satisfying an outgoing 
condition as in \eq{Uc} implies that they are not ABM, and therefore not included in the set of modes of \eq{lindec}. 
This remark applies to fermionic fields and implies that the set of ABM for the fermionic
dispersive field considered in~\cite{Corley98} and propagating in the BH-WH metric of \eq{vdparam}, 
is restricted to the continuous set of \eq{lindec}, {\it i.e.}, positive real frequency modes 
elastically scattered as given in \eq{RT}. Indeed, for fermions, \eq{Gddra} still applies, but using $| S_{22} |^2 = 1-  | S_{12} |^2 $ (coming from the conservation of the Dirac scalar product), a minus sign appears for the expression of $\Gamma_a$. To complete these remarks, one should notice that $\psi_a$, our decaying modes, 
are also ABM because they obey the incoming condition \eq{eqfpsi}.

Our treatment is similar to the interesting analysis presented in \cite{Damour76b}. In that work, the general solution is constructed in terms of WKB waves. Then the Bohr-Sommerfeld 
and the outgoing conditions are separately imposed in the small tunneling approximation, thereby 
fixing both the real part of the frequency $\om_a$ and its imaginary part $\Gamma_a$.
We followed another logic which leads to the same result, namely, 
the requirement that the mode be an ABM gives \eq{Uc}, which in turn gives
the complex equation $S_{22}= 1$ that encodes both conditions.
We note that $S_{22}= 1$ does not require that the tunneling amplitudes be small to be well defined.
We also note that our quantization scheme applies to the cases studied in \cite{Damour76b,Damour78}
and allows to remove the `formal trick' used in the second paper.

As a last remark, it is instructive to look how fast the laser instability disappears in the relativistic limit. As we discussed in \Sec{lasrootsbehav_Sec}, when $\Lambda \to \infty$, the growing rates $\Gamma_a$ of the complex frequencies should vanish. In \eq{Gddr}, one sees that the right hand side is basically independent of $\Lambda$ (at least when it is large enough), therefore, $\Gamma_a$ vanishes only through the bouncing time $T^b_\om$. A more precise computation leads to the asymptotic behavior 
\be
\Gamma_a \underset{\Lambda \to \infty}{\sim} \frac{\kappa}{2\ln(\Lambda/\om)} |S_{12}|^2. \label{singularGlimit}
\ee
Because of the log function, the vanishing of the growing rate in the limit $\Lambda \to \infty$ is \emph{very slow}. In other words, as long as Lorentz invariance is broken, $\Gamma_a$ is of the order of $\kappa$, even though $\Lambda$ is at the Planck scale or above. In that sense, the relativistic limit is almost singular, and as long as Lorentz invariance is lost, the laser instability is turned on. Moreover, we also notice that this log appears in $T^b_\om$ only as the time a negative frequency mode spends in the near horizon region. Hence, the transplanckian redshift is regulated by dispersion (see \Sec{eikonal_Sec}) for exactly the same reason that the laser instability appears, {\it i.e.}, the time spent in the near horizon region is finite. Note also that when $\Lambda$ is not infinite, the bouncing time $T_\om^b$ is essentially governed by the inter-horizon distance $L$. Hence, a better way to suppress the laser instability is to take $L\to \infty$, in which case\footnote{For example, in the numerical work of~\cite{Barbado11}, only the $L$ dependence of $\Gam$ was obtained. The logarithmic dependence was not seen, which is explained by the fact that $\Gam$ is almost insensitive to $\Lambda$.} $\Gamma = O(1/L)$.

\subsubsection{The $\Psi$ phase}
As obtained in \eq{Gddr}, $\Gamma_a$ is governed by the $\beta$ coefficients of the $S$-matrix. But there is also an interference term between contributions from both WH and BH horizons, which is determined by the phase $\Psi$. Interestingly, this $\Psi$ is best expressed in momentum space, using the Hamilton-Jacobi function $X_\om$ of \eq{Xomp}. However, because the flow profile $v$ is not monotonic in the black hole laser setup, $X_\om$ is no longer unique. In the chosen profile of \eq{vdparam}, $v$ has a single minimum at $x=0$ between the two horizons, we define $X_\om^W(p)$ (resp. $X_\om^B(p)$) as the solution of \eq{Xomp} 
of negative values describing the propagation toward the white hole 
(resp. positive values associated with the black hole). 
Both of these semi-classical trajectories run from a positive maximum value $p_{\rm max} = p_\om^{\rm in}(0)$ 
to a minimum negative value $p_{\rm min}  = p_{-\om}^{\rm in}(0)$. Therefore, the $\Psi$ phase reads 
\be
\Psi = \Re \left\{ \int_{p_{\rm min}}^{p_{\rm max}} [X_\om^{BH}(p) - X_\om^{WH}(p)] dp\right\} + \pi. \label{psiphase}
\ee
In this equation, we took the real part of the integral in order to remove the imaginary
contributions ($= i\pi \om/\kappa_W$ and $i\pi \om/\kappa_B$)  that arise when $p$ flips sign, 
see the discussion after \eq{theta0}. The contribution for $p > 0$ accounts for the propagation of the positive 
norm mode, whereas that with $p < 0$ for that of the trapped mode.
What is remarkable is that when they are combined, the net result for $S^{u}_{\om_a} + S^{(1)}_{-\om_a}$ takes the 
form of the first term of \eq{psiphase}. Notice again that this simple form is found only when using mode basis such that 
\be
\arg \left(z_{\om_a} w_{\om_a}^* \alpha_{\om_a}/\tilde \alpha_{\om_a}\right) = \pi.
\label{pii}
\ee 
Interestingly, \eq{psiphase} has the same structure as  Bohr-Sommerfeld condition 
of \eq{BS} with the role of $x$ and $p$ interchanged, {\it i.e.} \eq{psiphase} 
is a closed loop in $p$-space of some $X_\om(p)$. 
What is unusual is that the action receives imaginary contributions when $p$ changes sign.
In fact when removing the restriction to the real part of the action, one gets
\be
e^{i \int_{p_{\rm min}}^{p_{\rm max}} \left(X_\om^B(p) - X_\om^W(p) \right)dp + \pi} = - \frac{z_{\om_a} w_{\om_a}^* \alpha_{\om_a}}{\tilde \alpha_{\om_a}} e^{i \Psi }. 
\ee
In other words the Bogoliubov coefficients can be blamed on, and therefore computed from, the imaginary contributions of the action $S_\om$ of \eq{Som} that arise when $p$ flips sign. 
An early version of this relation was used for relativistic modes in~\cite{Damour76}, and it is at the core of the Unruh modes~\cite{Unruh76}, exposed in \Sec{eternalHR_Sec}. 
It was adapted to dispersive waves in~\cite{Brout95} and implicitly used above 
when computing the connection coefficients in \Sec{CF}. It was also recently exploited in~\cite{Scottthesis} in a similar context. 

\subsubsection{Numerical results}
Using the numerical work done in~\cite{Finazzi10}, we have compared the simulations with our theoretical predictions of Eqs.~\eqref{BS} and \eqref{Gddr}. In particular, we were able to clearly confirm the contribution of the extra phase shift $\eps$ appearing in \eq{BS}. This is a non trivial result, since it validates the treatment performed in Chapter \ref{LIV_Ch}, and especially the phases of the Bogoliubov coefficients. 

\begin{figure}[!ht]
\begin{subfigure}[b]{0.5\textwidth}
\includegraphics[scale=1]{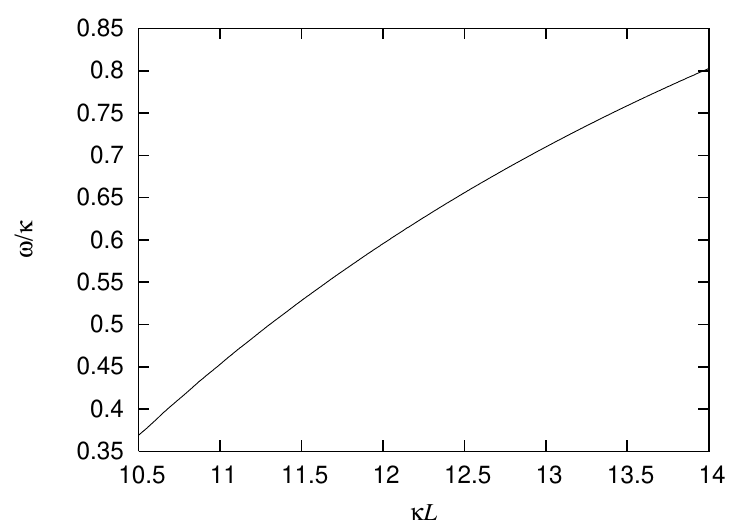}
\caption{Real part $\om = \Re(\lam)$.}
\end{subfigure}
\begin{subfigure}[b]{0.5\textwidth}
\includegraphics[scale=1]{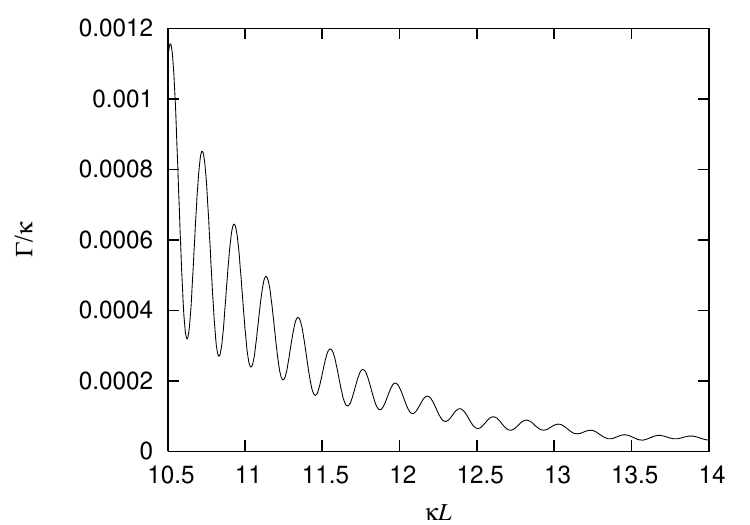}
\caption{Imaginary part $\Gamma = \Im(\lam)$.}
\end{subfigure}
\caption{Evolution of a complex frequency $\lam$ as a function of $L$, the distance between the 2 horizons. These curves have been obtained by making used of the numerical techniques described in~\cite{Finazzi10}. The parameters used are $\kappa_w/\kappa_b = 0.5$, $D=0.5$ and $\Lambda/\kappa_b = 8$, and we consider the $a = 22$ discrete mode.}
\label{omegalas} 
\end{figure}
\begin{figure}[!ht]
\begin{subfigure}[b]{0.5\textwidth}
\includegraphics[scale=1]{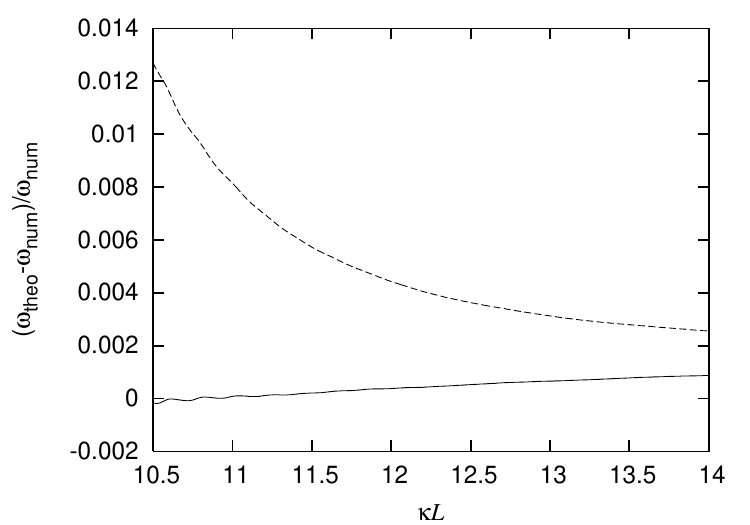}
\caption{Real part $\om = \Re(\lam)$.}
\end{subfigure}
\begin{subfigure}[b]{0.5\textwidth}
\includegraphics[scale=1]{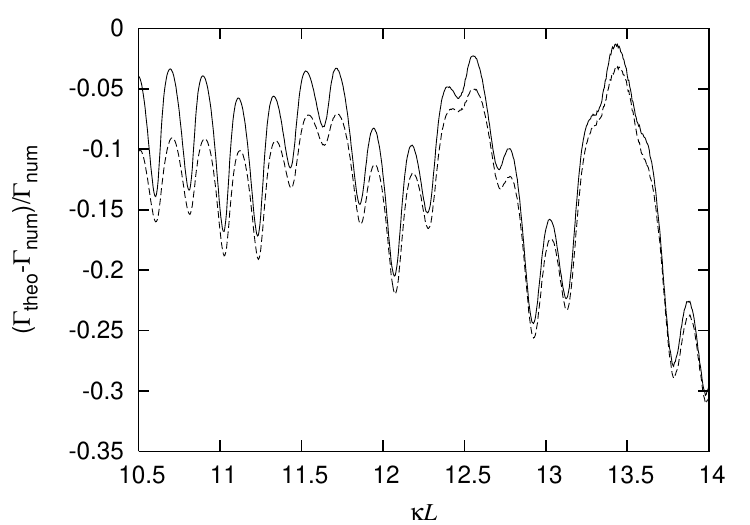}
\caption{Imaginary part $\Gamma = \Im(\lam)$.}
\end{subfigure}
\caption{Relative errors between the numerical results and our theoretical formulae \eqref{BS} and \eqref{Gddr} 
of the same complex frequency as in Fig.\ref{omegalas}.
The continuous lines take into account $\eps$ of \eq{epsphase}, while the dotted lines
are based on the standard expression $\eps = 0$. 
The improvement of the estimation is clear, and therefore 
the necessity of computing the phases in \eq{bogocoef} is established.}
\label{Gammalas} 
\end{figure}

\subsection{The density of unstable modes}
\label{density}

To further characterize the instability, 
it is of interest to inquire about the density of unstable modes, 
about the end of the set, its beginning, and about the most unstable mode.

The number of unstable modes will be either large, or small, depending on the value of $L \om_{\rm max}$. When $L \om_{\rm max}\gg 1$, the number is large because,
for $\om < \om_{\rm max}$, the gap between two neighboring modes roughly given  by $D/L$. 
The end of the set is controlled by $\om_{\rm max}$ as explained
after \eq{BS}. It does not significantly contribute to the instability when $\kappa/\om_{\rm max}\ll 1$,
since $z_\om$ and $w_\om$ tend to zero as $e^{-\pi \om/\kappa}\, (1 - \om / \om_{\rm max})^{1/4}$ for $\om \to \om_{\rm max}$. 
Thus the growth rate $\Gamma_a\to 0$ as $\om \to \om_{\rm max}$. 

The beginning of the set is governed by the first Bohr-Sommerfeld modes.
Their frequency is of the order of $D/L$.  A proper evaluation of $\Gamma_a$ is harder  
since, generically, the coefficients $\alpha_\om$ and $\gamma_\om$ both diverge as
$\sim \om^{-1/2}$. We thus conjecture that the most unstable mode is  the first mode, or one of the first ones, because the cosine
of \eq{Gddr} could in some cases lower the value of $\Gamma_1$ below that of the next ones.
We can also characterize the migration of the roots as $L$ increases.
The (total) variation of \eq{BS} with respect to $L$ and \eq{bouncet}
tells us that $\p \om_a /\p L> 0$. 
From this we conjecture that new unstable modes (obtained by increasing $L$)
appear with $\om_a = 0$. This is corroborated by the numerical simulations of \cite{Finazzi10}.
The determination of $\Gamma_a$ for $\om_a \to 0$ is difficult, 
and further study is needed to establish if $\Gamma_a $ follows the abrupt behavior present in Fig.1 of \cite{Cardoso04}.

\subsection{Taking into account the $u-v$ mixing}
\label{uvmix}

We now briefly discuss the modifications introduce when considering a nonvanishing $u-v$ mixing.
By $u-v$ mixing, we mean an extra elastic scattering that mixes left and right moving modes $\varphi^u_\om$ and $\varphi^v_\om$.
In fact, this hardly modifies the above results. However the algebra becomes 
more complicated because the two nontrivial matrices describing
the propagation across the WH and the BH horizon ($S_{\rm WH}$ and $S_{\rm BH}$) are now the $3\times 3$.
Hence we shall only sketch this enlarged case, but the reader can find more details in the numerical analysis of~\cite{Finazzi10}. 

When considering the {\it in} mode $\phi^{u,\, in}_\om$,
in the left region $L$, one has $\phi^{u,\, in}_\om = \varphi^u_\om + R_\om \varphi^v_\om$, see \eq{RT}. 
In the inside region $I$, one has
\be
\phi_\om^{u, {\rm in}} = {\cal A}_\om^u \, \varphi^u_\om +  {\cal A}_\om^v \, \varphi^v_\om +
{\cal B}^{(1)}_\om (\varphi^{(1)}_{-\om})^* + 
{\cal B}^{(2)}_\om (\varphi^{(2)}_{-\om})^* ,
\label{phiuin2}
\ee
and in the external $R$ region, the mode asymptotes to 
$T_\om \, \varphi^u_\om$. As before, the coefficients are determined by the matching conditions at the horizons.
Expressing $\phi_\om^{u, \,in}$ in terms of three coefficients associated with the 
three WKB waves, $(\varphi^u_\om, \varphi^v_\om, \varphi_{-\om}^*)$,
using the Bogoliubov matrices~\cite{Macher09} $S_{\rm WH},  S_{\rm BH}$ 
which give, for the WH and the BH respectively,
the coefficients after the scattering given the incident ones, one has
\be
\left(\begin{array}{c}  {\cal A}_\om^u \\ R_\om \\ {\cal B}^{(1)}_\om \end{array}\right)
= 
\begin{array}{cc} S_{\rm WH} \end{array}
\left(\begin{array}{c} 1 \\  {\cal A}_\om^v \\ {\cal B}^{(2)}_\om\end{array}\right), 
\label{WH}
\ee
and
\be
\left(\begin{array}{c}  T_\om \, e^{iS^u_\om}\\  {\cal A}_\om^v \, e^{iS^v_\om}  \\{\cal B}^{(2)}_\om \,
e^{- i S^{(2)}_{-\om}}
\end{array}\right)
= 
\begin{array}{cc} S_{\rm BH} \end{array}
\left(\begin{array}{c}  {\cal A}_\om^u \, e^{iS^u_\om} \\ 0
\\ 
{\cal B}^{(1)}_\om \, e^{-i S^{(1)}_{-\om}}
\end{array}\right). 
\label{BH}
\ee
In the second equation, the exponentials arise from the  
propagation from the WH to the BH horizon, and from our choice that the phase of  
WKB waves vanishes at the WH horizon.
The six relations fix the six coefficients. $T_\om$ and $R_\om$ characterize the asymptotic scattering, see \eq{RT},
whereas the four others determine $\phi_\om^{u, {\rm in}}$ in the inside region. 

To get the complex frequencies $\lam_a$, since the weight of the incoming $v$ mode 
is already set to $0$, one simply needs to replace the $1$ in the right hand side of \eq{WH} by a $0$, as done in \eq{Uc}.

\section{Physical predictions}
\label{predic2}

\subsection{Classical setting}
\label{clwp}

From an abstract point of view, any (bounded) initial condition, {\it i.e.}, the data of the field amplitude 
and its derivative with respect to time, can be translated
into the coefficients $a^i_\om,b_a,c_a$ of \eq{lindec}. This is guaranteed by the 
completeness of the mode basis. 
Yet, it is instructive to study more closely how initial conditions describing a wave
packet initially moving toward the WH-BH pair translate into these coefficients. 
This will allow us to relate our mode basis to the wave packet analysis of \cite{Corley98,Leonhardt08}.
Moreover it is a warming up exercise for the determination of the fluxes in quantum settings. 
For simplicity, we work in the regime in which the number of
unstable modes is large, {\it i.e.}, $L \om_{\rm max} \gg 1$, and with a wave packet of
mean frequency obeying $\bar \omega \ll \om_{\rm max}$  and $L \bar \om\gg 1$, {\it e.g.}, 
with $\bar \omega = \sqrt{ \om_{\rm max}/L}$.

We consider a unit norm wave packet which is initially in the left region and propagating to the right.
At early times, before it approaches the WH horizon, it can thus be expressed 
in terms of the WKB modes of \eq{WKBmodes} as 
\be
\bar \phi^u_{\bar \om}(t, x) = \int_0^\infty d\om \left(f_{\bar \om, x_0}(\om) 
\, e^{ -i\om t} \varphi^u_\om(x) + c.c. \right).
\label{wp}
\ee
As an example, one can consider the following Fourier components 
\be
f_{\bar \om, x_0}(\om) = \frac{e^{-(\om - \bar \om)^2/2\sigma^2}}{\pi^{1/4} \sigma^{1/2}}\,  e^{i \om t_0}\, 
e^{i S_\om^u(x_0,- L)} .
\ee
The first factor fixes the mean frequency $\bar \om $ and the spread $\sigma$ taken to satisfy $\sigma/\bar \om \ll 1$
and $\sigma L \gg 1$.
The last two exponentials ensure that at $t= t_0$ the incoming wave packet is centered around some $x_0 \ll - L$.
$S^u_\om(x, x')$ is the classical action of the $u$-mode evaluated from $x$ to $x'$, 
see the phase of $\varphi^u_\om$ in \eq{WKBmodes}.

When decomposing $\bar \phi^u_{\bar \om}$ as in \eq{lindec}, the coefficients are given by the overlaps
\bsub \label{overl}
\bea
\bar a^u_\om &=& (\varphi_\om^{u, \, in} | \bar \phi^u_{\bar \om}) = f_{\bar \om, x_0}(\om), \\
\bar a^v_\om &=&  (\varphi_\om^{v, \, in} | \bar \phi^u_{\bar \om}) = 0, \\
\bar b_a &=& (-i ) \, (\psi_a | \bar \phi^u_{\bar \om})  \neq 0 , \\ 
\bar c_a &=& (-i ) \,(\varphi_a | \bar \phi^u_{\bar \om}) = 0 . 
\eea 
\esub
Hence, at all times, the wave packet can be expressed as
\be
\bar \phi^u_{\bar \om}(t, x) = \int_0^\infty d\om  \left(f_{\bar \om, x_0}(\om) \, e^{ -i\om t} \phi^{u, \, in}_\om(x) +c.c. \right) 
+ \sum_a  \left(\bar  b_a\,  e^{- i \lam_a t} \varphi_a(x) + c.c. \right). 
\label{wp2}
\ee
The first two coefficients in \eq{overl} are easily computed and interpreted.
They fix the real frequency contribution of the wave packet.
The last two are very interesting and encode the instability. 
We first notice that, because of \eq{psivar}, they are given by the 
overlap with the partner wave. Hence the coefficients $\bar c_a$ identically vanish. Indeed the overlap of our  
wave packet with the 
{growing} modes $\varphi_a$ vanishes since these, by construction,  have no
incoming branch. On the contrary $ \bar b_a $, the amplitudes of the growing modes, 
do not vanish since the {\it decaying} modes $\psi_a$ contains an incoming branch, see \eq{eqfpsi}. 
It is thus through the nondiagonal character of \eq{psivar} that the instability enters in the game.
In this respect, our analysis differs from Ref.~\cite{Barcelo06}.
We do not understand the rules adopted in that work. 

To compute $\bar b_a$ we use the fact that at time $t_0$, the wave packet 
is localized around $x_0 \ll -L $. Thus we need the behavior of $\psi_a$ in this region. 
Using the results of Sec. \ref{zsmall} and \eq{WKBmodes} extended to complex frequencies, 
for $\Gamma_a T^b_{\om_a} \ll 1$, 
one gets
\bsub \label{51}
\bea
\psi_a(x) &=& \tilde \beta_a\,  \varphi^u_{\lam_a}(x)
= \tilde \beta_a\, \varphi^u_{\om_a}(x) \times \exp[{-\Gamma_a  \, t_{\om_a}(x, - L)}],
\eea 
\esub 
where $ t_{\om_a}(x,- L) = \p_\om S^u(x,-L)> 0$ 
is the time taken by a $u$-mode of frequency $\om_a$
to travel from $x$ to $-L$.\footnote{It is interesting to notice how $\psi_a$ and $\phi_a$ acquire 
a vanishing norm. Because the leaking wave has an amplitude $S_{21}\propto \Gamma_a^{1/2}$ 
and decreases in $x$ in $\Gamma_a^{-1}$, its positive contribution to the norm is 
independent of $\Gamma_a$ and cancels out that negative of the trapped mode.}
Inserting this result in  $(\psi_a | \bar \phi^u_{\bar \om})$, using the fact that 
the wave packet has a narrow spread in $x - x_0$ given by $1/(\sigma \p_\om k^u)$, 
the overlap is approximately 
\be
\bar b_a = \frac{ \tilde \beta_a^* 
}{\pi^{1/4} (2\sigma)^{1/2}}\, e^{-i(\bar \om - \om_a)t_0}\, e^{ i S^u_{\om_a}(x_0,-L)} \times e^{-\Gamma_a  t_{\om_a}(x_0, - L)},
\ee
The meaning of this result is clear. The spatial decay of $\psi_a$ governed by $\Gamma_a$
has the role to delay the growth of the amplitude of the wave packet until it reaches the WH-BH pair. The phase ensures the spatial coherence of the propagation. That is, the sum of $a$ in \eq{wp}
will give constructive interferences along the classical trajectory emerging from $x_0$ at $t_0$.
Let us also mention that, because of \eq{psivar}, the normalization of $\bar b_a$ is arbitrary,
but the product $\bar b_a \varphi_a$ is well defined and physically 
meaningful.

From this analysis, we see that at early times the wave packet $\bar \phi^u_{\bar \om}$ behaves as described 
in \cite{Corley98,Leonhardt08}. It propagates freely without any growth until
it reaches the horizon of the WH. Then, one piece is reflected and becomes a $v$ mode 
with amplitude $R_{\bar \om}$. The other piece enters in the inside region. 
At later times, a component stays trapped and bounces back and forth while its amplitude increases as $e^{\bar \Gamma t}$.
Every time it bounces there is leakage of a $u$ wave packet at the BH horizon, and a $v$ one at the WH horizon. 

This back and forth semiclassical movement goes on
until the most unstable mode, that with the largest $\Gamma_a$,
progressively dominates and therefore progressively destroys the coherence 
of the successive emissions. In \Sec{density} we saw that the most unstable mode is likely to be the one characterized by the smallest real frequency, $n=1$ in \eq{BS}. 
This implies that at late time, our wave packet will be completely governed
by the corresponding wave $\varphi_1(x)$ (unless of course $\bar b_1 = 0$)
\be
\bar \phi^u_{\bar \om} \to e^{\Gamma_1 (t-t_0)} 
 \times \, {\rm Re}[ e^{-i\om_1 t}\, \bar b_1 \varphi_1(x)]. 
\ee
In the late time limit, its behavior differs from the above semiclassical one. Indeed, in the inside region, one essentially has a standing wave whose amplitude exponentially grows
(it would have been one if the tunneling amplitudes were zero). 
Outside, for $x \gg L$, 
 using $\varphi_1(x) \sim \beta_1 \varphi^u_{\lam_1}(x)$ and the same approximation as in \eq{51},  
one has a modulated oscillatory pattern given by 
\be
\bar \phi^u_{\bar \om} \to e^{\Gamma_1 [(t -t_0) - t_{\om_1}(L,x)]}
\times {\rm Re}[ e^{-i\om_1 t}\, \bar b_1 \, \beta_1 \varphi^u_{\om_1}(x)]. 
\label{aswplt}
\ee 
It moves with a speed equal to $\sim 1$ in the dispersionless regime, when $\om_1/\Lambda \ll 1$. 
The energy flow is now given by a sine squared of period $2 \pi/\om_1$ 
rather than being composed of localized packets separated by the bounce time $T^b$.

\subsection{Quantum setting}
\label{QMsettings_Sec}
\subsubsection{The initial state}

Because of the instability, there is no clear definition of what the vacuum state should be.
Indeed, as can be seen from \eq{Hlindec},
the energy is unbounded from below. 
Therefore, to identify the physically relevant states, one should inquire what would be the state,
or better the subset of states, which would be obtained when the BH-WH pair is 
formed at some time $t_0$. If this formation is adiabatic, the initial state
would be close to a vacuum state at that time. 
That is, the expectation values of the square of the various field amplitudes 
would be close to their values in minimal uncertainty states, with no squeezing,  {\it i.e.}, no anisotropy
in $\phi_\om-\pi_\om$ plane, where $\pi_\om$ is the conjugated momentum of $\phi_\om$. 
Because of the orthogonality of the eigenmodes, this adiabatic state would be, and stay,  
a tensor product of states associated with each mode separately. 

There is no difficulty to apply these considerations 
to the real (positive) frequency oscillators which are described by 
standard destruction (creation) operators $\hat a_\om^u, \, \hat a_\om^v$ ($\hat a_\om^{u\, \dagger}, 
 \hat a_\om^{v\, \dagger}$). The adiabaticity guarantees that one obtains 
a state close to the ground state annihilated by the destruction operators. 
Because of the elastic character of \eq{RT}, for all 
real frequency oscillators, one gets stationary vacuum expectation values with no sign of instability.

There is no difficulty either for the complex frequency oscillators described by $\hat b_a$ and $\hat c_a$. 
Indeed, as shown in App.\ref{QHO_App}, one can define two destruction operators $\hat d_{a+}$ and $\hat d_{a-}$, and use them to 
define the state as that annihilated by them at $t_0$. In this state, we get the following 
{nonstationary} vacuum expectation values:
\bsub \bea 
\langle  \hat b_{a'}(t)\,  \hat b^\dagger_a(t') \rangle & =&
 \frac{ \delta_{a', \, a}}{2} \ e^{-i \om_a (t - t')} \, e^{\Gamma_a(t + t' - 2 t_0)}, \\
\langle  \hat c_{a'}(t)\,  \hat c^\dagger_a(t') \rangle & =&
 \frac{\delta_{a', \, a}}{2} \ e^{-i \om_a (t - t')} \, e^{- \Gamma_a(t + t' - 2 t_0)}  , \\
\langle  \hat b_{a'}(t)\,  \hat c^\dagger_a(t') \rangle & =& i \,  \frac{ \delta_{a', \, a}}{2} \
 e^{-i \lam_a (t - t')}. 
\eea \esub
As expected, the expression in $b b^\dagger$ ($c c^\dagger$) leads to an exponentially growing (decaying) contribution,
whereas the cross term is constant at equal time. This behavior is really peculiar to unstable systems.
Even though the metric is stationary, there is no normalizable state in which the expectation values of the $b,c$ operators are constant.
Stationary states do exist though, but they all have an infinite norm, see App.\ref{QHO_App}.

Using \eq{lindec}, and putting  $t_0 = 0$ for simplicity, 
the two-point function in this vacuum state is  
\be \bal
\langle \hat \phi(t,x)\,  \hat \phi(t',x') \rangle =& \int_0^\infty e^{- i \om (t- t')}\, [ \phi_{\om}^u(x) (\phi_{\om}^u(x'))^*  +  \phi_{\om}^v(x) (\phi_{\om}^v(x'))^*  ]  d\om \\
&+ \sum_{a=1}^N {\rm Re} \left( e^{- i \lam_a (t - t')} \varphi_a(x) \psi^*_a(x') -  e^{- i \lam^*_a (t - t')}  \psi_a(x)  \varphi^*_a(x') \right) \\
&+ \sum_{a=1}^N {\rm Re} \left( e^{- i (\lam_a t - \lam_a^* t')} \varphi_a(x) \varphi^*_a(x') + e^{- i (\lam^*_a t - \lam_a t')} \psi_a(x)  \psi^*_a(x') \right).
\eal \label{2ptf}
\ee
We notice that the last term, the second contribution of 
complex frequency modes is real, as a classical term (a stochastic noise) would be. We shall return to this point below.
We also notice that the second term is purely imaginary and, when evaluated at the 
same point $x=x'$, it is confined inside the horizons since $\varphi_a$ vanishes for $x\ll -L$
whereas $\psi_a$ does it for $x\gg L$. Thus it will give no asymptotic contribution to local observables.  

\subsubsection{The asymptotic fluxes}

Our aim is to characterize the asymptotic particle content 
encoded in the growing modes  $\varphi_a$. 
To this end
it is useful to introduce a particle detector localized far away from the BH-WH pair. We take
it to be sitting at $x \gg L$, in the $R$ region on the right of the BH horizon. We assume that 
it oscillates with a constant frequency $\om_0 > 0$, and that 
its coupling to $\phi$ is switched on at
 $t= - \infty$, and switched off suddenly at $t= T \gg t_0$ in order to see how the
response function is affected by the laser effect a finite time after the formation of the BH-WH pair at $t_0= 0$.

When the detector is initially in its ground state, applying the result of \Sec{UnruhDetector_Sec}, the probability to find it excited at time $T$ is given by 
\bsub \label{laser_rate}
\bea 
\mathcal P_e(T) &=& g^2 \int_{-\infty}^T \int_{-\infty}^T  \, e^{-i\om_0(t-t')} \, \langle \hat \phi(t,x)\, \hat \phi(t',x) \rangle dt' dt, \\
&=& g^2\,  \sum_a \, | \beta_a |^2 \, \,| \varphi^u_{\lam_a}(x)|^2 \, \, \left|  \int_{t_{\om_a}(L,x)}^T e^{- i(\om_0 - \lam_a)t} dt \right|^2 , \\
&=& g^2 \, \sum_a \, \frac{ | \varphi^u_{\om_a}(x)|^2}{(\om_0 - \om_a)^2 + (\Gamma_a)^2} \,\bar n_{\om_a}(T,x)  , 
\eea 
\esub 
where 
\be
\bar n_{\om_a}(T,x) = | \beta_a |^2 \, e^{2 \Gamma_a (T - t_{\om_a}(L,x))} | 1 - \exp  \{ - [\Gamma_a + i (\om_0 - \om_a)](T - t_{\om_a})\} \,  |^2 .
\ee
To get the second line of \eq{laser_rate}, we used $\varphi_a(x) \sim \beta_a \varphi^u_{\lam_a}(x)$, 
the fact that the BH-WH pair is formed at $t=0$,
and that it takes a time $t_{\om_a}(L,x)$ for the mode $\varphi^u_{\lam_a}$ to reach the detector at $x$. 
To get the third line, we used the inequality $\Gamma_a T^b_{\om_a}\ll 1$ 
as in \eq{51}. The meaning of the various factors appearing in $\mathcal P_e$ 
is the following. The sum over $a$ means that all unstable modes
contribute, but the Lorentz functions restrict the significant contributions to frequencies
$\om_a$ near $\om_0$, that of the detector. The prefactor $| \varphi^u_{\om_a}(x)|^2$ 
depends on the norm of the corresponding mode evaluated at the detector location, as in the usual case. 
The function $\bar n_{\om_a}(T,x)$ acts as 
the number of particles of frequency $\om_a$ received by the detector 
at time $T$, and at a distance $x-L \gg \kappa_B^{-1}$ from the BH horizon. 
It depends on the number initially emitted 
($=| \beta_a |^2$) multiplied by the exponential governed by $T - t_{\om_a}$, the 
lapse of time since the onset of the BH-WH pair minus the time needed to reach the detector at $x$.

From the response function of a localized detector, it is clear that  one cannot  
distinguish between the noise due to quanta of the real frequency modes $\phi^u_\om$ and 
that carried by the growing modes $\varphi_a$,  
because both modes asymptote to the WKB waves $\varphi^u_\om$ 
which are asymptotically complete. In this respect it is particularly 
interesting to compute the de-excitation probability $\mathcal P_d$ 
which governs the (spontaneous + induced) decay of the detector. 
It is obtained by replacing $\om_0$ by $- \om_0$ in the first line of \eq{laser_rate} \cite{Primer}.
In that case, one finds that the 
spontaneous decay only comes from the
$\phi^u_\om$ whereas the induced part only comes from the $\varphi_a$.
The induced part equals that of $\mathcal P_e$ since the asymptotic contribution of the $\varphi_a$ to \eq{2ptf} is real.
 We are not aware of other circumstances 
where orthogonal modes with different eigenfrequency (here $\phi^u_\om$ and $\varphi_a$)
are combined in this way in the spontaneous + induced de-excitation probability $\mathcal P_d$,
or equivalently, contribute in this way to the commutator and the anticommutator of the field,
{\it i.e.}, with the $\varphi_a$ only contributing to the latter. (For damped modes, the commutator and the anticommutator are related differently, see {\it e.g.}, Appendices A and B 
in \cite{Parentani07}.) The lesson we can draw is the following: even though the modes $\phi^u_\om$ are orthogonal to the growing 
modes $\varphi_a$, their respective contribution to the asymptotic particle content cannot be distinguished by external devices coupled to the field.

It is also interesting to compute the  asymptotic outgoing energy flux $ \langle \hat T_{uu}(t,x)  \rangle$, 
where $T_{uu} = [(\p_t - \p_x)\phi]^2$. At large distances in the $R$ region, using $\varphi_a \sim \beta_a \varphi^u_{\lam_a}$
and \eq{2ptf}, the renormalized value of the flux is 
\bsub \label{Tuu}
\bea 
 \langle \hat T_{uu}(t,x)  \rangle &=& \sum_a | (\p_t - \p_x)\,  e^{- i \lam_a t}\varphi_a(x)|^2 \\
&= & \sum_a \, | \beta_a |^2  \, 
| (\p_t - \p_x) \, e^{- i \lam_a t}\varphi^u_{\lam_a}(x)|^2 .
\eea 
\esub
It only depends on the discrete set of complex frequency modes.
Yet, because the imaginary part of $\lam_a$ defines a width equal to $\Gamma_a$,  the spectrum of real frequencies $\om$ is continuous,
as can be seen in \eq{laser_rate}. In fact the observables can either be written in terms of a discrete sum
over complex frequencies, or 
as a continuous integral of a sum of Lorentz functions
centered at $\om_a$ and of width $\Gamma_a$. 
However this second writing is only approximate and requires that the inequality $\Gamma_a T^b_a \ll 1$ 
of Sec. \ref{zsmall} be satisfied to provide a reliable approximation. 

\subsubsection{The correlation pattern}

As noticed in \cite{Corley98}, because of the bounces of the trapped modes,
the asymptotic fluxes possesses non-trivial correlations on the {\it same} side of the horizon,
and not across the horizon as in the case of Hawking radiation without 
dispersion~\cite{Schutzhold10,Parentani10,Massar96,Primer,Balbinot08},
or in a dispersive medium~\cite{Brout95}.
These new correlations 
are easily described in the wave packet 
language of that reference, or that of Sec. \ref{clwp}.
When using frequency eigenmodes, they can be recovered through constructive interferences, 
as in Sec.IV F. of \cite{Macher09b}.
Indeed, when the complex frequency modes
form a dense set so that the dispersion of the waves can be neglected, 
the sum over $a$ in \eq{2ptf} constructively interferes at equal time for two different positions $x$ and $x'$ separated by a propagation time 
\be
\p_\om S^u(x,x') = \int_x^{x'}\! dx\,  \p_\om k_\om^u(x) = T^b_\om,
\ee
where  $T^b_\om$ is the bounce time of \eq{bouncet}.
This is because the differences $\om_a - \om_{a+n}$ are equal $n\times 2\pi/T^b_\om$ 
since the $\om_a$ are solutions of \eq{BS}. 
With more precision, the conditions for having these multimodes 
interferences are, on one hand, $\om_{\rm max}/\kappa \gg 1$ so that dispersion hardly affects the modes
and, on the other hand, $\kappa L \gg 1$, so as to have many modes for $\om$ 
below the Hawking temperature $ \sim \kappa$. (For higher frequencies $| \beta_a |^2$, 
which governs the intensity of the correlations, is exponentially damped.) 

However, since dispersive effects grow  
and since the most unstable mode progressively dominates the two-point function of \eq{2ptf}, 
at sufficiently large time the above multimode coherence will be destroyed
and replaced by the single mode coherence of the most unstable one. 
This is unlike what is obtained when dealing with a single BH or WH horizon because in that case~\cite{Massar96}
the pattern is stationary, and all frequencies steadily contribute (significantly for $\om \leq \kappa$).
In the present case, at late time, if $\phi_1$ is the most unstable one,
the correlation pattern is given by 
\be
\langle \hat \phi(t,x)\,  \hat \phi(t',x') \rangle = e^{\Gamma_1 (t + t')}\times  {\rm Re}\left( e^{- i \om_1 (t - t')}\, \varphi_1(x) \varphi^*_1(x') \right). 
\label{2ptfbis}
\ee
The asymptotic pattern, for $x \gg L$, 
is obtained using $ \varphi_1(x) \sim \beta_1 e^{- \Gamma_1 t_{\om_1}(L,x)} \varphi^u_{\om_1}(x) $.
It is very similar to that of \eq{aswplt} found by sending 
a classical wave packet, see Appendix C of \cite{Macher09b} for a discussion of the correspondence 
between statistical correlations encoded in the two-point function and
deterministic correlations encoded in the mean value when dealing with a wave packet described by 
a coherent state.  

So far we worked under the assumption that the $u-v$ mixing coefficients are negligible.
When taking them into account, one obtains a richer pattern which is 
determined by the complex frequency mode solutions of Sec. \ref{uvmix}.

\subsubsection{The small supersonic region limit}

When $L$ of \eq{vdparam} (or $D= | v_+| /c - 1$)
decreases, the number of solutions of \eq{BS} diminishes.
Therefore, in the narrow supersonic limit $ \kappa L \to 0$,
there is a threshold value for $\kappa L$ given $D$, below which
there is no solution. In that case, there are no unstable mode, and
no laser effect. In fact no flux is emitted, and this even though the
surface gravity of the BH (and that of the WH) is not zero. The reason is that 
there is no room for the negative frequency modes $\phi_{-\om}$ to exist. 
In agreement with the absence of radiation, the entanglement entropy of the BH~\cite{Bombelli86} would vanish, because it accounts 
for the number of entangled modes across the horizon and thus of opposite frequency, see 
\cite{Jacobson07b} for the effects of dispersion on the entanglement entropy.

\subsubsection{Comparison with former works}

It is instructive to compare our expressions to those obtained in \cite{Corley98} and in \cite{Leonhardt08}. 
Our expressions differ from theirs because the discrete character of the set of 
complex frequency modes was ignored in these works. 
As a result, a continuous spectrum was obtained. Yet this spectrum 
possesses rapid superimposed oscillations stemming from the interferences that are
present in $S_{21}(\om)$ of \eq{S22}. 
{\it A priori} one might think that they could coincide with our
frequencies $\om_a = \Re(\lam_a)$.
However, as noticed in \cite{Leonhardt08}, their values are insensitive to the phase governed by 
$S^{(2)}_\om$, whereas it plays a crucial role in \eq{BS}. We found no regime in which the two sets could approximately agree.
Therefore, as far as the fine properties of the spectrum are concerned, 
the predictions of \cite{Corley98,Leonhardt08} are not trustworthy.

Nevertheless, when the density of complex frequency modes is high, 
and when ignoring these fine properties, the average properties derived using~\cite{Corley98,Leonhardt08} coincide with ours. Indeed when considering the mean flux 
in frequency intervals $\Delta \om \gg 1/L$, the rapid oscillations found in \cite{Corley98,Leonhardt08}
are averaged out. As a result the mean agrees with that over the contributions 
of complex frequency modes. This can be explicitly verified by comparing the 
norm of our discrete modes $\varphi_a, \psi_a$ with the continuous norm of the negative frequency modes used in \cite{Corley98,Leonhardt08}. 
In the limit of Sec. \ref{zsmall}
the relevant contribution to the overlap $(\psi_a| \varphi_a)$ 
comes from the negative frequency mode $\varphi_{-\om}$. 
Moreover $\varphi_a$ and $\psi_a$ are given by a sum of (normalized) WKB waves 
$\varphi^{(1)}_{-\om_a}$ and $\varphi^{(2)}_{-\om_a}$ times a prefactor $= \sqrt{2\pi/T^b_{\om_a}}$ where $T^b_{\om_a}$ is the bounce time
given in \eq{bouncet}. This is just what is needed for approximating
the discrete sum in \eq{2ptf} by a continuous integral $\int \! d\om$ with a measure equal to 1. 
 
In conclusion, when the number of bound modes becomes small, 
the difference between our description and the continuous one increases. This difference is maximal 
 when there are no solution of \eq{BS}. In this case, no radiation is emitted, something which 
cannot be derived by the continuous approach of \cite{Corley98,Leonhardt08}. 

\section{Conditions for having a laser effect}
\label{condi}
\subsection{General considerations}

Having understood the black hole laser effect, it is worth identifying in more general terms 
the conditions under which a laser effect would develop.
We define a `laser effect' by the fact that a free field possesses complex frequency ABM
while being governed by a quadratic hermitian Hamiltonian, as in \eq{Hphi}, and a conserved scalar product, 
as in \eq{KGnorm}. The field thus obeys an equation which is stationary, homogeneous, and second order in time. 
Let us note that this type of instability is often referred to as a \emph{dynamical instability}~\cite{Jain07,Barcelo06,Richartz09}, 
a denomination which indicates that quantum mechanics is not needed to describe or obtain it. Let us also note that we do not consider the  
case where the frequency of the ABM is purely imaginary. 
Such dynamical instability seems to belong to another class than that we are considering,
see the end of this subsection for more discussion on this. 

Using semiclassical concepts, the conditions for obtaining complex frequency ABM are the following :
\ben
\item For a finite
range of the real part of frequency $\om$,  
WKB solutions with both signs of norm should exist, 
or equivalently, positive norm WKB solutions should exist for both sign of $\om$. 
This is a rather strong condition which requires that the external field (gravitational or electric) 
must be strong enough for this level crossing to take place, {\it i.e.}, 
for the general solution be a superposition as in \eq{phiuin}.

\item These WKB solutions of opposite norm must mix when considering the exact solutions of the mode equation. 
In other words, they should be connected by a nonzero tunneling amplitude. This is a very weak condition
as different WKB branches are generally connected to each other.
In our case it means that $z_\om$ and $w_\om$ appearing in \eq{S22} should not vanish. 

\item One of these WKB solutions must be trapped so that the associated 
wave packets will bounce back and forth.
 This is also a rather strong condition.

\item The deepness of the potential trapping these modes should be deep enough so that at least
one pair of bound modes can exist, see \eq{BS}. This condition is rather mild once the first three are satisfied. 

\een

When these conditions are met for a sufficiently wide domain of frequency $\om$, 
they are sufficient to get a laser effect, and
they apply both when the external region is finite~\cite{Jain07} or infinite.
Being based on semiclassical concepts, {\it stricto sensu}, they cannot be considered as necessary.
But we are not here after mathematical rigor, rather we wish to identify
the relevant conditions in physically interesting situations.

In this respect, it should also be noticed that when only the first two conditions are satisfied, 
one obtains a {\it vacuum instability}~\cite{Primer}, also called a {\it superradiance} in the
context of rotating bodies~\cite{Bekenstein98,Richartz09}. 
Hence, whenever there is a vacuum instability, one can engender a laser effect by 
introducing a reflecting condition, as was done in \cite{Kang97,Cardoso04}, or by
modifying the potential, so that the last two conditions are also satisfied. It should be clear that 
when the laser effect takes place, it replaces the vacuum instability rather than occurs together with it. 
Indeed, as proven in Sec. \ref{zsmall}, the frequency of the trapped modes
are generically complex. The possibility of having a trapped mode
(subjected to a vacuum instability prior introducing the reflecting condition) 
with a real frequency is of measure zero,
as two conditions must be simultaneously satisfied.
To give an example of the replacement of a vacuum instability by a laser effect,
let us consider the archetypal case 
of pair production in a static electric field studied by Heisenberg~\cite{Heisenberg35} and Schwinger~\cite{Schwinger51}. 
In that case, one obtains a laser effect by replacing the Coulomb potential $A_0 = Ex$ by
$A_0 = E | x |$ which traps particles of charge $q$ for $qE < 0$, for frequencies $\om < - m$ where $m$ is their mass. 
We hope to return to such pedagogical examples in the near future.

\subsection{Stability of a single horizon}
\begin{figure} 
\centering
\includegraphics[height=80mm]{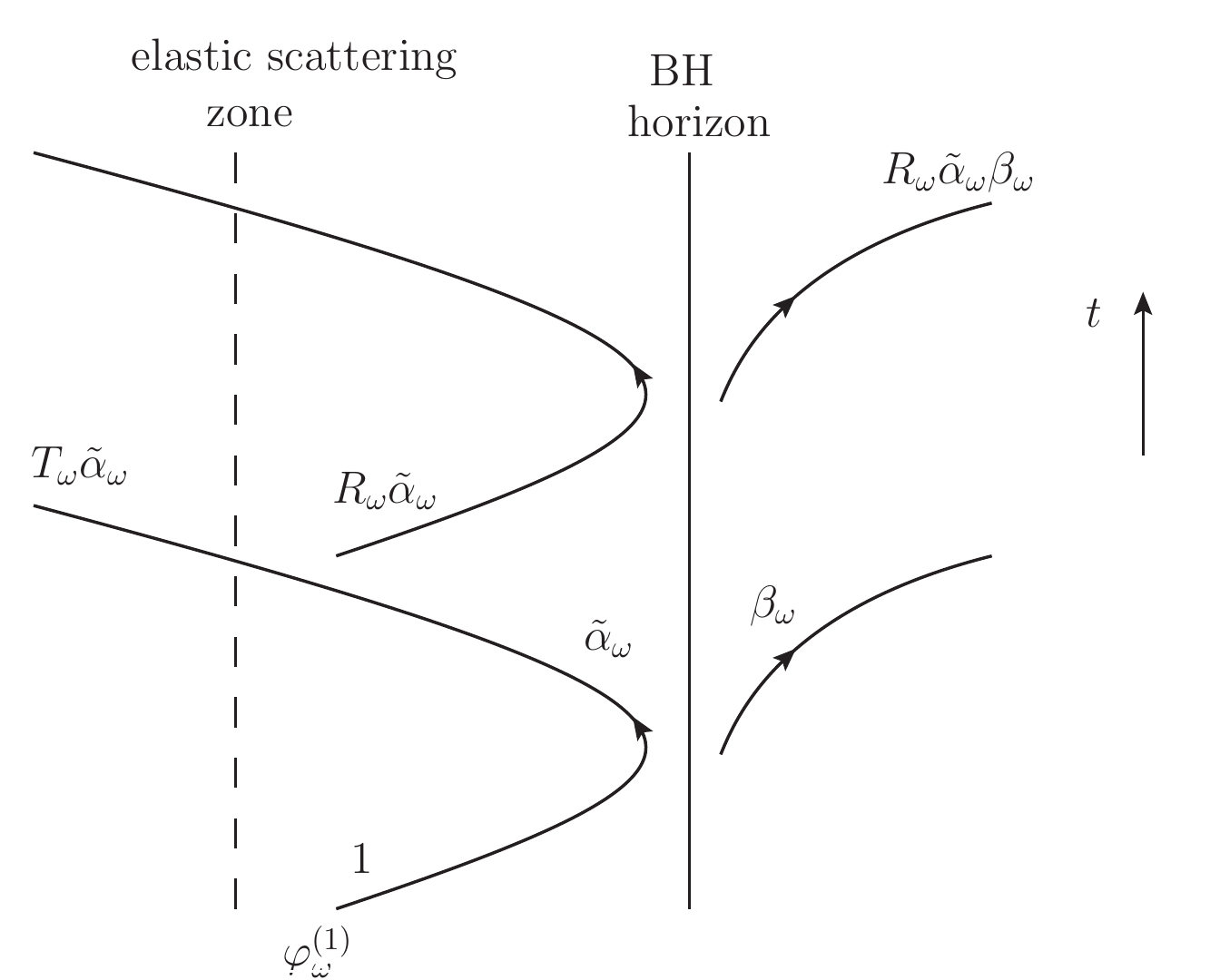}
\caption{Wave packet behavior of an incoming mode $\varphi_\om^{(1)}$. At each round trip, neglecting interferences, the outgoing flux is increased by a factor $| R_\om \tilde \alpha_\om|^2 $. If it is larger than 1, the system is unstable as shown by \eq{Gsingle}.}
\label{singlehorizonlaser_fig} 
\end{figure}

In this paragraph, we present a hand-waving argument, based on the $S$-matrix approach of \Sec{Sma}, to show that a laser effect \emph{may} also appear in the presence of a single horizon. This is due to the fact that  dispersion opens new channels through which Hawking radiation can be self-amplified. We consider a configuration where there is still a black hole horizon, but no white hole horizon. To simplify the discussion, we shall still assume that the $v$-mode decouples from the other ones. However, we now have 3 WKB modes in the asymptotic interior region (still with superluminal dispersion as on Fig.\ref{figdisp}), namely $\varphi_\om^u$, $\varphi_\om^{(1)}$ and $\varphi_\om^{(2)}$. Because the geometry is non homogeneous between $x\to -\infty$ and the horizon, these WKB modes will generally undergo some mixing. To further simplify the discussion, we assume that only $\varphi_\om^{(1)}$ and $\varphi_\om^{(2)}$ mix. Because both are of negative norm, this extra scattering is \emph{elastic}. As we shall argue, this is enough to generate a laser effect. Moreover, an instability can arise even though there is no classically closed orbit. It suffices that superradiant effects be large enough with respect to transmission coefficients. We parametrize the extra scattering by 
\be
S_{\rm el} = \bmat R_\om' & T_\om' \\ \tilde T_\om' & \tilde R_\om' \emat ,
\ee
where $S_{\rm el}$ is the $S$-matrix of the sector $(\varphi_\om^{(1)}, \varphi_\om^{(2)})$ on the left of the horizon (see Fig.\ref{singlehorizonlaser_fig}). Because this scattering is elastic, we have $|R_\om'|^2 = 1 - |T_\om'|^2$. (Note that these coefficients should not be confused with those of \eq{RT}, because we consider here another problem.) 

To keep the discussion close to that of \Sec{Sma}, we assume that this extra scattering is quite efficient, and hence $|R_\om'|^2 \sim 1$, whereas the tunneling coefficients $w_\om$ and $T_\om$ are small. The growth rate can be obtained as before, using semi-classical methods. Neglecting interference effects (the cos in \eq{Gddr}), we obtain 
\be
2 \Gamma_a T^b_\om \simeq  |w_\om|^2 - |T_\om'|^2 . \label{Gsingle}
\ee
We now see that the sign of $\Gamma$ is thus determined by a competition between the 2 kinds of scattering. If it is positive, {\it i.e.} superradiant effects due to the horizon are dominant, the mode is ABM and unstable. In the other case, if $\Gamma_a < 0$, it accounts for a resonant, or quasi-normal mode~\cite{Zapata11}. The conclusion of this little treatment is that dispersive effects open new possibilities to develop instabilities. This could put observational constraints on Lorentz violating theories, as was done in~\cite{Arvanitaki09}, where the possible masses of axions are constrained by the black hole bomb instability. This should be of particular concern in Lorentz violating theories with superluminal dispersion. Indeed, the dispersive modes then live inside the black hole, where the curvature is high. This potentially generates strong back scattering, and make $\Gamma$ of \eqref{Gsingle} positive. This might be the case {\it e.g.} in Ho\u rava-Lifshitz gravity, where stability of black holes has already been questioned~\cite{Blas11}.

\subsection{Quantum aspects}

We should also make several remarks concerning the quantum aspects of the laser effect. 
Even though the complex frequency ABM are orthogonal to the real frequency modes, as is guaranteed by \eq{ortho}, 
the {\it asymptotic} quanta associated with these modes are not of a new type but are, as we saw, superpositions of 
the standard ones associated with real frequency modes.
If laser instabilities can be studied in classical terms, the quantum aspects are not washed out. 
For instance, when considering a charged field, the charge received as infinity is
still quantized, albeit its mean value is described by a complex frequency ABM.
Moreover, in all cases, when the instability ceases,
the number of emitted quanta is a well-defined observable governed by standard destruction/creation operators as those appearing in \eq{HCa}.
From this, it appears that dynamical instabilities governed by an ABM with a purely imaginary 
frequency, as {\it e.g.}, the Gregory-Laflamme instability~\cite{Gregory93}, belong to another class since this asymptotic decomposition in terms of quanta does not seem available. Whether it could nevertheless make sense to quantize such instability is a moot point. 

\section{Main lessons of the analysis} 
\label{Cl_laser_Sec}
The work presented in this chapter provides a complete description, both at the classical and quantum level, of the black hole laser effect. Solving the wave equation \eqref{waveeq}, we showed that it should be described in terms of a finite and discrete set of complex frequency modes which asymptotically vanish~\cite{Coutant10}. We also showed that these modes are orthogonal to the continuous set of real frequency modes which are only elastically scattered, and which therefore play no role in the laser effect. In Sec. \ref{modeprop}, using the simplifying assumption that the near horizon regions are thin, we determined the set of complex frequencies and the properties of the modes using an approach that combines and generalizes  \cite{Leonhardt08} and \cite{Damour76b}. Comparing it with the numerical results of~\cite{Finazzi10}, we obtain quite good agreements,  shown in Fig.\ref{Gammalas}.

We described how an initial wave packet is amplified as it propagates in the BH-WH geometry. 
When the density of complex frequency modes is high we recovered the
picture of \cite{Corley98} at early times. Instead, at late time, or when the density is low, 
the successive emissions of distinct wave packets associated with the bouncing trajectories 
are replaced by an oscillating flux governed by the most unstable mode.

We then computed in quantum setting how the
growth of the complex frequency modes determine the asymptotic fluxes when the initial state at the formation of the BH-WH pair is vacuum. 
Because of the width associated with the instability, the spectral properties of
the fluxes are continuous albeit they arise from a discrete set of modes.
The properties we obtained significantly differ from those found in \cite{Corley98,Leonhardt08}.
We also briefly described the properties of the correlation pattern at early times when the number of 
complex frequency modes is large, and at late time when only the most unstable mode contributes. 
When the supersonic region between the two horizons is so small that 
there is no solution to \eq{BS}, we concluded that there is no instability,
that no flux is emitted, and that the entanglement entropy vanishes.
Finally, in Sec.~\ref{condi} we gave the general WKB conditions under which a laser effect would
obtain starting from the standard concepts that govern a vacuum instability in Quantum Field Theory.

It would be interesting to extend the present work to Fermionic fields. As noted after \eq{ortho}, the Hilbert structure of the Dirac equation prevents the appearance of laser modes. However, the equation \eqref{Upole} still makes sense, and picks up $N$ pairs of complex frequencies. They are not associated with ABM eigenmodes but with resonances. They indicate that the naive vacuum will decay in the lowest energy state vacuum plus $N$ asymptotic quanta by spontaneously emptying the $N$ Dirac holes that are trapped inside the horizons. A similar phenomenon occurs around a rotating black hole, when the bosonic field responsible for the black hole bomb effect is replaced by a fermionic one~\cite{Hartman09}. \\

In a more general context, this instability shows that the consequences of violating Lorentz invariance can be quite dramatic. Indeed, new instabilities can be triggered, just as the black hole laser. But as stressed in \Sec{condi}, it might be a concern for single horizons too. Moreover, as we explained after \eq{singularGlimit}, the rate of that instability is essentially insensitive to the dispersive scale. Therefore, even if Lorentz invariance were broken at an energy scale well above the Planck scale, it could have impacts at observable scales. In conclusion, the possible appearance of dynamical instabilities calls for caution when trying to build Lorentz violating theories~\cite{Jacobson01,Horava09} consistent with observational confirmations of general relativity.

\chapter*{Conclusions} 
\markboth{Conclusions}{Conclusions}
\phantomsection
\addcontentsline{toc}{chapter}{Conclusions}
In this thesis, we studied various implications of short distance dispersion on Hawking radiation. All along this work, our motivations were twofold. On one hand, in analog models, dispersive effects always exist. In the aim of detecting the analog of the Hawking effect, it is crucial to understand what properties are unaffected, but also what are the new features induced by dispersion. On the other hand, there is the `transplanckian question' of gravitational black holes. This question arises from the role of ultrahigh frequencies in the derivation of the Hawking effect. It is an open question whether these high frequencies can be described by the approximation of a free field propagating on a fixed background. Moreover, there is the possibility that quantum gravity effects might be approximately modeled by a modified dispersion relation when momenta approach the Planck length or any other `microscopic' scale. In our work, we assumed such a modified dispersion relation, to probe the sensitivity of Hawking effect to ultraviolet physics. 

Because of this double motivation, we shall draw general conclusions in both contexts. We do not recall the details of the obtained results, as this was done in the end of each chapter. In \Sec{Cl_LIV_Sec}, we exposed the parameters which govern the deviations from the Hawking spectrum, and in particular the role of the size of the near horizon region. In \Sec{Cl_mass_Sec}, we review the features of undulations, in the case of pure white holes, massive modes or double horizon configurations. In \Sec{Cl_laser_Sec} the main properties of the black hole laser effect were presented. Now, we shall discuss the general implications and lessons that we can draw from these studies. 

\section*{Analog gravity}
\addcontentsline{toc}{section}{Analog gravity}
The robustness of Hawking radiation against short distance dispersion is obviously a relevant fact for analog gravity. However, even in analog contexts, the Hawking effect is often very small, and hence quite hard to detect experimentally. Therefore, it is probably more valuable to understand that Hawking radiation could be detected and studied through the \emph{new} effects it generates, such as the laser effect, or the undulation. In fact, the possibility of having a laser arises in many setups, since it is most of the time easier to obtain two horizons rather than a single one. Moreover, the laser effect can appear to be either an instability that destroys the background, but also as an amplification of the radiation. Depending on what is aimed to be measured, this could be either a limitation or on the contrary an opportunity to study Hawking effect. In~\cite{Zapata11}, the proposal was literally to exploit a `pre-laser' setup, where there is no instability, but Hawking radiation is amplified by resonant effects. The lesson is similar for the undulation, if not stronger. Indeed, the undulation is less dangerous for the setup, but it is a clear consequence of the Hawking radiation emitted by white holes when using dispersive fields, and by black holes when using massive fields. In that sense, we believe that studying the undulation properties is part of the goal to detect analog Hawking effect, as we discussed in \Sec{WHundul_Sec}.

\section*{Quantum gravity}
\addcontentsline{toc}{section}{Quantum gravity}

As explained in \Sec{transpl_Sec}, the role of ultrahigh energy physics in the Hawking process is still unclear. Indeed, in the semi-classical approximation, ultraviolet frequencies seem to play a crucial role, even for large black holes. Taking quantum gravity into account could lead to small modifications of the Hawking effect, which decrease as the black hole becomes larger, but it could also generate new effects, that stay observable at low energies. In fact, the work performed and described in this thesis precisely show that both outcomes are possible. In our case, ultraviolet physics is modeled by short distance dispersion. On one hand, when analyzing the Hawking process for a single black hole, we obtain a phenomenology that is very close to the free field result of Hawking, as explained in chapter \ref{LIV_Ch}. This property is sometimes referred as the `robustness of Hawking radiation': effects due to dispersion are highly suppressed in the flux emitted by large black holes. On the other hand, in Chapters \ref{mass_Ch} and \ref{laser_Ch}, we analyzed effects that are generated by dispersion, but which are manifest at low energy scales. Indeed, both the undulation and the laser effect are phenomena that are induced by dispersion. However, as discussed in particular in \Sec{zsmall}, they are visible at a macroscopic scale. This provides a deep lesson concerning the physics of black hole radiation. It shows how black holes can act as `microscopes', in the sense that ultraviolet laws generate visible effects at very low energies. While in most physical systems, scales stay separated, in this case, the ultraviolet physics has an observable influence on infrared physics. \\

This work does not pretend to `solve' the transplanckian question, and it might well be that the Lorentz breaking hypothesis is never realized in nature. However, the present work shed some light on the importance of the transplanckian question. As we understood through these various effects, a complete understanding of Hawking radiation will be achieved only if one can provide a satisfactory way to incorporate ultraviolet gravitational physics. 


\begin{appendices}

\chapter{Useful properties of special functions}
\label{Specialfunction_App}
\section{Euler function}
The Euler $\Gam$ function is defined by its integral representation 
\be
\Gam(z) = \int_0^{+\infty} x^{z-1} e^{-x} dx.
\ee
When $\Re(z) > 0$, this integral is well-defined and $\Gam$ is analytic. Moreover, by an integration by part, we show the fundamental functional relation of the $\Gam$ function
\be
\Gam(z+1) = z\Gam(z).
\ee
Using this relation, we analytically extend the function for all complex values of $z$, except negative integers. At these latter values, $\Gam$ possesses a pole. The second most fundamental identity is the so-called `complement formula'
\be
\Gam(z) \Gam(1-z) = \frac{\pi}{\sin(\pi z)}.
\ee
We refer to~\cite{Olver} for its proof. These relations allow us to obtain several useful identities, which naturally arise in the study of Hawking radiation and especially in Sec.\ref{transfermatrix_Sec}. We now present a few of them.

$\forall x \in \mathbb R$,
\be
|\Gam(ix)|^2 = \frac{\pi}{x \sinh(\pi x)}.
\ee
We also have the asymptotic  
\be
\Gam(ix) = \sqrt{\frac{2\pi}x} e^{\frac{\pi x}2} e^{i(x \ln(x) - x) - i\frac\pi4} \left(1 - \frac i{12x} + O\left(\frac1{x^2}\right)\right),
\ee
or equivalently
\be
\Gam(1+ix) = \sqrt{2\pi x} e^{\frac{\pi x}2} e^{i(x \ln(x) - x) + i\frac\pi4} \left(1 - \frac i{12x} + O\left(\frac1{x^2}\right)\right).
\ee
Around 0, the behavior of $\Gam$ is much simpler, since for all $z \in \mathbb C$
\be
\Gam(z) \underset{0}{\sim} \frac1z.
\ee
This asymptotics are especially useful in Sec.\ref{massfields_Sec} in order to obtain asymptotic behaviors of the Bogoliubov coefficients.

\section{Hypergeometric functions}
\label{Hypergeo_App}
\subsection{Definitions}
An hypergeometric function is defined by the convergent series
\be
F(a,b;c;z) = \sum_{n \in \mathbb N} \frac{\Gamma(a+n) \Gamma(b+n)\Gamma(c)}{\Gamma(a)\Gamma(b) \Gamma(c+n)} \frac{z^n}{n!},
\ee
where $a$, $b$ and $c \notin \mathbb Z^-$ are complex parameters. The hypergeometric function obeys the second order differential equation
\be
z(1-z) \p_z^2 F + [c - (a+b+1)z] \p_z F - ab F = 0. \label{Hypergeoequ}
\ee
The other solution of that equation is, in fact, also an hypergeometric function, but with other values of the parameters. Many equations can be recasted under that form by a proper change of variable and therefore make the system completely integrable. Moreover, hypergeometric functions are powerful tools to solve scattering problems. Indeed, \eq{Hypergeoequ} has 3 regular singular points at $0$, $1$ and $\infty$. The asymptotic behavior of $F$ at these points can be fully derived. In scattering problems, this allows us to obtain analytical expressions for the scattering coefficients in terms of $\Gam$ functions~\cite{Gottfried}. For detailed explanations about the theory of hypergeometric functions, we refer to references~\cite{Olver,AbramoSteg}. We now use these hypergeometric functions to obtain the general solution of the massive field equation in the geometries of Sec.\ref{quattro}.

\subsection{Wave equation of massive modes}
In Sec.\ref{quattro}, we exploited hypergeometric functions to solve the mode equation \eqref{canmodequ}. As explained there, this is the mode equation of stationary modes in Schwarzschild-like coordinates. However, to solve them, it is easier to use another spatial coordinate
\be
x^* = 2\kappa \int \frac{dx}{1-v^2}, \label{xstar}
\ee
and a rescaled field 
\be
\psi = \frac{\varphi_\om}{\sqrt{|1-v^2|}}. \label{tortoisepsi}
\ee
The name of the new coordinate $x^*$ has been chosen because it is closely analogous to the `tortoise coordinate' in \Sch geometry. Note that it has to be defined separately on one side and the other of the horizon. Using it, the mode equation reads
 \be
- \partial_{x^*}^2 \psi + \frac{m^2}{4\kappa^2} (1-v^2) \psi =\frac{\om^2}{4\kappa^2} \psi \label{tortoisemodequ}.
\ee
Of course, $1-v^2$ has to be re-expressed as a function of $x^*$. Under that form, the wave equation is often easier to solve. In our case, even if it does not exactly correspond to the hypergeometric equation \eqref{Hypergeoequ}, it is close enough to easily obtain the general solution. It is important to notice that the $x^*$ coordinate is \emph{not} adapted to impose the regularity condition on the horizon, unlike $x$, see the discussion in Sec.\ref{NHRSec}.

\subsection{General solutions for massive modes}
To present the exact solution under a compact form, we shall give them in terms of the $x^*$ coordinate and field $\psi$. However, in Sec.\ref{quattro} what we need is the mode $\varphi_\om(x)$, and one must use Eqs. \eqref{xstar} and \eqref{tortoisepsi} to obtain it. (Note that in the Appendix of~\cite{Coutant12}, we gave them directly in $x$ coordinate.)

\subsubsection{Interior of the totally reflecting model}
We start with the totally reflecting model, described by the profile $v$ of \eq{Sandrov}, and use the notation of \eq{notations}.
For $x>0$, we solve \eq{tortoisemodequ} for 
\be \left\{ \bal
&1-v^2(x^*) = \frac{D e^{x^*}}{1- e^{x^*}},\\
&x^* \in ]-\infty;0[.
\eal \right. \ee
We then obtain the only ABM solution 
\be
\psi(x) = C (1-e^{x^*})(e^{x^*})^{-i\varpi} F\left(1 - i\varpi +i \bar \Om_+, 1 - i\varpi - i\bar \Om_+ ; 2 ; 1 - e^{x^*} \right),
\ee
where $C$ is an arbitrary constant. To go back to the $x$ coordinate, we use
\be
e^{x^*} = 1 - e^{-\frac{2\kappa x}D}.
\ee

\subsubsection{Exterior of the totally reflecting model}
For $x<0$, we have
\be \left\{ \bal
&1-v^2(x^*) = - \frac{D e^{x^*}}{1+ e^{x^*}},\\
&x^* \in ]-\infty;+\infty[.
\eal \right. \ee
And the general solution is
\bsub \label{Sandrosolution-}
\bea
\psi(x) &=& A (e^{x^*})^{i\varpi} F\left( i\varpi - i \bar \Om_+, i\varpi + i \bar \Om_+; 1 + 2i\varpi ;  -e^{x^*} \right),  \\
&& + B (e^{x^*})^{-i\varpi} F\left(- i\varpi - i \bar \Om_+, - i\varpi + i \bar \Om_+; 1 - 2i\varpi ; -e^{x^*} \right) ,
\eea 
\esub
with $A$ and $B$ arbitrary constants. The relation between $x$ and $x^*$ is here given by
\be
e^{x^*} = e^{-\frac{2\kappa x}D} - 1.
\ee

\subsubsection{Exterior of the CGHS model}
For the profile of the CGHS black hole, see \eq{CGHSv}, and for $x>0$, the conformal factor is
\be \left\{ \bal
&1-v^2(x^*) = \frac{D e^{x^*}}{1+ e^{x^*}},\\
&x^* \in ]-\infty;+\infty[,
\eal \right. \ee
leading to the general solution 
\bsub \label{CGHSsolution} 
\bea
\psi(x) &=& A (e^{x^*})^{i\varpi} F\left( i\varpi - i \bar \Om, i\varpi + i \bar \Om; 1 + 2i\varpi ; -e^{x^*} \right) ,\\
&+& B (e^{x^*})^{-i\varpi} F\left(- i\varpi - i \bar \Om, - i\varpi + i \bar \Om; 1 - 2i\varpi ; -e^{x^*} \right).  
\eea \esub
Here $A$ and $B$ are arbitrary constants and 
$\bar \Om = \bar \Om_>$ for $\om >\om_L$ and $ \bar \Om = i \bar \Om_<$ for $\om <\om_L$.
The definitions of these dimensionless quantities are given in Sec.~\ref{CGHSSec}. To express this in the $x$ coordinate, one must use 
\be
e^{x^*} = e^{\frac{2\kappa x}D} - 1.
\ee

\subsubsection{Analog model}
For the analog model, given by \eq{analogv} we combine the preceding two results. In the interior (resp. exterior), the solution is \eq{Sandrosolution-} (resp. \eq{CGHSsolution}) with $D \to D_L$ (resp. $D\to D_R$).

\chapter{Tools from the quantum harmonic oscillator}
\label{QHO_App}

\section{Harmonic oscillator}
In this section, we consider the quantum harmonic oscillator of mass $m=1$. The aim is to review some basic features and especially the notion of coherent states. For a more detailed description of this matter, we refer to~\cite{Gottfried,Leonhardt} At the end, we compare the two-point function obtained in the vacuum or in a coherent state. The presented result extended to quantum field theory, as shown {\it e.g.} in the appendix of~\cite{Macher09b}. This computation is used in Chapter \ref{mass_Ch} to interpret the role of the contribution of the undulation in the Wightman function. 

The harmonic oscillator is described by the Hamiltonian
\be
H = \frac12 \hat p^2 + \frac12 \om^2 \hat x^2.
\ee
We study the oscillator in the Heisenberg picture. Hence the operators evolve but not the state, which can represent initial conditions. As usual we define creation and annihilation operators such that
\bsub \bea
\hat x(t) &=& a \frac{e^{-i\om t}}{\sqrt{2\om}} + a^{\dagger} \frac{e^{i\om t}}{\sqrt{2\om}},\\
\hat p(t) &=& a \frac{-i\om}{\sqrt{2\om}} e^{-i\om t} + a^{\dagger} \frac{i\om}{\sqrt{2\om}} e^{i\om t}
\eea \esub
Using the creation and annihilation operators, the Hamiltonian reads
\be
\hat H = \om \left( \hat a^{\dagger}\hat a + \frac12 \right).
\ee
and the canonical commutation relations impose
\be
\left[ \hat a, \hat a^{\dagger} \right] = 1.
\ee
Note that this transformation is the one of Sec.\ref{QFTfst_Sec} applied to a single degree of freedom. The eigenstates of the Hamiltonian form a complete basis of the Hilbert space, and are noted $\{|n\rangle \}_{n \in \mathbb N}$. They obey the relation
\be
\hat H |n\rangle = \om \left(n+\frac12\right) |n\rangle.
\ee
Moreover, the annihilation and creation operators act as
\bsub \bea
\hat a |n\rangle &= \sqrt n |n-1\rangle ,\label{a_action} \\
\hat a^\dagger |n\rangle &= \sqrt{n+1} |n+1\rangle . \label{adag_action}
\eea \esub
The vacuum, or ground state is defined as the state of norm 1, annihilated by $\hat a$, {\it i.e.},
\be
\hat a \vac{} = 0.
\ee
It is also an eigenstate of the Hamiltonian, associated with the lowest eigenvalue $\om /2$.

\subsection{Coherent states}
We define a coherent state through the relation
\be
\hat a |\alpha\rangle = \alpha |\alpha\rangle \label{defcoh},
\ee
where $\alpha \in \mathbb C$. Hence $|\alpha\rangle$ is an eigenvector of the operator $\hat a$. Those states can be decomposed in the eigen-basis of $\hat H$. Combining Eqs.\eqref{a_action} and \eqref{defcoh}, we get
\be
|\alpha\rangle = \sum_{n \in \mathbb N} \frac{\alpha^n}{\sqrt{n!}}e^{-\frac{|\alpha|^2}2} |n\rangle.
\ee
This shows that the spectrum of $\hat a$ is the whole complex plane, $Sp(\hat a) = \mathbb C$. On the contrary, if one tries to do the same for $\hat a^\dagger$, one finds only non normalizable states. In fact, the spectrum of $\hat a^\dagger$ is also $\mathbb C$, but it consists of spectral values that are \emph{not} eigenvalues. We refer to~\cite{LevyBruhl,Reed} for proper distinctions between these notions. Here, we assume definition \eqref{defcoh}, and study the main features of these states. First, we see that the probability distribution to find $n$ excitations in a coherent state is a Poissonian distribution of mean value 
\be
\bar n = \langle \alpha | \hat a^\dagger \hat a |\alpha \rangle = |\alpha|^2. 
\ee
To further describe a coherent state, it is very convenient to define the displacement operator
\be
D(\alpha) = \exp (\alpha \hat a^\dagger - \alpha^* \hat a) = e^{-\frac{|\alpha|^2}2} e^{\alpha \hat a^\dagger} e^{-\alpha^* \hat a}.
\ee
The second expression for $D(\alpha)$ is deduced from the Baker-Campbell-Hausdorff formula. $D$ is unitary for any value of $\alpha$. Moreover it allows us to obtain any coherent state by acting on the vacuum state 
\be
|\alpha\rangle = D(\alpha) | 0 \rangle \label{Dcoh}.
\ee
We also recall the very useful identity
\be
D^\dagger(\alpha) \hat a D(\alpha) = \hat a + \alpha \label{disp}.
\ee
Coherent states are often called `classical' because they consist in a gaussian centered on the classical trajectory noted $x_{\rm cl}(t)$. In particular, we have
\be
\langle \alpha | \hat x_t |\alpha \rangle = \sqrt{\frac2{\om}} \Re(\alpha e^{-i\om t}) = x_{\rm cl}(t). \label{xcl_def}
\ee
In fact, the time evolution of $|\alpha\rangle$ is even more remarkable. Indeed, at all time $t$ we have
\be
|\alpha_t \rangle = U(t)|\alpha\rangle = |\alpha e^{-i\om t} \rangle.
\ee
This means that a coherent state stays coherent at all time and centered on the classical trajectory. Note also that the vacuum state $\vac{}$ is a particular coherent state, with $\alpha = 0$.

\subsection{Quantum fluctuations in the presence of a coherent state}
Quantum fluctuations can be characterized by the 2-point function
\be
G_0(t',t) = \langle 0 | \hat x_{t'}\hat x_t |0 \rangle = \frac{e^{-i\om(t'-t)}}{2\om}.
\ee
This function is the one degree of freedom version of the Wightman function $G_+$, defined in Sec.\ref{Greenfunctions_Sec}. Therefore, it can be seen as characterizing the `vacuum fluctuations'. When the system is in a coherent state, one can analyze the corresponding 2-point function
\be
G_\alpha(t',t) = \langle \alpha | \hat x_{t'} \hat x_t |\alpha \rangle.
\ee
Using the preceding section, and in particular Eq. (\ref{Dcoh}), (\ref{disp}), we obtain
\be
D^\dagger(\alpha) \hat x_t D(\alpha) = \hat x_t + x_{\rm cl}(t).
\ee
Moreover, since $\langle 0 | \hat x_t |0 \rangle = 0$, we get
\be
G_\alpha(t',t) = \langle 0 | \hat x_{t'} \hat x_t |0 \rangle + x_{\rm cl}(t')x_{\rm cl}(t).
\ee
Therefore, we see that the two-point function really consist in the vacuum fluctuations plus a classical contribution. This classical part is an exact product of the (real) classical solution \eqref{xcl_def}. The undulation studied in Chapter \ref{mass_Ch} contributes exactly the same way.

\section{The unstable quantum harmonic oscillator}
\label{IHO_App_Sec}
\subsection{Real upside down oscillator}

We review the quantization of upside down oscillators in the Heisenberg representation. To begin with, we start with a single real upside down harmonic oscillator.
Its Hamiltonian is 
\be
H = \frac12 (p^2 - \Gamma^2 q^2),
\label{Hho}
\ee
when written in terms of position $q$ and conjugated momentum $p$, obeying the standard equal time commutator (ETC) $[q, p] = i$. Introducing the `null' 
combinations 
\be
b = \frac1{\sqrt{2 \Gamma}} (p + \Gamma q), \quad c = \frac{1}{\sqrt{2 \Gamma}} (p - \Gamma q),
\label{realbc}
\ee
one gets 
\be
H = \frac{\Gamma}{2}( b c + c b). 
\label{Hho2}
\ee
One verifies that they obey the ETC $[b,c] = i$. The ordering of $b$ and $c$ in $H$ follows from that of \eq{Hho}. The equations of motions are
\be
\dot b = (-i) [b, H] = \Gamma\,  b , \quad \dot c = (-i) [c, H] = - \Gamma \,  c, 
\ee 
thereby establishing that $b$ (resp. $c$) is the growing (resp. decaying) mode $b= b_0 \, e^{\Gamma t}$ (resp. $c= c_0\,  e^{-\Gamma t}$). 

It is now relevant to look for stationary states. In the $b$-representation ($c = -i\partial_b$), 
the stationary Schröedinger equation $H \Psi_E = E \Psi_E$ reads
\be
- i b\partial_b \Psi_E(b) = (\frac{E}{\Gamma}- \frac{i}{2}) \Psi_E(b).
\ee
Solutions exist for all real values of $E$, and the general solution is 
\be
\Psi_E(b) = A_E \, \theta(b) (b)^{i E/\Gamma - 1/2} + B_E \, \theta(-b) (-b)^{i E/\Gamma - 1/2}.
\ee
Since the spectrum is continuum, one should adopt a Dirac delta normalization $\langle \Psi_{E'} \vert \Psi_E \rangle = \delta(E - E')$. This gives 
\be
\vert A_E \vert^2 + \vert B_E \vert^2 = \frac{1}{2 \pi \Gamma}. 
\ee
Imposing that the solution be even in $q$ ($p$, $b$, or $c$) imposes $A_E = B_E$. 
The important lesson one should retain is that there is no square integrable 
stationary states. Therefore, in all physically acceptable states ({\it i.e.}, square integrable) 
the expectation values of $q^2 + p^2$ will exponentially grow $\sim e^{2 \Gamma t}$ at late time. 

\subsection{Complex oscillators}

More relevant for the application of Chapter \ref{laser_Ch}, is the complex upside down harmonic oscillator. One can start from a 2 dimensional inverted oscillator $(q_1,q_2)$ conjugated with $(p_1,p_2)$, whose Hamiltonian is
\be
H = \frac12 \left[ (p_1^2 + p_2^2) - \Gamma^2 (q_1^2 + q_2^2) \right] + \om (q_1 p_2 - q_2 p_1). \label{Hrotinv}
\ee
This represents an unstable and rotating harmonic oscillator~\cite{Fulling}. The associated complex frequency is $\lam = \om + i\Gamma$. In fact, the real part term can always be gauged away, by a change of reference frame. However, for our purpose, we shall keep it explicitly in order to interpret \eqref{Hrotinv} as a standard harmonic oscillator in the interaction picture, see \eq{HCa}. Before that, we rewrite \eqref{Hrotinv} in a language adapted to Chapter \ref{laser_Ch}. We thus build the complex variables 
\be
q= q_1 + i q_2 \quad {\rm and} \quad p= p_1 + i p_2.
\ee
We then introduce the complex $b$ and $c$ variables
\be
b = \frac1{\sqrt{4 \Gamma}} (p + \Gamma q), \quad c = \frac{1}{\sqrt{4 \Gamma}} (p - \Gamma q),
\ee
which are normalized so that they obey the ETC
\be
[b,c^\dagger] = i.
\label{ETC2}
\ee
We then look for the (hermitian) Hamiltonian which gives the following equations 
\be
\dot b = (-i)[b,H] = - i \lam b, \quad \dot c = (-i)[c,H] = - i \lam^* c.
\label{bctev}
\ee
It is given by
\be
H = - i\lam \, c^\dagger b + i \lam^* \, b^\dagger c .
\label{HC}
\ee
It is instructive to reexpress this system in terms of a couple of destruction and creation operators
$d_-,d_-^\dagger$, $d_+,d_+^\dagger$ which are given, at a given time $t_0$, by
\be
b = \frac{1}{\sqrt{2}}(d_+ - i d_-^\dagger), \qquad 
c = \frac{1}{\sqrt{2}}(- i d_+ + d_-^\dagger).
\ee
They obey the standard commutation relations $[d_i,d_j^\dagger] = \delta_{ij}$, 
and \eq{HC} reads
\bsub \label{HCa} 
\bea 
H &=& \om \, (d_+^\dagger d_+ - d_-^\dagger d_-) + \Gamma \, ( d_- d_+ + d_+^\dagger d_-^\dagger), \\
&=& H_0 + H_{sq} .
\eea \esub
In the first term one recovers the standard form of an Hamiltonian is the presence of stationary modes
with opposite frequencies. The second term induces a squeezing of the state of the $d_-, d_+$ oscillators which grows linearly with time. 

To set initial conditions, and to be able to read the result of the instability in terms of quanta, it is appropriate to use this decomposition of $H$ and to work in the `interacting' picture
where the operators $b,c$ only evolve according to $H_0$, and where the squeezing operator acts on the state of the field. Indeed, in this picture the states can be expressed at any time as a superposition of states with a definite occupation numbers $n_-$ and $n_+$.


\end{appendices}

\bibliographystyle{utphys}

\cleardoublepage
\phantomsection
\addcontentsline{toc}{chapter}{Bibliography}
\bibliography{Thesis}

\providecommand{\href}[2]{#2}\begingroup\raggedright\begin{thebibliography}{10%
0}

\bibitem{Born26}
M.~Born, ``Zur quantenmechanik der sto{\ss}vorg{\"a}nge,''
  \href{http://dx.doi.org/10.1007/BF01397477}{{\em Zeitschrift f{\"u}r Physik A
  Hadrons and Nuclei} {\bfseries 37} no.~12, (1926) 863--867}.

\bibitem{Heisenberg27}
W.~Heisenberg, ``{\"U}ber den anschaulichen inhalt der quantentheoretischen
  kinematik und mechanik,'' \href{http://dx.doi.org/10.1007/BF01397280}{{\em
  Zeitschrift f{\"u}r Physik A Hadrons and Nuclei} {\bfseries 43} no.~3, (1927)
  172--198}.

\bibitem{Bohr49}
N.~Bohr, {\em Discussion with Einstein on epistemological problems in atomic
  physics}.
\newblock University of Copenhagen, 1949.

\bibitem{Wheeler}
J.~Wheeler and W.~Zurek, {\em Quantum theory and measurement}.
\newblock Princeton Univ. Press., 1983.

\bibitem{UnruhQG}
W.~Unruh, ``Steps toward a quantum theory of gravity,'' in {\em Quantum theory
  of gravity}.
\newblock Adam Hilger Ltd, 1984.
\newblock Ref.~\cite{Christensen}.

\bibitem{Christensen}
S.~Christensen, {\em Quantum theory of gravity}.
\newblock Adam Hilger Ltd, Bristol, 1984.

\bibitem{tHooft74}
G.~'t~Hooft and M.~Veltman, ``One loop divergencies in the theory of
  gravitation,'' {\em Annales Henri Poincare} {\bfseries A 20} (1974) 69--94.

\bibitem{Goroff85}
M.~H. Goroff and A.~Sagnotti, ``{Quantum Gravity at Two Loops},''
\href{http://dx.doi.org/10.1016/0370-2693(85)91470-4}{{\em Phys. Lett.}
  {\bfseries B160} (1985) 81}.

\bibitem{Goroff85b}
M.~H. Goroff and A.~Sagnotti, ``{The Ultraviolet Behavior of Einstein
  Gravity},''
\href{http://dx.doi.org/10.1016/0550-3213(86)90193-8}{{\em Nucl. Phys.}
  {\bfseries B 266} (1986) 709}.

\bibitem{Donoghue97}
J.~Donoghue, ``{Perturbative dynamics of quantum general relativity},''
\href{http://arxiv.org/abs/gr-qc/9712070}{{\ttfamily arXiv:gr-qc/9712070
  [gr-qc]}}.

\bibitem{Burgess}
C.~Burgess, ``{Quantum gravity in everyday life: General relativity as an
  effective field theory},'' {\em Living Rev. Rel.} {\bfseries 7} (2004) 5,
\href{http://arxiv.org/abs/gr-qc/0311082}{{\ttfamily arXiv:gr-qc/0311082
  [gr-qc]}}.

\bibitem{DeWitt67a}
B.~S. DeWitt, ``{Quantum Theory of Gravity. 1. The Canonical Theory},''
\href{http://dx.doi.org/10.1103/PhysRev.160.1113}{{\em Phys. Rev.} {\bfseries
  160} (1967) 1113--1148}.

\bibitem{DeWitt67b}
B.~S. DeWitt, ``{Quantum Theory of Gravity. 2. The Manifestly Covariant
  Theory},''
\href{http://dx.doi.org/10.1103/PhysRev.162.1195}{{\em Phys. Rev.} {\bfseries
  162} (1967) 1195--1239}.

\bibitem{DeWitt67c}
B.~S. DeWitt, ``{Quantum Theory of Gravity. 3. Applications of the Covariant
  Theory},''
\href{http://dx.doi.org/10.1103/PhysRev.162.1239}{{\em Phys. Rev.} {\bfseries
  162} (1967) 1239--1256}.

\bibitem{Zwiebach}
B.~Zwiebach, {\em A first course in string theory}.
\newblock Cambridge Univ Pr, 2004.

\bibitem{Thiemann}
T.~Thiemann, {\em Modern canonical quantum general relativity}.
\newblock Cambridge University Press, 2007.

\bibitem{Perez12}
A.~Perez, ``{The Spin Foam Approach to Quantum Gravity},''
\href{http://arxiv.org/abs/1205.2019}{{\ttfamily arXiv:1205.2019 [gr-qc]}}.

\bibitem{Geiller11}
S.~Alexandrov, M.~Geiller, and K.~Noui, ``{Spin Foams and Canonical
  Quantization},''
\href{http://arxiv.org/abs/1112.1961}{{\ttfamily arXiv:1112.1961 [gr-qc]}}.

\bibitem{Hawking74}
S.~Hawking, ``{Black hole explosions},''
\href{http://dx.doi.org/10.1038/248030a0}{{\em Nature} {\bfseries 248} (1974)
  30--31}.

\bibitem{Duff81}
M.~Duff, {\em Quantum Gravity 2: A Second Oxford Symposium}, pp.~81--105.
\newblock Oxford University Press, 1981.

\bibitem{FullingQG}
S.~Fulling, ``What have we learned from quantum field theory in curved
  space-time ?,'' in {\em Quantum theory of gravity}.
\newblock Adam Hilger Ltd, 1984.
\newblock Ref.~\cite{Christensen}.

\bibitem{Weinberg1}
S.~Weinberg, {\em The Quantum Theory of Fields, Vol. 1: Foundations}, vol.~1.
\newblock Cambridge University Press, 1995.

\bibitem{Parentani98}
R.~Parentani, ``{The Background field approximation in (quantum) cosmology},''
  \href{http://dx.doi.org/10.1088/0264-9381/17/6/314}{{\em Class. Quant. Grav.}
  {\bfseries 17} (2000) 1527--1547},
\href{http://arxiv.org/abs/gr-qc/9803045}{{\ttfamily arXiv:gr-qc/9803045
  [gr-qc]}}.

\bibitem{Massar98}
S.~Massar and R.~Parentani, ``{Unitary and nonunitary evolution in quantum
  cosmology},'' \href{http://dx.doi.org/10.1103/PhysRevD.59.123519}{{\em Phys.
  Rev.} {\bfseries D 59} (1999) 123519},
\href{http://arxiv.org/abs/gr-qc/9812045}{{\ttfamily arXiv:gr-qc/9812045
  [gr-qc]}}.

\bibitem{Bekenstein73}
J.~D. Bekenstein, ``{Black holes and entropy},''
\href{http://dx.doi.org/10.1103/PhysRevD.7.2333}{{\em Phys. Rev.} {\bfseries D
  7} (1973) 2333--2346}.

\bibitem{Jacobson91}
T.~Jacobson, ``{Black hole evaporation and ultrashort distances},''
\href{http://dx.doi.org/10.1103/PhysRevD.44.1731}{{\em Phys. Rev.} {\bfseries D
  44} (1991) 1731--1739}.

\bibitem{Unruh81}
W.~Unruh, ``{Experimental black hole evaporation},''
\href{http://dx.doi.org/10.1103/PhysRevLett.46.1351}{{\em Phys. Rev. Lett.}
  {\bfseries 46} (1981) 1351--1353}.

\bibitem{Coutant10}
A.~Coutant and R.~Parentani, ``{Black hole lasers, a mode analysis},''
  \href{http://dx.doi.org/10.1103/PhysRevD.81.084042}{{\em Phys. Rev.}
  {\bfseries D 81} (2010) 084042},
\href{http://arxiv.org/abs/0912.2755}{{\ttfamily arXiv:0912.2755 [hep-th]}}.

\bibitem{Coutant11}
A.~Coutant, R.~Parentani, and S.~Finazzi, ``{Black hole radiation with short
  distance dispersion, an analytical S-matrix approach},''
  \href{http://dx.doi.org/10.1103/PhysRevD.85.024021}{{\em Phys. Rev.}
  {\bfseries D 85} (2012) 024021},
\href{http://arxiv.org/abs/1108.1821}{{\ttfamily arXiv:1108.1821 [hep-th]}}.

\bibitem{aQuattro}
A.~Coutant, S.~Finazzi, S.~Liberati, and R.~Parentani, ``{On the impossibility
  of superluminal travel in Lorentz violating theories},''
  \href{http://dx.doi.org/10.1103/PhysRevD.85.064020}{{\em Phys. Rev.}
  {\bfseries D 85} (2012) 064020},
\href{http://arxiv.org/abs/1111.4356}{{\ttfamily arXiv:1111.4356 [gr-qc]}}.

\bibitem{Coutant12}
A.~Coutant, A.~Fabbri, R.~Parentani, R.~Balbinot, and P.~Anderson, ``{Hawking
  radiation of massive modes and undulations},''
  \href{http://dx.doi.org/10.1103/PhysRevD.86.064022}{{\em Phys.Rev.}
  {\bfseries D86} (2012) 064022},
\href{http://arxiv.org/abs/1206.2658}{{\ttfamily arXiv:1206.2658 [gr-qc]}}.

\bibitem{Einstein05}
A.~Einstein, ``Zur elektrodynamik bewegter k{\"o}rper,'' {\em Annalen der
  physik} {\bfseries 322} no.~10, (1905) 891--921.

\bibitem{Lafontaine}
J.~Lafontaine, {\em Introduction aux vari{\'e}t{\'e}s diff{\'e}rentielles}.
\newblock EDP Sciences, 2010.

\bibitem{Straumann}
N.~Straumann, {\em General relativity with applications to astrophysics}.
\newblock Springer, Berlin, 2004.

\bibitem{Wald}
R.~Wald, {\em General relativity}.
\newblock University of Chicago press, 1984.

\bibitem{Arnold}
V.~Arnold, {\em Mathematical methods of classical mechanics}, vol.~60.
\newblock Springer, 1989.

\bibitem{Kundt66}
W.~Kundt and M.~Trumper, ``{Orthogonal decomposition of axi-symmetric
  stationary spacetimes},''
{\em Z. Phys.} {\bfseries 192} (1966) 419--422.

\bibitem{Zegersthesis}
R.~Zegers, {\em Relativit{\'e} g{\'e}n{\'e}rale et dimensions
  suppl{\'e}mentaires}.
\newblock PhD thesis, Universit{\'e} Paris 7, 2006.

\bibitem{Frauendiener00}
J.~Frauendiener, ``{Conformal infinity},''
{\em Living Rev. Rel.} {\bfseries 3} (2000) 4.

\bibitem{Wald99}
R.~M. Wald, ``{Gravitation, thermodynamics, and quantum theory},''
  \href{http://dx.doi.org/10.1088/0264-9381/16/12A/309}{{\em Class. Quant.
  Grav.} {\bfseries 16} (1999) A177--A190},
\href{http://arxiv.org/abs/gr-qc/9901033}{{\ttfamily arXiv:gr-qc/9901033
  [gr-qc]}}.

\bibitem{Visser07}
M.~Visser, ``{The Kerr spacetime: A Brief introduction},''
\href{http://arxiv.org/abs/0706.0622}{{\ttfamily arXiv:0706.0622 [gr-qc]}}.

\bibitem{HawkingEllis}
S.~Hawking and G.~Ellis, {\em The large scale structure of space-time}, vol.~1.
\newblock Cambridge Univ Pr, 1975.

\bibitem{Poisson}
E.~Poisson, {\em A relativist's toolkit: the mathematics of black-hole
  mechanics}.
\newblock Cambridge University Press, 2004.

\bibitem{YorkQG}
J.~York, ``What happens to the horizon when a black hole radiates?,'' in {\em
  Quantum theory of gravity}.
\newblock Adam Hilger Ltd, 1984.
\newblock Ref.~\cite{Christensen}.

\bibitem{Emparan02}
R.~Emparan and H.~S. Reall, ``{A Rotating black ring solution in
  five-dimensions},''
  \href{http://dx.doi.org/10.1103/PhysRevLett.88.101101}{{\em Phys. Rev. Lett.}
  {\bfseries 88} (2002) 101101},
\href{http://arxiv.org/abs/hep-th/0110260}{{\ttfamily arXiv:hep-th/0110260
  [hep-th]}}.

\bibitem{Bardoux12}
Y.~Bardoux, M.~M. Caldarelli, and C.~Charmousis, ``{Shaping black holes with
  free fields},'' \href{http://dx.doi.org/10.1007/JHEP05(2012)054}{{\em JHEP}
  {\bfseries 1205} (2012) 054},
\href{http://arxiv.org/abs/1202.4458}{{\ttfamily arXiv:1202.4458 [hep-th]}}.

\bibitem{Jacobson07}
T.~Jacobson, ``{When is g(tt) g(rr) = -1?},''
  \href{http://dx.doi.org/10.1088/0264-9381/24/22/N02}{{\em Class. Quant.
  Grav.} {\bfseries 24} (2007) 5717--5719},
\href{http://arxiv.org/abs/0707.3222}{{\ttfamily arXiv:0707.3222 [gr-qc]}}.

\bibitem{Poisson00}
K.~Martel and E.~Poisson, ``{Regular coordinate systems for Schwarzschild and
  other spherical space-times},''
  \href{http://dx.doi.org/10.1119/1.1336836}{{\em Am. J. Phys.} {\bfseries 69}
  (2001) 476--480},
\href{http://arxiv.org/abs/gr-qc/0001069}{{\ttfamily arXiv:gr-qc/0001069
  [gr-qc]}}.

\bibitem{Gottfried}
K.~Gottfried and T.~Yan, {\em Quantum mechanics: fundamentals}.
\newblock Springer Verlag, 2nd~ed., 2003.

\bibitem{Page76}
D.~N. Page, ``{Particle Emission Rates from a Black Hole: Massless Particles
  from an Uncharged, Nonrotating Hole},''
\href{http://dx.doi.org/10.1103/PhysRevD.13.198}{{\em Phys. Rev.} {\bfseries D
  13} (1976) 198--206}.

\bibitem{Page76b}
D.~N. Page, ``{Particle Emission Rates from a Black Hole. 2. Massless Particles
  from a Rotating Hole},''
\href{http://dx.doi.org/10.1103/PhysRevD.14.3260}{{\em Phys. Rev.} {\bfseries D
  14} (1976) 3260--3273}.

\bibitem{Page77}
D.~N. Page, ``{Particle Emission Rates from a Black Hole. 3. Charged Leptons
  from a Nonrotating Hole},''
\href{http://dx.doi.org/10.1103/PhysRevD.16.2402}{{\em Phys. Rev.} {\bfseries D
  16} (1977) 2402--2411}.

\bibitem{Bardeen73}
J.~M. Bardeen, B.~Carter, and S.~Hawking, ``{The Four laws of black hole
  mechanics},''
\href{http://dx.doi.org/10.1007/BF01645742}{{\em Commun. Math. Phys.}
  {\bfseries 31} (1973) 161--170}.

\bibitem{Hawking71}
S.~Hawking, ``{Gravitational radiation from colliding black holes},''
\href{http://dx.doi.org/10.1103/PhysRevLett.26.1344}{{\em Phys. Rev. Lett.}
  {\bfseries 26} (1971) 1344--1346}.

\bibitem{Bekenstein74}
J.~D. Bekenstein, ``{Generalized second law of thermodynamics in black hole
  physics},''
\href{http://dx.doi.org/10.1103/PhysRevD.9.3292}{{\em Phys. Rev.} {\bfseries D
  9} (1974) 3292--3300}.

\bibitem{Unruh82}
W.~Unruh and R.~M. Wald, ``{Acceleration Radiation and Generalized Second Law
  of Thermodynamics},''
\href{http://dx.doi.org/10.1103/PhysRevD.25.942}{{\em Phys. Rev.} {\bfseries D
  25} (1982) 942--958}.

\bibitem{Unruh83}
W.~Unruh and R.~Wald, ``How to mine energy from a black hole,'' {\em General
  Relativity and Gravitation} {\bfseries 15} no.~3, (1983) 195--199.

\bibitem{Callan96}
C.~G. Callan and J.~M. Maldacena, ``{D-brane approach to black hole quantum
  mechanics},'' \href{http://dx.doi.org/10.1016/0550-3213(96)00225-8}{{\em
  Nucl. Phys.} {\bfseries B 472} (1996) 591--610},
\href{http://arxiv.org/abs/hep-th/9602043}{{\ttfamily arXiv:hep-th/9602043
  [hep-th]}}.

\bibitem{Engle09}
J.~Engle, A.~Perez, and K.~Noui, ``{Black hole entropy and SU(2) Chern-Simons
  theory},'' \href{http://dx.doi.org/10.1103/PhysRevLett.105.031302}{{\em Phys.
  Rev. Lett.} {\bfseries 105} (2010) 031302},
\href{http://arxiv.org/abs/0905.3168}{{\ttfamily arXiv:0905.3168 [gr-qc]}}.

\bibitem{Carlip96}
S.~Carlip, ``{The Statistical mechanics of the three-dimensional Euclidean
  black hole},'' \href{http://dx.doi.org/10.1103/PhysRevD.55.878}{{\em Phys.
  Rev.} {\bfseries D 55} (1997) 878--882},
\href{http://arxiv.org/abs/gr-qc/9606043}{{\ttfamily arXiv:gr-qc/9606043
  [gr-qc]}}.

\bibitem{Carlip93}
S.~Carlip and C.~Teitelboim, ``{The Off-shell black hole},''
  \href{http://dx.doi.org/10.1088/0264-9381/12/7/011}{{\em Class. Quant. Grav.}
  {\bfseries 12} (1995) 1699--1704},
\href{http://arxiv.org/abs/gr-qc/9312002}{{\ttfamily arXiv:gr-qc/9312002
  [gr-qc]}}.

\bibitem{tHooft96}
G.~'t~Hooft, ``The scattering matrix approach for the quantum black hole, an
  overview,'' \href{http://dx.doi.org/10.1142/S0217751X96002145}{{\em Int. J.
  Mod. Phys.} {\bfseries A 11} (1996) 4623--4688},
  \href{http://arxiv.org/abs/gr-qc/9607022}{{\ttfamily arXiv:gr-qc/9607022
  [gr-qc]}}.

\bibitem{Jacobson95}
T.~Jacobson, ``{Thermodynamics of space-time: The Einstein equation of
  state},'' \href{http://dx.doi.org/10.1103/PhysRevLett.75.1260}{{\em Phys.
  Rev. Lett.} {\bfseries 75} (1995) 1260--1263},
\href{http://arxiv.org/abs/gr-qc/9504004}{{\ttfamily arXiv:gr-qc/9504004
  [gr-qc]}}.

\bibitem{Delduc}
F.~Delduc, ``Introduction {\`a} la th{\'e}orie des champs,'' 2002.
\newblock \url{http://perso.ens-lyon.fr/francois.delduc/}.

\bibitem{BirrellDavies}
N.~Birrell and P.~Davies, {\em Quantum fields in curved space}.
\newblock Cambridge University Press, 1984.

\bibitem{Fulling}
S.~Fulling, {\em Aspects of Quantum Field Theory in Curved Space-Time},
  vol.~17.
\newblock Cambridge University Press, 1989.

\bibitem{Leonhardt}
U.~Leonhardt, {\em Essential quantum optics: from quantum measurements to Black
  Holes}, vol.~67.
\newblock Cambridge University Press, 2010.

\bibitem{Unruh76}
W.~G. Unruh, ``Notes on black-hole evaporation,''
  \href{http://dx.doi.org/10.1103/PhysRevD.14.870}{{\em Phys. Rev.} {\bfseries
  D 14} no.~4, (1976) 870--892}.

\bibitem{Primer}
R.~Brout, S.~Massar, R.~Parentani, and P.~Spindel, ``{A Primer for black hole
  quantum physics},''
  \href{http://dx.doi.org/10.1016/0370-1573(95)00008-5}{{\em Phys. Rept.}
  {\bfseries 260} (1995) 329--454},
\href{http://arxiv.org/abs/0710.4345}{{\ttfamily arXiv:0710.4345 [gr-qc]}}.

\bibitem{Unruh89}
W.~Unruh and W.~Zurek, ``{Reduction of a Wave Packet in Quantum Brownian
  Motion},''
\href{http://dx.doi.org/10.1103/PhysRevD.40.1071}{{\em Phys. Rev.} {\bfseries D
  40} (1989) 1071}.

\bibitem{Massar06}
S.~Massar and P.~Spindel, ``{Einstein-Podolsky-Rosen correlations between two
  uniformly accelerated oscillators},''
  \href{http://dx.doi.org/10.1103/PhysRevD.74.085031}{{\em Phys. Rev.}
  {\bfseries D 74} (2006) 085031},
\href{http://arxiv.org/abs/hep-th/0606174}{{\ttfamily arXiv:hep-th/0606174
  [hep-th]}}.

\bibitem{Parentani95}
R.~Parentani, ``{The Recoils of the accelerated detector and the decoherence of
  its fluxes},'' \href{http://dx.doi.org/10.1016/0550-3213(95)00452-X}{{\em
  Nucl. Phys.} {\bfseries B 454} (1995) 227--249},
\href{http://arxiv.org/abs/gr-qc/9502030}{{\ttfamily arXiv:gr-qc/9502030
  [gr-qc]}}.

\bibitem{Damour76}
T.~Damour and R.~Ruffini, ``{Black Hole Evaporation in the
  Klein-Sauter-Heisenberg-Euler Formalism},''
\href{http://dx.doi.org/10.1103/PhysRevD.14.332}{{\em Phys. Rev.} {\bfseries D
  14} (1976) 332--334}.

\bibitem{dInverno}
R.~d'Inverno, {\em Introducing Einstein's Relativity}.
\newblock Clarendon Press, Oxford, 1992.

\bibitem{Einstein17}
A.~Einstein, ``Zur quantentheorie der strahlung,'' {\em Physikalische
  Zeitschrift} {\bfseries 18} (1917) 121--128.

\bibitem{TerHaar}
D.~Ter~Haar, ``The old quantum theory,'' {\em American Journal of Physics}
  {\bfseries 35} (1967) 1098--1099.

\bibitem{WaldQ}
R.~Wald, {\em Quantum field theory in curved spacetime and black hole
  thermodynamics}.
\newblock University of Chicago Press, 1994.

\bibitem{Hawking75}
S.~Hawking, ``{Particle Creation by Black Holes},''
\href{http://dx.doi.org/10.1007/BF02345020, 10.1007/BF02345020}{{\em Commun.
  Math. Phys.} {\bfseries 43} (1975) 199--220}.

\bibitem{Massar96}
S.~Massar and R.~Parentani, ``{From vacuum fluctuations to radiation. 2. Black
  holes},''
\href{http://dx.doi.org/10.1103/PhysRevD.54.7444}{{\em Phys. Rev.} {\bfseries D
  54} (1996) 7444--7458}.

\bibitem{Macherthesis}
J.~Macher, {\em Brisure de l'invariance de Lorentz {\`a} haute {\'e}nergie :
  cons{\'e}quences pour l'inflation et le rayonnement des trous noirs}.
\newblock PhD thesis, Universit{\'e} Paris-Sud 11, 2009.

\bibitem{Hartle76}
J.~Hartle and S.~Hawking, ``{Path Integral Derivation of Black Hole
  Radiance},''
\href{http://dx.doi.org/10.1103/PhysRevD.13.2188}{{\em Phys. Rev.} {\bfseries D
  13} (1976) 2188--2203}.

\bibitem{Brout95}
R.~Brout, S.~Massar, R.~Parentani, and P.~Spindel, ``{Hawking radiation without
  transPlanckian frequencies},''
  \href{http://dx.doi.org/10.1103/PhysRevD.52.4559}{{\em Phys. Rev.} {\bfseries
  D 52} (1995) 4559--4568},
\href{http://arxiv.org/abs/hep-th/9506121}{{\ttfamily arXiv:hep-th/9506121
  [hep-th]}}.

\bibitem{tHooft85}
G.~'t~Hooft, ``On the quantum structure of a black hole,'' {\em Nucl. Phys.}
  {\bfseries B 256} (1985) 727--745.

\bibitem{Parentani02}
R.~Parentani, ``{Beyond the semiclassical description of black hole
  evaporation},'' \href{http://dx.doi.org/10.1023/A:1021133126804}{{\em Int. J.
  Theor. Phys.} {\bfseries 41} (2002) 2175--2200},
\href{http://arxiv.org/abs/0704.2563}{{\ttfamily arXiv:0704.2563 [hep-th]}}.

\bibitem{Barcelo05}
C.~Barcelo, S.~Liberati, and M.~Visser, ``{Analogue gravity},'' {\em Living
  Rev. Rel.} {\bfseries 8} (2005) 12,
\href{http://arxiv.org/abs/gr-qc/0505065}{{\ttfamily arXiv:gr-qc/0505065
  [gr-qc]}}.

\bibitem{Balbinot06}
R.~Balbinot, A.~Fabbri, S.~Fagnocchi, and R.~Parentani, ``{Hawking radiation
  from acoustic black holes, short distance and back-reaction effects},'' {\em
  Riv. Nuovo Cim.} {\bfseries 28} (2005) 1--55,
\href{http://arxiv.org/abs/gr-qc/0601079}{{\ttfamily arXiv:gr-qc/0601079
  [gr-qc]}}.

\bibitem{Visser12}
M.~Visser, ``{Survey of analogue spacetimes},''
\href{http://arxiv.org/abs/1206.2397}{{\ttfamily arXiv:1206.2397 [gr-qc]}}.

\bibitem{Schutzhold02}
R.~Schutzhold and W.~G. Unruh, ``{Gravity wave analogs of black holes},''
  \href{http://dx.doi.org/10.1103/PhysRevD.66.044019}{{\em Phys. Rev.}
  {\bfseries D 66} (2002) 044019},
\href{http://arxiv.org/abs/gr-qc/0205099}{{\ttfamily arXiv:gr-qc/0205099
  [gr-qc]}}.

\bibitem{Jacobson98}
T.~Jacobson and G.~Volovik, ``{Event horizons and ergoregions in He-3},''
\href{http://dx.doi.org/10.1103/PhysRevD.58.064021}{{\em Phys. Rev.} {\bfseries
  D 58} (1998) 064021}.

\bibitem{Macher09b}
J.~Macher and R.~Parentani, ``{Black hole radiation in Bose-Einstein
  condensates},'' \href{http://dx.doi.org/10.1103/PhysRevA.80.043601}{{\em
  Phys. Rev.} {\bfseries A 80} (2009) 043601},
\href{http://arxiv.org/abs/0905.3634}{{\ttfamily arXiv:0905.3634
  [cond-mat.quant-gas]}}.

\bibitem{Parentani02b}
R.~Parentani, ``{What did we learn from studying acoustic black holes?},''
  \href{http://dx.doi.org/10.1142/S0217751X02011679}{{\em Int. J. Mod. Phys.}
  {\bfseries A 17} (2002) 2721--2726},
\href{http://arxiv.org/abs/gr-qc/0204079}{{\ttfamily arXiv:gr-qc/0204079
  [gr-qc]}}.

\bibitem{Lahav09}
O.~Lahav, A.~Itah, A.~Blumkin, C.~Gordon, and J.~Steinhauer, ``{Realization of
  a sonic black hole analogue in a Bose-Einstein condensate},''
  \href{http://dx.doi.org/10.1103/PhysRevLett.105.240401}{{\em Phys. Rev.
  Lett.} {\bfseries 105} (2010) 240401},
\href{http://arxiv.org/abs/0906.1337}{{\ttfamily arXiv:0906.1337
  [cond-mat.quant-gas]}}.

\bibitem{Weinfurtner10}
S.~Weinfurtner, E.~W. Tedford, M.~C. Penrice, W.~G. Unruh, and G.~A. Lawrence,
  ``{Measurement of stimulated Hawking emission in an analogue system},''
  \href{http://dx.doi.org/10.1103/PhysRevLett.106.021302}{{\em Phys. Rev.
  Lett.} {\bfseries 106} (2011) 021302},
\href{http://arxiv.org/abs/1008.1911}{{\ttfamily arXiv:1008.1911 [gr-qc]}}.

\bibitem{Belgiorno10}
F.~Belgiorno, S.~Cacciatori, M.~Clerici, V.~Gorini, G.~Ortenzi, L.~Rizzi,
  E.~Rubino, V.~Sala, and D.~Faccio, ``Hawking radiation from ultrashort laser
  pulse filaments,''
  \href{http://dx.doi.org/10.1103/PhysRevLett.105.203901}{{\em Phys. Rev.
  Lett.} {\bfseries 105} no.~20, (2010) 203901},
  \href{http://arxiv.org/abs/1009.4634}{{\ttfamily arXiv:1009.4634 [gr-qc]}}.

\bibitem{Schutzhold10b}
R.~Schutzhold and W.~G. Unruh, ``{Comment on: Hawking Radiation from Ultrashort
  Laser Pulse Filaments},''
  \href{http://dx.doi.org/10.1103/PhysRevLett.107.149401}{{\em Phys. Rev.
  Lett.} {\bfseries 107} (2011) 149401},
\href{http://arxiv.org/abs/1012.2686}{{\ttfamily arXiv:1012.2686 [quant-ph]}}.

\bibitem{Volovik05}
G.~Volovik, ``{The Hydraulic jump as a white hole},''
  \href{http://dx.doi.org/10.1134/1.2166908, 10.1134/1.2166908}{{\em JETP
  Lett.} {\bfseries 82} (2005) 624--627},
\href{http://arxiv.org/abs/physics/0508215}{{\ttfamily arXiv:physics/0508215
  [physics]}}.

\bibitem{Jacobson93}
T.~Jacobson, ``{Black hole radiation in the presence of a short distance
  cutoff},'' \href{http://dx.doi.org/10.1103/PhysRevD.48.728}{{\em Phys. Rev.}
  {\bfseries D 48} (1993) 728--741},
\href{http://arxiv.org/abs/hep-th/9303103}{{\ttfamily arXiv:hep-th/9303103
  [hep-th]}}.

\bibitem{Jacobson01}
T.~Jacobson and D.~Mattingly, ``{Gravity with a dynamical preferred frame},''
  \href{http://dx.doi.org/10.1103/PhysRevD.64.024028}{{\em Phys. Rev.}
  {\bfseries D 64} (2001) 024028},
\href{http://arxiv.org/abs/gr-qc/0007031}{{\ttfamily arXiv:gr-qc/0007031
  [gr-qc]}}.

\bibitem{Horava09}
P.~Horava, ``{Quantum Gravity at a Lifshitz Point},''
  \href{http://dx.doi.org/10.1103/PhysRevD.79.084008}{{\em Phys. Rev.}
  {\bfseries D 79} (2009) 084008},
\href{http://arxiv.org/abs/0901.3775}{{\ttfamily arXiv:0901.3775 [hep-th]}}.

\bibitem{Libanov05}
M.~Libanov and V.~Rubakov, ``{More about spontaneous Lorentz-violation and
  infrared modification of gravity},''
  \href{http://dx.doi.org/10.1088/1126-6708/2005/08/001}{{\em JHEP} {\bfseries
  0508} (2005) 001},
\href{http://arxiv.org/abs/hep-th/0505231}{{\ttfamily arXiv:hep-th/0505231
  [hep-th]}}.

\bibitem{Barcelo01}
C.~Barcelo, M.~Visser, and S.~Liberati, ``{Einstein gravity as an emergent
  phenomenon?},'' \href{http://dx.doi.org/10.1142/S0218271801001591}{{\em Int.
  J. Mod. Phys.} {\bfseries D 10} (2001) 799--806},
\href{http://arxiv.org/abs/gr-qc/0106002}{{\ttfamily arXiv:gr-qc/0106002
  [gr-qc]}}.

\bibitem{Carlip12}
S.~Carlip, ``{Challenges for Emergent Gravity},''
\href{http://arxiv.org/abs/1207.2504}{{\ttfamily arXiv:1207.2504 [gr-qc]}}.

\bibitem{Oriti11}
D.~Oriti, ``{On the depth of quantum space},''
\href{http://arxiv.org/abs/1107.4534}{{\ttfamily arXiv:1107.4534
  [physics.pop-ph]}}.

\bibitem{Carrozza12}
S.~Carrozza, D.~Oriti, and V.~Rivasseau, ``Renormalization of tensorial group
  field theories: Abelian u(1) models in four dimensions,''
  \href{http://arxiv.org/abs/1207.6734}{{\ttfamily arXiv:1207.6734 [hep-th]}}.

\bibitem{Jacobson96}
T.~Jacobson, ``{On the origin of the outgoing black hole modes},'' {\em Phys.
  Rev.} {\bfseries D 53} (1996) 7082--7088,
\href{http://arxiv.org/abs/hep-th/9601064}{{\ttfamily arXiv:hep-th/9601064
  [hep-th]}}.

\bibitem{Kostelecky03}
V.~A. Kostelecky, ``{Gravity, Lorentz violation, and the standard model},''
  \href{http://dx.doi.org/10.1103/PhysRevD.69.105009}{{\em Phys. Rev.}
  {\bfseries D 69} (2004) 105009},
\href{http://arxiv.org/abs/hep-th/0312310}{{\ttfamily arXiv:hep-th/0312310
  [hep-th]}}.

\bibitem{Jacobson08}
T.~Jacobson, ``{Einstein-aether gravity: A Status report},'' {\em PoS}
  {\bfseries QG-PH} (2007) 020,
\href{http://arxiv.org/abs/0801.1547}{{\ttfamily arXiv:0801.1547 [gr-qc]}}.

\bibitem{Unruh95}
W.~Unruh, ``{Sonic analog of black holes and the effects of high frequencies on
  black hole evaporation},''
\href{http://dx.doi.org/10.1103/PhysRevD.51.2827}{{\em Phys. Rev.} {\bfseries D
  51} (1995) 2827--2838}.

\bibitem{Schutzhold08}
R.~Schutzhold and W.~G. Unruh, ``{On the origin of the particles in black hole
  evaporation},'' \href{http://dx.doi.org/10.1103/PhysRevD.78.041504}{{\em
  Phys. Rev.} {\bfseries D 78} (2008) 041504},
\href{http://arxiv.org/abs/0804.1686}{{\ttfamily arXiv:0804.1686 [gr-qc]}}.

\bibitem{Unruh12}
W.~Unruh, ``{Irrotational, two-dimensional Surface waves in fluids},''
\href{http://arxiv.org/abs/1205.6751}{{\ttfamily arXiv:1205.6751 [gr-qc]}}.

\bibitem{Parentani07}
R.~Parentani, ``{Constructing QFT's wherein Lorentz invariance is broken by
  dissipative effects in the UV},'' {\em PoS} {\bfseries QG-PH} (2007) 031,
\href{http://arxiv.org/abs/0709.3943}{{\ttfamily arXiv:0709.3943 [hep-th]}}.

\bibitem{Corley96}
S.~Corley and T.~Jacobson, ``{Hawking spectrum and high frequency
  dispersion},'' \href{http://dx.doi.org/10.1103/PhysRevD.54.1568}{{\em Phys.
  Rev.} {\bfseries D 54} (1996) 1568--1586},
\href{http://arxiv.org/abs/hep-th/9601073}{{\ttfamily arXiv:hep-th/9601073
  [hep-th]}}.

\bibitem{Macher09}
J.~Macher and R.~Parentani, ``{Black/White hole radiation from dispersive
  theories},'' \href{http://dx.doi.org/10.1103/PhysRevD.79.124008}{{\em Phys.
  Rev.} {\bfseries D 79} (2009) 124008},
\href{http://arxiv.org/abs/0903.2224}{{\ttfamily arXiv:0903.2224 [hep-th]}}.

\bibitem{Corley97}
S.~Corley, ``{Computing the spectrum of black hole radiation in the presence of
  high frequency dispersion: An Analytical approach},''
  \href{http://dx.doi.org/10.1103/PhysRevD.57.6280}{{\em Phys. Rev.} {\bfseries
  D 57} (1998) 6280--6291},
\href{http://arxiv.org/abs/hep-th/9710075}{{\ttfamily arXiv:hep-th/9710075
  [hep-th]}}.

\bibitem{Himemoto00}
Y.~Himemoto and T.~Tanaka, ``{A Generalization of the model of Hawking
  radiation with modified high frequency dispersion relation},''
  \href{http://dx.doi.org/10.1103/PhysRevD.61.064004}{{\em Phys. Rev.}
  {\bfseries D 61} (2000) 064004},
\href{http://arxiv.org/abs/gr-qc/9904076}{{\ttfamily arXiv:gr-qc/9904076
  [gr-qc]}}.

\bibitem{Unruh04}
W.~G. Unruh and R.~Schutzhold, ``{On the universality of the Hawking effect},''
  \href{http://dx.doi.org/10.1103/PhysRevD.71.024028}{{\em Phys. Rev.}
  {\bfseries D 71} (2005) 024028},
\href{http://arxiv.org/abs/gr-qc/0408009}{{\ttfamily arXiv:gr-qc/0408009
  [gr-qc]}}.

\bibitem{Rousseaux10}
G.~Rousseaux, P.~Maissa, C.~Mathis, P.~Coullet, T.~G. Philbin, and
  U.~Leonhardt, ``{Horizon effects with surface waves on moving water},''
  \href{http://dx.doi.org/10.1088/1367-2630/12/9/095018}{{\em New J. Phys.}
  {\bfseries 12} (2010) 095018},
\href{http://arxiv.org/abs/1004.5546}{{\ttfamily arXiv:1004.5546 [gr-qc]}}.

\bibitem{Parentani10}
R.~Parentani, ``{From vacuum fluctuations across an event horizon to long
  distance correlations},''
  \href{http://dx.doi.org/10.1103/PhysRevD.82.025008}{{\em Phys. Rev.}
  {\bfseries D 82} (2010) 025008},
\href{http://arxiv.org/abs/1003.3625}{{\ttfamily arXiv:1003.3625 [gr-qc]}}.

\bibitem{Busch12}
X.~Busch and R.~Parentani, ``Dispersive fields in de sitter space and event
  horizon thermodynamics,'' \href{http://arxiv.org/abs/1207.5961v1}{{\ttfamily
  arXiv:1207.5961v1 [hep-th]}}.

\bibitem{Finazzi10b}
S.~Finazzi and R.~Parentani, ``{Spectral properties of acoustic black hole
  radiation: broadening the horizon},''
  \href{http://dx.doi.org/10.1103/PhysRevD.83.084010}{{\em Phys. Rev.}
  {\bfseries D 83} (2011) 084010},
\href{http://arxiv.org/abs/1012.1556}{{\ttfamily arXiv:1012.1556 [gr-qc]}}.

\bibitem{Finazzi11}
S.~Finazzi and R.~Parentani, ``{On the robustness of acoustic black hole
  spectra},'' {\em J. Phys. Conf. Ser.} {\bfseries 314} (2011) 012030,
\href{http://arxiv.org/abs/1102.1452}{{\ttfamily arXiv:1102.1452 [gr-qc]}}.

\bibitem{Olver}
F.~Olver, {\em Asymptotics and special functions}, vol.~15.
\newblock Academic Press New York, 1974.

\bibitem{AbramoSteg}
M.~Abramowitz and I.~Stegun, {\em Handbook of mathematical functions with
  formulas, graphs, and mathematical tables}, vol.~55.
\newblock Dover publications, 1964.

\bibitem{Macher08}
J.~Macher and R.~Parentani, ``{Signatures of trans-Planckian dispersion in
  inflationary spectra},''
  \href{http://dx.doi.org/10.1103/PhysRevD.78.043522}{{\em Phys. Rev.}
  {\bfseries D 78} (2008) 043522},
\href{http://arxiv.org/abs/0804.1920}{{\ttfamily arXiv:0804.1920 [hep-th]}}.

\bibitem{Carusotto08}
I.~Carusotto, S.~Fagnocchi, A.~Recati, R.~Balbinot, and A.~Fabbri, ``{Numerical
  observation of Hawking radiation from acoustic black holes in atomic BECs},''
  {\em New J. Phys.} {\bfseries 10} (2008) 103001,
\href{http://arxiv.org/abs/0803.0507}{{\ttfamily arXiv:0803.0507
  [cond-mat.other]}}.

\bibitem{Balbinot08}
R.~Balbinot, A.~Fabbri, S.~Fagnocchi, A.~Recati, and I.~Carusotto, ``{Non-local
  density correlations as signal of Hawking radiation in BEC acoustic black
  holes},'' \href{http://dx.doi.org/10.1103/PhysRevA.78.021603}{{\em Phys.
  Rev.} {\bfseries A 78} (2008) 021603},
\href{http://arxiv.org/abs/0711.4520}{{\ttfamily arXiv:0711.4520
  [cond-mat.other]}}.

\bibitem{Schutzhold10}
R.~Schutzhold and W.~G. Unruh, ``{On Quantum Correlations across the Black Hole
  Horizon},'' \href{http://dx.doi.org/10.1103/PhysRevD.81.124033}{{\em Phys.
  Rev.} {\bfseries D 81} (2010) 124033},
\href{http://arxiv.org/abs/1002.1844}{{\ttfamily arXiv:1002.1844 [gr-qc]}}.

\bibitem{Scottthesis}
S.~Robertson, {\em Hawking Radiation in Dispersive Media}.
\newblock PhD thesis, University of St Andrews, 2011.

\bibitem{Zapata11}
I.~Zapata, M.~Albert, R.~Parentani, and F.~Sols, ``{Resonant Hawking radiation
  in Bose-Einstein condensates},''
  \href{http://dx.doi.org/10.1088/1367-2630/13/6/063048}{{\em New J. Phys.}
  {\bfseries 13} (2011) 063048},
\href{http://arxiv.org/abs/1103.2994}{{\ttfamily arXiv:1103.2994
  [cond-mat.quant-gas]}}.

\bibitem{Jacobson99}
T.~Jacobson, ``{Trans Planckian redshifts and the substance of the space-time
  river},'' \href{http://dx.doi.org/10.1143/PTPS.136.1}{{\em Prog. Theor. Phys.
  Suppl.} {\bfseries 136} (1999) 1--17},
\href{http://arxiv.org/abs/hep-th/0001085}{{\ttfamily arXiv:hep-th/0001085
  [hep-th]}}.

\bibitem{Mayoral11}
C.~Mayoral, A.~Recati, A.~Fabbri, R.~Parentani, R.~Balbinot, and I.~Carusotto,
  ``{Acoustic white holes in flowing atomic Bose-Einstein condensates},'' {\em
  New J. Phys.} {\bfseries 13} (2011) 025007,
\href{http://arxiv.org/abs/1009.6196}{{\ttfamily arXiv:1009.6196
  [cond-mat.quant-gas]}}.

\bibitem{Chanson95}
H.~Chanson and J.~Montes, ``Characteristics of undular hydraulic jumps:
  experimental apparatus and flow patterns,'' {\em Journal of hydraulic
  engineering} {\bfseries 121} no.~2, (1995) 129--144.

\bibitem{Wolsthesis}
B.~Wols, ``Undular hydraulic jumps,'' Master's thesis, Delft University of
  Technology, 2005.

\bibitem{Carusotto12}
I.~Carusotto and G.~Rousseaux, ``{The Cerenkov effect revisited: From swimming
  ducks to zero modes in gravitational analogs},''
\href{http://arxiv.org/abs/1202.3494}{{\ttfamily arXiv:1202.3494
  [physics.class-ph]}}.

\bibitem{Pitaevskii84}
L.~P. Pitaevskii, ``Layered structure of superfluid 4 he with supercritical
  motion,'' {\em JETP Lett} {\bfseries 39} no.~9, (1984) 423--425.

\bibitem{Baym12}
G.~Baym and C.~J. Pethick, ``Landau critical velocity in weakly interacting
  bose gases,'' \href{http://arxiv.org/abs/1206.7066}{{\ttfamily 1206.7066}}.
  \url{http://arxiv.org/abs/1206.7066}.

\bibitem{Rousseaux07}
G.~Rousseaux, C.~Mathis, P.~Maissa, T.~G. Philbin, and U.~Leonhardt,
  ``{Observation of negative phase velocity waves in a water tank: A classical
  analogue to the Hawking effect?},''
  \href{http://dx.doi.org/10.1088/1367-2630/10/5/053015}{{\em New J.Phys.}
  {\bfseries 10} (2008) 053015},
\href{http://arxiv.org/abs/0711.4767}{{\ttfamily arXiv:0711.4767 [gr-qc]}}.

\bibitem{Frolov}
V.~Frolov and I.~Novikov,
``{Black hole physics: Basic concepts and new developments},''.

\bibitem{Leonhardt02}
U.~Leonhardt, T.~Kiss, and P.~Ohberg, ``{Intrinsic instability of sonic white
  holes},''
\href{http://arxiv.org/abs/gr-qc/0211069}{{\ttfamily arXiv:gr-qc/0211069
  [gr-qc]}}.

\bibitem{Unruh11}
W.~Unruh, ``{Quantum Noise in Amplifiers and Hawking/Dumb-Hole Radiation as
  Amplifier Noise},''
\href{http://arxiv.org/abs/1107.2669}{{\ttfamily arXiv:1107.2669 [gr-qc]}}.

\bibitem{Campo05}
D.~Campo and R.~Parentani, ``{Inflationary spectra and partially decohered
  distributions},'' \href{http://dx.doi.org/10.1103/PhysRevD.72.045015}{{\em
  Phys. Rev.} {\bfseries D 72} (2005) 045015},
\href{http://arxiv.org/abs/astro-ph/0505379}{{\ttfamily arXiv:astro-ph/0505379
  [astro-ph]}}.

\bibitem{Coutant13}
A.~Coutant and R.~Parentani, ``{Undulations from amplified low frequency
  surface waves},'' \href{http://dx.doi.org/10.1063/1.4872025}{{\em Phys.
  Fluids} {\bfseries 26} (2014) 044106},
\href{http://arxiv.org/abs/1211.2001}{{\ttfamily arXiv:1211.2001
  [physics.flu-dyn]}}.

\bibitem{Witten91}
E.~Witten, ``{On string theory and black holes},''
\href{http://dx.doi.org/10.1103/PhysRevD.44.314}{{\em Phys. Rev.} {\bfseries D
  44} (1991) 314--324}.

\bibitem{Callan92}
C.~G.~J. Callan, S.~B. Giddings, J.~A. Harvey, and A.~Strominger, ``{Evanescent
  black holes},'' \href{http://dx.doi.org/10.1103/PhysRevD.45.R1005}{{\em Phys.
  Rev.} {\bfseries D 45} (1992) 1005--1009},
\href{http://arxiv.org/abs/hep-th/9111056}{{\ttfamily arXiv:hep-th/9111056
  [hep-th]}}.

\bibitem{to_appear}
P.~Anderson {\em et~al.}
\newblock Work in progress.

\bibitem{Jacobson07b}
T.~Jacobson and R.~Parentani, ``{Black hole entanglement entropy regularized in
  a freely falling frame},''
  \href{http://dx.doi.org/10.1103/PhysRevD.76.024006}{{\em Phys. Rev.}
  {\bfseries D 76} (2007) 024006},
\href{http://arxiv.org/abs/hep-th/0703233}{{\ttfamily arXiv:hep-th/0703233
  [hep-th]}}.

\bibitem{Jannes11}
G.~Jannes, P.~Maissa, T.~G. Philbin, and G.~Rousseaux, ``{Hawking radiation and
  the boomerang behaviour of massive modes near a horizon},''
  \href{http://dx.doi.org/10.1103/PhysRevD.83.104028}{{\em Phys. Rev.}
  {\bfseries D 83} (2011) 104028},
\href{http://arxiv.org/abs/1102.0689}{{\ttfamily arXiv:1102.0689 [gr-qc]}}.

\bibitem{Jannes11b}
G.~Jannes, ``{Hawking radiation of $E<m$ massive particles in the tunneling
  formalism},'' \href{http://dx.doi.org/10.1134/S0021364011130091}{{\em JETP
  Lett.} {\bfseries 94} (2011) 18--21},
\href{http://arxiv.org/abs/1105.1656}{{\ttfamily arXiv:1105.1656 [gr-qc]}}.

\bibitem{Barcelo04}
C.~Barcelo, S.~Liberati, S.~Sonego, and M.~Visser, ``{Causal structure of
  acoustic spacetimes},''
  \href{http://dx.doi.org/10.1088/1367-2630/6/1/186}{{\em New J. Phys.}
  {\bfseries 6} (2004) 186},
\href{http://arxiv.org/abs/gr-qc/0408022}{{\ttfamily arXiv:gr-qc/0408022
  [gr-qc]}}.

\bibitem{Finazzi12}
S.~Finazzi and R.~Parentani, ``{Hawking radiation in dispersive theories, the
  two regimes},'' \href{http://dx.doi.org/10.1103/PhysRevD.85.124027}{{\em
  Phys.Rev.} {\bfseries D85} (2012) 124027},
\href{http://arxiv.org/abs/1202.6015}{{\ttfamily arXiv:1202.6015 [gr-qc]}}.

\bibitem{Balbinot07}
R.~Balbinot, S.~Fagnocchi, and A.~Fabbri, ``{The Depletion in Bose Einstein
  condensates using quantum field theory in curved space},''
  \href{http://dx.doi.org/10.1103/PhysRevA.75.043622}{{\em Phys. Rev.}
  {\bfseries A 75} (2007) 043622},
\href{http://arxiv.org/abs/cond-mat/0610367}{{\ttfamily arXiv:cond-mat/0610367
  [cond-mat.other]}}.

\bibitem{Alcubierre94}
M.~Alcubierre, ``{The Warp drive: Hyperfast travel within general
  relativity},'' \href{http://dx.doi.org/10.1088/0264-9381/11/5/001}{{\em
  Class. Quant. Grav.} {\bfseries 11} (1994) L73--L77},
\href{http://arxiv.org/abs/gr-qc/0009013}{{\ttfamily arXiv:gr-qc/0009013
  [gr-qc]}}.

\bibitem{Everett95}
A.~E. Everett, ``{Warp drive and causality},''
\href{http://dx.doi.org/10.1103/PhysRevD.53.7365}{{\em Phys. Rev.} {\bfseries D
  53} (1996) 7365--7368}.

\bibitem{Pfenning97}
M.~J. Pfenning and L.~Ford, ``{The Unphysical nature of 'warp drive'},''
  \href{http://dx.doi.org/10.1088/0264-9381/14/7/011}{{\em Class. Quant. Grav.}
  {\bfseries 14} (1997) 1743--1751},
\href{http://arxiv.org/abs/gr-qc/9702026}{{\ttfamily arXiv:gr-qc/9702026
  [gr-qc]}}.

\bibitem{VanDenBroeck99}
C.~Van Den~Broeck, ``{A 'Warp drive' with reasonable total energy
  requirements},'' \href{http://dx.doi.org/10.1088/0264-9381/16/12/314}{{\em
  Class. Quant. Grav.} {\bfseries 16} (1999) 3973--3979},
\href{http://arxiv.org/abs/gr-qc/9905084}{{\ttfamily arXiv:gr-qc/9905084
  [gr-qc]}}.

\bibitem{Lobo04}
F.~S. Lobo and M.~Visser, ``{Fundamental limitations on 'warp drive'
  spacetimes},'' \href{http://dx.doi.org/10.1088/0264-9381/21/24/011}{{\em
  Class. Quant. Grav.} {\bfseries 21} (2004) 5871--5892},
\href{http://arxiv.org/abs/gr-qc/0406083}{{\ttfamily arXiv:gr-qc/0406083
  [gr-qc]}}.

\bibitem{Lobo04b}
F.~S. Lobo and M.~Visser, ``{Linearized warp drive and the energy
  conditions},''
\href{http://arxiv.org/abs/gr-qc/0412065}{{\ttfamily arXiv:gr-qc/0412065
  [gr-qc]}}.

\bibitem{Finazzi09}
S.~Finazzi, S.~Liberati, and C.~Barcelo, ``{Semiclassical instability of
  dynamical warp drives},''
  \href{http://dx.doi.org/10.1103/PhysRevD.79.124017}{{\em Phys. Rev.}
  {\bfseries D 79} (2009) 124017},
\href{http://arxiv.org/abs/0904.0141}{{\ttfamily arXiv:0904.0141 [gr-qc]}}.

\bibitem{Hiscock97}
W.~A. Hiscock, ``{Quantum effects in the Alcubierre warp drive space-time},''
  \href{http://dx.doi.org/10.1088/0264-9381/14/11/002}{{\em Class. Quant.
  Grav.} {\bfseries 14} (1997) L183--L188},
\href{http://arxiv.org/abs/gr-qc/9707024}{{\ttfamily arXiv:gr-qc/9707024
  [gr-qc]}}.

\bibitem{Liberati09}
S.~Liberati and L.~Maccione, ``{Lorentz Violation: Motivation and new
  constraints},''
  \href{http://dx.doi.org/10.1146/annurev.nucl.010909.083640}{{\em Ann. Rev.
  Nucl. Part. Sci.} {\bfseries 59} (2009) 245--267},
\href{http://arxiv.org/abs/0906.0681}{{\ttfamily arXiv:0906.0681
  [astro-ph.HE]}}.

\bibitem{Gambini98}
R.~Gambini and J.~Pullin, ``{Nonstandard optics from quantum space-time},''
  \href{http://dx.doi.org/10.1103/PhysRevD.59.124021}{{\em Phys.Rev.}
  {\bfseries D59} (1999) 124021},
\href{http://arxiv.org/abs/gr-qc/9809038}{{\ttfamily arXiv:gr-qc/9809038
  [gr-qc]}}.

\bibitem{Roman05}
T.~Roman, ``Proceedings of the tenth marcel grossmann meeting on general
  relativity,''.

\bibitem{Kay96}
B.~S. Kay, M.~J. Radzikowski, and R.~M. Wald, ``{Quantum field theory on
  space-times with a compactly generated Cauchy horizon},''
  \href{http://dx.doi.org/10.1007/s002200050042}{{\em Commun. Math. Phys.}
  {\bfseries 183} (1997) 533--556},
\href{http://arxiv.org/abs/gr-qc/9603012}{{\ttfamily arXiv:gr-qc/9603012
  [gr-qc]}}.

\bibitem{Chaline12}
J.~Chaline, G.~Jannes, P.~Maissa, and G.~Rousseaux, ``{Some aspects of
  dispersive horizons: lessons from surface waves},''
\href{http://arxiv.org/abs/1203.2492}{{\ttfamily arXiv:1203.2492
  [physics.flu-dyn]}}.

\bibitem{Corley98}
S.~Corley and T.~Jacobson, ``{Black hole lasers},''
  \href{http://dx.doi.org/10.1103/PhysRevD.59.124011}{{\em Phys. Rev.}
  {\bfseries D 59} (1999) 124011},
\href{http://arxiv.org/abs/hep-th/9806203}{{\ttfamily arXiv:hep-th/9806203
  [hep-th]}}.

\bibitem{Leonhardt08}
U.~Leonhardt and T.~G. Philbin, ``{Black Hole Lasers Revisited},''
\href{http://arxiv.org/abs/0803.0669}{{\ttfamily arXiv:0803.0669 [gr-qc]}}.

\bibitem{Jain07}
P.~Jain, A.~S. Bradley, and C.~W. Gardiner, ``The quantum de laval nozzle:
  stability and quantum dynamics of sonic horizons in a toroidally trapped bose
  gas containing a superflow,''
  \href{http://dx.doi.org/10.1103/PhysRevA.76.023617}{{\em Phys. Rev.}
  {\bfseries A 76} (05, 2007) 023617},
  \href{http://arxiv.org/abs/0705.3093v1}{{\ttfamily arXiv:0705.3093v1
  [cond-mat.other]}}.

\bibitem{Garay00}
L.~Garay, J.~Anglin, J.~Cirac, and P.~Zoller, ``{Sonic black holes in dilute
  Bose-Einstein condensates},''
  \href{http://dx.doi.org/10.1103/PhysRevA.63.023611}{{\em Phys. Rev.}
  {\bfseries A 63} (2001) 023611},
\href{http://arxiv.org/abs/gr-qc/0005131}{{\ttfamily arXiv:gr-qc/0005131
  [gr-qc]}}.

\bibitem{Damour76b}
T.~Damour, N.~Deruelle, and R.~Ruffini, ``{On Quantum Resonances in Stationary
  Geometries},''
{\em Lett. Nuovo Cim.} {\bfseries 15} (1976) 257--262.

\bibitem{Damour78}
T.~Damour and N.~Deruelle, ``{Dressing up a Reissner Naked Singularity},''
\href{http://dx.doi.org/10.1016/0370-2693(78)90736-0}{{\em Phys. Lett.}
  {\bfseries B 72} (1978) 471--476}.

\bibitem{Kang97}
G.~Kang, ``{Quantum aspects of ergoregion instability},''
  \href{http://dx.doi.org/10.1103/PhysRevD.55.7563}{{\em Phys. Rev.} {\bfseries
  D 55} (1997) 7563--7573},
\href{http://arxiv.org/abs/gr-qc/9701040}{{\ttfamily arXiv:gr-qc/9701040
  [gr-qc]}}.

\bibitem{Cardoso04}
V.~Cardoso, O.~J. Dias, J.~P. Lemos, and S.~Yoshida, ``{The Black hole bomb and
  superradiant instabilities},''
  \href{http://dx.doi.org/10.1103/PhysRevD.70.044039,
  10.1103/PhysRevD.70.049903}{{\em Phys. Rev.} {\bfseries D 70} (2004) 044039},
\href{http://arxiv.org/abs/hep-th/0404096}{{\ttfamily arXiv:hep-th/0404096
  [hep-th]}}.

\bibitem{Greiner}
W.~Greiner, ``Quantum electrodynamics of strong fields,'' {\em Hadrons and
  Heavy Ions} (1985) 95--226.

\bibitem{Barcelo06}
C.~Barcelo, A.~Cano, L.~Garay, and G.~Jannes, ``{Stability analysis of sonic
  horizons in Bose-Einstein condensates},''
  \href{http://dx.doi.org/10.1103/PhysRevD.74.024008}{{\em Phys. Rev.}
  {\bfseries D 74} (2006) 024008},
\href{http://arxiv.org/abs/gr-qc/0603089}{{\ttfamily arXiv:gr-qc/0603089
  [gr-qc]}}.

\bibitem{Itzykson}
C.~Itzykson and J.~Zuber, {\em Quantum field theory}, vol.~59.
\newblock Mcgraw-hill New York, 1980.

\bibitem{LevyBruhl}
P.~L{\'e}vy-Bruhl, {\em Introduction {\`a} la th{\'e}orie spectrale}.
\newblock Dunod, 2003.

\bibitem{Reed}
M.~Reed and B.~Simon, {\em Methods of modern mathematical physics. IV. Analysis
  of operators.}, vol.~4.
\newblock Academic Press, 1978.

\bibitem{Langer82}
H.~Langer, ``Spectral functions of definitizable operators in krein spaces,''
  {\em Functional analysis} (1982) 1--46.

\bibitem{Bognar}
J.~Bogn{\'a}r, {\em Indefinite inner product spaces}.
\newblock Springer New York, 1974.

\bibitem{Gerard11}
C.~Gerard, ``{Scattering theory for Klein-Gordon equations with non-positive
  energy},'' \href{http://dx.doi.org/10.1007/s00023-011-0138-8}{{\em Annales
  Henri Poincare} {\bfseries 13} (2012) 883--941},
\href{http://arxiv.org/abs/1101.2145}{{\ttfamily arXiv:1101.2145 [math-ph]}}.

\bibitem{Berti09}
E.~Berti, V.~Cardoso, and A.~O. Starinets, ``{Quasinormal modes of black holes
  and black branes},''
  \href{http://dx.doi.org/10.1088/0264-9381/26/16/163001}{{\em Class. Quant.
  Grav.} {\bfseries 26} (2009) 163001},
\href{http://arxiv.org/abs/0905.2975}{{\ttfamily arXiv:0905.2975 [gr-qc]}}.

\bibitem{Finazzi10}
S.~Finazzi and R.~Parentani, ``{Black-hole lasers in Bose-Einstein
  condensates},'' \href{http://dx.doi.org/10.1088/1367-2630/12/9/095015}{{\em
  New J.Phys.} {\bfseries 12} (2010) 095015},
\href{http://arxiv.org/abs/1005.4024}{{\ttfamily arXiv:1005.4024
  [cond-mat.quant-gas]}}.

\bibitem{Mayoral10}
C.~Mayoral, A.~Fabbri, and M.~Rinaldi, ``{Step-like discontinuities in
  Bose-Einstein condensates and Hawking radiation: dispersion effects},''
  \href{http://dx.doi.org/10.1103/PhysRevD.83.124047}{{\em Phys. Rev.}
  {\bfseries D 83} (2011) 124047},
\href{http://arxiv.org/abs/1008.2125}{{\ttfamily arXiv:1008.2125 [gr-qc]}}.

\bibitem{Nardin09}
J.-C. Nardin, G.~Rousseaux, and P.~Coullet, ``{Wave-Current Interaction as a
  Spatial Dynamical System: Analogies with Rainbow and Black Hole Physics},''
\href{http://dx.doi.org/10.1103/PhysRevLett.102.124504}{{\em Phys. Rev. Lett.}
  {\bfseries 102} (2009) 124504}.

\bibitem{Barbado11}
L.~C. Barbado, C.~Barcelo, L.~J. Garay, and G.~Jannes, ``{The Trans-Planckian
  problem as a guiding principle},''
  \href{http://dx.doi.org/10.1007/JHEP11(2011)112}{{\em JHEP} {\bfseries 1111}
  (2011) 112},
\href{http://arxiv.org/abs/1109.3593}{{\ttfamily arXiv:1109.3593 [gr-qc]}}.

\bibitem{Bombelli86}
L.~Bombelli, R.~K. Koul, J.~Lee, and R.~D. Sorkin, ``{A Quantum Source of
  Entropy for Black Holes},''
\href{http://dx.doi.org/10.1103/PhysRevD.34.373}{{\em Phys. Rev.} {\bfseries D
  34} (1986) 373--383}.

\bibitem{Richartz09}
M.~Richartz, S.~Weinfurtner, A.~Penner, and W.~Unruh, ``{General universal
  superradiant scattering},''
  \href{http://dx.doi.org/10.1103/PhysRevD.80.124016}{{\em Phys. Rev.}
  {\bfseries D 80} (2009) 124016},
\href{http://arxiv.org/abs/0909.2317}{{\ttfamily arXiv:0909.2317 [gr-qc]}}.

\bibitem{Bekenstein98}
J.~D. Bekenstein and M.~Schiffer, ``{The Many faces of superradiance},''
  \href{http://dx.doi.org/10.1103/PhysRevD.58.064014}{{\em Phys. Rev.}
  {\bfseries D 58} (1998) 064014},
\href{http://arxiv.org/abs/gr-qc/9803033}{{\ttfamily arXiv:gr-qc/9803033
  [gr-qc]}}.

\bibitem{Heisenberg35}
W.~Heisenberg and H.~Euler, ``{Consequences of Dirac's theory of positrons},''
  {\em Z. Phys.} {\bfseries 98} (1936) 714--732,
\href{http://arxiv.org/abs/physics/0605038}{{\ttfamily arXiv:physics/0605038
  [physics]}}.

\bibitem{Schwinger51}
J.~Schwinger, ``On gauge invariance and vacuum polarization,'' {\em Phys. Rev.}
  {\bfseries 82} no.~5, (1951) 664.

\bibitem{Arvanitaki09}
A.~Arvanitaki, S.~Dimopoulos, S.~Dubovsky, N.~Kaloper, and J.~March-Russell,
  ``{String Axiverse},''
  \href{http://dx.doi.org/10.1103/PhysRevD.81.123530}{{\em Phys. Rev.}
  {\bfseries D 81} (2010) 123530},
\href{http://arxiv.org/abs/0905.4720}{{\ttfamily arXiv:0905.4720 [hep-th]}}.

\bibitem{Blas11}
D.~Blas and S.~Sibiryakov, ``{Horava gravity versus thermodynamics: The Black
  hole case},'' \href{http://dx.doi.org/10.1103/PhysRevD.84.124043}{{\em Phys.
  Rev.} {\bfseries D 84} (2011) 124043},
\href{http://arxiv.org/abs/1110.2195}{{\ttfamily arXiv:1110.2195 [hep-th]}}.

\bibitem{Gregory93}
R.~Gregory and R.~Laflamme, ``{Black strings and p-branes are unstable},''
  \href{http://dx.doi.org/10.1103/PhysRevLett.70.2837}{{\em Phys. Rev. Lett.}
  {\bfseries 70} (1993) 2837--2840},
\href{http://arxiv.org/abs/hep-th/9301052}{{\ttfamily arXiv:hep-th/9301052
  [hep-th]}}.

\bibitem{Hartman09}
T.~Hartman, W.~Song, and A.~Strominger, ``{The Kerr-Fermi Sea},''
\href{http://arxiv.org/abs/0912.4265}{{\ttfamily arXiv:0912.4265 [hep-th]}}.

\end{thebibliography}\endgroup

\end{document}